\documentclass[12pt,a4paper]{report}

\usepackage{amsthm,amsxtra,amsfonts,amssymb,latexsym}

\begin{document}

\newcommand{\TeXButton}[2]{#2}
\newcommand{\stackunder}[2]{\underset{#1}{#2}}
\newcommand{\NEG}{\not}
\newcommand{\QATOP}[2]{#1 \atop #2}
\newcommand{\intmul}{\vdash}

\setcounter{secnumdepth}{3}
\setcounter{tocdepth}{1}

\title{Spacetime Topology, Conserved Charges, and the Splitting of Classical
Multiplets}

\vspace{3ex}

\author{Hanno Hammer}

\date{
\vspace{4ex}
A dissertation submitted for the degree of Doctor of Philosophy\\
\vspace{1ex}
University of Cambridge\\
\vspace{1ex}
April 1999}

\maketitle

\pagebreak
\tableofcontents
\pagebreak

\chapter{Introduction}

Collective excitations of physical systems such as extended objects are
expressed by the conservation of charges which have the interpretation of
overall Poincare momenta and, in a supersymmetry context, their associated
supercharges. Generally, these overall charges do not describe intrinsic
excitations of the system; rather, their algebra reflects the symmetries and
the topology of a fixed ''background'' where the system is propagating in.
For genuine Noether charges, this is true independent of whether the charges
are conserved or not (see chapter \ref{TopEx}, section \ref{NoethChar}); if
they are conserved, then the Lagrangian describing the system shares the
symmetry of the background in the sense that the action of the symmetry
group leaves the Lagrangian invariant. It is for this reason that large
classes of dynamical systems have isomorphic algebras of overall Noether
charges.

It need not be true, however, that the algebra of conserved charges
coincides with the algebra of Noether charges. If a Lagrangian transforms
under the group only semi-invariantly, this being a term that will be
explained in chapter \ref{TopEx}, the conserved charges extend the Noether
charge algebra by topological, central or non-central, charges. These
charges measure the non-trivial homology of the background spacetime as well
as non-trivial topological configurations of gauge fields in the Lagrangian.
Now, on every configuration of the system, i.e. on every solution of the
underlying dynamical equations, each charge takes on a certain real or
Grassmann-value. This specific $n$-tupel of values in turn distinguishes a
class of configurations of the underlying theory, namely those for which the
collection of conserved charges produces the same $n$-tupel of real or
Grassmann-values. If the algebra of charges determines (on exponentiation) a
(super-)Lie group $G$, say, then we will refer to these distinguished
classes of configurations as $G${\it -states} (see section \ref{GStateSpaces}
in chapter \ref{MomentMaps}), or simply states. We can now state that the
major topic of this thesis is the interplay of states that transform under a
given (super-)Lie group, and the geometry and topology of the background in
which the physical system is evolving. In chapter \ref{MomentMaps} we will
see that an important tool in handling these problems in a classical context
will be a mapping of phase spaces of different topologies, which
nevertheless ''look locally the same''; technically speaking, we shall
relate the dynamics on a topologically non-trivial phase space to the
dynamics on a covering phase space, by preserving the local structure of the
dynamics involved, ultimately in order to gain a better understanding of the
multiplets of states on the topologically nontrivial phase space.

Aspects of these ideas are worked out in chapters \ref{TopEx}--\ref
{MomentMaps}. The line of reasoning connecting the different chapters is as
follows:

In chapter \ref{TopEx} we study classical solutions of $D=10$ type II
supergravity describing extended objects commonly called $D$-$p$-branes. $p$%
-branes are solitons carrying conserved charges that act as sources for
antisymmetric gauge fields in the underlying supergravity theory. $D$-$p$%
-branes arise from mixed boundary conditions on open strings in type II
string theory; they are introduced as $\left( p+1\right) $-dimensional
hypersurfaces in a spacetime, where open strings are constrained to end on.
A spacelike section of a $D$-brane can acquire a finite volume in a
spacetime with compact dimensions by wrapping around non-trivial homology
cycles of the spacetime. In this case the supertranslation algebra of
conserved charges carried by the $D$-brane is extended by topological
charges. In chapter \ref{TopEx} we perform a ''naive'' analysis of these
topological extensions of the modified Noether charge algebras carried by $D$%
-branes. This analysis is called ''naive'', as, in the first place, we are
assuming that the brane is propagating in a flat 10-dimensional
super-Minkowski space, whose super-isometry group is maximal and coincides
with the $N=2$ super-Poincare group in ten dimensions. This super-Minkowski
space is known to be a solution of massless type IIA supergravity in ten
dimensions. Propagation of the brane in this fixed background is described
by a Lagrangian which is semi-invariant under the supertranslation group;
this means that apart from a supersymmetric kinetic term, the Lagrangian
contains a term ${\cal L}_{WZ}$ which arises from the pull-back of a $\left(
p+1\right) $-form $\beta $ on the target space whose differential $d\beta $
is a nontrivial cocycle in the Chevalley-Eilenberg cohomology on the
super-translation group; consequently, the form $\beta $, which is usually
called a Wess-Zumino form,  is not left-invariant. Physically, the
Wess-Zumino term in the Lagrangian describes the coupling of the extended
object to the various antisymmetric tensor fields in the underlying
supergravity. The fact that $\beta $ is not left-invariant gives rise to the
algebra extensions encountered in the sequel. This follows, since subjecting 
${\cal L}_{WZ}$ to supertranslations produces de Rham cocycles on the
worldvolume of the brane which can be non-trivial, i.e. closed but not
exact, provided that both the worldvolume and the background space time have
nontrivial homology, and the brane is wrapping around such homology cycles.
These de Rham cocycles make themselves apparent by central charges in the
modified Noether charge algebra which extend the supertranslation algebra on
the super-spacetime. Technically, these central charges arise as integrals
of the non-trivial de Rham cocycles over non-trivial homology cycles in the
worldvolume of the extended object. Further extensions can arise from gauge
fields on the worldvolume of the brane that are forced to transform under
the supersymmetry group on the super-Minkowski space in order to keep the
kinetic term in the Lagrangian of the brane supertranslation-invariant. The
analysis in chapter \ref{TopEx} shows how these extensions appear in
principal by the mechanism just described; in actuality, the central charges
that describe non-trivial homology of the target space all vanish in this
case, since a flat super-Minkowski space is topologically trivial, and has
no nontrivial homology cycles (The extensions that are solely due to the
gauge field might still be non-zero, however). Put it another way, assuming
that some of the central charges pertaining to spacetime homology that
appear in the extension \ref{bos10} are indeed nonzero implies that the
actual target space cannot be a flat super-Minkowski space, but rather must
be a compactified, and possibly non-flat, supergravity solution. In this
case, the isometry group of the actual target space will be only a proper
subgroup of the maximal supertranslation group in ten flat spacetime
dimensions. Therefore, the analysis of this chapter gives only a general
pattern, which has to be refined, once a compactified version of the
ten-dimensional target space has been chosen, which is why we have called
this approach ''naive''.

These considerations naturally lead us to the question how a compactified
spacetime can arise as a solution of an underlying supergravity; in
particular, is it possible to obtain a compactified spacetime from a
non-compactified one, without violating the property of being a solution to
some underlying field theory, say, a supergravity? Here a crucial point is
that the supergravity field equations are purely local; this means that once
we have found a solution to these equations, constructing identification
spaces, typically orbit spaces of the non-compactified solution under the
free, properly discontinuous action of a discrete subgroup of the isometry
group of the non-compactified version, must yield another solution to the
underlying supergravity, simply because identifying points of the spacetime
under such a group action preserves all local properties of the spacetime,
changing only the global topology. If the non-compactified solution was a
manifold (and not itself a more complicated topological space) then the
compactified orbit space is again a manifold. More generally, the
non-compactified version is a covering space of the compactified one. It is
important to understand in which sense this construction changes the global
topology of the spacetime: The main statement here is that the fundamental
group (and hence the first homology group) becomes larger in the process of
identification; the fundamental group of the non-compactified spacetime
becomes a normal subgroup of the fundamental group of the compactified
version; and the quotient of the compactified fundamental group over this
normal factor basically gives the discrete group whose orbits define the
compactification. Another important classical result is that the higher
homotopy groups of the spacetime are left invariant in this process.
However, higher homology groups can become non-trivial as well. As an
example, think of an $r$-dimensional torus factor in a compactified
spacetime, which gives rise to a non-trivial homology group $H_r$. Although
toroidal compactifications of a flat non-compactified spacetime are
conceptually very simple, they are nevertheless of central importance, since
an important result by \cite{Wolf} states that, in the flat case, it is
precisely the class of toroidally compactified spacetimes that are
geodesically complete homogeneous spaces; in this case the discrete group
acting on the non-compactified flat version is a group of pure translations.
This result is a special case of a more general beautiful structure theorem
about flat connected homogeneous geodesically complete pseudo-Riemannian
manifolds proven on p. 135 in \cite{Wolf}.

If a compactified spacetime arises as an orbit space from an $n$-dimensional
covering spacetime as discussed above, it is comparatively easy to obtain
the precise form of the (bosonic part of the super-) isometry group of the
compactified version from the larger isometry group of the covering space;
in particular this means that global aspects of the residual isometry group
can be computed. It it for this reason that we have mainly worked with
Riemann-flat spacetimes in chapters \ref{TopEx} and \ref{semiGroup}. In
chapter \ref{MomentMaps} we made no assumptions about curvature properties
of the underlying manifolds, however, so that the results derived in this
chapter hold in full generality.

It is a well-known result that, if $\Gamma $ denotes a discrete subgroup of
the larger isometry group which acts on a covering spacetime, then the
isometry group of the orbit space, i.e. the compactified version, is given
by the quotient $N\left( \Gamma \right) /\Gamma $, where $N\left( \Gamma
\right) $ denotes the normalizer of the group $\Gamma $ in the larger
isometry group (for an explanation of the term ''normalizer'' see chapter 
\ref{semiGroup}). The bosonic part of the isometry group of the compactified
spacetime in this construction typically has the structure of a product of a
lower-dimensional (super-)Poincare- or Galilei-group with a discrete or Lie
group which describes internal symmetries. In particular, this is true if
the group $\Gamma $ is a discrete group of translations acting on a flat
Minkowski-type covering space. In this case the resulting compactification
is a Cartesian product of a flat $\left( n-m\right) $-dimensional space with
an internal $m$-dimensional torus. -- This approach via the normalizer does
not explain the reduction of the number of supersymmetry generators,
however. A precise investigation of this point must involve global
computations of Killing spinors equations, and has to examine whether the
bosonic part of the supermanifold admits a spin structure. These
considerations are outside the scope of this work, however.

In a torus-compactification of a Riemann-flat spacetime, one can
distinguish two major cases, according to whether all basic lattice
vectors of the group $\Gamma $ are spacelike, or whether one of
these lattice vectors is lightlike (timelike lattice vectors are not
considered in this work, as they give rise to a compact time
direction; see, for example, \cite{Hull} and the references
therein). It is fairly well-known that if all $m$ basis vectors are
spacelike, the ''external'' spacetime is an $\left( n-m\right)$
-dimensional (super-) Minkowski spacetime, and the restriction of
the metric to the internal torus is Euclidean. Furthermore, the
associated isometry group is a direct product of a lower-dimensional
$\left( n-m\right) $-dimensional ''external'' (super-) Poincare
group and a semidirect product of a discrete rotation group with a
product $U\left( 1\right) ^m$, this last factor describing the
translational symmetries of the internal torus. In chapter
\ref{semiGroup} we have performed an analogous investigation for the
case that one lattice vector in the group $\Gamma $ is lightlike,
with the purpose to understand the precise structure of the
resulting isometries, including possible discrete
transformations. The analysis reveals the somewhat unexpected result
that the resulting isometry group admits a natural extension to a
semigroup. The semigroup transformations form a discrete set
isomorphic to the positive natural numbers with multiplication as
composition, and arise from the fact that the metric of the covering
spacetime when restricted to the subspace spanned by the lattice
vectors generating $\Gamma $ is degenerate. In this case the
compactified spacetime is again a Cartesian product of some
$\TeXButton{R}{\mathbb{R}}^{n-m}$ times an internal torus, but the
restriction of the metric to both the external and the internal
factor is now degenerate. The Lie algebra of the isometry group of
the compactified spacetime turns out to be the direct sum of an
Abelian Lie algebra isomorphic to $\TeXButton{R}{\mathbb{R}}^{m-1}$
and a centrally extended Galilei algebra in $\left( n-m-1\right) $
dimensions. The semigroup transformations correspond to discrete
transformations generated by the ''mass'' generator which spans the
central extension of the extended Galilei group. As the mass
generator commutes with all other generators of the Lie algebra, its
quantum version should provide a superselection operator for the
spacetime degrees of freedom of the theory. This means that the
discrete winding numbers of the lightlike torus around itself 
should label noncoherent, i.e. non-superposable, subspaces of
physical states in the overall Hilbert space of a system defined on
this compactified spacetime.

The motivation for the developments in chapter \ref{MomentMaps} lies again
in the results of chapter \ref{TopEx} on topological extensions of Noether
charge algebras: We may naturally pose the question on which kind of space
such an extended algebra is represented or realized. In a quantum context
the answer would normally be straightforward; we would assume that the
overall Hilbert space of states of the theory containes physical subspaces,
each of which is labelled by the collection of (conserved) values of the
topological generators. This would be even more natural, as the topological
charges typically take on discrete values, according to the fact that they
represent winding numbers, degrees of maps, or are associated with Chern
classes of gauge fields, as in formula \ref{bos10} of chapter \ref{TopEx}.
But what would be the analogous classical construction? Before we can think
about this problem we must understand what we mean by saying that a physical
state transforms under a symmetry group $G$. Again, in the quantum context
the answer is comparatively simple. A state transforming under a group $G$,
or a $G$-state, is an element of a subvector space of an overall Hilbert
space which serves as a carrier space for an irreducible representation of $G
$. If $e^{iB}$ is an irreducible representation of $G$ on this carrier
space, and $\left| \psi \right\rangle $ is a state vector in this subspace,
then the mean value of the observable $A$, being an element of the Lie
algebra of $G$, on the transformed state $e^{iB}\left| \psi \right\rangle $
is 
\begin{equation}
\label{intro1}\left\langle A\right\rangle _{e^{iB}\left| \psi \right\rangle
}=\left\langle Ad\left( e^{-iB}\right) A\right\rangle _{\left| \psi
\right\rangle }\quad .
\end{equation}
The last two statements can be transferred into the classical domain as
follows: Irreducibility means that the group acts transitively on a certain
subspace of states (we learned this from \cite{Woodhouse}); but transitivity
of a group action is a meaningful concept in the classical context as well.
Also formula (\ref{intro1}) has a classical counterpart, given by the $%
Ad\left( G\right) $-transformation behaviour of global moment maps
associated with the action of $G$, as defined in section \ref
{SymplecticGActionsMoMaps} of chapter \ref{MomentMaps}. With the help of a
global moment map one can partition the underlying phase space into subsets
such that the dynamics of every physical system that is $G$-invariant (which
means that the Hamiltonian Poisson-commutes with all phase space functions
representing the Lie algebra of the group $G$) lies completely in one of
these subsets; this follows, since the moment map is constant on each of
these subsets, and hence all generators of $G$ are conserved on trajectories
that lie entirely in one of these subsets. These subsets coincide with the $G
$-states described in the second paragraph of this introduction. Now the
following problem arises: On a phase space, or more general, on a symplectic
manifold, a global moment map associated with a symplectic group action is
available only if the phase space is simply connected. But this is not true
in general for the phase spaces we are concerned in this work, as these
spaces will have non-trivial topology in general. However, symplectic group
actions can be defined on non-simply connected phase spaces as well, and
clearly, the concept of conserved charges makes sense also on a phase space
with non-trivial topology. In chapter \ref{MomentMaps} we therefore take up
the problem of how one can define $G$-states as defined by moment maps on
such a space. The non-simply connectedness is an obstruction to the
existence of global moment maps; we will show, however, how one can remove
this obstruction be transferring the dynamics to a universal symplectic
covering manifold, on which a global moment map exists. This construction is
based on the fact that a covering projection is a local diffeomorphism which
can be regarded as a local symplectomorphism between simply connected and
non-simply connected phase spaces. In turn, one can define $G$-states on the
compactified phase space via the notion of $G$-states on the covering space.
At the end of chapter \ref{MomentMaps} it is shown, then, that the $G$%
-states on the compactified phase space and its covering space are related
by an identification map, where states on the covering space are identified
under the action of the deck transformation group of the covering. This
works out successfully some aspects of the idea that was alluded to at the
beginning of this introduction, which proposed that the multiplet structure
of states living on a topologically non-trivial phase space can be related
to multiplets on a phase space whose topological structure is simpler:
Namely, such a simplifying construction is available whenever the 
non-trivial homology of a phase space can be traced back to the fact that
the phase space is an orbit space of a topologically simpler symplectic
covering manifold.

\chapter{Topological extensions of the algebra of Noether charges}
\label{TopEx}

\section*{Introduction}

Topological extensions of the algebra of Noether currents and corresponding
Noether charges have been studied in the past by a number of authors \cite
{Azca,Soro}. In \cite{Azca} the extensions of the algebra of Noether
and modified Noether charges carried by supersymmetric extended objects have
been examined; furthermore, it has been pointed out that the origin of these
modifications is the Wess-Zumino term in the Lagrangian of the extended
object. In \cite{Soro}
the algebra of the Noether supercharges of the M-$5$-brane was derived, and
it was observed that not all central charges occuring in the superalgebra
extension are entirely due to the Wess-Zumino term; it was shown that
another contribution to the central charges originates in the presence of a
gauge field potential on the worldvolume of the M-$5$-brane, which takes
part in the action of the superPoincare group acting on the target space. In
this chapter we shall prove a theorem explaining that this is a general feature
for a whole class of Lagrangians containing a gauge potential on the
worldvolume which is forced to transform under a Lie group that acts on the
target space of the theory; in this case even the algebra of the
(unmodified) Noether charges suffers modifications, which otherwise would
close into the original algebra of the group that acts on the target space,
possibly up to a sign, which is a consequence of whether the group acts from
the left or from the right.

In this chapter we have performed an analysis of the extensions of the
superalgebra of Noether and modified Noether charges carried by D-$p$-branes
in a IIA superspace. In doing so, however, we have faced a number of
difficulties which could not be illuminated by consulting the literature; in
section \ref{NoethChar} we therefore provide an introduction to the basic
concepts of the algebra of Noether currents, Noether charges, and associated
modified currents and charges, which might arise from a Lagrangian
transforming as a semi-invariant under the action of the group. We show that
for a left action the algebra of the associated Noether charges always
closes to the original algebra, regardless of whether the Lagrangian is
invariant under the transformation or not; we derive the action of the
Noether charges on the currents and show that for a left action the Noether
currents transfom in the adjoint representation of the group. We then extend
this analysis to the case of semi-invariant Lagrangians and examine
carefully under which circumstances certain contributions to the Poisson
brackets of the modified currents vanish or may be neglected; this question
is not always fully adressed in the literature, and becomes even more
non-trivial in the case of having a gauge field present on the worldvolume,
since the gauge field degrees of freedom are subject to primary and
secondary constraints (in Dirac's terminology). We analyze the constraint
structure of a theory possessing such a gauge field on the world-volume; the
results apply to D-branes and M-branes as well. We derive conditions under
which the additional charges obtained so far are conserved and central. We
finish the first section with showing how modifications of the Noether
charge algebra arise from the presence of such a gauge field, even if the
modifications of the charge algebra due to semi-invariant pieces in the
Lagrangian are not yet taken into account.

In section \ref{SuperAlgebra} we apply these ideas to derive the extensions
of the algebra of modified Noether charges for D-$p$-brane Lagrangians in
IIA superspace. We first derive a general form of these extensions
applicable for the most general forms of gauge fields (NS-NS and RR) on the
superspace; then we choose a particular background in putting all bosonic
components of the gauge fields to zero, and taking into account that the
remainder are subject to superspace constraints which allow to reconstruct
the leading components of the RR gauge field strengths unambiguously. In
doing so we must check whether the RR field strengths thus derived actually
satisfy the appropriate Bianchi identities; we find that this question can
be traced back to the validity of a set of generalized $\Gamma $-matrix
identities; it is known that the first two members in this set are actually
valid; as for the rest we derive a necessary condition using the technique
that has been applied in similar circumstances previously, see \cite
{Achu,PKT1}, and find that it is satisfied. The extended algebra thus
derived contains topological charges that probe the existence of compact
spacetime dimensions the brane is wrapping around; furthermore, which is a
new feature here, we find that central charges show up that probe the
non-triviality of the worldvolume regarded as a $U\left( 1\right) $-bundle
of the gauge field $A_\mu $; we find that for the D-$4$-, $6$- and $8$%
-branes there exist central charges originating in the Wess-Zumino term that
can be interpreted as probing the coupling of these non-trivial gauge-field
configurations to compact dimensions in the spacetime; they are zero if
either there are no compact dimensions, or the brane is not wrapping around
them, or the $U\left( 1\right) $-bundle is trivial, which requires the gauge
field configuration to be trivial. As for the D-$2$-brane such a coupling of
spacetime topology to the gauge field is only present in the central charges
that stem from the fact that the gauge field transforms under supersymmetry;
they have nothing to do with the Wess-Zumino term; for the special case of a
D-$2$-brane given by ${\bf R}\times S^2$, where ${\bf R}$ denotes the time
dimension, we find that the algebra can contain the charge of a Dirac
monopole of the gauge field; this result is very neat, so we present it here:%
$$
\left\{ Q_\alpha ,Q_\beta \right\} =2\left( C\Gamma ^m\right) _{\alpha \beta
}\cdot P_m\;-\;2i\left( C\Gamma _{11}\Gamma _m\right) _{\alpha \beta }\cdot
Y^m\;- 
$$
$$
-\;i\left( C\Gamma _{m_2m_1}\right) _{\alpha \beta }\cdot
T^{m_1m_2}\;-\;2i\left( C\Gamma _{11}\right) _{\alpha \beta }\cdot 4\pi
g\quad . 
$$
Here $Y^m$ is a central charge that couples the canonical gauge field
momentum to compact dimensions in the spacetime allowing for $1$-cycles in
the brane wrapping around them; $T^{m_1m_2}$ probes the presence of compact
dimensions in spacetime the brane wraps around, i.e. allowing for $2$-cycles
wrapping around them, and $g$ is the quantized charge of a Dirac monopole
resulting from the gauge field.

\section{Actions of a Lie group and associated Noether charges\quad \quad 
\label{NoethChar}}

\subsection{Noether currents}

Let $\left( x^\mu \right) =\left( t,\sigma ^r\right) $, $\mu =0,\ldots ,p$; $%
r=1,\ldots ,p$ denote coordinates on a $\left( p+1\right) $-dimensional
manifold (''worldvolume'') $W$. Here $t$ refers to a ''timelike''
coordinate, $\sigma ^r$ refers to ''spacelike'' coordinates. Let $W\left(
t\right) $ denote the hypersurfaces in $W$ with constant $t$. Let ${\cal L}=%
{\cal L}\left( \phi ,\partial _\mu \phi \right) $ be a Lagrangian of a field
multiplet multiplet $\phi =\left( {\phi ^i} \right)$ defined on $W$ , with unspecified
dimension. The objects $\left( \phi ^i\right) $ are regarded to be
coordinates on a target space $\Sigma $; at present we do not make any
further assumptions on the precise nature of $\Sigma $. Let $G$ be a Lie
group with generators $T_M\in Lie\left( G\right) $, where $Lie\left(
G\right) $ is the Lie algebra of $G$; the generators $T_M$ act on $\phi ^i$
according to $\phi \mapsto \delta _M\phi =\left( {\delta _M\phi ^i} \right)$%
; here $\delta _M\phi ^i$ are the components of the vector field $\widetilde{%
T_M}$ induced by the generator $T_M$ on $\Sigma $, i.e., the action of $%
e^{t\cdot T_M}$ defines a flow $\left( \phi ,t\right) \mapsto \left(
e^{t\cdot T_M}\phi \right) ^i$, which generates the vector field \cite
{cram/pir} 
\begin{equation}
\label{gract0.0}\left. \frac d{dt}\left( e^{t\cdot T_M}\phi \right)
^i\right| _{t=0}\frac \partial {\partial \phi ^i}=\left. \left( \widetilde{%
T_M}\right) ^i\right| _\phi \frac \partial {\partial \phi ^i}=\left( \delta
_M\phi ^i\right) \frac \partial {\partial \phi ^i}=\widetilde{T_M}\quad . 
\end{equation}
For a right action the map $Lie\left( G\right) \ni X\mapsto \tilde X$, which
sends an element of the Lie algebra of $G$ to an induced vector field on $%
\Sigma $, is a Lie algebra homomorphism into the set of all vector fields on 
$\Sigma $ endowed with the Lie bracket as multiplication: 
\begin{equation}
\label{zw1}\widetilde{\left[ X,Y\right] }=\left[ \widetilde{X},\widetilde{Y}%
\right] \quad . 
\end{equation}
For a left action this is true for the map $Lie\left( G\right) \ni X\mapsto
-\tilde X$, since in this case 
\begin{equation}
\label{zw2}\widetilde{\left[ X,Y\right] }=-\left[ \widetilde{X},\widetilde{Y}%
\right] \quad . 
\end{equation}
Now denote the expression for the equations of motion for the fields $\phi
^i $ by 
\begin{equation}
\label{gract1}\left( eq,{\cal L}\right) _i:=\frac{\partial {\cal L}}{%
\partial \phi ^i}-\partial _\mu \frac{\partial {\cal L}}{\partial \partial
_\mu \phi ^i}\quad , 
\end{equation}
then the action of the generator $T_M$ on ${\cal L}$ takes the form 
\begin{equation}
\label{gract2}\delta _M{\cal L}=\delta _M\phi ^i\cdot \left( eq,{\cal L}%
\right) _i+\partial _\mu j_M^\mu \quad , 
\end{equation}
where 
\begin{equation}
\label{gract3}j_M^\mu =\delta _M\phi ^i\cdot \frac{\partial {\cal L}}{%
\partial \partial _\mu \phi ^i} 
\end{equation}
is the {\it Noether current associated with} $T_M$. If $sol\left( {\cal L}%
\right) $ denotes a solution to the {\it equations of motion} $\left( eq,%
{\cal L}\right) _i=0$, we have 
\begin{equation}
\label{gract4}\left[ \delta _M{\cal L}=\partial _\mu j_M^\mu \right]
_{sol\left( {\cal L}\right) }\quad , 
\end{equation}
i.e. on the solution $sol\left( {\cal L}\right) $. Now assume that ${\cal L}=%
{\cal L}_0+{\cal L}_1$, where ${\cal L}_0$ is invariant under $G$, $\delta _M%
{\cal L}_0=0$. Then 
\begin{equation}
\label{gract5}\delta _M{\cal L}_1=\delta _M\phi ^i\cdot \left( eq,{\cal L}_0+%
{\cal L}_1\right) _i+\partial _\mu j_M^\mu \quad , 
\end{equation}
where 
\begin{equation}
\label{gract6}j_M^\mu =\delta _M\phi ^i\cdot \frac{\partial {\cal L}_0}{%
\partial \partial _\mu \phi ^i}+\delta _M\phi ^i\cdot \frac{\partial {\cal L}%
_1}{\partial \partial _\mu \phi ^i}=:j_{0,M}^\mu +J_M^\mu \quad . 
\end{equation}
Therefore, 
\begin{equation}
\label{gract7}\delta _M{\cal L}_1=\delta _M\phi ^i\cdot \left( eq,{\cal L}_0+%
{\cal L}_1\right) _i+\partial _\mu \left( j_{0,M}^\mu +J_M^\mu \right) \quad
, 
\end{equation}
and 
\begin{equation}
\label{gract8}\left[ \delta _M{\cal L}_1=\partial _\mu \left( j_{0,M}^\mu
+J_M^\mu \right) \right] _{sol\left( {\cal L}_0+{\cal L}_1\right) }\quad . 
\end{equation}
This is to be compared with 
\begin{equation}
\label{gract9}0=\delta _M\phi ^i\cdot \left( eq,{\cal L}_0\right)
_i+\partial _\mu j_{0,M}^\mu \quad , 
\end{equation}
and 
\begin{equation}
\label{gract10}\left[ 0=\partial _\mu j_{0,M}^\mu \right] _{sol\left( {\cal L%
}_0\right) }\quad . 
\end{equation}
Since%
$$
\left[ \partial _\mu j_{0,M}^\mu \right] _{sol\left( {\cal L}_0+{\cal L}%
_1\right) }\neq \left[ \partial _\mu j_{0,M}^\mu \right] _{sol\left( {\cal L}%
_0\right) }=0 
$$
in general, we see that $j_{0,M}^\mu $ is no longer conserved in the
presence of ${\cal L}_1$, although it is conserved on the critical
trajectories of ${\cal L}_0$. Neither is the total current conserved,%
$$
\left[ \delta _M{\cal L}_1=\partial _\mu j_M^\mu \right] _{sol\left( {\cal L}%
_0+{\cal L}_1\right) }\quad . 
$$

To proceed, we now specify the action of $G$ on ${\cal L}_1$: We assume
that, under the action of $G$, ${\cal L}_1$ transforms as a total derivative 
{\bf on- and off-shell}, i.e. without using the equations of motion. Then $%
{\cal L}_1$ is said to be {\it semi-invariant} under the action of $G$. This
means that 
\begin{equation}
\label{gract11}\delta _M{\cal L}_1=\partial _\mu U_M^\mu \quad , 
\end{equation}
for some functions $U_M^\mu $ of the fields and its derivatives. This gives,
using (\ref{gract7}),%
$$
0=\delta _M\phi ^i\cdot \left( eq,{\cal L}_0+{\cal L}_1\right) _i+\partial
_\mu \left( j_M^\mu -U_M^\mu \right) \quad , 
$$
and we see that the modified current 
\begin{equation}
\label{gract12}\widetilde{j_M^\mu }:=j_M^\mu -U_M^\mu 
\end{equation}
is conserved on the critical trajectories of ${\cal L}_0+{\cal L}_1$, i.e. 
\begin{equation}
\label{gract12.1}\partial _\mu \widetilde{j_M^\mu }=0\quad . 
\end{equation}
Note that in this case the {\bf conserved current is no longer a Noether
current}.

\subsection{Algebra of Poisson brackets\quad \quad \label{PB Algebra}}

The Poisson brackets of the zero components $j_M^0$ of the {\bf total}
Noether currents $j_M^\mu $ associated with the action of $T_M$ on some
Lagrangian ${\cal L}$ satisfy the Lie algebra of $G$, possibly up to a sign,
regardless of whether ${\cal L}$ is invariant or not. This can be proven by
introducing canonical momenta%
$$
\Lambda _i:=\frac{\partial {\cal L}}{\partial \dot \phi ^i}\quad , 
$$
so that 
\begin{equation}
\label{gract13}j_M^0=\delta _M\phi ^i\cdot \Lambda _i\quad .
\end{equation}
Let $C_{MN}^{\;K}$ denote the structure constants of the Lie algebra $%
Lie\left( G\right) $ of $G$, i.e.%
$$
\left[ T_M,T_N\right] =C_{MN}^{\;K}\cdot T_K\quad , 
$$
where $\left[ \cdot ,\cdot \right] $ denotes a (graded) commutator. Working
out the Poisson bracket we get 
\begin{equation}
\label{gr01}\left\{ j_M^0\left( t,\sigma \right) ,j_N^0\left( t,\sigma
^{\prime }\right) \right\} _{PB}=-\delta _M\phi ^i\frac{\partial \delta
_N\phi ^j}{\partial \phi ^i}\Lambda _j\,\delta \left( \sigma -\sigma
^{\prime }\right) +\delta _N\phi ^j\frac{\partial \delta _M\phi ^i}{\partial
\phi ^j}\Lambda _i\,\delta \left( \sigma -\sigma ^{\prime }\right) \quad .
\end{equation}
where $\left\{ \cdot ,\cdot \right\} _{PB}$ denotes a (graded) Poisson
bracket with canonical variables $\phi ^i$, $\Lambda _i$. If we now use the
results from (\ref{gract0.0}) we find%
$$
\left( \ref{gr01}\right) =-\left[ \widetilde{T_M},\widetilde{T_N}\right]
^i\Lambda _i\,\delta \left( \sigma -\sigma ^{\prime }\right) \quad . 
$$
Taking account of (\ref{zw1}, \ref{zw2}) this gives 
\begin{equation}
\label{gra131}\left\{ j_M^0\left( t,\sigma \right) ,j_N^0\left( t,\sigma
^{\prime }\right) \right\} _{PB}\;=\;\pm C_{MN}^{\;K}\cdot j_K^0\left(
t,\sigma \right) \cdot \delta \left( \sigma -\sigma ^{\prime }\right) \quad ,
\end{equation}
where $"+/-"$ refers to a {\bf left / right} action. If we define associated 
{\it Noether charges} 
\begin{equation}
\label{gract25}Q_M\left( t\right) :=\int\limits_{W\left( t\right) }d^p\sigma
\cdot j_M^0\left( t,\sigma \right) \quad ,
\end{equation}
then the once integrated version of (\ref{gra131}) is 
\begin{equation}
\label{gra132}\left\{ Q_M,j_N^0\right\} _{PB}=\pm \;j_K^0\cdot ad\left(
T_M\right) _{\;N}^K\quad ,
\end{equation}
where $ad\left( T\right) $ denotes the adjoint representation of the Lie
algebra element $T$. This now means that the total Noether currents span the
adjoint representation of $G$ in the case of a left action. Moreover, in
this case a further integration of (\ref{gra132}) yields back the algebra we
have started with, 
\begin{equation}
\label{gra133}\left\{ Q_M,Q_N\right\} _{PB}=\pm C_{MN}^{\;K}\cdot Q_K\quad .
\end{equation}

We omit the subscript $PB$ in what follows, and reintroduce it only when
there is danger of confusion with an anticommutator.

Now we look at the situation when the Lagrangian contains a semi-invariant
piece ${\cal L}_1$. In this case the conserved currents are $\widetilde{%
j_M^\mu }=j_M^\mu -U_M^\mu $, and their zero components have Poisson bracket
relations 
\begin{equation}
\label{gract15}\left\{ \widetilde{j_M^0}\left( t,\sigma \right) ,\widetilde{%
j_N^0}\left( t,\sigma ^{\prime }\right) \right\} =\left\{
j_M^0,j_N^0\right\} -\left\{ j_M^0,U_N^0\right\} -\left\{
U_M^0,j_N^0\right\} +\left\{ U_M^0,U_N^0\right\} \quad . 
\end{equation}
The brackets $\left\{ j_M^0,U_N^0\right\} =\left\{ \delta _M\phi ^i\Lambda
_i,U_N^0\right\} $ are always unequal zero when $U_N^0$ is not a constant,
since the presence of the canonical momenta amounts to derivatives with
respect to the fields on $U_N^0$. However, the brackets $\left\{
U_M^0,U_N^0\right\} $ are also non-vanishing in general; although in the
Lagrangian description they contained only fields $\phi ^i$ and their
derivatives, the shift to the Hamiltonian (first order) picture amounts to
inverting the relations%
$$
\Lambda _i=\frac{\partial {\cal L}}{\partial \dot \phi ^i}\left( \phi
,\partial _0\phi ,\partial _r\phi \right) 
$$
for $\partial _0\phi ^i$, which gives $\dot \phi ^i=\Phi ^i\left( \phi
,\partial _r\phi ,\Lambda \right) $, where $r,s=1,\ldots ,p$ refers to the
''spatial'' coordinates on $W$. Therefore,%
$$
U_M^0\left( \phi ,\partial _\mu \phi \right) \longrightarrow U_M^0\left(
\phi ,\Phi \left( \phi ,\partial _r\phi ,\Lambda \right) ,\partial _s\phi
\right) =\widehat{U_M^0}\left( \phi ,\partial _r\phi ,\Lambda \right) \quad
, 
$$
so that {\bf after Legendre transforming} the $U_M^0$ {\bf do} depend on the
canonical momenta, which makes their mutual Poisson brackets in general
non-vanishing. Note that the ''hatted'' $\widehat{U_M^0}$ is of course a
different function of its arguments than $U_M^0$ which makes itself manifest
when we are performing partial or functional derivatives, respectively.
Therefore, in a Poisson bracket we are always dealing with $\widehat{U_M^0}$%
; outside a Poisson bracket we can replace $\widehat{U_M^0}$ by $U_M^0$, as
we shall do in the following.

Now we subtract and add $\pm C_{MN}^{\;K}\cdot U_K^0\left( t,\sigma \right)
\cdot \delta \left( \sigma -\sigma ^{\prime }\right) $ on the right hand
side of (\ref{gract15}), and use (\ref{gra131}). This gives us 
\begin{equation}
\label{gract15.1}\left\{ \widetilde{j_M^0}\left( t,\sigma \right) ,%
\widetilde{j_N^0}\left( t,\sigma ^{\prime }\right) \right\} =\pm
C_{MN}^{\;K}\cdot \widetilde{j_K^0}\left( t,\sigma \right) \cdot \delta
\left( \sigma -\sigma \right) \;+\;\widetilde{S_{MN}^0}\;+\left\{ \widehat{%
U_M^0}\left( t,\sigma \right) ,\widehat{U_N^0}\left( t,\sigma ^{\prime
}\right) \right\} \quad , 
\end{equation}
with the ''anomalous'' piece%
$$
\widetilde{S_{MN}^0}+\left\{ \widehat{U_M^0}\left( t,\sigma \right) ,%
\widehat{U_N^0}\left( t,\sigma ^{\prime }\right) \right\} =-\left\{
j_M^0\left( t,\sigma \right) ,\widehat{U_N^0}\left( t,\sigma ^{\prime
}\right) \right\} -\left\{ \widehat{U_M^0}\left( t,\sigma \right)
,j_N^0\left( t,\sigma ^{\prime }\right) \right\} \;\pm 
$$
\begin{equation}
\label{gract16}\pm \;C_{MN}^{\;K}\cdot U_K^0\left( t,\sigma \right) \cdot
\delta \left( \sigma -\sigma ^{\prime }\right) \;+\left\{ \widehat{U_M^0}%
\left( t,\sigma \right) ,\widehat{U_N^0}\left( t,\sigma ^{\prime }\right)
\right\} . 
\end{equation}

Let us compute the brackets $\left\{ j_M^0,\widehat{U_N^0}\right\} $ for the
special case that the action of $G$ on covectors $\Lambda _i$ is specified
so as to make expressions like $\dot \phi ^i\Lambda _i$ transforming as
scalars under the group operation; this means that $\Lambda _i$ transform
contragrediently to $\phi ^i$, 
\begin{equation}
\label{gract17}\delta _M\Lambda _i=-\Lambda _j\frac{\partial \delta _M\phi ^j%
}{\partial \phi ^i}\quad .
\end{equation}
To see that this specification leaves $\dot \phi ^i\Lambda _i$ invariant we
apply $\delta _M$,%
$$
\delta _M\left( \dot \phi ^i\Lambda _i\right) =\left( \delta _M\frac
d{dt}\phi ^i-\dot \phi ^j\frac{\partial \delta _M\phi ^i}{\partial \phi ^j}%
\right) \Lambda _i\quad ; 
$$
if we assume now, as usual, that $\delta _M$ commutes with $\partial _\mu $
the expression in the bracket vanishes. We find 
\begin{equation}
\label{gr6}\left\{ j_M^0\left( t,\sigma \right) ,\widehat{U_N^0}\left(
t,\sigma ^{\prime }\right) \right\} =-\delta _MU_N^0\cdot \delta \left(
\sigma ^{\prime }-\sigma \right) \;+\frac \partial {\partial \sigma
^r}\left[ \delta _M\phi ^i\frac{\partial \widehat{U_N^0}}{\partial \partial
_r\phi ^i}\cdot \delta \left( \sigma ^{\prime }-\sigma \right) \right] \quad
.
\end{equation}
Double integration of the second term over $W\left( t\right) $, $t=const$.,
yields 
\begin{equation}
\label{gr3}\int\limits_{W\left( t\right) }d^p\sigma \cdot \frac \partial
{\partial \sigma ^r}\left[ \delta _M\phi ^i\frac{\partial \widehat{U_N^0}}{%
\partial \partial _r\phi ^i}\right] =\int\limits_{\partial W\left( t\right)
}d{\cal A}_r^{p-1}\cdot \delta _M\phi ^i\frac{\partial \widehat{U_N^0}}{%
\partial \partial _r\phi ^i}\quad ,
\end{equation}
where $d{\cal A}_r^{p-1}$ is a $\left( p-1\right) $-dimensional area
element. We must deal with this surface term appropriately. The manifold $%
W\left( t\right) $ can be infinitely extended in all spatial directions, or
some of these spatial directions may be compact. To avoid bothering with the
surface terms we assume from now on that the integrands of surface
contributions vanish sufficiently strong at the boundary $\partial W\left(
t\right) $, i.e. at points which lie at infinite values of the non-compact
coordinates. As a special case this includes the possibility that $W\left(
t\right) $ is closed, which implies that {\bf all} spatial coordinates $%
\sigma ^{\mu \,}$ are compact.Furthermore we assume that all expressions in
a total derivative, such as on the left hand side of (\ref{gr3}), are smooth
and defined {\bf globally} on $W$. (The emphasis on being globally defined
is of course to prevent us from situations where Stokes' theorem is not
applicable, i.e. ''surface terms cannot be integrated away''; this can be
true for the topological current to be defined below). Under these
circumstances {\bf all surface terms vanish}, and we obtain for the current
algebra 
\begin{equation}
\label{al}\left\{ \widetilde{j_M^0}\left( t,\sigma \right) ,\widetilde{j_N^0}%
\left( t,\sigma ^{\prime }\right) \right\} =\pm C_{MN}^{\;K}\cdot \widetilde{%
j_K^0}\left( t,\sigma \right) \cdot \delta \left( \sigma -\sigma ^{\prime
}\right) \;+\;\widetilde{S_{MN}^0}\;+\left\{ \widehat{U_M^0}\left( t,\sigma
\right) ,\widehat{U_N^0}\left( t,\sigma ^{\prime }\right) \right\} \quad ,
\end{equation}
\begin{equation}
\label{gract23}\widetilde{S_{MN}^0}\left( t,\sigma ,\sigma ^{\prime }\right)
=\left[ \delta _MU_N^0-\delta _NU_M^0\pm C_{MN}^{\;K}\cdot U_K^0\right]
\cdot \delta \left( \sigma -\sigma ^{\prime }\right) \quad +\quad \left( 
\mbox{total derivatives}\right) \quad ,
\end{equation}
with the total derivatives from (\ref{gr6}). Let us now define 
\begin{equation}
\label{gract251}Q_M\left( t\right) :=\int\limits_{W\left( t\right) }d^p\sigma
\cdot \widetilde{j_M^0}\left( t,\sigma \right) \quad ,
\end{equation}
\begin{equation}
\label{gract24}S_{MN}^\mu \left( t,\sigma \right) =\delta _MU_N^\mu -\delta
_NU_M^\mu \pm C_{MN}^{\;K}\cdot U_K^\mu \quad ,
\end{equation}
\begin{equation}
\label{gract26}Z_{MN}\left( t\right) :=\int\limits_{W\left( t\right)
}d^p\sigma \cdot S_{MN}^0\left( t,\sigma \right) \quad .
\end{equation}
Note that the charge $Q_M\left( t\right) =Q_M$ is no longer a Noether
charge, since it is defined through the conserved current $\widetilde{j_M^0}$
rather than the Noether current $j_M^0$. It is conserved, however, due to (%
\ref{gract12.1}). We show now that $Z_{MN}$ is conserved as well.

\subsection{Conservation of the new charges}

To prove this, observe that $\delta _M$ commutes with $\partial _\mu $;
therefore we can write%
$$
\partial _\mu S_{MN}^\mu =\delta _M\partial _\mu U_N^\mu -\delta _N\partial
_\mu U_M^\mu \pm C_{MN}^{\;K}\cdot \partial _\mu U_K^\mu \quad = 
$$
\begin{equation}
\label{gract27}=\quad \left( \delta _M\delta _N-\delta _N\delta _M\pm
C_{MN}^{\;K}\cdot \delta _K\right) {\cal L}_1\quad . 
\end{equation}
If we work out the double variation we find that the last expression
vanishes due to%
$$
\left[ \delta _M,\delta _N\right] {\cal L}=\left[ \delta _M,\delta _N\right]
\phi ^i\cdot {\cal L}_{\phi ^i}+\partial _\mu \left[ \delta _M,\delta
_N\right] \phi ^i\cdot {\cal L}_{\partial _\mu \phi ^i}\quad . 
$$
This can be seen yet in another way: On account of $\left[ \partial _\mu
,\delta _M\right] =0$, $\delta _N=\widetilde{T_N}$ acts on coordinates $\phi
^i$ in the same way as it acts on $\partial _\mu \phi ^i$. Therefore we can
replace the $\delta ^{\prime }$s in the round bracket in (\ref{gract27}) by
vector fields $\widetilde{T_N}$, which yields%
$$
\delta _M\delta _N-\delta _N\delta _M\pm C_{MN}^{\;K}\cdot \delta _K=\left[ 
\widetilde{T_M},\widetilde{T_N}\right] \pm C_{MN}^{\;K}\cdot \widetilde{T_K}%
\;= 
$$
$$
=\;\mp \widetilde{\left( \left[ T_M,T_N\right] -C_{MN}^{\;K}\cdot T_K\right) 
}=0\quad , 
$$
according to the algebra of the generators $\left( T_M\right) $. What we
have shown is 
\begin{equation}
\label{gract28}\partial _\mu S_{MN}^\mu =0\quad , 
\end{equation}
which is the local conservation law for the charge $Z_{MN}\left( t\right) $
defined in (\ref{gract26}).

Using the definitions (\ref{gract251}, \ref{gract26}), we find on double
integration of (\ref{al}) (and on assumption that this integration is
defined) 
\begin{equation}
\label{gract29}\left\{ Q_M,Q_N\right\} _{PB}=\pm C_{MN}^{\;K}\cdot
Q_K\;+\;Z_{MN}\;+\int\limits_{W\left( t\right) }d^p\sigma \,d^p\sigma
^{\prime }\cdot \left\{ \widehat{U_M^0}\left( t,\sigma \right) ,\widehat{%
U_N^0}\left( t,\sigma ^{\prime }\right) \right\} \quad .
\end{equation}
We see that our original algebra has been extended by conserved charges $%
Z_{MN}$; however, unless the Poisson brackets $\left\{ \widehat{U_M^0},%
\widehat{U_N^0}\right\} $ vanish, this extension does not close to a new
algebra!

\section{Coset spaces of Lie groups as target spaces\quad \quad \label
{CosetSpaces}}

\subsection{Closure of the algebra extension\quad \quad \label{Closure}}

In order to proceed further we now make more detailed assumptions about the
structure of the target space and the geometric origin of the invariant and
semi-invariant pieces in the Lagrangian. We assume that the target space is
now the group $G$ itself, with coordinates $\phi ^i$. More generally, we
could have that $G$ is a subgroup of a larger group $\tilde G$, which
contains yet another subgroup $H$ : $G,H\subset \tilde G$. Then $\Sigma $
could be the coset space $\tilde G/H$, and $G$ would act on elements of $%
\Sigma =\tilde G/H$ by left or right multiplication. This is the situation
we shall consider later, where $\tilde G=\,${\rm super}$Poincare$ in $D=10$
spacetime dimensions, $G$ is the subgroup generated by $\left\{ P_m,Q_\alpha
\right\} $, i.e. the generators of Poincare- and super-translations, and $H$
is the subgroup $SO\left( 1,9\right) $. If the objects $Q_\alpha $ build two 
$16$-component spinors with opposite chirality, then the coset space $\Sigma 
$ is type IIA superspace. However, for the purpose of illustrating of how
topological currents emerge we shall in the following refrain from any
graded groups, algebras, or whatsoever, and restrict ourselves to the
simpler case of $\Sigma =G$.

The fields $\phi ^i$ on $W$ accomplish an embedding $emb:W\rightarrow \Sigma 
$ of $W$ into $\Sigma $ by $emb\left( x\right) =\left( \phi ^1\left(
x\right) ,\ldots ,\phi ^{\dim G}\left( x\right) \right) $. From now on we
call $W$ the ''worldvolume'', following standard conventions. If the
hypersurfaces $W\left( t\right) $ are closed then the same holds for their
images in $\Sigma $, since $\partial \left[ embW\left( t\right) \right]
=emb\left[ \partial W\left( t\right) \right] =\emptyset $. In other words,
the images $embW\left( t\right) $ are $p${\it -cycles}$\ $ in $\Sigma $ in
this case. We assume that the previously made assumptions concerning surface
terms in integrands still hold, and that those spatial dimensions of $%
W\left( t\right) $ which are not infinitely extended are closed. Furthermore
we {\bf assume} that the semi-invariant piece ${\cal L}_1$ or Wess-Zumino
(WZ) term, as it will be called in the sequel, is the result of the
pull-back of a target space $\left( p+1\right) $-form $\left( WZ\right) $ to
the worldvolume $W$; from now on, we write ${\cal L}_1=:{\cal L}_{WZ}$ for
the semi-invariant piece. Its construction proceeds as follows:

Let $\left( \Pi ^A\right) _{A=1,\ldots ,\dim G}$ be left-invariant (LI) $1$%
-forms on $\Sigma =G$; this means, that at every point in $\Sigma $ they
span the cotangent space to $\Sigma $ at this point, {\bf and} they are
invariant under the action of the group, 
\begin{equation}
\label{gract30}\delta _M\Pi ^A={\cal \not L}_{\widetilde{T_M}}\Pi ^A=0\quad
, 
\end{equation}
where ${\cal \not L}_{\widetilde{T_M}}$ denotes the Lie derivative with
respect to the induced vector field $\widetilde{T_M}$. The WZ-form $\left(
WZ\right) $ on $\Sigma $ can be expanded in this basis,%
$$
\left( WZ\right) =\frac 1{\left( p+1\right) !}\Pi ^{A_1}\cdots \Pi
^{A_{p+1}}\cdot \left( WZ\right) _{A_{p+1}\cdots A_1}\left( \phi \right)
\quad , 
$$
with pull-back 
\begin{equation}
\label{gr333}emb^{*}\left( WZ\right) =\frac 1{\left( p+1\right) !}dx^{\mu
_1}\cdots dx^{\mu _{p+1}}\cdot \Pi _{,\mu _1}^{A_1}\cdots \Pi _{,\mu
_{p+1}}^{A_{p+1}}\cdot \left( WZ\right) _{A_{p+1}\cdots A_1}\left( \phi
\right) \quad ; 
\end{equation}
since $dx^{\mu _1}\cdots dx^{\mu _{p+1}}$ is proportional to the canonical
volume form $\omega _0$ with respect to the coordinates $\left( x^\mu
\right) $ on $W$,%
$$
dx^{\mu _1}\cdots dx^{\mu _{p+1}}=\epsilon ^{\mu _1\cdots \mu _{p+1}}\cdot
\omega _0\quad ,\quad \omega _0=dx^0\cdots dx^p\quad , 
$$
we find that $emb^{*}\left( WZ\right) =\omega _0\cdot {\cal L}_{WZ}$, where 
\begin{equation}
\label{gract319}{\cal L}_{WZ}=\frac 1{\left( p+1\right) !}\,\epsilon ^{\mu
_1\cdots \mu _{p+1}}\cdot \,\Pi _{,\mu _1}^{A_1}\cdots \Pi _{,\mu
_{p+1}}^{A_{p+1}}\cdot \,\left( WZ\right) _{A_{p+1}\cdots A_1}\left( \phi
\right) \quad . 
\end{equation}
Note that we have tacitly used the superspace summation conventions on the
indices $M_i$, which, of course, does not affect the validity of the results
to be shown.

Semi-invariance of the WZ-term then implies that for every generator $T_M$
of $G$ there exists a $p$-form $\Delta _M$ on $\Sigma $ such that 
\begin{equation}
\label{gract31}\delta _M\left( WZ\right) =d\Delta _M\quad . 
\end{equation}
This implies that%
$$
\omega _0\cdot \delta _M{\cal L}_{WZ}=\delta _M\left[ emb^{*}\left(
WZ\right) \right] \;=\;emb^{*}\delta _M\left( WZ\right) =emb^{*}d\Delta _M= 
$$
\begin{equation}
\label{gract32}=d\left( emb^{*}\Delta _M\right) \quad . 
\end{equation}
Expanding $\Delta _M$ in the LI-basis we can compute $d\left( emb^{*}\Delta
_M\right) =$%
$$
=\omega _0\cdot \frac 1{p!}\epsilon ^{\mu _1\cdots \mu _{p+1}}\cdot \partial
_{\mu _1}\,\left[ \Pi _{,\mu _2}^{A_2}\cdots \Pi _{,\mu
_{p+1}}^{A_{p+1}}\cdot \,\Delta _{MA_{p+1}\cdots A_2}\right] \quad , 
$$
and comparison with (\ref{gract32}) then yields%
$$
\delta _M{\cal L}_{WZ}=\partial _\mu U_M^\mu \quad , 
$$
\begin{equation}
\label{gract33}U_M^\mu =\frac 1{p!}\epsilon ^{\mu \mu _2\cdots \mu
_{p+1}}\cdot \,\left[ \Pi _{,\mu _2}^{A_2}\cdots \Pi _{,\mu
_{p+1}}^{A_{p+1}}\cdot \,\Delta _{MA_{p+1}\cdots A_2}\right] \quad . 
\end{equation}
In particular, for $\mu =0$ we obtain%
$$
U_M^0=\frac 1{p!}\epsilon ^{0\mu _2\cdots \mu _{p+1}}\cdot \,\left[ \Pi
_{,\mu _2}^{A_2}\cdots \Pi _{,\mu _{p+1}}^{A_{p+1}}\cdot \,\Delta
_{MA_{p+1}\cdots A_2}\right] \quad , 
$$
from which it is seen that $U_M^0$ {\bf cannot} contain $\Pi _{,0}^A$, due
to the antisymmetry of the $\epsilon $-tensor. Reexpanding the forms $\Pi ^A$
in the coordinate basis $d\phi ^M$ gives%
$$
\Pi ^A=\Pi _N^Ad\phi ^N\quad ,\quad \Pi _{,\mu }^A=\Pi _N^A\phi _{,\mu
}^N\quad , 
$$
from which we see that $U_M^0$ cannot contain $\phi _{,0}^N=\dot \phi ^N$
either. This point is crucial in light of our previous considerations, of
course, since, if we now assume, that the equations%
$$
\Lambda _M=\frac{\partial {\cal L}}{\partial \dot \phi ^M}\quad ,\quad \mbox{%
for\quad }{\cal L}={\cal L}_0+{\cal L}_{WZ} 
$$
are invertible with respect to $\dot \phi ^N$, then $\dot \phi ^M=\Phi
^M\left( \phi ,\partial _r\phi ,\Lambda \right) $ for $r=1,\ldots ,p$, and
after performing the Legendre transformation $\left( \phi ^M,\dot \phi
^N\right) \rightarrow \left( \phi ^M,\Lambda _N\right) $ we have%
$$
U_M^s=U_M^s\left( \phi ,\Phi ^M\left( \phi ,\partial _r\phi ,\Lambda \right)
,\partial _t\phi \right) =\widehat{U_M^s}\left( \phi ,\partial _t\phi
,\Lambda ,\right) \quad ;\quad s=1,\ldots ,p\,;\;r,t\neq s\quad , 
$$
but 
\begin{equation}
\label{gr001}U_M^0=\widehat{U_M^0}\left( \phi ,\partial _r\phi \right) \quad
;\quad r=1,\ldots ,p\quad . 
\end{equation}
Therefore we now have Poisson brackets 
\begin{equation}
\label{gract34}\left\{ j_M^0,\widehat{U_N^0}\right\} =\delta _M\phi ^K\cdot
\left\{ \Lambda _K,\widehat{U_N^0}\right\} \quad , 
\end{equation}
\begin{equation}
\label{gract35}\left\{ \widehat{U_M^0},\widehat{U_N^0}\right\} =0\quad . 
\end{equation}
This point being clarified we omit the ''hats'' on $\widehat{U_N^0}$ from
now on, it being understood that it is the ''hatted'' version that appears
in a Poisson bracket.

Referring to (\ref{gract29}) we can now state that the algebra of the
charges $Q_M$ closes to a linear combination of the $Q_M$ and the new
charges $Z_{MN}$, 
\begin{equation}
\label{cl1}\left\{ Q_M,Q_N\right\} _{PB}=\pm C_{MN}^{\;K}\cdot
Q_K\;+\;Z_{MN}\quad .
\end{equation}
Furthermore, due to (\ref{gract35}), we have 
\begin{equation}
\label{cll}\left\{ S_{MN}^0,S_{M^{\prime }N^{\prime }}^0\right\} =0\quad ,
\end{equation}
and therefore 
\begin{equation}
\label{cl2}\left\{ Z_{MN},Z_{M^{\prime }N^{\prime }}\right\} =0\quad ,
\end{equation}
i.e. the mutual algebra of the new charges $Z_{MN}$ also closes into the
extension generated by $\left( Q_M,Z_{MN}\right) $. But what about the
algebra of $\left\{ Q_K,Z_{MN}\right\} _{PB}$ ? We now examine under which
conditions this expression yields a linear combination of $\left\{
Q_M,Z_{MN}\right\} _{PB}$.

\subsection{Topological currents\quad \quad \label{TopCurrents}}

Consider the object $S_{MN}^\mu $ defined in (\ref{gract24}), 
\begin{equation}
\label{sch}S_{MN}^\mu \left( t,\sigma \right) =\delta _MU_N^\mu -\delta
_NU_M^\mu \pm C_{MN}^{\;K}\cdot U_K^\mu \quad . 
\end{equation}
Using the form of $U_M^\mu $ given in (\ref{gract33}) and taking into
account that $\delta _M\Pi _{,\nu }^N=0$ we have%
$$
S_{MN}^\mu =\frac 1{p!}\epsilon ^{\mu \mu _1\cdots \mu _p}\,\,\Pi _{,\mu
_1}^{A_1}\cdots \Pi _{,\mu _p}^{A_p}\;\times 
$$
\begin{equation}
\label{gract36}\times \;\left[ \,\delta _M\Delta _{NA_p\cdots A_1}-\delta
_N\Delta _{MA_p\cdots A_1}\pm C_{MN}^{\;K}\cdot \Delta _{KA_p\cdots
A_1}\right] \quad . 
\end{equation}
For the sake of simplicity we define the expression 
\begin{equation}
\label{gract37}\tilde R_{MNA_p\cdots A_1}:=\left[ \,\delta _M\Delta
_N-\delta _N\Delta _M\pm C_{MN}^{\;K}\cdot \Delta _K\right] _{A_p\cdots
A_1}\quad , 
\end{equation}
so that%
$$
S_{MN}^\mu =\frac 1{p!}\epsilon ^{\mu \mu _1\cdots \mu _p}\,\,\Pi _{,\mu
_1}^{A_1}\cdots \Pi _{,\mu _p}^{A_p}\cdot \tilde R_{MNA_p\cdots A_1}\quad , 
$$
and rewrite this form in the coordinate basis $\left( d\phi ^N\right) $, $%
\Pi ^A=\Pi _N^Ad\phi ^N$, which yields 
\begin{equation}
\label{gract38}S_{MN}^\mu =\frac 1{p!}\epsilon ^{\mu \mu _1\cdots \mu
_p}\,\,\phi _{,\mu _1}^{N_1}\cdots \phi _{,\mu _p}^{N_p}\cdot R_{MNN_p\cdots
N_1}\quad , 
\end{equation}
with the new components 
\begin{equation}
\label{gract39}R_{MNN_p\cdots N_1}=\Pi _{N_1}^{A_1}\cdots \Pi
_{N_p}^{A_p}\cdot \tilde R_{MNA_p\cdots A_1}\quad . 
\end{equation}
We note that $R_{MNN_p\cdots N_1}$ is a function of the fields $\phi $ only.
Appealing to (\ref{gract38}) we now define the identically conserved \cite
{Azca} {\it topological currents}

\begin{equation}
\label{gract40}j_T^{\mu M_1\cdots M_p}:=\epsilon ^{\mu \mu _1\cdots \mu
_p}\,\,\phi _{,\mu _1}^{M_1}\cdots \phi _{,\mu _p}^{M_p}\quad , 
\end{equation}
and the {\it topological charges} 
\begin{equation}
\label{gract401}T^{M_1\cdots M_p}:=\int\limits_{W\left( t\right) }d^p\sigma
\cdot j_T^{0M_1\cdots M_p}\quad , 
\end{equation}
which are conserved due to $\partial _\mu j_T^{\mu M_1\cdots M_p}=0$. The
topological charges $T^{M_1\cdots M_p}$ are invariant under the group
action, 
\begin{equation}
\label{da}\delta _KT^{M_1\cdots M_p}=0\quad , 
\end{equation}
see (\ref{ko5}) below. We write (\ref{gract38}) as 
\begin{equation}
\label{gract41}S_{MN}^\mu =\frac 1{p!}j_T^{\mu N_1\cdots N_p}\cdot
R_{MNN_p\cdots N_1}=:\,j_T^\mu \bullet R_{MN}\quad , 
\end{equation}
then the charges $Z_{MN}$ take the form 
\begin{equation}
\label{gr401}Z_{MN}=\int\limits_{W\left( t\right) }d^p\sigma \cdot
S_{MN}^0=\int\limits_{W\left( t\right) }d^p\sigma \;j_T^0\bullet R_{MN}\quad
; 
\end{equation}
for {\bf constant} $R_{MNN_p\cdots N_1}$ this is 
\begin{equation}
\label{gract262}Z_{MN}=T\bullet R_{MN}\quad . 
\end{equation}

Now we can turn to the bracket $\left\{ Q_K,Z_{MN}\right\} $; a computation
yields 
\begin{equation}
\label{gract49}\left\{ Q_K,Z_{MN}\right\} =-\,\int\limits_{W\left( t\right)
}d^p\sigma \cdot \delta _K\left[ j_T^0\bullet R_{MN}\right] \quad . 
\end{equation}
It is clear that this can never close into an expression involving the
charges $Q_M$, since this would require the occurence of $j_M^0=\delta
_M\phi ^N\Lambda _N$ in the integrand, but the integrand contains no
canonical momenta (recall that $j_T^0$ contains no time derivatives of
fields, and $R_{MN}$ contains no field derivatives at all). Hence, at best
the left hand side can close into a linear combination of the new charges $%
Z_{M^{\prime }N^{\prime }}$. If we now look at (\ref{gr401}) we see that
requiring that (\ref{gract49}) be a linear combination of $Z_{M^{\prime
}N^{\prime }}$ is equivalent to demanding that 
\begin{equation}
\label{gract505}\delta _K\left[ j_T^0\bullet R_{MN}\right] =-\frac
12B_{KMN}^{M^{\prime }N^{\prime }}\cdot j_T^0\bullet R_{M^{\prime }N^{\prime
}}\;+\;\cdots \quad , 
\end{equation}
where $\cdots $ denote possible surface terms, and where $B_{KMN}^{M^{\prime
}N^{\prime }}$ are {\bf constant}; the factor $\frac 12$ is due to the
antisymmetry of $R_{MN}$ in $M$ and $N$. (\ref{gract49}) then reads 
\begin{equation}
\label{gract501}\left\{ Q_K,Z_{MN}\right\} =\frac 12B_{KMN}^{M^{\prime
}N^{\prime }}\cdot Z_{M^{\prime }N^{\prime }}\quad . 
\end{equation}
(\ref{cl1}, \ref{cl2}) and (\ref{gract49} - \ref{gract501}) now tell us that
the algebra of the conserved charges $Q_K,Z_{MN}$ closes if and only if (\ref
{gract505}) holds.

\subsection{When are the charges $Z_{MN}$ central ?}

This can be read off from (\ref{gract501}): The charges $Z_{MN}$ are {\it %
central}, i.e. they commute with all other elements in the algebra, {\bf iff}
all coefficients $B_{KMN}^{M^{\prime }N^{\prime }}$ vanish; according to (%
\ref{gract505}) this is true {\bf iff} 
\begin{equation}
\label{ko2}\delta _K\left[ j_T^0\bullet R_{MN}\right] =\left( \mbox{globally
defined smooth surface term}\right) \quad . 
\end{equation}
Let us now examine 
\begin{equation}
\label{ko3}\delta _Kj_T^{0M_1\cdots M_p}=\sum_{k=1}^p\partial _{\mu
_k}\left[ \frac 1{p!}\epsilon ^{0\mu _1\cdots \mu _k\cdots \mu _p}\cdot \phi
_{,\mu _1}^{M_1}\cdots \delta _K\phi ^{M_k}\cdots \phi _{,\mu
_p}^{M_p}\right] \quad . 
\end{equation}
We take the point of view that the expression in square brackets is smooth
and globally defined (since $\delta _K\phi ^{M_k}$ amounts to a derivative
of the field $\phi ^{M_k}$ which can be smoothy continued over the whole of $%
W$) so that its integral over $W\left( t\right) $ indeed vanishes, on using
Stokes' theorem. This means that $\delta _K\left[ j_T^0\bullet R_{MN}\right] 
$ is a surface term, {\bf provided} that $R_{MN}$ are constant. A {\bf %
sufficient} condition for the charges $Z_{MN}$ to be {\bf central} is
therefore that 
\begin{equation}
\label{ko4}R_{MNN_p\cdots N_1}\left( \phi \right) =const.=R_{MNN_p\cdots
N_1}\quad , 
\end{equation}
where $R_{MNN_p\cdots N_1}$ are the components of $R_{MN}$ in the coordinate
basis $\left( d\phi ^N\right) $.

As an aside we remark that (\ref{ko3}) implies that the topological charges $%
T^{M_1\cdots M_p}$ are invariant under the group action, 
\begin{equation}
\label{ko5}\delta _KT^{M_1\cdots M_p}=0\quad . 
\end{equation}

We now have (see (\ref{gract262})) $Z_{MN}=T\bullet R_{MN}$, and the
non-vanishing brackets of our extended algebra then read%
$$
\left\{ Q_M,Q_N\right\} =C_{MN}^{\;K}\cdot Q_K\;+\;T\bullet R_{MN}\quad , 
$$
\begin{equation}
\label{gract53}\left\{ Q_K,T\bullet R_{MN}\right\} =\frac
12B_{KMN}^{M^{\prime }N^{\prime }}\cdot T\bullet R_{M^{\prime }N^{\prime
}}\quad , 
\end{equation}
and the charges $T\bullet R_{MN}$ are all {\bf central}.

\section{Summary}

At this point it is appropriate to summarize the results we have obtained so
far in the form of three theorems.

\subsection{Theorem 1}

The Noether currents satisfy the Poisson bracket algebra, possibly up to a
sign, 
\begin{equation}
\label{th1}\left\{ j_M^0\left( t,\sigma \right) ,j_N^0\left( t,\sigma
^{\prime }\right) \right\} _{PB}\;=\;\pm C_{MN}^{\;K}\cdot j_K^0\left(
t,\sigma \right) \cdot \delta \left( \sigma -\sigma ^{\prime }\right) \quad ,
\end{equation}
{\bf regardless} of whether the Lagrangian ${\cal L}$ is {\bf invariant or
not}. $\pm $ refers to a left/right action. The once integrated version is 
\begin{equation}
\label{th2}\left\{ Q_M,j_N^0\right\} _{PB}=\;\pm j_K^0\cdot ad\left(
T_M\right) _{\;N}^K\quad ,
\end{equation}
where $ad\left( T\right) $ denotes the adjoint representation of the Lie
algebra element $T$. This implies that the total Noether currents span the
adjoint representation of $G$ in the case of a left action.

Double integration of the current algebra yields the algebra of the
generators of $G$, possibly up to a sign, 
\begin{equation}
\label{th3}\left\{ Q_M,Q_N\right\} _{PB}=\pm C_{MN}^{\;K}\cdot Q_K\quad .
\end{equation}

\subsection{Theorem 2}

Assume that the Lagrangian ${\cal L}$ is {\bf semi-invariant} under the
action of $G$, i.e. $\delta _M{\cal L}=\partial _\mu U_M^\mu $ for functions 
$U_M^\mu =U_M^\mu \left( \phi ,\partial _\nu \phi \right) $ of the fields
and its derivatives {\bf on-shell and off-shell}; that the action of $G$ on
canonical momenta $\Lambda _i={\cal L}_{\dot \phi ^i}$ is defined by (\ref
{gract17}); and that surface integrals with smooth integrands may be
neglected. Then

\begin{enumerate}
\item  The modified currents 
\begin{equation}
\label{th5}\widetilde{j_M^\mu }=j_M^\mu -U_M^\mu \quad ,
\end{equation}
where $j_M^\mu \,$ are the Noether currents associated with ${\cal L}$, are
conserved, 
\begin{equation}
\label{th6}\partial _\mu \widetilde{j_M^\mu }=0\quad .
\end{equation}

\item  Double integration of the Poisson bracket algebra yields 
\begin{equation}
\label{th7}\left\{ Q_M,Q_N\right\} _{PB}=\pm C_{MN}^{\;K}\cdot
Q_K\;+\;Z_{MN}\;+\int\limits_{W\left( t\right) }d^p\sigma \,d^p\sigma
^{\prime }\cdot \left\{ \widehat{U_M^0}\left( t,\sigma \right) ,\widehat{%
U_N^0}\left( t,\sigma ^{\prime }\right) \right\} \quad ,
\end{equation}
where 
\begin{equation}
\label{th8}Z_{MN}\left( t\right) :=\int\limits_{W\left( t\right) }d^p\sigma
\cdot S_{MN}^0\left( t,\sigma \right) \quad ,
\end{equation}
and 
\begin{equation}
\label{th9}S_{MN}^\mu \left( t,\sigma \right) =\delta _MU_N^\mu -\delta
_NU_M^\mu \pm C_{MN}^{\;K}\cdot U_K^\mu \quad .
\end{equation}
Due to 
\begin{equation}
\label{th10}\partial _\mu S_{MN}^\mu =0
\end{equation}
the ''charges'' $Z_{MN}$ are {\bf conserved}.
\end{enumerate}

\subsection{Theorem 3}

Let those directions of the hypersurfaces $W\left( t\right) $ which are not
infinitely extended be closed. Let the target space $\Sigma $ be the group $%
G $ itself; let the semi-invariant piece ${\cal L}_1={\cal L}_{WZ}$ in the
Lagrangian be the pull-back of a target space $\left( p+1\right) $-form to
the worldvolume $W$, which transforms under $G$ according to $\delta _M{\cal %
L}_{WZ}=\partial _\mu U_M^\mu $, with 
\begin{equation}
\label{th11}U_M^\mu =\frac 1{p!}\epsilon ^{\mu \mu _2\cdots \mu _{p+1}}\cdot
\,\left[ \Pi _{,\mu _2}^{M_2}\cdots \Pi _{,\mu _{p+1}}^{M_{p+1}}\cdot
\,\Delta _{MM_{p+1}\cdots M_2}\right] \quad , 
\end{equation}
where $\Delta _{MM_{p+1}\cdots M_2}$ are the components of $\dim G$ $p$%
-forms $\Delta _M$ in a left-invariant basis $\left( \Pi ^M\right) $. Let
the action of $G$ on canonical momenta $\Lambda _i={\cal L}_{\dot \phi ^i}$
be defined according to (\ref{gract17}). Then

\begin{enumerate}
\item  The Poisson bracket algebra of the Noether charges $Q_M$ and the
charges $Z_{MN}$ closes {\bf iff} 
\begin{equation}
\label{th12}\delta _K\left[ j_T^0\bullet R_{MN}\right] =-\frac
12B_{KMN}^{M^{\prime }N^{\prime }}\cdot j_T^0\bullet R_{M^{\prime }N^{\prime
}}\;+\;\cdots \quad ,
\end{equation}
where $\cdots $ denote possible surface terms, $B_{KMN}^{M^{\prime
}N^{\prime }}$ are {\bf constant}, and where 
\begin{equation}
\label{th13}R_{MNN_p\cdots N_1}=\Pi _{N_1}^{A_1}\cdots \Pi _{N_p}^{A_p}\cdot
\left[ \,\delta _M\Delta _N-\delta _N\Delta _M\pm C_{MN}^{\;K}\cdot \Delta
_K\right] _{A_p\cdots A_1}\quad .
\end{equation}
The extended algebra then reads 
\begin{equation}
\label{th14}\left\{ Q_M,Q_N\right\} _{PB}=\pm C_{MN}^{\;K}\cdot
Q_K\;+\;Z_{MN}\quad ,
\end{equation}
\begin{equation}
\label{th15}\left\{ Q_K,Z_{MN}\right\} =\frac 12B_{KMN}^{M^{\prime
}N^{\prime }}\cdot Z_{M^{\prime }N^{\prime }}\quad ,
\end{equation}
\begin{equation}
\label{th16}\left\{ Z_{MN},Z_{M^{\prime }N^{\prime }}\right\} _{PB}=0\quad .
\end{equation}

\item  A {\bf sufficient} condition for the charges $Z_{MN}$ to be {\bf %
central} is that 
\begin{equation}
\label{th17}R_{MNN_p\cdots N_1}\left( \phi \right) =const.=R_{MNN_p\cdots
N_1}\quad .
\end{equation}

\item  The topological charges $T^{M_1\cdots M_p}$ are invariant under the
group action, 
\begin{equation}
\label{th19}\delta _KT^{M_1\cdots M_p}=0\quad .
\end{equation}
\end{enumerate}

\subsection{Corollary}

If $R_{MNN_p\cdots N_1}=const.$, then all charges $Z_{MN}$ are central, and
are linear combinations of the topological charges $T^{M_1\cdots M_p}$, 
\begin{equation}
\label{th18}Z_{MN}=T\bullet R_{MN}\quad . 
\end{equation}
\pagebreak

\section{Lagrangians including (Abelian) Gauge fields \label{Eich}}

\subsection{Structure of the Lagrangian}

Now let us study the case when the Lagrangian ${\cal L}$ contains additional
degrees of freedom in the form of an Abelian $\left( q-1\right) $-form gauge
potential $A_{\mu _1\ldots \mu _{q-1}}$, $q\le p$, that is defined on the 
{\bf worldvolume}. A priori, the group $G$ acts on the target space $\Sigma $
and there is no reason why $A$ should be involved in the transformation of
fields on $\Sigma $, but that is what we now impose on $A$, since it is the
situation that occurs when the Lagrangian describes $D$-$p$-branes, which we
want to study later. To this end, we assume that on the target space there
exists a $q$-form potential $B=\frac 1{q!}\Pi ^{C_q}\cdots \Pi
^{C_1}B_{C_1\cdots C_q}$, with an associated $\left( q+1\right) $-form field
strength $H=dB$. The field strength $H$ is taken to be invariant under the
action of $G$, i.e. $\delta _MH=0$. This implies that locally 
\begin{equation}
\label{ga0}\delta _MB=d\Delta _M\quad , 
\end{equation}
with $\dim G$ $\left( q-1\right) $-forms%
$$
\Delta _M=\frac 1{\left( q-1\right) !}\Pi ^{A_q}\cdots \Pi ^{A_2}\widetilde{%
\Delta _{MA_2\cdots A_q}}=\frac 1{\left( q-1\right) !}d\phi ^{A_q}\cdots
d\phi ^{A_2}\,\Delta _{MA_2\cdots A_q}\quad , 
$$
where we have used a tilde to distinguish the components of $\Delta _M$ with
respect to the LI-basis $\left( \Pi ^A\right) $ from the components in the
coordinate basis $\left( d\phi ^M\right) $, which we shall need later. $%
A_{\mu _1\ldots \mu _{q-1}}$ are therefore
$\left( \begin{array}{c}
{p+1} \\
{q-1}
\end{array}  \right)$
additional
degrees of freedom involved in the dynamics; it is assumed, however, that $%
A_{\mu _1\ldots \mu _{q-1}}$ enters the Lagrangian {\bf only} via the field
strengths $F_{\mu _1\ldots \mu _q}=q\cdot \partial _{[\mu _1}A_{\mu _2\ldots
\mu _q]}$. The Lagrangian again splits into an invariant piece ${\cal L}_0$
and a semi-invariant piece ${\cal L}_{WZ}$, where ${\cal L}_0$ takes the
form 
\begin{equation}
\label{ga2}{\cal L}_0={\cal L}_0\left( \phi ,\partial _\mu \phi ,\widehat{%
F_{\mu _1\ldots \mu _q}}\right) \quad ,\quad \widehat{F_{\mu _1\ldots \mu _q}%
}=F_{\mu _1\ldots \mu _q}-\left( emb^{*}B\right) _{\mu _1\ldots \mu _q}\quad
; 
\end{equation}
in order to have ${\cal L}_0$ invariant we {\bf impose} the transformation
behaviour 
\begin{equation}
\label{ga3}\delta _MA=emb^{*}\Delta _M\quad ,\mbox{\quad }\left( \delta
_MA\right) _{\mu _2\ldots \mu _q}=\phi _{,\mu _2}^{A_q}\cdots \phi _{,\mu
_q}^{A_2}\,\Delta _{MA_2\cdots A_q}\left( \phi \right) 
\end{equation}
on $A$. Since 
\begin{equation}
\label{ga4}\delta _M\widehat{F}=\delta _M\left[ dA-emb^{*}B\right] =\left[
d\delta _MA-emb^{*}d\Delta _M\right] =0\quad , 
\end{equation}
this is sufficient to have an invariant ${\cal L}_0$. As for the
semi-invariant part ${\cal L}_{WZ}$, we assume the following: 
\begin{equation}
\label{ga5}{\cal L}_{WZ}={\cal L}_{WZ}\left( \phi ,\partial _\mu \phi ,%
\widehat{F_{\mu _1\ldots \mu _q}}\right) \quad , 
\end{equation}
with transformation behaviour $\delta _M{\cal L}_{WZ}=\partial _\mu U_M^\mu $%
, where $U_M^\mu =U_M^\mu \left( \phi ,\partial _\mu \phi ,F_{\mu \nu
}\right) $, {\bf but} 
\begin{equation}
\label{ga6}\frac{\partial U_M^\mu }{\partial \partial _\nu \phi }=0\quad
\quad \mbox{if\quad }\mu =\nu \quad ;\quad \frac{\partial U_M^\mu }{\partial
F_{\nu _1\ldots \nu _q}}=0\quad \quad \mbox{if\quad }\mu \in \left\{ \nu
_1,\ldots ,\nu _q\right\} \quad . 
\end{equation}
Note the absence of a hat in the field $F$ in the definition of the field
content of $U_M^\mu $.

Now we define canonical momenta 
\begin{equation}
\label{ga61}\Lambda _N=\frac{\partial {\cal L}}{\partial \partial _0\phi ^N}%
\quad ,\quad \Lambda ^{\nu _2\ldots \nu _q}=\frac{\partial {\cal L}}{%
\partial \partial _0A_{\nu _2\ldots \nu _q}}\quad . 
\end{equation}

\subsection{Constraints on the gauge field degrees of freedom}

The fact that we are dealing with a gauge field $A_{\mu _1\ldots \mu _{q-1}}$
as dynamical degrees of freedom makes itself manifest in the form of {\bf %
constraints} that are imposed on the dynamics \cite{Govaerts}: Using the
formula 
\begin{equation}
\label{ga7}\frac{\partial {\cal L}}{\partial \partial _{\nu _1}A_{\nu
_2\ldots \nu _q}}=\frac 1{\left( q-1\right) !}\frac{\partial {\cal L}}{%
\partial F_{\nu _1\ldots \nu _q}} 
\end{equation}
we see that, due to the antisymmetry of $F$, ${\cal L}$ cannot contain $%
\partial _0A_{\nu _2\ldots \nu _q}$, whenever one of the $\nu _2,\ldots ,\nu
_q$ is zero; this implies that the canonical momenta 
\begin{equation}
\label{ga8}\Lambda ^{\nu _2\ldots \nu _q}=0\quad \quad \mbox{for\quad }0\in
\left\{ \nu _2,\ldots \nu _q\right\} \quad , 
\end{equation}
i.e. they vanish identically. The number of independent constraints (\ref
{ga8}) is 
$ \left( \begin{array}{c}
p \\
{q-2} \end{array}  \right)$. The second set of constraints follows from the
equations of motion for $A_{\nu _2\ldots \nu _q}$: They are given by%
$$
\left( eq\right) ^{\nu _2\ldots \nu _q}:=\frac{\partial {\cal L}}{\partial
A_{\nu _2\ldots \nu _q}}-\partial _0\Lambda ^{\nu _2\ldots \nu _q}-\partial
_r\frac{\partial {\cal L}}{\partial \partial _rA_{\nu _2\ldots \nu _q}}%
=0\quad , 
$$
where the sum over $r$ ranges from $1$ to $p$. The first term on the RHS
vanishes since ${\cal L}$ contains no $A_{\nu _2\ldots \nu _q}$; the second
one vanishes if we choose one of the $\nu $'s to be equal to zero, say $\nu
_2$. On using (\ref{ga7}) we have $\frac{\partial {\cal L}}{\partial
\partial _rA_{0\nu _3\ldots \nu _q}}=-\frac{\partial {\cal L}}{\partial
\partial _0A_{r\nu _3\ldots \nu _q}}$, so we get 
\begin{equation}
\label{ga9}\partial _r\Lambda ^{r\nu _3\ldots \nu _q}=0\quad ,\quad \nu
_3,\ldots ,\nu _q\mbox{\quad arbitrary.} 
\end{equation}
This yields a number of another 
$ \left( \begin{array}{c}
p \\
{q-2} \end{array}  \right)$
 constraints.

These constraints are not on an equal footing, however; as can be seen from
the above arguments, the first set (\ref{ga8}) holds before any equations of
motion are considered, and therefore amounts to a reduction of phase space
to a submanifold of the original phase space of codimension 
$ \left( \begin{array}{c}
p \\
{q-2} \end{array}  \right)$
;
in Dirac's terminology this is a set of {\it primary constraints}. The
second set (\ref{ga9}) comes into play only {\bf on-shell}, i.e. on using
equations of motion, and is called a set of {\it secondary constraints}. We
shall not use Dirac's machinery for handling these constraints here, but
shall work with Poisson brackets instead; in this case, however, it is
crucial to impose (\ref{ga8}, \ref{ga9}) {\bf not before} all Poisson
brackets have been worked out, otherwise we would obtain wrong results.

After these remarks let us now study the Poisson bracket algebra of the
Noether currents. The Noether currents are 
\begin{equation}
\label{ga10}j_M^\mu =\delta _M\phi ^K\cdot \frac{\partial {\cal L}}{\partial
\partial _\mu \phi ^K}+\frac 1{\left( q-1\right) !}\delta _MA_{\nu _2\ldots
\nu _q}\cdot \frac{\partial {\cal L}}{\partial \partial _\mu A_{\nu _2\ldots
\nu _q}}\quad , 
\end{equation}
\begin{equation}
\label{ga11}j_M^0=\delta _M\phi ^K\cdot \Lambda _K+\frac 1{\left( q-1\right)
!}\delta _MA_{\nu _2\ldots \nu _q}\cdot \Lambda ^{\nu _2\ldots \nu _q}\quad
. 
\end{equation}

\subsection{Algebra of Noether currents}

In working out brackets $\left\{ j_M^0,j_N^0\right\} $ we make use of the
fact that $\delta _M\phi ^K$ is a function of the fields $\phi $ only,
therefore the brackets $\left\{ \delta _M\phi ^K,\Lambda ^{\nu _2\ldots \nu
_q}\right\} $ vanish; and that $\delta _MA_{\nu _2\ldots \nu _q}$ is a
function of the fields $\phi $ and their derivatives $\partial _\mu \phi $
only, see (\ref{ga3}), therefore the brackets $\left\{ \delta _MA_\nu
,\Lambda ^{\nu _2\ldots \nu _q}\right\} $ vanish. The computation then yields%
$$
\left\{ j_M^0\left( t,\sigma \right) ,j_N^0\left( t,\sigma ^{\prime }\right)
\right\} =\pm C_{MN}^{\;K}\cdot j_K^0\cdot \delta \left( \sigma -\sigma
^{\prime }\right) \;+\;\frac 1{\left( q-1\right) !}\left[ \,\mp
C_{MN}^{\;K}\,\delta _KA_{\nu _2\ldots \nu _q}\cdot \delta \left( \sigma
-\sigma ^{\prime }\right) \;+\right. 
$$
\begin{equation}
\label{ga12}\left. +\;\delta _M\phi ^K\cdot \left\{ \Lambda _K,\delta
_NA_{\nu _2\ldots \nu _q}\right\} \;-\;\delta _N\phi ^K\cdot \left\{ \Lambda
_K,\delta _MA_{\nu _2\ldots \nu _q}\right\} \right] \cdot \Lambda ^{\nu
_2\ldots \nu _q}\quad . 
\end{equation}
This can be written as%
$$
\left\{ j_M^0,j_N^0\right\} =\pm C_{MN}^{\;K}\cdot j_K^0\cdot \delta \left(
\sigma -\sigma ^{\prime }\right) \;+\; 
$$
\begin{equation}
\label{ga13}+\;\frac 1{\left( q-1\right) !}\left[ \,\left( -\delta _M\delta
_N+\delta _N\delta _M\mp C_{MN}^{\;K}\,\cdot \delta _K\right) A_{\nu
_2\ldots \nu _q}\right] \Lambda ^{\nu _2\ldots \nu _q}\cdot \delta \left(
\sigma -\sigma ^{\prime }\right) \quad \cdots \quad , 
\end{equation}
where $\cdots $ denotes surface terms.

The first term is just what we have expected; $\pm $ again refers to a
left/right action. Since $\delta _MA=emb^{*}\Delta _M$ we have $\left(
-\delta _M\delta _N+\delta _N\delta _M\mp C_{MN}^{\;K}\,\cdot \delta
_K\right) A=$%
$$
=\;emb^{*}\left( -\delta _M\Delta _N+\delta _N\Delta _M\mp
C_{MN}^{\;K}\,\Delta _K\right) \quad , 
$$
where the expression in the brackets 
\begin{equation}
\label{ga14}-\delta _M\Delta _N+\delta _N\Delta _M\mp C_{MN}^{\;K}\,\Delta
_K\;=:\;S\left( \Delta \right) _{MN}\; 
\end{equation}
measures the {\bf deviation} of the forms $\Delta _M$ from transforming as a
multiplet under the adjoint representation of the group $G$; this is seen
from 
\begin{equation}
\label{ga15}T_M\cdot \Delta _N=\mp \,\Delta _K\cdot ad\left( T_M\right)
_{\;N}^K\;-\;S\left( \Delta \right) _{MN}\quad , 
\end{equation}
where the point denotes the action of the ''abstract'' generator $T_M$ on
the component $\Delta _N$ according to $T_M\cdot \Delta _N=\left[ \delta
_M,\Delta _N\right] $.

We now introduce the notation 
\begin{equation}
\label{ga16}\left[ emb^{*}S\left( \Delta \right) _{MN}\right] _{\nu _2\ldots
\nu _q}=:S\left( \Delta \right) _{MN\nu _2\ldots \nu _q}\quad ,
\end{equation}
\begin{equation}
\label{ga17}\frac 1{\left( q-1\right) !}S\left( \Delta \right) _{MN\nu
_2\ldots \nu _q}\Lambda ^{\nu _2\ldots \nu _q}=:S\left( \Delta \right)
_{MN}\bullet \Lambda ^{gauge}\quad ,
\end{equation}
then (\ref{ga13}) reads 
\begin{equation}
\label{ga18}\left\{ j_M^0,j_N^0\right\} =\left[ \pm C_{MN}^{\;K}\cdot
j_K^0\;+\;S\left( \Delta \right) _{MN}\bullet \Lambda ^{gauge}\right] \cdot
\delta \left( \sigma -\sigma ^{\prime }\right) \quad .
\end{equation}
Let us define 
\begin{equation}
\label{ga19}Q_M=\int\limits_{W\left( t\right) }d^p\sigma \cdot j_M^0\quad
,\quad Y_{MN}\left( t\right) =\int\limits_{W\left( t\right) }d^p\sigma \cdot
S\left( \Delta \right) _{MN}\bullet \Lambda ^{gauge}\quad ,
\end{equation}
then the once integrated version of (\ref{ga18}) is 
\begin{equation}
\label{ga20}\left\{ Q_M,j_N^0\right\} =\pm j_K^0\cdot ad\left( T_M\right)
_{\;N}^K\;+\;S\left( \Delta \right) _{MN}\bullet \Lambda ^{gauge}\quad ,
\end{equation}
which defines the action of the generator $T_M$ on the Noether current $j_N^0
$. We see that due to the presence of the $S$-term on the right hand side
the Noether currents now {\bf fail} to transform as a multiplet in the
adjoint representation, as was the case previously.

The twice integrated version is 
\begin{equation}
\label{ga21}\left\{ Q_M,Q_N\right\} =\pm C_{MN}^{\;K}\cdot
Q_K\;+\;Y_{MN}\quad .
\end{equation}
$Q_M$ are conserved when the Lagrangian is invariant under $G$; we need to
check when $Y_{MN}\left( t\right) $ are conserved. To this end we perform $%
\frac d{dt}$ on $Y_{MN}$ in (\ref{ga19}) and assume, for the sake of
convenience, that the hypersurfaces $W\left( t\right) $ do not change shape
as $t$ varies; then the only contribution to $\frac{dY_{MN}}{dt}$ comes from 
$\frac d{dt}\left[ S\left( \Delta \right) _{MN}\bullet \Lambda
^{gauge}\right] $. A calculation then shows that a {\bf sufficient}
condition for the charge $Y_{MN}\left( t\right) $ to be conserved is 
\begin{equation}
\label{ga23}S\left( \Delta \right) _{MNN_q\ldots N_2}\left( \phi \right)
=const.=S\left( \Delta \right) _{MNN_q\ldots N_2}\quad .
\end{equation}
Under the same condition the charges $Y_{MN}$ are seen to be central.

The complete algebra is then 
\begin{equation}
\label{ga25}\left\{ Q_M,Q_N\right\} =\pm C_{MN}^{\;K}\cdot
Q_K\;+\;Y_{MN}\quad ,\quad \left\{ Q_K,Y_{MN}\right\} =\left\{
Y_{MN},Y_{M^{\prime }N^{\prime }}\right\} =0\quad .
\end{equation}

\subsection{Algebra of modified currents}

At last then let us determine the general structure of the extended algebra
of the charges associated with the {\bf modified} currents $\widetilde{%
j_M^\mu }=j_M^\mu -U_M^\mu $, given that the WZ-term ${\cal L}_{WZ}$ behaves
as in (\ref{ga5}, \ref{ga6}). We again find that%
$$
\left\{ \widetilde{j_M^0},\widetilde{j_N^0}\right\} =\left\{
j_M^0,j_N^0\right\} -\left\{ U_M^0,j_N^0\right\} -\left\{
j_M^0,U_N^0\right\} \quad , 
$$
with $\left\{ j_M^0,j_N^0\right\} $ given in (\ref{ga18}). $\left\{
j_M^0,U_N^0\right\} $ can be determined using the properties of $U_N^0\,$
given in (\ref{ga6}). Up to surface terms we then find%
$$
\left\{ \widetilde{j_M^0},\widetilde{j_N^0}\right\} \approx \pm
C_{MN}^{\;K}\cdot \widetilde{j_K^0}\cdot \delta \left( \sigma -\sigma
^{\prime }\right) \;+\;\left[ \pm C_{MN}^{\;K}\cdot U_K^0\;+\;S\left( \Delta
\right) _{MN}\bullet \Lambda ^{gauge}\;-\right.  
$$
\begin{equation}
\label{ga26}\left. -\delta _N\phi ^K\frac{\partial U_M^0}{\partial \phi ^K}%
\;+\;\delta _M\phi ^K\frac{\partial U_N^0}{\partial \phi ^K}\right] \cdot
\delta \left( \sigma -\sigma ^{\prime }\right) \quad .
\end{equation}
Analogous to (\ref{sch}) we define 
\begin{equation}
\label{ga27}S\left( U\right) _{MN}=\delta _M\phi ^K\cdot \frac{\partial U_N^0%
}{\partial \phi ^K}-\delta _N\phi ^K\cdot \frac{\partial U_M^0}{\partial
\phi ^K}\pm C_{MN}^{\;K}\cdot U_K^0
\end{equation}
and its integral 
\begin{equation}
\label{ga28}Z_{MN}=\int\limits_{W\left( t\right) }d^p\sigma \cdot S\left(
U\right) _{MN}\quad ,
\end{equation}
and $Y_{MN}\,$ as the integral of $S\left( \Delta \right) _{MN}\bullet
\Lambda ^{gauge}$ over $W\left( t\right) $, according to (\ref{ga19}). Then
double integration of (\ref{ga26}) yields 
\begin{equation}
\label{ga29}\left\{ Q_M,Q_N\right\} =\pm C_{MN}^{\;K}\cdot
Q_K\;+\;Y_{MN}\;+\;Z_{MN}\quad .
\end{equation}
In the case of constant $S\left( \Delta \right) _{MNN_q\ldots N_2}$ (see (%
\ref{ga23})) we can write%
$$
Y_{MN}=\frac 1{\left( q-1\right) !}S\left( \Delta \right) _{MNN_q\ldots
N_2}\int\limits_{W\left( t\right) }d^p\sigma \cdot \phi _{,r_2}^{N_2}\cdots
\phi _{,r_q}^{N_q}\cdot \Lambda ^{r_2\ldots r_q}\,\quad ; 
$$
on the right hand side now there appear charges 
\begin{equation}
\label{ga30}Y_{MN}^{N_2\ldots N_q}:=\int\limits_{W\left( t\right) }d^p\sigma
\cdot \phi _{,r_2}^{N_2}\cdots \phi _{,r_q}^{N_q}\cdot \Lambda ^{r_2\ldots
r_q}\quad ,
\end{equation}
and due to the first class constraints the summation in the integrand runs
over ''spatial'' indices $r_2,\ldots ,r_q\in \left\{ 1,\ldots ,p\right\} $
only, so that finally%
$$
Y_{MN}=\frac 1{\left( q-1\right) !}S\left( \Delta \right) _{MNN_q\ldots
N_2}\cdot Y_{MN}^{N_2\ldots N_q}\quad , 
$$
and (\ref{ga29}) becomes now 
\begin{equation}
\label{ga31}\left\{ Q_M,Q_N\right\} =\pm C_{MN}^{\;K}\cdot Q_K\;+\;\frac
1{\left( q-1\right) !}S\left( \Delta \right) _{MNN_q\ldots N_2}\cdot
Y_{MN}^{N_2\ldots N_q}\;+Z_{MN}\quad .
\end{equation}

Yet another expression for the above relations can be obtained \cite{Soro}
if we regard the worldvolume as a pseudo-Riemannian manifold with an
(auxiliary) metric which is diagonal in the coordinate system $\left(
t,\sigma \right) $, 
$$
-dt\otimes dt+\delta _{rs}\cdot d\sigma ^r\otimes d\sigma ^s\quad . 
$$
Then the restriction of this metric to the hypersurfaces $W\left( t\right) $
is a Euclidean metric, and $W\left( t\right) $ become Riemannian manifolds,
on which we can introduce a Hodge star operator with respect to this metric.
We need not distinguish between upper and lower indices here, so that $%
\Lambda ^{r_2\ldots r_q}$ for $r_2,\ldots ,r_q\in \left\{ 1,\ldots
,p\right\} $ can be regarded as components of a $\left( q-1\right) $-form $%
\Lambda ^{gauge}$ on $W\left( t\right) $. Its Hodge dual is then 
\begin{equation}
\label{ho1}\left( *\Lambda ^{gauge}\right) _{s_q\ldots s_p}=\frac 1{\left(
q-1\right) !}\,\epsilon _{t_2\ldots t_qs_q\ldots s_p}\cdot \Lambda
^{t_2\ldots t_q}\quad , 
\end{equation}
where all indices are taken from the set $\left\{ 1,\ldots ,p\right\} $. On
using 
\begin{equation}
\label{ho2}\left( **\Lambda ^{gauge}\right) =\left( -1\right) ^{\left(
q-1\right) \left( p-q+1\right) }\cdot \Lambda ^{gauge} 
\end{equation}
we can write%
$$
d\sigma ^1\cdots d\sigma ^p\cdot \phi _{,r_2}^{N_2}\cdots \phi
_{,r_q}^{N_q}\cdot \Lambda ^{r_2\ldots r_q}= 
$$
\begin{equation}
\label{ho3}=\frac 1{\left( p-q+1\right) !}\,\left( *\Lambda ^{gauge}\right)
\cdot emb^{*}d\phi ^{N_2}\cdots emb^{*}d\phi ^{N_q}\quad ; 
\end{equation}
in what follows we shall omit the ''$emb^{*}$'' for the sake of simplicity.
Multiplication of (\ref{ho3}) by $\frac 1{\left( q-1\right) !}S\left( \Delta
\right) _{MNN_q\ldots N_2}$ gives%
$$
d\sigma ^1\cdots d\sigma ^p\cdot S\left( \Delta \right) _{MN}\bullet \Lambda
^{gauge}\;= 
$$
\begin{equation}
\label{ho4}=\;\frac 1{\left( p-q+1\right) !}\,\left( *\Lambda
^{gauge}\right) \cdot S\left( \Delta \right) _{MN}\quad , 
\end{equation}
where $S\left( \Delta \right) _{MN}$ now denotes the pullback of this form
to $W\left( t\right) $, 
\begin{equation}
\label{ho5}S\left( \Delta \right) _{MN}=emb^{*}\,\frac 1{\left( q-1\right)
!}\,d\phi ^{N_2}\cdots d\phi ^{N_q}\cdot S\left( \Delta \right)
_{MNN_q\ldots N_2}\quad . 
\end{equation}
Thus we can rewrite (\ref{ga19}) as 
\begin{equation}
\label{ho6}Y_{MN}\left( t\right) =\frac 1{\left( p-q+1\right)
!}\int\limits_{W\left( t\right) }\,\left( *\Lambda ^{gauge}\right) \cdot
S\left( \Delta \right) _{MN}\quad , 
\end{equation}

We again summarize this section in the form of a theorem.

\subsection{Theorem}

Let an Abelian $\left( q-1\right) $-form gauge potential $A_{\mu _2\ldots
\mu _q}$, $q\le p$, be defined on the worldvolume. On the target space a $q$%
-form potential $B$ transforms according to $\delta _MB=d\Delta _M$ under $G$%
. We impose a transformation behaviour $\delta _MA=emb^{*}\Delta _M$ on $A$.
The Lagrangian splits into an invariant piece ${\cal L}_0$ and a
semi-invariant piece ${\cal L}_{WZ}$, as described above. Then

\begin{enumerate}
\item  The Poisson bracket algebra of the {\bf Noether} currents is 
\begin{equation}
\label{th1x1}\left\{ j_M^0,j_N^0\right\} \approx \left[ \pm C_{MN}^{\;K}\cdot
j_K^0\;+\;S\left( \Delta \right) _{MN}\bullet \Lambda ^{gauge}\right] \cdot
\delta \left( \sigma -\sigma ^{\prime }\right) \quad ,
\end{equation}
where $S\left( \Delta \right) _{MN\nu _2\ldots \nu _q}=\left[ emb^{*}S\left(
\Delta \right) _{MN}\right] _{\nu _2\ldots \nu _q}$, and 
\begin{equation}
\label{th2x}S\left( \Delta \right) _{MN}=-\delta _M\Delta _N+\delta _N\Delta
_M\mp C_{MN}^{\;K}\,\Delta _K\;\;
\end{equation}
measures the deviation of the forms $\Delta _M$ from transforming as a
multiplet under the adjoint representation of the group $G$: 
\begin{equation}
\label{th3x}T_M\cdot \Delta _N=\mp \,\Delta _K\cdot ad\left( T_M\right)
_{\;N}^K\;-\;S\left( \Delta \right) _{MN}\quad .
\end{equation}
The once integrated version of (\ref{th1x1}) is 
\begin{equation}
\label{th4x}\left\{ Q_M,j_N^0\right\} =\pm j_K^0\cdot ad\left( T_M\right)
_{\;N}^K\;+\;S\left( \Delta \right) _{MN}\bullet \Lambda ^{gauge}\quad ,
\end{equation}
which defines the action of the generator $T_M$ on the Noether current $j_N^0
$, and $\pm $ refers to a left/right action. The twice integrated version is 
\begin{equation}
\label{th5x}\left\{ Q_M,Q_N\right\} =\pm C_{MN}^{\;K}\cdot
Q_K\;+\;Y_{MN}\quad ,
\end{equation}
where 
\begin{equation}
\label{th6x}Y_{MN}\left( t\right) =\int\limits_{W\left( t\right) }d^p\sigma
\cdot S\left( \Delta \right) _{MN}\bullet \Lambda ^{gauge}\quad .
\end{equation}
If $S\left( \Delta \right) _{MNN_q\ldots N_2}=const.$, then the charges $%
Y_{MN}$ are conserved and central. (\ref{th6x}) can be rewritten in the form 
\begin{equation}
\label{th61}Y_{MN}\left( t\right) =\frac 1{\left( p-q+1\right)
!}\int\limits_{W\left( t\right) }\,\left( *\Lambda ^{gauge}\right) \cdot
S\left( \Delta \right) _{MN}\quad ,
\end{equation}
with $*\Lambda ^{gauge}$ being the Hodge dual of the form $\Lambda
^{gauge}=\frac 1{\left( q-1\right) !}d\sigma ^{r_2}\cdots d\sigma
^{r_q}\cdot \Lambda ^{r_2\ldots r_q}$ on the worldvolume.

\item  The Poisson bracket algebra of the {\bf modified} currents is 
\begin{equation}
\label{th7x}\left\{ \widetilde{j_M^0},\widetilde{j_N^0}\right\} \approx \pm
C_{MN}^{\;K}\cdot \widetilde{j_K^0}\cdot \delta \left( \sigma -\sigma
^{\prime }\right) \;+\;\left[ S\left( U\right) _{MN}\;+\;S\left( \Delta
\right) _{MN}\bullet \Lambda ^{gauge}\;\right] \quad ,
\end{equation}
with 
\begin{equation}
\label{th8x}S\left( U\right) _{MN}=\delta _M\phi ^K\cdot \frac{\partial U_N^0%
}{\partial \phi ^K}-\delta _N\phi ^K\cdot \frac{\partial U_M^0}{\partial
\phi ^K}\pm C_{MN}^{\;K}\cdot U_K^0
\end{equation}
and its integral 
\begin{equation}
\label{th9x}Z_{MN}=\int\limits_{W\left( t\right) }d^p\sigma \cdot S\left(
U\right) _{MN}\quad .
\end{equation}
Double integration of the current algebra yields the charge algebra 
\begin{equation}
\label{th10x}\left\{ Q_M,Q_N\right\} =\pm C_{MN}^{\;K}\cdot
Q_K\;+\;Y_{MN}\;+\;Z_{MN}\quad .
\end{equation}
In the case of constant $S\left( \Delta \right) _{MN}$ this can be written
as 
\begin{equation}
\label{th11x}\left\{ Q_M,Q_N\right\} =C_{MN}^{\;K}\cdot Q_K\;+\;\frac
1{\left( q-1\right) !}\,S\left( \Delta \right) _{MNN_q\ldots N_2}\cdot
Y_{MN}^{N_2\ldots N_q}\;+Z_{MN}\quad .
\end{equation}
\end{enumerate}
\pagebreak

\section{Extended superalgebras carried by $D$-$p$-branes in IIA superspace 
\label{SuperAlgebra}}

Now we apply the ideas we have developed in the previous sections to the
case of $D$-$p$-branes in IIA superspace. This restricts $p$ to be even, $%
p=0,2,4,6,8$. However, before doing so, we first discuss our superspace
conventions, and our definitions of graded Poisson brackets. Then we first
compute the algebra of Noether charges and of modified Noether charges
resulting from a D-brane Lagrangian without specific assumptions on the
background the brane propagates in, or on the specific form of the various
gauge fields occuring in the Lagrangian. Then we recapitulate how
supergravity determines the background in which the branes propagate, and
the relation of superspace constraints with $\kappa $-symmetry of the
branes. Then we study the Bianchi identities associated with a specific
choice of background gauge fields in superspace; and only then we work out
the explicit superalgebra extensions carried by D-branes in this particular $%
D=10$ vacuum.

\section{Conventions}

\subsection{Superspace conventions \label{conventions}}

The target space $\Sigma $ is now the coset space%
$$
\mbox{IIA-superMinkowski}=\mbox{IIA-superPoincare}/SO\left( 1,9\right)  
$$
with coordinates $\left( X,\theta \right) $ that label the coset
representative $e^{iX\cdot P+\theta Q}$. Adopting the convention that the
complex conjugate of a product of two spinors reverses their order this
implies that in an operator realization the coset representatives are mapped
to unitary operators, provided that $P$ and $Q$ are hermitian. The
assumption of IIA superspace means that we have two $16$-component spinor
generators of opposite chirality which transform under the two irreducible $%
\left( 16\times 16\right) $-dimensional spin representations of $SO\left(
1,9\right) $; but effectively, this yields one non-chiral $32$-component
spinor, transforming under the direct sum of the two irreducible spin
representations, which is just the representation of $SO\left( 1,9\right) $
obtained from the $32$-component $\Gamma $-matrices. The metric $\eta _{mn}$
on the target space is flat $10$-dimensional ''mostly plus'' Minkowski
metric. Spinor components occur with natural index up; an inner product
between spinors is provided by the bilinear form $\left( \chi ,\theta
\right) \mapsto \chi ^\alpha C_{\alpha \beta }\theta ^\beta $, where $C$ is
a charge conjugation matrix. In $D=10$ and with the Minkowski metric as
specified above we can choose a Majorana-Weyl representation for the spinors
and the $\Gamma $-matrices, respectively, in which spinors have real
Grassmann-odd components, the matrices $C\Gamma _m$ are real and symmetric,
and $C$ is real and antisymmetric. In such a representation we can choose $%
C=\pm \Gamma _0$. Altogether we have $32$ real fermion degrees of freedom a
priori. Spinor indices are lowered and raised {\bf from the left} with the
charge conjugation matrix and its inverse, respectively; e.g., raising is
accomplished with the inverse of $C$, the components of which are denoted by 
$C^{\alpha \beta }$, by $\theta _\beta \mapsto \theta ^\alpha =C^{\alpha
\beta }\theta _\beta $. By definition, $C^{\alpha \beta }C_{\beta \gamma
}=\delta _\gamma ^\alpha $. An expression like $\bar \epsilon \Gamma
_m\theta $ therefore means%
$$
\bar \epsilon \Gamma _m\theta =\epsilon ^\alpha C_{\alpha \beta }\left(
\Gamma _m\right) _{\;\gamma }^\beta \theta ^\gamma \quad , 
$$
etc. Our supertranslation algebra is 
\begin{equation}
\label{ap1}\left\{ Q_\alpha ,Q_\beta \right\} =2\Gamma _{\alpha \beta
}^m\cdot P_m\quad .
\end{equation}
The action of $e^{iY\cdot P+\epsilon Q}$ on $\left( X,\theta \right) $
yields $\left( X^{\prime },\theta ^{\prime }\right) $, where $\left(
X^{\prime },\theta ^{\prime }\right) $ is implicitly defined by 
\begin{equation}
\label{ap2}e^{iY\cdot P+\epsilon Q}e^{iX\cdot P+\theta Q}=e^{iX^{\prime
}\cdot P+\theta ^{\prime }Q}\quad ;
\end{equation}
for infinitesimal $\epsilon $ this yields $\left( X^{\prime },\theta
^{\prime }\right) =\left( X+Y+i\bar \epsilon \Gamma \theta ,\theta +\epsilon
\right) $. From (\ref{ap2}) it can be seen that this is a left action. The
vector fields $\widetilde{T_\alpha }$, $i\widetilde{T_m}$ induced by the
generators $Q_\alpha $, $iP_m$ on $\Sigma $ are therefore 
\begin{equation}
\label{ap3}\widetilde{T_\alpha }=\left( i\Gamma ^m\theta \right) _\alpha
\cdot \frac \partial {\partial X^m}+\frac \partial {\partial \theta ^\alpha
}=:\delta _\alpha \quad ,
\end{equation}
\begin{equation}
\label{ap4}i\widetilde{T_m}=\frac \partial {\partial X^m}=:\delta _m\quad .
\end{equation}
By construction they are right-invariant vector fields. The corresponding
left-invariant vector fields are obtained by replacing $\theta \mapsto
-\theta $ in (\ref{ap3}, \ref{ap4}). Their duals are the left invariant $1$%
-forms $\Pi ^M=\left( \Pi ^m,\Pi ^a\right) $ on superspace, where
\begin{equation}
\label{ap44}\Pi ^m=dX^m+id\bar \theta \Gamma ^m\theta \quad ,\quad \Pi
^\alpha =d\theta ^\alpha \quad .
\end{equation}
From (\ref{ap4}) we see that $\delta _m$ is strictly speaking ''$i\times $
Poincare-translation with generator $P_m$''.

The graded Lie-bracket of $\widetilde{T_\alpha }$, $\widetilde{T_\beta }$ is 
$\left\{ \delta _\alpha ,\delta _\beta \right\} =\left[ \widetilde{T_\alpha }%
,\widetilde{T_\beta }\right] _{graded\;Lie}=$%
\begin{equation}
\label{ap5}=\left( -2\Gamma _{\alpha \beta }^m\right) \cdot \left( -i\frac
\partial {\partial X^m}\right) =-2\Gamma _{\alpha \beta }^m\cdot \widetilde{%
T_m}=2i\Gamma _{\alpha \beta }^m\cdot \delta _m\quad ,
\end{equation}
i.e. the algebra (\ref{ap1}) is satisfied up to a sign, which is in accord
with (\ref{zw2}) in the first section, since the action of the supergroup on 
$\Sigma $ is from the left, or equivalently, since the $\widetilde{T_M}$ are
right-invariant.

Summation of superspace indices is defined according to%
$$
\omega =\frac 1{r!}dZ^{M_1}\ldots dZ^{M_r}\cdot \omega _{M_r\ldots M_1}\quad
, 
$$
where $\omega $ is a superspace $p$-form. The forms on the worldvolume obey
the usual summation conventions, however; e.g. for the pull-back of the
above $r$-form to the worldvolume we write%
$$
emb^{*}\omega =\frac 1{r!}\partial _{\mu _1}Z^{M_1}\ldots \partial _{\mu
_r}Z^{M_r}\cdot \omega _{M_r\ldots M_1}\cdot dx^{\mu _1}\ldots dx^{\mu
_r}\quad . 
$$
Exterior derivative $d$ is defined to act from the right on superspace forms
as well as on worldvolume forms,%
$$
d\left( \omega \chi \right) =\omega d\chi +\left( -1\right) ^qd\omega \cdot
\chi \quad , 
$$
where $\chi $ is a $q$-form.

\subsection{Graded Poisson brackets}

Given two (possibly graded) functionals $F$, $G$ of the (possibly graded)
time-dependent fields $\phi ^i\left( t,\sigma \right) $ and their canonical
conjugate momenta $\Lambda _i\left( t,\sigma \right) $ that are defined on a 
$p$-dimensional manifold $S$ with coordinates $\left( \sigma ^1,\ldots
,\sigma ^p\right) $, their Poisson bracket is defined by (see, e.g., \cite
{Govaerts}) 
\begin{equation}
\label{ap6}\left\{ F,G\right\} _{PB}=\int\limits_Sd^p\sigma \,\sum_i\left[
\left( -1\right) ^{\phi ^i}F\frac{\overleftarrow{\delta }}{\delta \phi
^i\left( \sigma \right) }\frac{\overrightarrow{\delta }}{\delta \Lambda
_i\left( \sigma \right) }G-F\frac{\overleftarrow{\delta }}{\delta \Lambda
_i\left( \sigma \right) }\frac{\overrightarrow{\delta }}{\delta \phi
^i\left( \sigma \right) }G\right] \quad ; 
\end{equation}
this can be expressed in terms of derivatives acting solely from the left by 
\begin{equation}
\label{ap7}\left\{ F,G\right\} _{PB}=\int\limits_Sd^p\sigma \,\sum_i\left[
\left( -1\right) ^{F\phi ^i}\frac{\delta F}{\delta \phi ^i\left( \sigma
\right) }\frac{\delta G}{\delta \Lambda _i\left( \sigma \right) }-\left(
-1\right) ^{\left( F+1\right) \phi ^i}\frac{\delta F}{\delta \Lambda
_i\left( \sigma \right) }\frac{\delta G}{\delta \phi ^i\left( \sigma \right) 
}\right] \quad , 
\end{equation}
where $\left( -1\right) ^{F\phi ^i}=1$ {\bf iff} both $F$ and $\phi ^i$ are
Grassmann-odd. With these definitions the following rules are satisfied:

\begin{enumerate}
\item  Graded Antisymmetry, 
\begin{equation}
\label{ap8}\left\{ F,G\right\} =-\left( -1\right) ^{FG}\left\{ G,F\right\}
\quad .
\end{equation}

\item  Graded Leibnitz rule, 
\begin{equation}
\label{ap9}\left\{ F,GH\right\} =\left\{ F,G\right\} H+\left( -1\right)
^{FG}G\left\{ F,H\right\} \quad .
\end{equation}

\item  Graded Jacobi identity, 
\begin{equation}
\label{ap10}\left( -1\right) ^{FH}\left\{ F,\left\{ G,H\right\} \right\}
+\left( -1\right) ^{GF}\left\{ G,\left\{ H,F\right\} \right\} +\left(
-1\right) ^{HG}\left\{ H,\left\{ F,G\right\} \right\} =0\quad .
\end{equation}
\end{enumerate}

\section{D-$p$ brane Lagrangians}

\subsection{Structure of the Lagrangian}

The kinetic supertranslation-invariant part ${\cal L}_0$ in the D-$p$-brane
Lagrangian is given by 
\begin{equation}
\label{an1}{\cal L}_0=\sqrt{-\det \left( g_{\mu \nu }+\widehat{F}_{\mu \nu
}\right) }\quad ,
\end{equation}
where $g_{\mu \nu }=\Pi _\mu ^m\Pi _\nu ^n\eta _{nm}$ is the pull-back of
the $10$-dimensional ''mostly plus'' Minkowski metric $\eta _{nm}$ to the
worldvolume $W$ of the D-$p$-brane using left-invariant (LI) $1$-forms $\Pi
^A$, and 
\begin{equation}
\label{an2}\widehat{F}_{\mu \nu }=\partial _\mu A_\nu -\partial _\nu A_\mu
-\Pi _\mu ^{A_1}\Pi _\nu ^{A_2}\cdot B_{A_2A_1}\quad ,
\end{equation}
where $F_{\mu \nu }:=\partial _\mu A_\nu -\partial _\nu A_\mu $ are the
components of the field strength $F$ of the gauge potential $A$ defined on
the worldvolume, and $B_{A_2A_1}$ are the components of the superspace $2$%
-form potential $B$ in the LI-basis whose leading component in a $\theta $%
-expansion is the NS-NS gauge potential. In the discussion below we shall
assume that its bosonic components are zero, but at present $B$ could be
quite arbitrary. Under supertranslations $\delta _\alpha $ the field
strength $H=dB$ is assumed to be invariant,
\begin{equation}
\label{an20}\delta _\alpha H=0\quad .
\end{equation}
This implies that $B$ transforms locally as a differential, 
\begin{equation}
\label{an3}\quad \delta _\alpha B=d\Delta _\alpha \quad .
\end{equation}
Therefore it transforms as a differential under Poincare translations as
well: To see this compare%
$$
2i\Gamma _{\alpha \beta }^m\cdot \delta _mB=\left\{ \delta _\alpha ,\delta
_\beta \right\} B=d\left( \delta _\alpha \Delta _\beta +\delta _\beta \Delta
_\alpha \right) \quad , 
$$
where in the first equation (\ref{ap5}) has been used. If we multiply with
another $\Gamma $-matrix and take the trace we find that
\begin{equation}
\label{an310}\delta _mB=d\Delta _m
\end{equation}
with
\begin{equation}
\label{an4}\Delta _m=\frac{\Gamma _m^{\alpha \beta }}{i\cdot tr\left( {\bf 1}%
_{32}\right) }\,\delta _\alpha \Delta _\beta \quad .
\end{equation}
Since $\left[ \delta _m,\delta _n\right] =0$ it follows from (\ref{an310})
that $d\left( \delta _m\Delta _n-\delta _n\Delta _m\right) =0$, which
implies, that locally 
\begin{equation}
\label{an5}\delta _m\Delta _n-\delta _n\Delta _m=df_{mn}
\end{equation}
for some function $f_{mn}$. This function need not be defined globally,
however.

Although $A_{\mu \,}$ is a worldvolume field it is defined to transform
under these translations according to 
\begin{equation}
\label{an6}\delta _mA=emb^{*}\Delta _m\quad ,\quad \delta _\alpha
A=emb^{*}\Delta _\alpha \quad , 
\end{equation}
where $emb:W\rightarrow \Sigma $ denotes the embedding of the worldvolume
into the target space, and $emb^{*}$ denotes the associated pull-back. With
this definition the quantity $\widehat{F}$ is invariant under Poincare- and
supertranslations, as explained in section \ref{Eich}. Since the same is
true for $g_{\mu \nu }$ we see that therefore ${\cal L}_0$ is invariant as
well.

\subsection{Wess-Zumino term}

The Wess-Zumino form $\left( WZ\right) $ in the D-$p$-brane Lagrangian is
the $\left( p+1\right) $-form on the worldvolume 
\begin{equation}
\label{an7}\left( WZ\right) =\sum\limits_{n=0}^{\left[ \frac{p+1}2\right]
}\frac 1{n!}\,emb^{*}C^{\left( p+1-2n\right) }\cdot \widehat{F}^n\quad ,
\end{equation}
where $C^{\left( r\right) }$ are the superspace potentials whose leading
components in a $\theta $-expansion are the usual bosonic RR gauge
potentials \cite{BergTown1}; in the discussion below the bosonic components
of its field strengths will be set to zero, which then amounts to the choice
of a particular background, but here we make no specific assumptions on the
form of $C^{\left( r\right) }$. The pull-back of $\left( WZ\right) $ to the
worldvolume gives the Wess-Zumino term ${\cal L}_{WZ}$ in the Lagrangian, 
\begin{equation}
\label{an8}{\cal L}_{WZ}=\sum\limits_{n=0}^{\left[ \frac{p+1}2\right] }\frac{%
\epsilon ^{\lambda _1\ldots \lambda _{p+1-2n}\nu _1\ldots \nu _{2n}}}{\left(
p+1-2n\right) !\cdot 2^n\cdot n!}\,\left[ emb^{*}C^{\left( p+1-2n\right)
}\right] _{\lambda _1\ldots \lambda _{p+1-2n}}\widehat{F}_{\nu _1\nu
_2}\cdots \widehat{F}_{\nu _{2n-1}\nu _{2n}}\quad .
\end{equation}
Since we are in IIA superspace we have actually $p=2q$, $q=0,\ldots ,4$.

If $\delta $ denotes either a supertranslation or a Poincare translation
then invariance of $\widehat{F}$ under either of these transformations
implies that 
\begin{equation}
\label{lag1}\delta {\cal L}_{WZ}=\sum\limits_{n=0}^{\left[ \frac{p+1}%
2\right] }\frac{\epsilon ^{\lambda _1\ldots \lambda _{p+1-2n}\nu _1\ldots
\nu _{2n}}}{\left( p+1-2n\right) !\cdot 2^n\cdot n!}\,\left[ emb^{*}\delta
C^{\left( p+1-2n\right) }\right] _{\lambda _1\ldots \lambda _{p+1-2n}}%
\widehat{F}_{\nu _1\nu _2}\cdots \widehat{F}_{\nu _{2n-1}\nu _{2n}}\quad .
\end{equation}
The field strengths associated with the RR-potentials $C^{\left( r\right) }$
are defined to be 
\begin{equation}
\label{lag2}R^{\left( r+1\right) }=\left\{ 
\begin{array}{ccc}
dC^{\left( r\right) } & ; & r=0,1\; \\ 
dC^{\left( r\right) }-C^{\left( r-2\right) }H & ; & r=2,\ldots ,10
\end{array}
\right. \quad ;
\end{equation}
they obey the Bianchi identities 
\begin{equation}
\label{lag3}\left\{ 
\begin{array}{ccc}
dR^{\left( r+1\right) }=0 & ; & r=0,1\; \\ 
dR^{\left( r+1\right) }-R^{\left( r-1\right) }H=0 & ; & r=2,\ldots ,10
\end{array}
\right. \quad .
\end{equation}
It is now assumed that the field strengths (\ref{lag2}) are supertranslation
invariant, 
\begin{equation}
\label{lag4}\delta _\alpha R^{\left( r+1\right) }=0\quad ;\quad r=0,\ldots
,10\quad .
\end{equation}
Then it follows from 
\begin{equation}
\label{lag5}\delta _m=\frac{\Gamma _m^{\alpha \beta }}{2i\cdot tr\left( {\bf %
1}_{32}\right) }\cdot \left\{ \delta _\alpha ,\delta _\beta \right\} 
\end{equation}
that it is invariant under Poincare-translations $-i\delta _m$ as well. From
(\ref{an20}, \ref{lag3}, \ref{lag4}) we can construct the general form of $%
\delta _\alpha C$ for both the IIA and IIB case recursively by starting with
the lowest rank form $C^{\left( 1\right) }$ or $C^{\left( 0\right) }$,
respectively. For the IIA case the result is that there exist superspace
forms 
\begin{equation}
\label{lag6}D_\alpha ^{\left( 2r\right) }\quad ;\quad r=0,\ldots ,4\quad
;\quad \alpha =1,\ldots ,32
\end{equation}
such that 
\begin{equation}
\label{lag7}\delta _\alpha C^{\left( 2q+1\right)
}=\sum\limits_{k=0}^qdD_\alpha ^{\left( 2q-2k\right) }\cdot \frac{B^k}{k!}%
\quad .
\end{equation}
The superscript $\left( 2r\right) $ in (\ref{lag6}) refers to the fact that
the index $\alpha $ does not take part in a summation, but labels one of $32$
components of a spinor-valued $\left( 2r\right) $-form $D^{\left( 2r\right)
}=\left( D_\alpha ^{\left( 2r\right) }\right) _{\alpha =1,\ldots ,32}$.
Using (\ref{lag5}) we find that 
\begin{equation}
\label{lag8}\delta _mC^{\left( 2q+1\right) }=\sum\limits_{k=0}^qdD_m^{\left(
2q-2k\right) }\cdot \frac{B^k}{k!}\quad ,
\end{equation}
where 
\begin{equation}
\label{lag9}D_m^{\left( 0\right) }=\frac{\Gamma _m^{\alpha \beta }}{2i\cdot
tr\left( {\bf 1}_{32}\right) }\cdot \left[ \delta _\alpha D_\beta ^{\left(
0\right) }+\delta _\beta D_\alpha ^{\left( 0\right) }\right] \;=\;\frac{%
\Gamma _m^{\alpha \beta }}{i\cdot tr\left( {\bf 1}_{32}\right) }\cdot \delta
_\alpha D_\beta ^{\left( 0\right) }\quad ,
\end{equation}
$$
D_m^{\left( 2q\right) }=\frac{\Gamma _m^{\alpha \beta }}{2i\cdot tr\left( 
{\bf 1}_{32}\right) }\cdot \left[ \delta _\alpha D_\beta ^{\left( 2q\right)
}+\delta _\beta D_\alpha ^{\left( 2q\right) }+D_\alpha ^{\left( 2q-2\right)
}\cdot d\Delta _\beta +D_\beta ^{\left( 2q-2\right) }\cdot d\Delta _\alpha
\right] \;= 
$$
\begin{equation}
\label{lag10}=\;\frac{\Gamma _m^{\alpha \beta }}{i\cdot tr\left( {\bf 1}%
_{32}\right) }\cdot \left[ \delta _\alpha D_\beta ^{\left( 2q\right)
}+D_\alpha ^{\left( 2q-2\right) }\cdot d\Delta _\beta \right] \quad .
\end{equation}

Now let us return to the supertranslation variation of $C^{\left( r\right) }$
in (\ref{lag7}). If we insert (\ref{lag7}) in (\ref{lag1}) we find that all
terms involving $B$ cancel, 
\begin{equation}
\label{lag11}\delta _\alpha {\cal L}_{WZ}=\partial _\mu U_\alpha ^\mu \quad
, 
\end{equation}
with 
\begin{equation}
\label{lag12}U_\alpha ^\mu =\sum\limits_{n=0}^q\frac{\epsilon ^{\mu \mu
_2\ldots \mu _{2q+1-2n}\nu _1\ldots \nu _{2n}}}{\left( 2q-2n\right) !\cdot
2^n\cdot n!}\,\left[ emb^{*}D_\alpha ^{\left( 2q-2n\right) }\right] _{\mu
_2\ldots \mu _{2q+1-2n}}F_{\nu _1\nu _2}\cdots F_{\nu _{2n-1}\nu _{2n}}\quad
. 
\end{equation}
Similarly, application of (\ref{lag5}) gives 
\begin{equation}
\label{lag13}U_m^\mu =-i\sum\limits_{n=0}^q\frac{\epsilon ^{\mu \mu _2\ldots
\mu _{2q+1-2n}\nu _1\ldots \nu _{2n}}}{\left( 2q-2n\right) !\cdot 2^n\cdot n!%
}\,\left[ emb^{*}D_m^{\left( 2q-2n\right) }\right] _{\mu _2\ldots \mu
_{2q+1-2n}}F_{\nu _1\nu _2}\cdots F_{\nu _{2n-1}\nu _{2n}}\quad , 
\end{equation}
with the $D_m$ given in (\ref{lag9}, \ref{lag10}). We observe that $U_\alpha
^0,U_m^0$ contain neither of $\dot X,\dot \theta ,F_{0r}$.

\subsection{Noether currents and Noether charges}

We now want to compute the algebra of Noether currents and Noether charges
resulting from these currents; as discussed in section \ref{Eich} we may
expect that due to the fact that the worldvolume gauge field $A_\mu \,$
takes part in the supersymmetry transformations on the target space even the
current algebra of the {\bf Noether} currents fails to close in the ordinary
form, but will be extended by central pieces. All the more this will be true
for the algebra of the modified currents and charges, respectively.

Our degrees of freedom are now $\left( X^m,\theta ^\alpha ,A_\nu \right) $
with $\nu =0,\ldots ,p$; the associated canonical conjugate momenta are $%
\left( \Lambda _m,\Lambda _\alpha ,\Lambda ^\nu \right) $, respectively. As
discussed in section \ref{Eich} we have a primary constraint $\Lambda ^0=0$
which amounts to a reduction of phase space, and a secondary constraint $%
\sum_{r=1}^p\partial _r\Lambda ^r=0$, which holds only on-shell, i.e. on
using the equations of motion. The zeroth components of the Noether currents
associated with the generators $Q_\alpha $ are 
\begin{equation}
\label{an9}j_\alpha ^0=\left( i\Gamma ^m\theta \right) _\alpha \cdot \Lambda
_m+\Lambda _\alpha +\left( emb^{*}\Delta _\alpha \right) _\nu \cdot \Lambda
^\nu \quad ; 
\end{equation}
as explained in section \ref{Eich} the primary constraint $\Lambda ^0=0$
must not be taken into account before all Poisson brackets have been worked
out. The zeroth components of the Noether currents associated with the
generators $P_m$ are 
\begin{equation}
\label{an10}j_m^0=-i\Lambda _m-i\left( emb^{*}\Delta _m\right) _\nu \cdot
\Lambda ^\nu \quad . 
\end{equation}
Then the Poisson bracket of the currents $\left\{ j_\alpha ^0,j_\beta
^0\right\} $ is%
$$
\left\{ j_\alpha ^0\left( t,\sigma \right) ,j_\beta ^0\left( t,\sigma
^{\prime }\right) \right\} \approx -2i\Gamma _{\alpha \beta }^m\cdot \Lambda
_m\cdot \delta \left( \sigma -\sigma ^{\prime }\right) \;- 
$$
\begin{equation}
\label{an11}-\;\left[ \delta _\alpha \left( emb^{*}\Delta _\beta \right)
_r+\delta _\beta \left( emb^{*}\Delta _\alpha \right) _r\right] \cdot
\Lambda ^r\cdot \delta \left( \sigma -\sigma ^{\prime }\right) \quad , 
\end{equation}
where $"\approx "$ means ''on using all constraints and equations of motion
and on discarding surface terms''. If we insert (\ref{an10}) into the last
equation we get 
\begin{equation}
\label{an12}\left\{ j_\alpha ^0,j_\beta ^0\right\} \approx \left[ 2\Gamma
_{\alpha \beta }^m\cdot j_m^0\;+\left( emb^{*}S_{\alpha \beta }\left( \Delta
\right) \right) _r\cdot \Lambda ^r\right] \cdot \delta \left( \sigma -\sigma
^{\prime }\right) \quad , 
\end{equation}
where we have used the primary constraint $\Lambda ^0=0$ at last. $S_{\alpha
\beta }\left( \Delta \right) $ is given by 
\begin{equation}
\label{an13}S_{\alpha \beta }\left( \Delta \right) =2i\Gamma _{\alpha \beta
}^n\cdot \Delta _m-\delta _\alpha \Delta _\beta -\delta _\beta \Delta
_\alpha \quad . 
\end{equation}
We see that the presence of the gauge field $A_\mu $ taking part in the
supersymmetry variation of the Lagrangian alters the form of the algebra
even of the {\bf Noether} currents, i.e. before taking into account the
possible modifications of the Noether currents by terms originating in the
Wess-Zumino term.

The once integrated version of (\ref{an12}) defines the action of the
generator $Q_\alpha $ on the current $j_\beta ^0$, 
\begin{equation}
\label{an14}\left\{ Q_\alpha ,j_\beta ^0\right\} =2\Gamma _{\alpha \beta
}^m\cdot j_m^0\;+\left( emb^{*}S_{\alpha \beta }\left( \Delta \right)
\right) _r\cdot \Lambda ^r\quad . 
\end{equation}
As explained in section \ref{NoethChar} the presence of the Wess-Zumino term
in the Lagrangian implies that the Noether charges $Q_\alpha $ which are
obtained by integrating the zero components $j_\alpha ^0$ over the
hypersurface $W\left( t\right) $ are no longer conserved; however, if the
Wess-Zumino term ${\cal L}_{WZ}$ and the NS-NS gauge potential are
translational invariant, as will be the case below, the Noether charges $P_m$
obtained by integrating $j_m^0$ are still conserved.

The twice integrated version of (\ref{an12}) describes the modified algebra
of the Noether charges, 
\begin{equation}
\label{an15}\left\{ Q_\alpha ,Q_\beta \right\} =2\Gamma _{\alpha \beta
}^m\cdot P_m+\int\limits_{W\left( t\right) }d^p\sigma
\sum\limits_{r=1}^p\left( emb^{*}S_{\alpha \beta }\left( \Delta \right)
\right) _r\cdot \Lambda ^r\quad ;
\end{equation}
here $\left( emb^{*}S_{\alpha \beta }\left( \Delta \right) \right) _r=$ $%
\partial _rZ^M\cdot S_{\alpha \beta M}$, when $S_{\alpha \beta }$ is
expanded in the coordinate basis $\left( dZ^M\right) =\left( dX^m,d\theta
^\alpha \right) $. Now let us assume (see section \ref{Eich}) that $%
S_{\alpha \beta M}$ are constants; we want to find extensions of the current
and charge algebra by topological charges carried by the brane; but since it
is only the bosonic coordinates $X^m$ and the pull-back of their
differentials to the worldvolume that describe the topology of the image $%
embW\left( t\right) $ of the brane in the spacetime $\Sigma $ we need only
consider the terms involving bosonic $1$-forms $dX^m$, i.e. $\partial _rX^m$%
, in the above pull-back; therefore if we now define the charge 
\begin{equation}
\label{an16}Y^m=\int\limits_{W\left( t\right) }d^p\sigma
\sum\limits_{r=1}^p\partial _rX^m\cdot \Lambda ^r\quad ,
\end{equation}
then the algebra of the Noether charges in (\ref{an15}) becomes 
\begin{equation}
\label{an17x}\left\{ Q_\alpha ,Q_\beta \right\} =2\Gamma _{\alpha \beta
}^m\cdot P_m+S_{\alpha \beta m}\cdot Y^m\quad ,\quad S_{\alpha \beta m}\quad 
\mbox{constant.}
\end{equation}
If we think of $W\left( t\right) $ as being endowed with an auxiliary
Euclidean metric which is diagonal in the coordinates $\left( \sigma
^r\right) $ then we can introduce the Hodge dual of the $1$-form $\Lambda
^{gauge}=\sum_{r=1}^pd\sigma ^r\Lambda ^r$ \cite{Soro}, which is given by 
\begin{equation}
\label{an18x}\left( *\Lambda ^{gauge}\right) _{s_2\ldots s_p}=\epsilon
_{rs_2\ldots s_p}\Lambda ^r\quad ,
\end{equation}
and rewrite (\ref{an16}) as 
\begin{equation}
\label{an17}Y^m=\frac 1{\left( p-1\right) !}\int\limits_{W\left( t\right)
}\left( *\Lambda ^{gauge}\right) \,dX^m\quad ,
\end{equation}
where $dX^m$ now denotes the pull-back $emb^{*}dX^m$, and exterior product
of forms is understood in the integrand. Furthermore we note that $*\Lambda
^{gauge}$ is {\bf closed} on the physical trajectories, since 
\begin{equation}
\label{an171}d*\Lambda ^{gauge}=\partial _r\Lambda ^r\cdot d\sigma ^1\cdots
d\sigma ^p\quad .
\end{equation}

A similar computation now shows that 
\begin{equation}
\label{an18}\left\{ j_\alpha ^0\left( t,\sigma \right) ,j_m^0\left( t,\sigma
^{\prime }\right) \right\} \approx \left( emb^{*}S_{\alpha m}\left( \Delta
\right) \right) _r\cdot \Lambda ^r\cdot \delta \left( \sigma -\sigma
^{\prime }\right) \quad ,
\end{equation}
with 
\begin{equation}
\label{an19}S_{\alpha m}\left( \Delta \right) =i\left( \delta _\alpha \Delta
_m-\delta _m\Delta _\alpha \right) \quad ;
\end{equation}
the factor of $i$ comes from our parametrising of the coset elements of
superMinkowski space, see section \ref{conventions}.

Finally, we find 
\begin{equation}
\label{an20x}\left\{ j_m^0\left( t,\sigma \right) ,j_n^0\left( t,\sigma
^{\prime }\right) \right\} \approx \left( emb^{*}S_{mn}\left( \Delta \right)
\right) _r\cdot \Lambda ^r\cdot \delta \left( \sigma -\sigma ^{\prime
}\right) \quad ,
\end{equation}
where 
\begin{equation}
\label{an21}S_{mn}\left( \Delta \right) =\delta _m\Delta _n-\delta _n\Delta
_m\quad .
\end{equation}
Double integration of (\ref{an20x}) using (\ref{an5}) then yields 
\begin{equation}
\label{an22}\left[ P_m,P_n\right] =\int\limits_{W\left( t\right) }d^p\sigma
\cdot \partial _r\left( f_{mn}\Lambda ^r\right) \quad ,
\end{equation}
where we have used the secondary constraint $\partial _r\Lambda ^r=0$ and
the equations of motion. We see that the momenta can be {\bf non-commuting}
in the case that the functions $f_{mn}$ are not globally defined; this could
happen if some of the dimensions of $W\left( t\right) $ are compact, and
their images in the spacetime under the embedding describe a closed but
non-contractible cycle.

\subsection{Modified currents and charges \label{Modificatio}}

The modified currents are 
\begin{equation}
\label{an23}\widetilde{j_\alpha ^0}=j_\alpha ^0-U_\alpha ^0\quad ,\quad 
\widetilde{j_m^0}=j_m^0-U_m^0\quad . 
\end{equation}
Their Poisson brackets are found to be 
\begin{equation}
\label{an24}\left\{ \widetilde{j_\alpha ^0},\widetilde{j_\beta ^0}\right\}
\approx \left[ 2\Gamma _{\alpha \beta }^n\cdot \widetilde{j_m^0}+\left(
emb^{*}S_{\alpha \beta }\left( \Delta \right) \right) _r\cdot \Lambda
^r+S_{\alpha \beta }\left( U\right) \right] \cdot \delta \left( \sigma
-\sigma ^{\prime }\right) \quad , 
\end{equation}
with 
\begin{equation}
\label{an25}S_{\alpha \beta }\left( U\right) =\delta _\alpha U_\beta
^0+\delta _\beta U_\alpha ^0+2\Gamma _{\alpha \beta }^n\cdot U_n^0\quad . 
\end{equation}
$S_{\alpha \beta }$ is given in (\ref{an13}). Furthermore, 
\begin{equation}
\label{an26}\left\{ \widetilde{j_\alpha ^0},\widetilde{j_m^0}\right\}
\approx \left[ \left( emb^{*}S_{\alpha m}\left( \Delta \right) \right)
_r\cdot \Lambda ^r+S_{\alpha m}\left( U\right) \right] \cdot \delta \left(
\sigma -\sigma ^{\prime }\right) \quad , 
\end{equation}
\begin{equation}
\label{an27}S_{\alpha m}\left( U\right) =\delta _\alpha U_m^0+i\delta
_mU_\alpha ^0\quad , 
\end{equation}
with $S_{\alpha m}$ given in (\ref{an19}); and finally, 
\begin{equation}
\label{an28}\left\{ \widetilde{j_m^0},\widetilde{j_n^0}\right\} \approx
\left[ \left( emb^{*}S_{mn}\left( \Delta \right) \right) _r\cdot \Lambda
^r+S_{mn}\left( U\right) \right] \cdot \delta \left( \sigma -\sigma ^{\prime
}\right) \quad , 
\end{equation}
\begin{equation}
\label{an29x}S_{mn}\left( U\right) =-i\left[ \delta _mU_n^0-\delta
_nU_m^0\right] \quad , 
\end{equation}
with $S_{mn}$ from (\ref{an21}).

We can work out the expressions for $S_{MN}\left( U\right) $ using (\ref
{lag12}, \ref{lag13}), which yields%
$$
S_{\alpha \beta }\left( U\right) \;=\;\frac{\epsilon ^{0\nu _1\ldots \nu
_{2q}}}{2^q\cdot q!}\,emb^{*}\left[ \delta _\alpha D_\beta ^{\left( 0\right)
}+\delta _\beta D_\alpha ^{\left( 0\right) }+2\Gamma _{\alpha \beta }^n\cdot
D_n^{\left( 0\right) }\right] \cdot F_{\nu _1\nu _2}\cdots F_{\nu _{2q-1}\nu
_{2q}}\;+ 
$$
$$
+\;\sum\limits_{k=0}^{q-1}\frac{\epsilon ^{0\mu _2\ldots \mu _{2q+1-2k}\nu
_1\ldots \nu _{2k}}}{\left( 2q-2k\right) !\,2^k\cdot k!}\,\left\{
emb^{*}\left[ \delta _\alpha D_\beta ^{\left( 2q-2k\right) }+\delta _\beta
D_\alpha ^{\left( 2q-2k\right) }+\right. \right. 
$$
$$
\left. +\;2\Gamma _{\alpha \beta }^n\cdot D_n^{\left( 2q-2k\right) }\right]
_{\mu _2\ldots \mu _{2q+1-2k}}\;-\;\left( 2q-2k\right) \left( 2q-2k-1\right)
\cdot 
$$
$$
\cdot \left[ \left( emb^{*}D_\beta ^{\left( 2q-2k-2\right) }\right) _{\mu
_2\ldots \mu _{2q-1-2k}}\cdot \partial _{\mu _{2q-2k}}\left( emb^{*}\Delta
_\alpha \right) _{\mu _{2q+1-2k}}\right. \;+ 
$$
$$
\left. +\;\left. \left( emb^{*}D_\alpha ^{\left( 2q-2k-2\right) }\right)
_{\mu _2\ldots \mu _{2q-1-2k}}\cdot \partial _{\mu _{2q-2k}}\left(
emb^{*}\Delta _\beta \right) _{\mu _{2q+1-2k}}\right] \right\} \cdot \; 
$$
\begin{equation}
\label{an29}\cdot F_{\nu _1\nu _2}\cdots F_{\nu _{2k-1}\nu _{2k}}\quad . 
\end{equation}
\pagebreak[3]
The appropriate expression for $S_{\alpha m}$ is%
$$
S_{\alpha m}\left( U\right) \;=\;\frac{\epsilon ^{0\nu _1\ldots \nu _{2q}}}{%
2^q\cdot q!}\,emb^{*}\left[ \delta _\alpha D_m^{\left( 0\right) }+i\delta
_mD_\alpha ^{\left( 0\right) }\right] \cdot F_{\nu _1\nu _2}\cdots F_{\nu
_{2q-1}\nu _{2q}}\;+ 
$$
$$
+\;\sum\limits_{k=0}^{q-1}\frac{\epsilon ^{0\mu _2\ldots \mu _{2q+1-2k}\nu
_1\ldots \nu _{2k}}}{\left( 2q-2k\right) !\,2^k\cdot k!}\,\left\{
emb^{*}\left[ \delta _\alpha D_m^{\left( 2q-2k\right) }+i\delta _mD_\alpha
^{\left( 2q-2k\right) }\right] _{\mu _2\ldots \mu _{2q+1-2k}}\right. \;+ 
$$
$$
+\;\left( 2q-2k\right) \left( 2q-2k-1\right) \cdot 
$$
$$
\cdot \left[ \left( emb^{*}D_m^{\left( 2q-2k-2\right) }\right) _{\mu
_2\ldots \mu _{2q-1-2k}}\cdot \partial _{\mu _{2q-2k}}\left( emb^{*}\Delta
_\alpha \right) _{\mu _{2q+1-2k}}\right. \;+ 
$$
$$
\left. +\;i\cdot \left. \left( emb^{*}D_\alpha ^{\left( 2q-2k-2\right)
}\right) _{\mu _2\ldots \mu _{2q-1-2k}}\cdot \partial _{\mu _{2q-2k}}\left(
emb^{*}\Delta _m\right) _{\mu _{2q+1-2k}}\right] \right\} \cdot \; 
$$
\begin{equation}
\label{an30}\cdot F_{\nu _1\nu _2}\cdots F_{\nu _{2k-1}\nu _{2k}}\quad . 
\end{equation}
The expression for $S_{mn}$ is%
$$
S_{mn}\left( U\right) \;=\;-i\cdot \frac{\epsilon ^{0\nu _1\ldots \nu _{2q}}%
}{2^q\cdot q!}\,emb^{*}\left[ \delta _mD_n^{\left( 0\right) }-\delta
_nD_m^{\left( 0\right) }\right] \cdot F_{\nu _1\nu _2}\cdots F_{\nu
_{2q-1}\nu _{2q}}\;- 
$$
$$
-i\cdot \sum\limits_{k=0}^{q-1}\frac{\epsilon ^{0\mu _2\ldots \mu
_{2q+1-2k}\nu _1\ldots \nu _{2k}}}{\left( 2q-2k\right) !\,2^k\cdot k!}%
\,\left\{ emb^{*}\left[ \delta _mD_n^{\left( 2q-2k\right) }-\delta
_nD_m^{\left( 2q-2k\right) }\right] _{\mu _2\ldots \mu _{2q+1-2k}}\right.
\;+ 
$$
$$
\;+\;\left( 2q-2k\right) \left( 2q-2k-1\right) \cdot 
$$
$$
\cdot \left[ \left( emb^{*}D_n^{\left( 2q-2k-2\right) }\right) _{\mu
_2\ldots \mu _{2q-1-2k}}\cdot \partial _{\mu _{2q-2k}}\left( emb^{*}\Delta
_m\right) _{\mu _{2q+1-2k}}\right. \;- 
$$
$$
\left. -\;\left. \left( emb^{*}D_m^{\left( 2q-2k-2\right) }\right) _{\mu
_2\ldots \mu _{2q-1-2k}}\cdot \partial _{\mu _{2q-2k}}\left( emb^{*}\Delta
_n\right) _{\mu _{2q+1-2k}}\right] \right\} \cdot \; 
$$
\begin{equation}
\label{an31}\cdot F_{\nu _1\nu _2}\cdots F_{\nu _{2k-1}\nu _{2k}}\quad . 
\end{equation}
\pagebreak

\section{D-$p$-branes in IIA supergravity backgrounds}

\subsection{Superspace constraints}

$D=10$ type II supergravity theories are the low-energy effective field
theories of type II superstring theories \cite{Thor}. These theories have
classical solutions which describe extended objects called $p$-branes. The $%
p $-branes are solitons carrying conserved charges that act as sources for
the various anti-symmetric gauge fields of the underlying supergravity
theory, i.e. RR-gauge fields $C^{\left( r\right) }$ and the NS-NS $2$-form
potential $B$.

In superspace all ordinary components of RR and NS-NS gauge fields are
introduced as first components of their corresponding superfields. From the
gauge fields one can derive field strengths with associated Bianchi
identities. In order to reduce the enormous field content of these
superfields down to the on-shell content one introduces {\it constraints} on
some of the components of the superfield field strengths. When these
constraints are inserted into the Bianchi identities the latter cease to be
identities, but rather become equations the consistency of which has to be
examined separately. If the constraints are properly chosen the equations so
obtained are just the supergravity equations of motion.

D-branes arise from prescribing mixed (Neumann- and Dirichlet) boundary
conditions on open strings in type II string theory. They are introduced as $%
\left( p+1\right) $-dimensional hypersurfaces in spacetime where open
strings are constrained to end on, but the ends are free to move on this
submanifold. A spacelike section of a D-brane can be given a finite volume
in a spacetime with compact dimensions by wrapping around topologically
non-trivial cycles in the spacetime. In this case the supertranslation
algebra of Noether charges or modified charges carried by the brane is
extended by topological charges, which we derive below.

We want to consider D-branes in a flat IIA background. This condition
requires the underlying supergravity theory to be massless, $m=0$, since it
is known that $D=10$ Minkowski spacetime is {\bf not} a solution to the
field equations of massive IIA supergravity \cite{Berg2}. The massless
theory allows a flat solution, however; its constraints, i.e. the
constraints on the massless IIA supergravity background, can be obtained by
dimensional reduction of the standard $D=11$ superspace constraints \cite
{Berg3}; in particular, they imply the field equations of massless IIA
supergravity. Moreover, we have the observation that, once the constraints
on the NS-NS fields coupling to the kinetic (supersymmetry invariant) term
in the D-brane action are given, the constraints on the RR-fields coupling
to the brane via the Wess-Zumino term can be read off from $\kappa $%
-symmetry, see \cite{Ceder1}; thus, consistent propagation of $D$-branes
demands a background solving the equations of motion of the appropriate
supergravity theory.

\subsection{Superspace background and Bianchi identities \label{Vakuum}}

In the following we choose a massless flat background vacuum with Dilaton $%
\phi =0$, Dilatino $D\phi =0$, where $D$ denotes a supercovariant
derivative. Moreover, we assume that all bosonic components of the field
strengths associated with the NS-NS fields and RR gauge fields,
respectively, are zero; the non-bosonic components of these field strengths
as well as the non-bosonic torsion components are uniquely determined by the
superspace constraints, see \cite{BergTown1}. The RR superfield potentials
are usually collected in a formal sum $C=\sum_{r=0}^{10}C^{\left( r\right) }$%
, where the ordinary RR gauge potentials are just the leading components of
the $C^{\left( r\right) }$ in a $\theta $-expansion; for the IIA case only
the odd forms are relevant. Their field strengths are defined in (\ref{lag2}%
). The Bianchi identities associated with these field strengths are given in
(\ref{lag3}). The Bianchi identities for the field strength $H$ of the NS-NS
field $B$ is $dH=0$.

Let us now define a family of superspace forms $K^{\left( p+2\right) }\left(
S\right) $ by 
\begin{equation}
\label{ap13}K^{\left( p+2\right) }\left( S\right) =\frac i{p!}\Pi
^{m_p}\cdots \Pi ^{m_1}\cdot d\bar \theta S\Gamma _{m_1\ldots m_p}d\theta
\quad , 
\end{equation}
where $p=0,\ldots ,9$, $S\in \left\{ {\bf 1}_{32},\Gamma _{11}\right\} $,
and $\Gamma _{m_1\ldots m_p}$ is the usual antisymmetrised product of $%
\Gamma $-matrices. Then our choice of vacuum determines the field strengths
to be \cite{BergTown1} 
\begin{equation}
\label{ap14}
\begin{array}{c}
R^{\left( 2\right) }=K^{\left( 2\right) }\left( \Gamma _{11}\right) \quad ,
\\ 
R^{\left( 4\right) }=K^{\left( 4\right) }\left( 
{\bf 1}\right) \quad , \\ R^{\left( 6\right) }=K^{\left( 6\right) }\left(
\Gamma _{11}\right) \quad , \\ 
R^{\left( 8\right) }=K^{\left( 8\right) }\left( 
{\bf 1}\right) \quad , \\ R^{\left( 10\right) }=K^{\left( 10\right) }\left(
\Gamma _{11}\right) \quad ; 
\end{array}
\end{equation}
furthermore, the field strength $H$ must take the form 
\begin{equation}
\label{ap15}H=-K^{\left( 3\right) }\left( \Gamma _{11}\right) \quad . 
\end{equation}
These field strengths are determined by superspace constraints; the Bianchi
identities (\ref{lag3}) and the relation $dH=0$ are therefore identities no
longer, and we must check whether they are actually satisfied.

\subsection{Explicit form of $B$}

Let us first consider the field strength $H$ in (\ref{ap15}); the $3$-form $%
K^{\left( 3\right) }\left( \Gamma _{11}\right) $ has a potential 
\begin{equation}
\label{ap151}B=\left( -\Pi ^m+\frac i2d\bar \theta \Gamma ^m\theta \right)
\cdot \left( id\bar \theta \Gamma _{11}\Gamma _m\theta \right) \quad , 
\end{equation}
which yields $H=dB=-K^{\left( 3\right) }\left( \Gamma _{11}\right) $ on
account of the identity 
\begin{equation}
\label{ap152}d\bar \theta \Gamma ^n\theta \cdot d\bar \theta \Gamma
_{11}\Gamma _nd\theta +d\bar \theta \Gamma _{11}\Gamma _n\theta \cdot d\bar
\theta \Gamma ^nd\theta =0\quad ; 
\end{equation}
this in turn is a consequence of the identity 
\begin{equation}
\label{ap153}\Gamma _{(\alpha \beta }^n\left( \Gamma _{11}\Gamma _n\right)
_{\gamma \delta )}=0\quad , 
\end{equation}
which is known to to hold in $D=10$. Therefore (\ref{ap15}) actually is a
consistent choice for $H$. Since $H=-K^{\left( 3\right) }\left( \Gamma
_{11}\right) $ is indeed supertranslation invariant we have $\delta _\alpha
B=d\Delta _\alpha $ for some $1$-form $\Delta _\alpha $; this can be
computed to be 
\begin{equation}
\label{ap154}\Delta _\alpha =dX^m\cdot \left( i\Gamma _{11}\Gamma _m\theta
\right) _\alpha -\frac 16\left[ d\bar \theta \Gamma _m\theta \cdot \left(
\bar \theta \Gamma _{11}\Gamma _m\right) _\alpha +d\bar \theta \Gamma
_{11}\Gamma _m\theta \cdot \left( \bar \theta \Gamma _m\right) _\alpha
\right] \quad . 
\end{equation}
Moreover we note that%
$$
\delta _\alpha \Delta _\beta +\delta _\beta \Delta _\alpha =dX^m\cdot \left(
2i\Gamma _{11}\Gamma _m\right) _{\alpha \beta }\;+ 
$$
\begin{equation}
\label{ap155}+\;\frac 12\cdot d\left[ \left( \Gamma _{11}\Gamma _m\theta
\right) _\alpha \cdot \left( \Gamma ^m\theta \right) _\beta +\left( \Gamma
_{11}\Gamma _m\theta \right) _\beta \cdot \left( \Gamma ^m\theta \right)
_\alpha \right] \quad . 
\end{equation}
Taking the trace with $\Gamma _n^{\alpha \beta }$ of this expression yields
zero: due to the tracelessness of products of $\Gamma $-matrices the first
contribution vanishes, and the terms in the square bracket yield zero since 
$$
\bar \theta \Gamma _{11}\Gamma _m\Gamma _n\Gamma ^m\theta =\left( 2-D\right)
\bar \theta \Gamma _{11}\Gamma _n\theta =0\quad , 
$$
for $C\Gamma _{11}\Gamma _n$ is {\bf symmetric}, see Table \ref{Tabelle2}.
By (\ref{an4}) this implies that 
\begin{equation}
\label{ap156}\Delta _m=0\quad , 
\end{equation}
as can be seen directly from (\ref{ap151}), since $B$ is translation
invariant. From definitions (\ref{an13}, \ref{an19}, \ref{an21}) we now see
that 
\begin{equation}
\label{for1}S_{\alpha \beta }\left( \Delta \right) =-dX^m\cdot \left(
2i\Gamma _{11}\Gamma _m\right) _{\alpha \beta }\;+\;\cdots \quad , 
\end{equation}
\begin{equation}
\label{for2}S_{\alpha m}\left( \Delta \right) =S_{mn}\left( \Delta \right)
=0\quad , 
\end{equation}
where $"\cdots "$ denotes terms that involve only fermionic $1$-forms.
Finally, note that $d\Delta _m=0$.

\subsection{Generalized $\Gamma $-matrix identities}

Now let us turn attention to the RR field strengths. If we insert (\ref{ap14}%
) into the Bianchi identities (\ref{lag3}) we obtain 
\begin{equation}
\label{ap16}dK^{\left( 2\right) }\left( \Gamma _{11}\right) =0\quad ,
\end{equation}
\begin{equation}
\label{ap17}
\begin{array}{c}
dK^{\left( 4\right) }\left( 
{\bf 1}\right) +K^{\left( 2\right) }\left( \Gamma _{11}\right) K^{\left(
3\right) }\left( \Gamma _{11}\right) =0\quad , \\ dK^{\left( 6\right)
}\left( \Gamma _{11}\right) +K^{\left( 4\right) }\left( 
{\bf 1}\right) K^{\left( 3\right) }\left( \Gamma _{11}\right) =0\quad , \\ 
dK^{\left( 8\right) }\left( 
{\bf 1}\right) +K^{\left( 6\right) }\left( \Gamma _{11}\right) K^{\left(
3\right) }\left( \Gamma _{11}\right) =0\quad , \\ dK^{\left( 10\right)
}\left( \Gamma _{11}\right) +K^{\left( 8\right) }\left( {\bf 1}\right)
K^{\left( 3\right) }\left( \Gamma _{11}\right) =0\quad .
\end{array}
\end{equation}
Here (\ref{ap16}) is trivially satisfied due to $d\left( id\bar \theta
\Gamma _{11}d\theta \right) =0$. The equations in (\ref{ap17}) can be
written as 
\begin{equation}
\label{ap18}dK^{\left( 2q+2\right) }\left( S\right) +K^{\left( 2q\right)
}\left( S\Gamma _{11}\right) K^{\left( 3\right) }\left( \Gamma _{11}\right)
=0\quad ;\quad q=1,2,3,4,
\end{equation}
and $p=2q$ is related to $S$ by Table \ref{Tabelle1}. {%
\begin{table} \centering
   \begin{tabular}{|c||c|c|c|c|c|} \hline
   ${\bf {p=2q}}$  &  $0$  &  $2$  &  $4$  &  $6$  &  $8$  \\ \hline
   ${\bf S}$  &  $\Gamma _{11}$  &  ${\bf 1}_{32}$  & $\Gamma _{11}$  & ${\bf 1}_{32}$  &  $\Gamma _{11}$    \\ \hline
  \end{tabular}
 \caption{Relation between $p$ and $S$.  \label{Tabelle1}}
\end{table}} If we now use the explicit definitions of $K^{\left( 2q\right)
}\left( S\right) $ as given in (\ref{ap13}) we find that equations (\ref
{ap17}) are satisfied {\bf iff} 
\begin{equation}
\label{ap19}\Gamma _{(\alpha \beta }^n\left( S\Gamma _{nm_1\ldots
m_{2q-1}}\right) _{\gamma \delta )}\;+\;\left( 2q-1\right) \cdot \left(
\Gamma _{11}\Gamma _{[m_1}\right) _{(\alpha \beta }\left( S\Gamma
_{11}\Gamma _{m_2\ldots m_{2q-1}]}\right) _{\gamma \delta )}\;=\;0\quad .
\end{equation}
In the first term the symmetrisation involves spinor indices $\alpha ,\beta
,\gamma ,\delta $, but {\bf no} covector indices $n,m_1,\ldots ,m_{2q-1}$,
of course. In the second term we have a symmetrisation over $\alpha ,\beta
,\gamma ,\delta $, and independently, an antisymmetristion over $m_1,\ldots
m_{2q-1}$. (\ref{ap19}) is a set of {\it generalized }$\Gamma ${\it -matrix
identities}. We shall derive a necessary condition for them to hold, and
show, that it is indeed satisfied. Before we do so, however, let us examine
the special case of (\ref{ap19}) when $q=1$. In this case (\ref{ap19})
becomes (see table \ref{Tabelle1} for the choice of $S$) 
\begin{equation}
\label{ap20}\Gamma _{(\alpha \beta }^n\left( \Gamma _{nm}\right) _{\gamma
\delta )}+\left( \Gamma _{11}\right) _{(\alpha \beta }\left( \Gamma
_{11}\Gamma _m\right) _{\gamma \delta )}\;=\;0\quad .
\end{equation}
This is just the dimensional reduction to $D=10$ of the $D=11$ identity
required for $\kappa $-symmetry of the $D=11$ supermembrane \cite{Berg3},
and is known to hold in $D=11$. This means that the validity of at least the
first equation in (\ref{ap17}) is assured.

To examine the validity of the other cases we reexpress (\ref{ap19}) as%
$$
\Gamma _{(\alpha \beta }^n\left( S\Gamma _{nm_1\ldots m_{2q-1}}\right)
_{\gamma \delta )}+\left( \Gamma _{11}\Gamma _{m_1}\right) _{(\alpha \beta
}\left( S\Gamma _{11}\Gamma _{m_2\ldots m_{2q-1}}\right) _{\gamma \delta )}+ 
$$
\begin{equation}
\label{ap21}+\;\left( \mbox{cycl. }m_1\rightarrow m_2\rightarrow \cdots
\right) \;+\;\cdots \;=0\quad , 
\end{equation}
where ''cyc.'' denotes a sum over all cyclic permutations of $m_i$-indices
in the second term of the first line. Now we multiply (\ref{ap21}) by $%
\Gamma _{\alpha \beta }^l$; this yields%
$$
tr\left( \Gamma ^n\Gamma _l\right) \cdot \left( CS\Gamma _{nm_1\ldots
m_{2q-1}}\right) +\left( C\Gamma ^n\right) \cdot tr\left( \Gamma _lS\Gamma
_{nm_1\ldots m_{2q-1}}\right) + 
$$
$$
+4\left( C\Gamma ^n\Gamma _lS\Gamma _{nm_1\ldots m_{2q-1}}\right) _{\left(
sym\right) }+ 
$$
$$
+\left\{ tr\left( \Gamma _{11}\Gamma _{m_1}\Gamma _l\right) \cdot \left(
CS\Gamma _{11}\Gamma _{m_2\ldots m_{2q-1}}\right) +\left( C\Gamma
_{11}\Gamma _{m_1}\right) \cdot tr\left( \Gamma _lS\Gamma _{11}\Gamma
_{m_2\ldots m_{2q-1}}\right) +\right. 
$$
\begin{equation}
\label{ap22}+\left. 4\left( C\Gamma _{11}\Gamma _{m_1}\Gamma _lS\Gamma
_{11}\Gamma _{m_2\ldots m_{2q-1}}\right) _{\left( sym\right) }+\left( \mbox{%
cycl. }m_1\rightarrow m_2\rightarrow \cdots \right) \;+\;\cdots \right\}
=0\quad . 
\end{equation}
Here $\left( sym\right) $ denotes the symmetric part of the matrix in
brackets, i.e. $M_{\left( sym\right) }=\frac{1}{2}\left( M+M^T\right) 
$. We list the contributions to (\ref{ap22}):%
$$
tr\left( \Gamma ^n\Gamma _l\right) \cdot \left( CS\Gamma _{nm_1\ldots
m_{2q-1}}\right) =tr\left( {\bf 1}_{32}\right) \cdot \left( CS\Gamma
_{lm_1\ldots m_{2q-1}}\right) \quad , 
$$
$$
tr\left( \Gamma _lS\Gamma _{nm_1\ldots m_{2q-1}}\right) =0\quad \mbox{for
all\quad }q=1,\ldots ,4\,;\;S=S\left( q\right) \mbox{, see Table (\ref
{Tabelle1})\quad }, 
$$
\begin{equation}
\label{ap23}\left( C\Gamma ^n\Gamma _lS\Gamma _{nm_1\ldots m_{2q-1}}\right)
_{\left( sym\right) }=-\left( D-2q-1\right) \cdot \left( CS\Gamma
_{lm_1\ldots m_{2q-1}}\right) \quad , 
\end{equation}
where we have used the fact that if $\left( CS\Gamma _{lm_1\ldots
m_{2q-1}}\right) $ is symmetric then $\left( CS\Gamma _{m_2\ldots
m_{2q-1}}\right) $ is always antisymmetric, see Table \ref{Tabelle2}.{%
\begin{table} \centering
  \begin{tabular}{||c||c|c||c|c||} \hline
     ${\bf p}$  &  $S={\bf 1}_{32}$  &  type  &  $S=\Gamma _{11}$  &  type  \\ \hline \hline
     $0$  &  $C$  &  $-$  &  $C\Gamma _{11}$  &  $+$  \\  \hline 
     $1$  &  $C\Gamma _{m_1}$  &  $+$  &  $C\Gamma _{11}\Gamma _{m_1}$
  &  $+$  \\  \hline
     $2$  &  $C\Gamma _{m_1m_2}$  &  $+$  &  $C\Gamma _{11}\Gamma _{m_1m_2}$  &  $-$  \\  \hline
     $3$  &  $C\Gamma _{m_{1\ldots }m_3}$  &  $-$  &  $C\Gamma _{11}\Gamma _{m_{1\ldots }m_3}$  &  $-$  \\  \hline
     $4$  &  $C\Gamma _{m_{1\ldots }m_4}$  &  $-$  &  $C\Gamma _{11}\Gamma _{m_{1\ldots }m_4}$  &  $+$  \\  \hline
     $5$  &  $C\Gamma _{m_{1\ldots }m_5}$  &  $+$  &  $C\Gamma _{11}\Gamma _{m_{1\ldots }m_5}$  &  $+$  \\  \hline
     $6$  &  $C\Gamma _{m_{1\ldots }m_6}$  &  $+$  &  $C\Gamma _{11}\Gamma _{m_{1\ldots }m_6}$  &  $-$  \\  \hline
     $7$  &  $C\Gamma _{m_{1\ldots }m_7}$  &  $-$  &  $C\Gamma _{11}\Gamma _{m_{1\ldots }m_7}$  &  $-$  \\  \hline
     $8$  &  $C\Gamma _{m_{1\ldots }m_8}$  &  $-$  &  $C\Gamma _{11}\Gamma _{m_{1\ldots }m_8}$  &  $+$  \\  \hline
     $9$  &  $C\Gamma _{m_{1\ldots }m_9}$  &  $+$  &  $C\Gamma _{11}\Gamma _{m_{1\ldots }m_9}$  &  $+$  \\  \hline
     $10$  &  $C\Gamma _{m_{1\ldots }m_{10}}$  &  $+$  &  $C\Gamma _{11}\Gamma _{m_{1\ldots }m_{10}}$  &  $-$  \\  \hline
  \end{tabular}
  \caption{Symmetry and Antisymmetry of products of $\Gamma$-matrices in $D=10$. $+/-$ denotes Symmetry/Antisymmetry; $C$ is a charge conjugation matrix.\label{Tabelle2}}
\end{table}} The last three contributions to (\ref{ap22}) are%
$$
tr\left( \Gamma _{11}\Gamma _{m_1}\Gamma _l\right) =0\quad , 
$$
$$
tr\left( \Gamma _lS\Gamma _{11}\Gamma _{m_2\ldots m_{2q-1}}\right) =0\quad , 
$$
$$
\left( CS\Gamma _{m_1}\Gamma _l\Gamma _{m_2\ldots m_{2q-1}}\right) _{\left(
sym\right) }\;=\;-\left( CS\Gamma _{lm_1\ldots m_{2q-1}}\right) - 
$$
\begin{equation}
\label{ap24}-\left( 2q-1\right) \left( 2q-3\right) \cdot \eta
_{m_1[m_2}\cdot \eta _{\left| l\right| m_3}\left( CS\Gamma _{m_4\ldots
m_{2q-1}]}\right) \quad . 
\end{equation}
Now we must perform the cyclic sum $\left( \mbox{cycl. }m_1\rightarrow
m_2\rightarrow \cdots \right) $ in (\ref{ap24}). Since this is equal to $%
\left( 2q-1\right) \times $ ''antisymmetrisation of (\ref{ap24}) over $%
\left( m_1,\ldots ,m_{2q-1}\right) $'' we see that the second contribution
on the right hand side of (\ref{ap24}) must vanish, since it involves
antisymmetrisation over $\eta _{m_1m_2}$, and therefore the total
contribution from this term is
\begin{equation}
\label{ap25}\left( 2q-1\right) \cdot \left( CS\Gamma _{[m_1}\Gamma _{\left|
l\right| }\Gamma _{m_2\ldots m_{2q-1}]}\right) _{\left( sym\right) }=-\left(
2q-1\right) \cdot \left( CS\Gamma _{lm_1\ldots m_{2q-1}}\right) \quad . 
\end{equation}
Altogether, (\ref{ap22}) leads to the condition 
\begin{equation}
\label{ap251}\left[ tr\left( {\bf 1}_{32}\right) -4\left( D-2q-1\right)
-4\left( 2q-1\right) \right] \cdot \left( CS\Gamma _{lm_1\ldots
m_{2q-1}}\right) =0\quad ; 
\end{equation}
remarkably, the contributions involving $q$ cancel each other in this
equation, so we arrive at 
\begin{equation}
\label{ap26}tr\left( {\bf 1}_{32}\right) -4\left( D-2\right) =0 
\end{equation}
as a necessary condition for the $\Gamma $-matrix identities (\ref{ap19}) to
hold; but this is satisfied precisely in $D=10$, {\bf independent} of $q$.

We do not know whether (\ref{ap26}) is also sufficient to ensure (\ref{ap19}%
); in the past, sufficiency of a similar condition to (\ref{ap26}) to
establish the well-known $\Gamma $-matrix identity $\Gamma _{\left( \alpha
\beta \right) }^n\left( \Gamma _n\right) _{\gamma \delta )}=0$ in $D=10$
could be established only via computer \cite{PKT1}. In the following we
shall assume that (\ref{ap26}) is sufficient and therefore (\ref{ap19})
holds for all allowed values of $q$; if this assumption should turn out to
be wrong, then at least our analysis is valid for $q=1$, since in this case
the validity of (\ref{ap20}) is known; our results then would be restricted
to the D-$2$-brane in a IIA superspace.

\subsection{Constructing the leading terms of $C^{\left( r\right) }$}

Provided that (\ref{ap19}) is valid we show that under these circumstances
we can construct the potentials $C^{\left( 3\right) },C^{\left( 5\right)
},C^{\left( 7\right) },C^{\left( 9\right) }$ recursively from $C^{\left(
1\right) }$. From (\ref{ap13}, \ref{ap14}) we see that, up to a gauge
transformation, we have 
\begin{equation}
\label{ap27}C^{\left( 1\right) }=id\bar \theta \Gamma _{11}\theta \quad . 
\end{equation}
Now assuming that we have constructed $C^{\left( 2q-1\right) }$ we can use (%
\ref{lag2}) to give 
\begin{equation}
\label{ap28}dC^{\left( 2q+1\right) }=K^{\left( 2q+2\right) }\left( S\right)
-C^{\left( 2q-1\right) }K^{\left( 3\right) }\left( \Gamma _{11}\right) \quad
, 
\end{equation}
where $S$ is chosen according to Table \ref{Tabelle1}. A nessecary and
sufficient condition for the existence of a (local) $\left( 2q+1\right) $%
-form $C^{\left( 2q+1\right) }$ that satisfies (\ref{ap28}) is that the
differential of the right hand side of (\ref{ap28}) vanishes; but since $%
dC^{\left( 2q-1\right) }=K^{\left( 2q\right) }\left( S\Gamma _{11}\right)
-C^{\left( 2q-3\right) }K^{\left( 3\right) }\left( \Gamma _{11}\right) $ by
assumption, this is 
\begin{equation}
\label{ap29}d\left[ K^{\left( 2q+2\right) }\left( S\right) -C^{\left(
2q-1\right) }K^{\left( 3\right) }\left( \Gamma _{11}\right) \right]
=dK^{\left( 2q+2\right) }\left( S\right) +K^{\left( 2q\right) }\left(
S\Gamma _{11}\right) K^{\left( 3\right) }\left( \Gamma _{11}\right) \quad , 
\end{equation}
where we have used the fact that $H=-K^{\left( 3\right) }\left( \Gamma
_{11}\right) $ is a closed $3$-form. But the right hand side of (\ref{ap29})
are just the Bianchi identities (\ref{ap18}), which are identically zero
provided that (\ref{ap19}) holds; the Bianchi identities are therefore
integrability conditions for the forms $C^{\left( 2q+1\right) }$ in (\ref
{ap28}). The existence of $C^{\left( r\right) }$ is therefore guaranteed at
least for $r=1,3$.

We have solved (\ref{ap28}) for $C^{\left( 3\right) }$ explicitly; the
result is%
$$
C^{\left( 3\right) }=\frac i2\Pi ^m\Pi ^n\cdot d\bar \theta \Gamma
_{nm}\theta \;+ 
$$
$$
+\;\frac 12\Pi ^m\cdot \left[ d\bar \theta \Gamma _n\theta \cdot d\bar
\theta \Gamma _{nm}\theta -d\bar \theta \Gamma _{11}\theta \cdot d\bar
\theta \Gamma _{11}\Gamma _m\theta \right] \;+ 
$$
\begin{equation}
\label{ap30}+\;\frac i6d\bar \theta \Gamma _m\theta \cdot \left[ d\bar
\theta \Gamma _{11}\theta \cdot d\bar \theta \Gamma _{11}\Gamma _m\theta
-d\bar \theta \Gamma ^n\theta \cdot d\bar \theta \Gamma _{nm}\theta \right]
\quad . 
\end{equation}
In proving that (\ref{ap30}) is actually a solution to (\ref{ap28}) for $q=1$
one has to make use of the identities 
\begin{equation}
\label{ap31}\left( Id\,1\right) :=d\bar \theta \Gamma ^nd\theta \cdot d\bar
\theta \Gamma _{11}\Gamma _n\theta +d\bar \theta \Gamma ^n\theta \cdot d\bar
\theta \Gamma _{11}\Gamma _nd\theta =0\quad , 
\end{equation}
and%
$$
\left( Id\,2\right) _m:=d\bar \theta \Gamma ^nd\theta \cdot d\bar \theta
\Gamma _{nm}\theta +d\bar \theta \Gamma ^n\theta \cdot d\bar \theta \Gamma
_{nm}d\theta \;+ 
$$
\begin{equation}
\label{ap32}+\;d\bar \theta \Gamma _{11}d\theta \cdot d\bar \theta \Gamma
_{11}\Gamma _m\theta +d\bar \theta \Gamma _{11}\theta \cdot d\bar \theta
\Gamma _{11}\Gamma _md\theta \;=\;0\quad , 
\end{equation}
where (\ref{ap31}) is a consequence of (\ref{ap153}), and (\ref{ap32})
follows from (\ref{ap20}). Then%
$$
dC^{\left( 3\right) }=K^{\left( 4\right) }\left( {\bf 1}_{32}\right)
-C^{\left( 1\right) }K^{\left( 3\right) }\left( \Gamma _{11}\right) \;+ 
$$
$$
+\;\left( \frac 12\Pi ^m-\frac i6d\bar \theta \Gamma ^m\theta \right) \cdot
\left( Id\,2\right) _m\;-\;\left( \frac i3d\bar \theta \Gamma _{11}\theta
\right) \cdot \left( Id\,1\right) \quad , 
$$
and (\ref{ap28}) is fulfilled.

In principle we could apply the same procedure to construct the other
potentials $C^{\left( 5\right) },C^{\left( 7\right) },C^{\left( 9\right) }$.
But for the purpose we are pursuing here, namely the determination of the
topological extensions of Noether algebras, we do not need to know the full
expression for $C^{\left( 2q+1\right) }$; as mentioned earlier, these
algebra extensions come into play when the D-$p$-brane wraps around compact
dimensions in the spacetime; but the topology of this configuration is
entirely determined by the bosonic coordinates $X$ on the superspace, and
the pull-back of the differentials $dX^m$ to the worldvolume of the brane,
respectively. In evaluating the anomalous contributions to the charge
algebra as far as they origin in the WZ-term we therefore can restrict
attention to those components of the $C$'s which have only bosonic indices.
The strategy is as follows:

From section \ref{Modificatio} we see that all we need are the components of
the forms $S_{MN}\left( U\right) $, $M=\left( m,\alpha \right) $, carrying
the maximum number of bosonic indices; since we shall work with the LI-basis
now, this means that we need only consider terms involving the maximum
number of bosonic basis-$1$-forms $\Pi ^m$; in the following we shall refer
to such terms simply as ''leading terms''; furthermore we shall call the
number of bosonic indices in the leading term as the ''order'' of the term.
From (\ref{an29})-(\ref{an31}) we see that $S_{MN}\left( U\right) $ is
composed of terms $\delta _MD_N$, $D_M$ and $D_Md\Delta _N$. Since $\delta
_M $ leaves the number of LI-$1$-forms invariant we see that in order to
construct the leading terms of $S_{MN}\left( U\right) $ we need only
construct the leading terms of $D_M$. Now let us look back at formula (\ref
{lag7}), 
\begin{equation}
\label{bos2}\delta _\alpha C^{\left( 2q+1\right)
}=\sum\limits_{k=0}^qdD_\alpha ^{\left( 2q-2k\right) }\cdot \frac{B^k}{k!}%
\quad . 
\end{equation}
From our choice of $B$ in (\ref{ap151}) we see that $B$ contains only one
bosonic $1$-form $\Pi ^m$; the order of the terms in the sum in (\ref{bos2})
therefore decreases by $1$ as $k$ increases by $1$; this means that in order
to construct the leading term of $dD_\alpha ^{\left( 2q\right) }$ we need
only construct the leading term in $\delta _\alpha C^{\left( 2q+1\right) }$;
but this can be done using (\ref{ap28}) recursively: 
\begin{equation}
\label{bos2x}dC^{\left( 2q+1\right) }=K^{\left( 2q+2\right) }\left( S\right)
-C^{\left( 2q-1\right) }K^{\left( 3\right) }\left( \Gamma _{11}\right) \quad
. 
\end{equation}
From (\ref{ap13}) we see that the order of $K^{\left( 3\right) }\left(
\Gamma _{11}\right) $ is one, and that of $K^{\left( 2q+2\right) }\left(
S\right) $ is $\left( 2q\right) $; from (\ref{ap27}) and (\ref{ap30}) we
deduce that the order of $C^{\left( 2q-1\right) }$ is $\left( 2q-2\right) $,
therefore the first term on the right hand side of (\ref{bos2x}) is the
leading term, and we must construct a $C^{\left( 2q+1\right) }$ such that%
$$
dC^{\left( 2q+1\right) }=K^{\left( 2q+2\right) }\left( S\right) +\cdots
\quad . 
$$
Thus we find the leading term of $C^{\left( 2q+1\right) }$ to be 
\begin{equation}
\label{bos3}C^{\left( 2q+1\right) }=\frac i{\left( 2q\right) !}\Pi
^{m_{2q}}\cdots \Pi ^{m_1}\cdot d\bar \theta S\Gamma _{m_1\ldots
m_{2q}}\theta \quad , 
\end{equation}
with $S$ given in Table \ref{Tabelle1}. Therefore the leading term of $%
dD_\alpha ^{\left( 2q\right) }$ is 
\begin{equation}
\label{bos4}dD_\alpha ^{\left( 2q\right) }=-\frac i{\left( 2q\right) !}\Pi
^{m_{2q}}\cdots \Pi ^{m_1}\cdot \left( d\bar \theta S\Gamma _{m_1\ldots
m_{2q}}\right) _\alpha \quad , 
\end{equation}
and, up to a differential, we have 
\begin{equation}
\label{bos5}D_\alpha ^{\left( 2q\right) }=-\frac i{\left( 2q\right) !}\Pi
^{m_{2q}}\cdots \Pi ^{m_1}\cdot \left( \bar \theta S\Gamma _{m_1\ldots
m_{2q}}\right) _\alpha \quad . 
\end{equation}
This gives 
\begin{equation}
\label{bos6}\delta _\alpha D_\beta ^{\left( 2q\right) }+\delta _\beta
D_\alpha ^{\left( 2q\right) }=\frac{-2i}{\left( 2q\right) !}\Pi
^{m_{2q}}\cdots \Pi ^{m_1}\cdot \left( S\Gamma _{m_1\ldots m_{2q}}\right)
_{\alpha \beta }\quad = 
\end{equation}
$$
=\;\frac{-2i}{\left( 2q\right) !}dX^{m_1}\cdots dX^{m_{2q}}\cdot \left(
S\left( 2q\right) \Gamma _{m_{2q}\ldots m_1}\right) _{\alpha \beta
}\;+\;\cdots \quad . 
$$
multiplying (\ref{bos6}) with $\Gamma _n^{\alpha \beta }$ then yields a
vanishing result due to the vanishing of the trace%
$$
tr\left( S\Gamma _{m_1\ldots m_{2q}}\Gamma _n\right) =0 
$$
for all allowed values of $q$, $n$ and $S$. But since the order of the
leading term of $D_\alpha ^{\left( 2q-2\right) }\cdot d\Delta _\beta $ is $%
\left( 2q-1\right) $, and the order of $\delta _\alpha D_\beta ^{\left(
2q\right) }$ is $\left( 2q\right) $, as can be seen from (\ref{ap154}) and (%
\ref{bos5}), we infer from (\ref{lag10}) that indeed 
\begin{equation}
\label{bos7}D_m^{\left( 0\right) }=0\quad ,\quad D_m^{\left( 2q\right)
}=0\quad ; 
\end{equation}
these equations will actually hold in a rigorous sense, not only to leading
order; from (\ref{ap27}) and (\ref{ap30}) we see that at least $C^{\left(
1\right) }$ and $C^{\left( 3\right) }$ are strictly translation invariant,
and this will be true for the others as well, since the higher rank
potentials are constructed recursively from the lower rank ones.
Furthermore, from (\ref{bos5}) we infer that $\delta _mD_\alpha ^{\left(
2q\right) }=0$ for all $q$.

\subsection{Extended superalgebras for D-$2q$-branes}

Now we can turn to evaluating the expressions $S_{MN}\left( U\right) $ as
given in (\ref{an29})-(\ref{an31}). Since expressions involving $d\Delta _M$
have leading order smaller than the leading order of $\delta _\alpha D_\beta
^{\left( 2q-2k\right) }$, see (\ref{for1}, \ref{for2}), they can be omitted
in the discussion. The final expression for $S_{\alpha \beta }\left(
U\right) $ is therefore%
$$
S_{\alpha \beta }\left( U\right) \;=\;-2i\cdot \sum\limits_{k=0}^q\left[
S\left( 2q-2k\right) \Gamma _{m_{2q-2k}\ldots m_1}\right] _{\alpha \beta
}\cdot \frac{\epsilon ^{0\mu _1\ldots \mu _{2q-2k}\nu _1\ldots \nu _{2k}}}{%
\left( \left( 2q-2k\right) !\right) ^2\,2^k\cdot k!}\,\cdot 
$$
\begin{equation}
\label{bos8}\cdot \partial _{\mu _1}X^{m_1}\cdots \partial _{\mu
_{2q-2k}}X^{m_{2q-2k}}\cdot F_{\nu _1\nu _2}\cdots F_{\nu _{2k-1}\nu
_{2k}}\quad . 
\end{equation}
Let us now write $dX^m:=emb^{*}dX^m$ for the sake of convenience; then we
have%
$$
dX^{m_1}\cdots dX^{m2q-2k}\cdot \frac{\left( dA\right) ^k}{k!}= 
$$
$$
=\;\omega _0\cdot \frac{\epsilon ^{0\mu _1\ldots \mu _{2q-2k}\nu _1\ldots
\nu _{2k}}}{\left( 2q-2k\right) !\,2^k\cdot k!}\cdot \partial _{\mu
_1}X^{m_1}\cdots \partial _{\mu _{2q-2k}}X^{m_{2q-2k}}\cdot F_{\nu _1\nu
_2}\cdots F_{\nu _{2k-1}\nu _{2k}}\quad , 
$$
where $\omega _0=d\sigma ^1\cdots d\sigma ^{2q}$; therefore 
\begin{equation}
\label{bos9}\omega _0\cdot S_{\alpha \beta }\left( U\right) =-2i\cdot
\sum\limits_{k=0}^q\frac{\left[ S\left( 2q-2k\right) \Gamma
_{m_{2q-2k}\ldots m_1}\right] _{\alpha \beta }}{\left( 2q-2k\right) !}\cdot
dX^{m_1}\cdots dX^{m2q-2k}\cdot \frac{\left( dA\right) ^k}{k!}\quad . 
\end{equation}

Furthermore, from (\ref{an30}) and (\ref{an31}) we infer that both $%
S_{\alpha m}\left( U\right) $ and $S_{mn}\left( U\right) $ are zero.

Now we can collect everything together to write down the general structure
of the modified charge algebra; we assume that double integration is
defined, so that we get%
$$
\left\{ Q_\alpha ,Q_\beta \right\} =2\Gamma _{\alpha \beta }^n\cdot
P_m\;-\;2i\left( \Gamma _{11}\Gamma _m\right) _{\alpha \beta }\cdot Y^m\;- 
$$
\begin{equation}
\label{bos10}-\;2i\cdot \sum\limits_{k=0}^q\frac{\left[ S\left( 2q-2k\right)
\Gamma _{m_{2q-2k}\ldots m_1}\right] _{\alpha \beta }}{\left( 2q-2k\right) !}%
\cdot Z^{m_1\ldots m_{2q-2k}}\quad , 
\end{equation}
with 
\begin{equation}
\label{bos11}Y^m=\frac 1{\left( 2q-1\right) !}\int\limits_{W\left( t\right)
}\left( *\Lambda ^{gauge}\right) \,dX^m\quad , 
\end{equation}
which was defined in (\ref{an17}), and 
\begin{equation}
\label{bos12}Z^{m_1\ldots m_{2q-2k}}=\int\limits_{W\left( t\right)
}dX^{m_1}\cdots dX^{m2q-2k}\cdot \frac{\left( dA\right) ^k}{k!}\quad = 
\end{equation}
\begin{equation}
\label{bos121}=\int\limits_{W\left( t\right) }d^{2q}\sigma \,\frac{\epsilon
^{0\mu _1\ldots \mu _{2q-2k}\nu _1\ldots \nu _{2k}}}{\left( 2q-2k\right)
!\,2^k\cdot k!}\cdot \partial _{\mu _1}X^{m_1}\cdots \partial _{\mu
_{2q-2k}}X^{m_{2q-2k}}\cdot F_{\nu _1\nu _2}\cdots F_{\nu _{2k-1}\nu
_{2k}}\quad . 
\end{equation}
Moreover, the relation between $q$ and $S\left( 2q\right) $ is given in
Table \ref{Tabelle1}. Note that the integrand of the charge $Y^m$ is closed
on the physical trajectories, see (\ref{an171}).

At last, from (\ref{an26}) and (\ref{an28}) we learn that 
\begin{equation}
\label{bos13}\left[ Q_\alpha ,P_m\right] =0\quad ,\quad \left[
P_m,P_n\right] =0\quad . 
\end{equation}
To avoid confusion we emphasize that in (\ref{bos10}) the bracket $\left\{
Q_\alpha ,Q_\beta \right\} $ denotes a graded {\bf Poisson}-bracket between
two Grassmann-odd quantities, but in (\ref{bos13}) we have chosen a square
bracket to denote the {\bf Poisson} bracket between quantities of which at
least one of them is Grassmann-even.

\section{Interpretation of the central charges}

Let us try to interprete the structure of the charges $Z^{m_1\ldots
m_{2q-2k}}$ in (\ref{bos12}). Let us fix $q$ and first of all look at the
extreme values of $k$, i.e. $k=0$ and $k=q$. For $k=0$ we find 
\begin{equation}
\label{bos14}Z^{m_1\ldots m_{2q}}=\int\limits_{W\left( t\right)
}dX^{m_1}\cdots dX^{m_{2q}}\quad ; 
\end{equation}
from (\ref{gract40}) we see that this is just the integral over the
topological current $j_T^{0m_1\cdots m_{2q}}$, i.e. the topological charge 
\begin{equation}
\label{bos15}Z^{m_1\ldots m_{2q}}=T^{m_1\ldots m_{2q}} 
\end{equation}
from (\ref{gract401}). This charge will not be defined if the brane $W\left(
t\right) $ is infinitely extended in one of the spatial directions $X^{m_i}$
occuring in $T^{m_1\ldots m_{2q}}$. On the other hand, if all spacetime
directions occuring in $T^{m_1\ldots m_{2q}}$ are compact, but the brane is
not wrapped around {\bf all} of them then this charge will be zero. It will
be non-zero {\bf only} if the brane wrappes around all these compact
dimensions; consider, for example, a compact $U\left( 1\right) $-factor in
the spacetime, which may be taken as direction $m=1$, and a closed string
that is wrapped around this $n$ times \cite{Azca} (the string is not a IIA
brane, of course, but that does not affect the discussion here); then $T^1$
is proportional to $2n\pi $, where $n$ is an integer. On the other hand, if
the string is closed in a flat spacetime, then $T^1=0$.

For $k=q$ we find that 
\begin{equation}
\label{bos16}Z=\frac 1{q!}\int\limits_{W\left( t\right) }\left( dA\right)
^q\quad . 
\end{equation}
This can be given a simple interpretation in the case of $q=1$, $p=2q$, i.e.
the D-$2$-brane: In this case 
\begin{equation}
\label{bos17}Z=\int\limits_{W\left( t\right) }dA 
\end{equation}
is just the flux of the field strength $F$ of the gauge potential through
the brane $W\left( t\right) $. The worldvolume is now to be regarded as a $%
U\left( 1\right) $-bundle $P\left( W,U\left( 1\right) \right) $. If the
section $W\left( t\right) $ is infinitely extended then $Z$ will vanish
provided that the gauge field vanishes sufficiently fast at infinity, and
the bundle is trivial. If, however, $W\left( t\right) $ describes a $S^2$,
say, then we have the possibility that the gauge potential is no longer
defined globally on $S^2$; if two gauge patches are necessary to cover $S^2$
then the flux integral (\ref{bos17}) yields 
\begin{equation}
\label{bos18}Z=4\pi g\quad ,\quad 2g\in {\bf Z}\quad , 
\end{equation}
where $g$ now is the charge of a {\it Dirac monopole} of the gauge field
sitting ''in the centre of $S^2$'', and the quantization condition $2g\in 
{\bf Z}$ comes from the requirement that the transition function between the
two gauge patches be unique, see for example \cite{Nakahara}. It is not
clear to us whether this interpretation extends to all possible values of $q$%
; we might conjecture that the $U\left( 1\right) $-bundle can always be
non-trivial, in which case similar arguments apply to (\ref{bos16}), since
then we must cover $W\left( t\right) $ by more than one gauge patch, which
should yield analogous results.

As for the values $1\le k\le q$ we see that the currents $dX^{m_1}\cdots
dX^{m2q-2k}$ in (\ref{bos12}) now probe whether the brane has subcycles of
dimension $\left( 2q-2k\right) $ embedded in it that wrap around $\left(
2q-2k\right) $ compact dimensions of the spacetime. Only in this case the
charges $Z^{m_1\ldots m_{2q-2k}}$ will be non-vanishing. Furthermore we see
that the $U\left( 1\right) $-bundle defined by the gauge field must be
non-trivial in order to having a non-vanishing charge. To see this we can
choose a static gauge $\sigma ^\mu =X^\mu $, then from (\ref{bos121}) we
have that 
\begin{equation}
\label{bos19}Z^{m_1\ldots m_{2q-2k}}=\int\limits_{W\left( t\right)
}d^{2q}\sigma \,\frac{\epsilon ^{0m_1\ldots m_{2q-2k}\nu _1\ldots \nu _{2k}}%
}{\left( 2q-2k\right) !\cdot k!}\cdot \partial _{\nu _1}A_{\nu _2}\cdots
\partial _{\nu _{2k-1}}A_{\nu _{2k}}\quad . 
\end{equation}
(This static gauge will be allowed at least on a certain coordinate patch on
the worldvolume; in this case we have to sum over contributions from the
different patches). We see that similar considerations concerning the
non-triviality of the $U\left( 1\right) $-bundle should apply here. In
particular, (\ref{bos19}) will vanish if the bundle is trivial, since in
this case the gauge potential $A_\mu $ is globally defined, and then (\ref
{bos19}) yields a surface term. A tentative interpretation of the charges (%
\ref{bos12}) therefore would be that they measure the coupling of compact
spacetime dimensions the brane or some directions of the brane wrap around
to non-trivial gauge field configurations on the brane.

We have not found an easy interpretation for $Y^m$; from its structure we
see that the $dX^m$-factor together with the fact that $*\Lambda ^{gauge}$
is closed on the physical trajectories will make this charge non-vanishing
only when the brane contains a $1$-cycle wrapping around a compact spacetime
dimension, e.g. a $S^1$-factor. This charge then describes the coupling of
the canonical gauge field momentum to this particular topological
configuration.

We finally present the modified charge algebra in the case of the D-$2$%
-brane with worldvolume ${\bf R}\times S^2$, since this allows for an easy
interpretation, as we have seen above:%
$$
\left\{ Q_\alpha ,Q_\beta \right\} =2\left( C\Gamma ^m\right) _{\alpha \beta
}\cdot P_m\;-\;2i\left( C\Gamma _{11}\Gamma _m\right) _{\alpha \beta }\cdot
Y^m\;- 
$$
\begin{equation}
\label{bos20}-\;i\left( C\Gamma _{m_2m_1}\right) _{\alpha \beta }\cdot
T^{m_1m_2}\;-\;2i\left( C\Gamma _{11}\right) _{\alpha \beta }\cdot 4\pi
g\quad , 
\end{equation}
where $T^{m_1m_2}$ probes the presence of compact dimensions in spacetime
the brane wraps around, and $g$ is the quantized charge of a possible Dirac
monopole resulting from the gauge field.

\chapter{Lightlike Compactifications}
\label{semiGroup}

\section{Introduction}

As described in the introduction, in this chapter we examine the structure of maps, in particular of
elements of the isometry group $E_t^n$ of a flat pseudo-Euclidean space $%
\TeXButton{R}{\mathbb{R}}_t^n$, that preserve the points of a lattice $lat$,
whose associated real vector space $\left[ lat\right] $, called the $%
\TeXButton{R}{\mathbb{R}}$-linear enveloppe of the lattice, is lightlike. In
this case, the restriction of the metric $\eta $ with signature $\left(
-t,+s\right) $ to $\left[ lat\right] $ is no longer definite. This gives
rise to the possibility of having lattice-preserving transformations in the
overall pseudo-Euclidean group $E_t^n$ that are injective, but no longer
surjective on the lattice; in other words, their inverses do not preserve $%
lat$. The set of all these transformations therefore will no longer be a
group, but only a semigroup. Since it is precisely the lattice preserving
transformations that descend to the quotient of $\TeXButton{R}{\mathbb{R}}%
_t^n$ over the discrete group of primitive lattice translations $\Gamma $,
i.e. to the ''compactified'' spacetime $\TeXButton{R}{\mathbb{R}}_t^n/\Gamma
=\TeXButton{R}{\mathbb{R}}^{n-m}\times T^m$, these semigroup elements
constitute an extension of the isometry group $I\left( \TeXButton{R}
{\mathbb{R}}_t^n/\Gamma \right) $, which act non-invertibly on the
compactified space. We present in detail the case of the compactification of
a  Lorentzian spacetime over a lightlike lattice $\Gamma $, where it is
shown that the non-invertible elements wind the lightlike circle $k$ times
around itself. We argue that this should map the different sectors of a
Lagrange theory on $\TeXButton{R}{\mathbb{R}}_t^n/\Gamma $, as labelled by
the lightlike compactification radius, in a one-way process into each other.
In the case under consideration, this map will be accomplished by finite
discrete transformations generated by the ''mass'' generator of the
centrally extended Galilei group; since it is known that the eigenvalues of
this generator label different superselection sectors of a theory, we argue
that our semigroup transformations connect different superselection sectors
of any Lagrange theory on a lightlike compactified spacetime.

The plan of this chapter is as follows: In section \ref{Sc.1} we provide some
background on orbit spaces and the associated fibre preserving sets. In
section \ref{Sc.2} we introduce our notation conventions. In section \ref
{Sc.3} and \ref{Sc.4} we introduce our concepts of lattice preserving
transformations, and examine some of their structure. In particular, we show
in a theorem why no semigroup extensions are available in a Euclidean
background space, or more generally, for a lattice whose $\TeXButton{R}
{\mathbb{R}}$-linear enveloppe $\left[ lat\right] $ is spacelike. In section 
\ref{Sc.5} we construct the sets normalizing a lightlike lattice in a
Minkowski spacetime. In section \ref{Sc.6} we show how the semigroup
transformations act on the lightlike circle of the spacetime, and how they
relate theories belonging to different compactification radii.

\section{Orbit spaces and normalizing sets \label{Sc.1}}

Assume that a group $G$ has a left action on a topological space $X$ such
that the map $G\times X\ni \left( g,x\right) \mapsto gx$ is a homeomorphism.
When a discrete subgroup $\Gamma \subset G$ acts properly discontinuously
and freely on $X$, then the natural projection $p:X\rightarrow X/\Gamma $ of 
$X$ onto the space of orbits, $X/\Gamma $, can be made into a covering map,
and $X$ becomes a covering space of $X/\Gamma $ (e.g. \cite
{Fulton,Jaehnich,Massey}) . More specially, if $X=M$ is a connected
pseudo-Riemannian manifold with a metric $\eta $, and $G=I\left( M\right) $
is the group of isometries of $M$, so that $\Gamma $ is a discrete subgroup
of isometries acting on $M$, then there is a unique way to make the quotient 
$M/\Gamma $ a pseudo-Riemannian manifold (e.g. \cite{Wolf,ONeill}); in this
construction one stipulates that the projection $p$ be a local isometry,
which determines the metric on $M/\Gamma $. In such a case, we speak of $%
p:M\rightarrow M/\Gamma $ as a pseudo-Riemannian covering.

In both cases, the quotient $p:X\rightarrow $ $X/\Gamma $ can be regarded as
a principal fibre bundle with bundle space $X$, base $X/\Gamma $, and $%
\Gamma $ as structure group, the fibre over $m\in X/\Gamma $ being the orbit
of any element $x\in p^{-1}\left( m\right) $ under $\Gamma $, i.e. $%
p^{-1}\left( m\right) =\Gamma x=\left\{ \gamma x\mid \gamma \in \Gamma
\right\} $. If $g\in G$ induces the homeomorphism $x\mapsto gx$ of $X$ (or
an isometry of $M$), then $g$ gives rise to a well-defined map $%
g_{\#}:X/\Gamma \rightarrow X/\Gamma $ {\bf only} when $g$ preserves all
fibres, i.e. when $g\left( \Gamma x\right) \subset \Gamma \left( gx\right) $
for all $x\in X$. This is equivalent to saying that $g\Gamma g^{-1}\subset
\Gamma $. If this relation is replaced by the stronger condition $g\Gamma
g^{-1}=\Gamma $, then $g$ is an element of the {\it normalizer }$N\left(
\Gamma \right) $ of $\Gamma $ in $G$, where 
\begin{equation}
\label{pp3fo1}N\left( \Gamma \right) =\left\{ g\in G\mid g\Gamma
g^{-1}=\Gamma \right\} \quad . 
\end{equation}
The normalizer is a group by construction. It contains all fibre preserving
elements $g$ of $G$ such that $g^{-1}$ is fibre preserving as well. In
particular, it contains the group $\Gamma $, which acts trivially on the
quotient space; this means, that for any $\gamma \in \Gamma $, the induced
map $\gamma _{\#}:X/\Gamma \rightarrow X/\Gamma $ is the identity on $%
X/\Gamma $, $\gamma _{\#}=\left. id\right| _{X/\Gamma }$. This follows,
since the action of $\gamma _{\#}$ on the orbit $\Gamma x$, say, is defined
to be $\gamma _{\#}\left( \Gamma x\right) =\Gamma \left( \gamma x\right)
=\Gamma x$, where the last equality holds, since $\Gamma $ is a group.

In this work we are interested in relaxing the equality in the condition
defining $N\left( \Gamma \right) $; to this end we introduce what we call
the {\it extended normalizer}, denoted by $eN\left( \Gamma \right) $,
through 
\begin{equation}
\label{pp3fo2}eN\left( \Gamma \right) :=\left\{ g\in G\mid g\Gamma
g^{-1}\subset \Gamma \right\} \quad . 
\end{equation}
The elements $g\in G$ which give rise to well-defined maps $g_{\#}$ on $%
X/\Gamma $ are therefore precisely the elements of the extended normalizer $%
eN\left( \Gamma \right) $, as we have seen in the discussion above. Such
elements $g$ are said to {\it descend} to the quotient space $X/G$. Hence $%
eN\left( \Gamma \right) $ contains all homeomorphisms of $X$ (isometries of $%
M$) that descend to the quotient space $X/\Gamma $ ($M/\Gamma $); the
normalizer $N\left( \Gamma \right) $, on the other hand, contains all those $%
g$ for which $g^{-1}$ descends to the quotient as well. Thus, $N\left(
\Gamma \right) $ is the group of all $g$ which descend to {\bf invertible}
maps $g_{\#}$ (homeomorphisms; isometries) on the quotient space. In the
case of a semi-Riemannian manifold $M$, for which the group $G$ is the
isometry group $I\left( M\right) $, the normalizer $N\left( \Gamma \right) $
therefore contains all isometries of the quotient space, the only point
being that the action of $N\left( \Gamma \right) $ is not effective, since $%
\Gamma \subset N\left( \Gamma \right) $ acts trivially on $M/\Gamma $.
However, $\Gamma $ is a normal subgroup of $N\left( \Gamma \right) $, so
that the quotient $N\left( \Gamma \right) /\Gamma $ is a group again, which
is now seen to act effectively on $M/\Gamma $, and the isometries of $%
M/\Gamma $ which descend from isometries of $M$ are in a 1--1 relation to
elements of this group. Thus, denoting the isometry group of the quotient
space $M/\Gamma $ as $I\left( M/\Gamma \right) $, we have the well-known
result that 
\begin{equation}
\label{pp3fo3}I\left( M/\Gamma \right) =N\left( \Gamma \right) /\Gamma \quad
. 
\end{equation}

Now we turn to the extended normalizer. For a general element $g\in eN\left(
\Gamma \right) $ the induced map $g_{\#}$ is no longer injective on $%
X/\Gamma $. To see this assume that for a fixed element $g\in G$, the
inclusion in definition (\ref{pp3fo2}) is proper, i.e. $g\Gamma g^{-1}%
\stackunder{\neq }{\subset }\Gamma $. Take an arbitrary $x\in M$, then $%
g\left( \Gamma x\right) \stackunder{\neq }{\subset }\Gamma \left( gx\right) $%
; this means that there exists an element $z\in \Gamma \left( gx\right) $
that is not the image under $g$ of any $\gamma x$ in the orbit $\Gamma x$.
Hence, since $g$ is invertible on $X$, there exists an $x^{\prime }\in X$,
whose orbit $\Gamma x^{\prime }$ is different from the orbit $\Gamma x$,
such that $z=gx^{\prime }$. Since $g$ preserves fibres we have $g\left(
\Gamma x^{\prime }\right) \stackunder{\neq }{\subset }\Gamma \left(
gx^{\prime }\right) =\Gamma z=\Gamma \left( gx\right) $, the last equality
following, since $z$ lies in the orbit of $gx$. This implies that the
induced map $g_{\#}$ maps the distinct orbits $\Gamma x\neq \Gamma x^{\prime
}$ into the same orbit $\Gamma \left( gx\right) $, which expresses that $%
g_{\#}$ is not injective. In particular, if $g$ was an isometry of $M$, then 
$g_{\#}$ can no longer be an isometry on the quotient space, since it is not
invertible. It also follows that $eN\left( \Gamma \right) $ will not be a
group, in general. For this reason, the elements of the extended normalizer
seem to have attracted limited attention in the literature so far.

In this work, however, we will show that the extended normalizer naturally
emerges when we study identification spaces $M/\Gamma $, where $M=\left( 
\TeXButton{R}{\mathbb{R}}^n,\eta \right) $ is flat $\TeXButton{R}{\mathbb{R}}%
^n$ endowed with a symmetric bilinear form $\eta $ with signature $\left(
-t,+s\right) $ or index $t$; to such a space $M$ we will also refer to as $M=%
\TeXButton{R}{\mathbb{R}}_t^n$. The group $\Gamma $ will be realized as a
discrete group of translations in $M$, the elements being in 1--1
correspondence with the points of a {\it lattice }$lat\subset \TeXButton{R}
{\mathbb{R}}_t^n$, which is regarded as a subset of $\TeXButton{R}
{\mathbb{R}}_t^n$. We will find that in the Lorentzian case, the fact that
the identity component $SO_{1,n-1}^{+}\subset O_{1,n-1}$ is no longer
compact will give rise to a natural extension of the isometry group $N\left(
\Gamma \right) /\Gamma $ of the quotient $M/\Gamma $ to the set $eN\left(
\Gamma \right) /\Gamma $, provided that the $\TeXButton{R}{\mathbb{R}}$%
-linear enveloppe of the lattice is a {\bf lightlike} subvector space (we do
not consider lattices whose associated $\TeXButton{R}{\mathbb{R}}$-linear
vector space is timelike; this would give rise to ''compactifications along
a time direction''). We will show that $eN\left( \Gamma \right) /\Gamma $ in
general has the structure of a {\it semigroup}, naturally containing the
isometry group $N\left( \Gamma \right) /\Gamma $ as a subgroup. This will be
compared with the orthogonal case, and it will be shown that the compactness
of $SO_n$ obstructs such an extension. That is probably why such extensions
have not been studied in crystallography in the past.

\section{Notations and conventions \label{Sc.2}}

--- If a subgroup $H$ of a group $G$ is normal in $G$, we denote this fact
by $H\lhd G$.

--- The isometry group $I\left( \TeXButton{R}{\mathbb{R}}_t^n\right) $ of $%
\TeXButton{R}{\mathbb{R}}_t^n$ is the semi-direct product 
\begin{equation}
\label{pp3not1}I\left( \TeXButton{R}{\mathbb{R}}_t^n\right) =E_t^n=%
\TeXButton{R}{\mathbb{R}}^n\odot O_{t,n-t}\quad , 
\end{equation}
called {\it pseudo-Euclidean group}, where the translational factor $%
\TeXButton{R}{\mathbb{R}}^n$ is normal in $E_t^n$, $\TeXButton{R}{\mathbb{R}}%
^n\lhd E_t^n$. Elements of $E_t^n$ will be denoted by $\left( t,R\right) $
with group law $\left( t,R\right) \left( t^{\prime },R^{\prime }\right)
=\left( Rt^{\prime }+t,RR^{\prime }\right) $. Projections onto the first and
second factor of $E_t^n$ are defined as $p_1:E_t^n\rightarrow \TeXButton{R}
{\mathbb{R}}^n$, $p_1\left( t,R\right) =\left( t,1\right) $; $%
p_2:E_t^n\rightarrow O_{t,n-t}$, $p_2\left( t,R\right) =\left( 0,R\right) $.
Elements of the form $\left( 0,R\right) $ will be referred to as
''rotations'', although in general they are pseudo-orthogonal
transformations. The Lie algebra of the pseudo-Euclidean group $E_t^n$ will
be denoted by 
\begin{equation}
\label{pp3not2}euc_t^n\equiv Lie\left( E_t^n\right) 
\end{equation}
henceforth. -- For $t=1$, $E_1^n$ is the Poincare group.

--- Given a subset $S\subset \TeXButton{R}{\mathbb{R}}_t^n$, the $%
\TeXButton{R}{\mathbb{R}}${\it -linear span} of $S$ is the vector subspace
of $\TeXButton{R}{\mathbb{R}}_t^n$ generated by elements of $S$, i.e. the
set of all finite linear combinations of elements in $S$ with coefficients
in $\TeXButton{R}{\mathbb{R}}$; we denote the $\TeXButton{R}{\mathbb{R}}$%
-linear span of $S$ by $\left[ S\right] _{\TeXButton{R}{\mathbb{R}}}$ or
simply $\left[ S\right] $, if no confusion is likely. If $S=\left\{
u_1,\ldots ,u_m\right\} $ is finite, one also writes $\left[ S\right]
=\sum_{i=1}^m\TeXButton{R}{\mathbb{R}}\cdot u_i$. This contains the $%
\TeXButton{Z}{\mathbb{Z}}${\it -linear span }$\left[ S\right] _{\TeXButton{Z}
{\mathbb{Z}}}=\sum_{i=1}^m\TeXButton{Z}{\mathbb{Z}}\cdot u_i$ as a proper
subset.

--- The {\it index} $ind\left( W\right) $ of a vector subspace $W\subset 
\TeXButton{R}{\mathbb{R}}_t^n$ is the maximum in the set of all integers
that are the dimensions of $\TeXButton{R}{\mathbb{R}}$-vector subspaces $%
W^{\prime }\subset W$ on which the restriction of the metric $\left. \eta
\right| W^{\prime }$ is negative definite, see e.g. \cite{ONeill}. Hence $%
0\le ind\left( W\right) \le m$, and $ind\left( W\right) =0$ if and only if $%
\left. \eta \right| W$ is positive definite. In the Lorentzian case, i.e. $M=%
\TeXButton{R}{\mathbb{R}}_1^n$, we call $W$ timelike $\Leftrightarrow $ $%
\left. \eta \right| W$ nondegenerate, and $ind\left( W\right) =1$; $W$
lightlike $\Leftrightarrow $ $\left. \eta \right| W$ degenerate, and $W$
contains a $1$-dimensional lightlike vector subspace, but no timelike
vector; and $W$ spacelike $\Leftrightarrow $ $\left. \eta \right| W$ is
positive definite and hence $ind\left( W\right) =0$.

\section{Lattices and their symmetries \label{Sc.3}}

In this section we introduce our conventions of lattices in $\TeXButton{R}
{\mathbb{R}}_t^n$ and their associated sets of symmetries. These notions
will be appropriate to examine the peculiarities that arise when the vector
space $\TeXButton{R}{\mathbb{R}}_t^n$ is non-Euclidean. On the one hand,
they extend the usual terminology encountered in crystallography. On the
other hand, our definition of a lattice is adapted to the purposes of this
paper, and therefore somewhat simplified compared with the most general
definitions possible in crystallography. This means that a full adaption of
our terminology introduced here with well-established crystallographic
notions would have been a tedious task with no contribution to deeper
understanding; we therefore have made no attempt to do so.

In this work we restrict attention to lattices that contain the origin $0\in 
\TeXButton{R}{\mathbb{R}}_t^n$ as a lattice point, which suffices for our
purposes. Let $1\le m\le n$, let $\underline{u}\equiv \left( u_1,\ldots
,u_m\right) $ be a set of $m$ linearly independent vectors in $\TeXButton{R}
{\mathbb{R}}_t^n$; then the $\TeXButton{Z}{\mathbb{Z}}$-linear span of $%
\underline{u}$, 
\begin{equation}
\label{pp3fo4}lat\equiv \sum_{i=1}^m\TeXButton{Z}{\mathbb{Z}}\cdot
u_i=\left\{ \sum_{i=1}^mz_i\cdot u_i\mid z_i\in \TeXButton{Z}{\mathbb{Z}}%
\right\} \quad , 
\end{equation}
is called the set of {\it lattice points} with respect to $\underline{u}$.
Elements of $lat$ are regarded as points in $\TeXButton{R}{\mathbb{R}}_t^n$
as well as vectors on $T\TeXButton{R}{\mathbb{R}}_t^n$. Let $\left[
lat\right] \equiv \left[ \underline{u}\right] _{\TeXButton{R}{\mathbb{R}}}$
denote the $\TeXButton{R}{\mathbb{R}}$-linear span of $lat$. We define the 
{\it index of the lattice }as the index of its $\TeXButton{R}{\mathbb{R}}$%
-linear span $\left[ lat\right] $, $ind\left( lat\right) \equiv ind\left(
\left[ lat\right] \right) $. In the Lorentzian case, $M=\TeXButton{R}
{\mathbb{R}}_1^n$, the lattice $lat$ is called {\it timelike / lightlike /
spacelike} if the enveloppe $\left[ lat\right] $ is.

The subset $T_{lat}\subset E_t^n$ is the subgroup of all translations in $%
E_t^n$ through elements of $lat$, 
\begin{equation}
\label{pp3fo5}T_{lat}=\left\{ \left( t_z,0\right) \in E_t^n\mid t_z\in
lat\right\} \quad . 
\end{equation}
Elements of $T_{lat}$ are called {\it primitive translations}.

Now we introduce the set $pres\left( lat\right) $, which is defined to be
the set of all diffeomorphisms on $\TeXButton{R}{\mathbb{R}}_t^n$ preserving 
$lat$, i.e. 
\begin{equation}
\label{pp3fo6}pres\left( lat\right) \equiv \left\{ \phi \in diff\left( 
\TeXButton{R}{\mathbb{R}}_t^n\right) \mid \phi \,lat\subset lat\right\}
\quad . 
\end{equation}
Every such $\phi $ is invertible, hence surjective, on $\TeXButton{R}
{\mathbb{R}}_t^n$; however, it need not be surjective on $lat$, which means
that inverses in this set do not necessarily exist. Neither is it required
that $\phi $ be linear. $pres\left( lat\right) $ is therefore a semigroup
with composition of maps as multiplication, and $\left. id\right| _{%
\TeXButton{R}{\mathbb{R}}_t^n}$ as unit element.

We also define the associated set 
\begin{equation}
\label{pp3fo6a}pres^{\times }\left( lat\right) \equiv \left\{ \phi \in
diff\left( \TeXButton{R}{\mathbb{R}}_t^n\right) \mid \phi \,lat=lat\right\}
\subset pres\left( lat\right) \quad . 
\end{equation}
$pres^{\times }\left( lat\right) $ contains all diffeomorphisms of $%
pres\left( lat\right) $ whose restriction to $lat$ is invertible, and hence
is a group.

The intersection 
\begin{equation}
\label{pp3fo7}sym\equiv pres\left( lat\right) \cap E_t^n 
\end{equation}
we term the {\it set of symmetries} of the lattice $lat$, and 
\begin{equation}
\label{pp3fo7a}sym^{\times }\equiv pres^{\times }\left( lat\right) \cap
E_t^n 
\end{equation}
we call the {\it set of invertible symmetries} of $lat$. $sym^{\times }$ is
a group by construction. On the other hand, $sym$ is only a semigroup, since
inverses \underline{in $sym$} do not necessarily exist. Cleary, $sym^{\times
}\subset sym$. This inclusion is not always proper. Below we prove a theorem
that explains the details. First, however, we examine the structure of $%
sym^{\times }$ and $sym$ more closely.

\section{The structure of $sym^{\times }$ and $sym$ \label{Sc.4}}

An immediate statement is

\subsection{Proposition \label{sym}}

\begin{enumerate}
\item  All translations in $sym$ belong to $sym^{\times }$.

\item  All translations in $T_{lat}$ belong to $sym^{\times }$.

\item  There are no pure translations in $sym^{\times }$ other than
primitive translations from $T_{lat}$.

\item  $T_{lat}$ is normal in $sym^{\times }$.
\end{enumerate}

\TeXButton{Beweis}{\raisebox{-1ex}{\it Proof :}
\vspace{1ex}}

The fist two statements are obvious. As for the third, assume $\left(
t,1\right) \in sym^{\times }$ were no element of $T_{lat}$; then it would
map the lattice point $0$ into the lattice point $t$, which is a
contradition. Now prove $\left( 4\right) $: Since, by $\left( 2\right) $,
translations $\left( t_z,0\right) $ belong to $sym^{\times }$, we have that
for a given $\left( t,R\right) \in sym^{\times }$, the product $\left(
t,R\right) \left( t_z,0\right) \left( t,R\right) ^{-1}\in sym^{\times }$ as
well. But this product equals $\left( Rt_z,1\right) $, hence is a pure
translation, hence, by $\left( 3\right) $, must belong to $T_{lat}$, which
proves $T_{lat}\lhd sym^{\times }$. \TeXButton{BWE}
{\hfill
\vspace{2ex}
$\blacksquare$}

These results do not a priori imply, however, that $p_1\left( sym\right)
=T_{lat}$. We show below that this is true for $sym^{\times }$.

We now examine the projection of $sym^{\times }$ onto the second factor $%
p_2\left( sym^{\times }\right) $. A priori it is not clear whether this
projection is a subset of $sym$, $sym^{\times }$, or not. We will see
shortly that indeed $p_2\left( sym^{\times }\right) \subset sym^{\times }$.
We start with observing that the elements $\left( 0,R\right) $ of $p_2\left(
sym^{\times }\right) $ are in 1--1 correspondence with the left cosets $%
T_{lat}\cdot \left( t,R\right) $, where $\left( t,R\right) $ is in the
inverse image $p_2^{-1}\left( 0,R\right) $. This follows, since $T_{lat}$ is
a subgroup of $sym^{\times }$, so that $T_{lat}\cdot \left( t,R\right) $ is
certainly in the inverse image; and furthermore, any two elements in this
coset must differ by a primitive translation, since for $\left( t,R\right) $%
, $\left( t^{\prime },R\right) $ we have $\left( t,R\right) ^{-1}\in
sym^{\times }$, hence $\left( t^{\prime },R\right) \left( t,R\right)
^{-1}=\left( t^{\prime }-t,1\right) \in sym^{\times }$. It is at this point
that we need the condition that $\left( t,R\right) $ be in $sym^{\times }$
rather in $sym$. The last equation says that $\left( t^{\prime }-t,1\right) $
must be a primitive translation, since $sym^{\times }$ contains no other
translations than these. From these considerations we conclude that there
must exist a coset representative, denoted by $\left( \tau _R,R\right) $ of
the coset $T_{lat}\cdot \left( t,R\right) $ such that 
\begin{equation}
\label{pp3fo7b}\tau _R=\sum_{i=1}^mq_i\cdot u_i\quad ,\quad 0\le q_i<1\quad
. 
\end{equation}
This defines a map 
\begin{equation}
\label{pp3fo7c}\tau :\left\{ 
\begin{array}{c}
sym^{\times }\rightarrow W \\ 
\left( t,R\right) =\left( t_z+\tau _R,R\right) \mapsto \tau _R 
\end{array}
\right. , 
\end{equation}
which is constant on the cosets. We remark that if the map $\tau \equiv 0$
is identically zero, then the associated group $sym^{\times }$ is called 
{\it symmorphic} in crystallography (see, e.g., \cite{Corn1}).

We now define the following subgroups of $sym$, $sym^{\times }:$%
\begin{equation}
\label{pp3fo8}rot=\left\{ \Lambda \in sym\mid \Lambda =\left( 0,R\right)
\right\} \quad , 
\end{equation}
\begin{equation}
\label{pp3fo9}rot^{\times }=rot\cap sym^{\times }\quad , 
\end{equation}
i.e. these are the subsets of $sym$, $sym^{\times }$, respectively, that are
pure (pseudo-Euclidean) ''rotations''. We can now prove the result announced
above:

\subsection{Proposition}

The projection of $sym^{\times }$ onto the ''rotational'' factor coincides
with the set of all pure ''rotations'' in $sym^{\times }$, i.e. 
\begin{equation}
\label{pp3fo10}p_2\left( sym^{\times }\right) =rot^{\times }\quad . 
\end{equation}

\TeXButton{Beweis}{\raisebox{-1ex}{\it Proof :}
\vspace{1ex}}

The inclusion $"\supset "$ is trivial. We prove $"\subset ":$\ If $\left(
0,R\right) \in p_2\left( sym^{\times }\right) $, then there exists a vector $%
t\in W$ (not necessarily in $lat$) so that $\left( 0,R\right) =p_2\left(
t,R\right) $, \underline{and} $\left( t,R\right) $ as well as $\left(
t,R\right) ^{-1}$ are elements of $sym^{\times }$. Now let $t_z\in lat$
arbitrary, then $\left( t_z,0\right) \in sym^{\times }$, and so is the
product $\left( t,R\right) \left( t_z,0\right) \left( t,R\right)
^{-1}=\left( Rt_z,1\right) $. The RHS must be a primitive translation, hence 
$Rt_z\in lat$ for all $t_z$, or $Rlat\subset lat$. The same argument holds
for $R^{-1}$, which says that $Rlat=lat$, or $\left( 0,R\right) \in
rot^{\times }$. \TeXButton{BWE}{\hfill
\vspace{2ex}
$\blacksquare$}

We next show

\subsection{Proposition \label{Tau}}

The restriction of the $\tau $-map to $sym^{\times }$ vanishes identically.

\TeXButton{Beweis}{\raisebox{-1ex}{\it Proof :}
\vspace{1ex}}

Let $\left( t,R\right) \in sym^{\times }$, then the projection onto the
second factor is $\left( 0,R\right) \in p_2\left( sym^{\times }\right)
=rot^{\times }\subset sym^{\times }$, where we have used (\ref{pp3fo10}).
Therefore $\left( t,R\right) $ and $\left( 0,R\right) $ both lie in the same
coset $T_{lat}\cdot \left( t,R\right) $; but this means that $\left(
0,R\right) $ is the unique coset representative that determines the value of
the $\tau $-map on the argument $\left( t,R\right) $. Hence $\tau _R=0$. 
\TeXButton{BWE}{\hfill
\vspace{2ex}
$\blacksquare$}

From proposition \ref{Tau} we infer that every element of $sym^{\times }$
has the form $\left( t_z,R\right) $, where $t_z\in lat$. For every $\left(
t,R\right) \in sym^{\times }$ lies in the coset $T_{lat}\cdot \left(
t,R\right) $, which contains the coset representative $\left( 0,R\right) $.
Hence $\left( t,R\right) $ and $\left( 0,R\right) $ must differ by a
primitive translation, which says that $t=t_z\in lat$.

As a corollary we infer the result we have announced after proposition \ref
{sym}, namely 
\begin{equation}
\label{pp3fo11}p_1\left( sym^{\times }\right) =T_{lat}\quad . 
\end{equation}
Thus, all elements of $sym^{\times }$ have a standard decomposition $\left(
t_z,R\right) =\left( t_z,1\right) \left( 0,R\right) \in T_{lat}\times
rot^{\times }$ according to (\ref{pp3fo10}). Furthermore, the groups $%
T_{lat} $ and $rot^{\times }$ have only the unit element $\left( 0,1\right) $
in common, and $T_{lat}$ is normal in $sym^{\times }$. Thus, we have proven
the

\subsection{Proposition}

$sym^{\times }$ is the semidirect product 
\begin{equation}
\label{pp3fo12}sym^{\times }=T_{lat}\odot rot^{\times }\quad . 
\end{equation}

As an immediate conclusion we see that we must have 
\begin{equation}
\label{pp3fo13}T_{lat}\odot rot\subset sym\quad . 
\end{equation}
We have not examined, however, whether this inclusion is proper, or if $sym$
can contain elements $\left( t,R\right) $ for which $t\not \in lat$.

Finally, we present a condition under which $rot$ coincides with $%
rot^{\times }$; this sheds some light on the question under which
circumstances $sym^{\times }$ is actually a proper subset of $sym$.

\subsection{Theorem \label{Bedingung}}

If $ind\left( lat\right) =0$ or $ind\left( lat\right) =m$ (i.e. minimal or
maximal), then $rot=rot^{\times }$.

\TeXButton{Beweis}{\raisebox{-1ex}{\it Proof :}
\vspace{1ex}}

We first assume that $ind\left( W\right) =0$, i.e. $\left. \eta \right| W$
is positive definite. Let $\Lambda \in rot$. Since $\Lambda $
preserves $lat$, it also preserves its $\TeXButton{R}{\mathbb{R}}$-linear
enveloppe $W$, i.e. $\Lambda W\subset W$. Let $x,y\in W$ arbitrary, then $%
\Lambda x,\Lambda y\in W$. This says that%
$$
\left( \left. \eta \right| W\right) \left[ \left( \left. \Lambda \right|
W\right) x,\left( \left. \Lambda \right| W\right) y\right] =\eta \left(
\Lambda x,\Lambda y\right) =\eta \left( x,y\right) =\left( \left. \eta
\right| W\right) \left[ x,y\right] \quad , 
$$
which says that the restriction $\left. \Lambda \right| W$ of $\Lambda $ to
the subvector space $W$ preserves the bilinear form $\left. \eta \right| W$
on this space. But $\left. \eta \right| W$ is positive definite by
assumption, hence $\left. \Lambda \right| W\in O\left( W\right) $, where $%
O\left( W\right) $ denotes the orthogonal group of $W$.

Now we assume that $\Lambda $ has the property 
\begin{equation}
\label{pp3fo14}\Lambda \in sym\quad ,\quad \text{and\quad }\Lambda ^{-1}\NEG
\in sym\quad , 
\end{equation}
in other words, $\Lambda lat\stackrel{\subset }{\neq }lat$. This means that $%
\left. \Lambda \right| lat$ is not surjective. Hence $\exists \;x\in
lat:\Lambda u\neq x$ for all $u\in lat$. $x$ cannot be zero, since $0\in lat$%
, and $\Lambda $ is linear. Hence $r\equiv \left\| x\right\| >0$, where $%
\left\| x\right\| =\eta \left( x,x\right) $ denotes the Euclidean norm on $W$%
. Now let $S_{m-1}$ be the $\left( m-1\right) $-dimensional sphere 
\underline{in $W$}, centered at $0$. Consider the intersection $sct=lat\cap
r\cdot S_{m-1}$, where $r\cdot S_{m-1}$ is the $\left( m-1\right) $%
-dimensional sphere with radius $r$ in $W$. Note that this set coincides
with the orbit $O_{m-1}\cdot x$ of $x$ under the action of the orthogonal
group $O_{m-1}$, which is a compact subset of $W\simeq \TeXButton{R}
{\mathbb{R}}^m$. From the compactness of $r\cdot S_{m-1}$ it follows that
the number of elements $\#sct$ of $sct$ is {\bf finite}, $0\le \#sct<\infty $%
. Then

\begin{enumerate}
\item  $\left. \Lambda \right| W$ orthogonal$\quad \Rightarrow $\quad $%
\Lambda \left( sct\right) \subset r\cdot S_{n-1}$;

\item  $\Lambda $ lattice preserving\quad $\Rightarrow $\quad $\Lambda
\left( sct\right) \subset lat$;

\item  $\Lambda $ injective\quad $\Rightarrow $\quad $\#\Lambda \left(
sct\right) =\#\left( sct\right) $.
\end{enumerate}

The first two statements imply that $\left. \Lambda \right| W$ preserves $%
sct $, $\left( \left. \Lambda \right| W\right) \left( sct\right) \subset sct$%
; from the third we deduce that $\left( \left. \Lambda \right| W\right)
\left( sct\right) =sct$. But this says that all elements of $sct$ are in the
image of $\left( \left. \Lambda \right| W\right) $, hence $x=\left( \left.
\Lambda \right| W\right) \left( x^{\prime }\right) $ for some $x^{\prime
}\in sct$, which is a contradiction to the result above. This says that our
initial assumption (\ref{pp3fo14}) concerning $\Lambda $ was wrong.

Now assume that $ind\left( lat\right) $ is maximal. Then $\left. \eta
\right| W$ is negative definite, but the argument given above clearly still
applies, since $O_{0,m-1}\simeq O_{m-1,0}$, and the only point in the proof
was the compactness of the $O_{m-1}$-orbits. This completes our proof. 
\TeXButton{BWE}{\hfill
\vspace{2ex}
$\blacksquare$}

We see that the structure of the proof relies on the compactness of orbits $%
O\cdot x$ of $x$ under the orthogonal group, which, in turn, comes from the
fact that the orthogonal groups $O$ are compact. If the metric restricted to 
$\left[ lat\right] $ were pseudo-Euclidean instead, we could have
non-compact orbits, related to the non-compactness of the groups $O_{t,s}$.
We have not proved this in full, but we conjecture that the converse of
theorem \ref{Bedingung} should read:

''If $0<ind\left( lat\right) <m=\dim _{\TeXButton{R}{\mathbb{R}}}\left[
lat\right] $, then $rot^{\times }\stackrel{\subset }{\neq }rot$''.

An explicit example of this situation will be constructed now.

\section{Identifications over a lightlike lattice \label{Sc.5}}

Given a lattice $lat$ in a pseudo-Euclidean space $M=\TeXButton{R}
{\mathbb{R}}_t^n$, we have the associated group of primitive translations $%
\Gamma =T_{lat}$. We want to study the quotient space $M/\Gamma $, its
isometry group $I\left( M/\Gamma \right) =N\left( \Gamma \right) /\Gamma $,
and the possible extension $eN\left( \Gamma \right) /\Gamma $ of this
isometry group. We now show how $N\left( \Gamma \right) $ and $eN\left(
\Gamma \right) $ are related to the sets $rot^{\times }$ and $rot$,
respectively.

An element $\left( t,R\right) \in E_t^n$ is in the extended normalizer $%
eN\left( \Gamma \right) $ iff $\left( t,R\right) \Gamma \left( t,R\right)
^{-1}\subset \Gamma $, where $\left( t,R\right) ^{-1}$ is the inverse of $%
\left( t,R\right) $ in $E_t^n$. This is true iff $\left( Rt_z,1\right) \in
\Gamma $ for all $t_z\in lat$, hence iff $Rlat\subset lat$, hence iff $%
\left( 0,R\right) \in sym$, hence iff $\left( 0,R\right) \in rot$. On the
other hand, $\left( t,R\right) $ is in the normalizer $N\left( \Gamma
\right) $ if the same condition holds for $R^{-1}$ as well, i.e. $R^{-1}\in
rot$. But $R,R^{-1}\in rot$ is true iff $R\in rot^{\times }$. Hence 
\begin{equation}
\label{pp3fo15}eN\left( \Gamma \right) =\left\{ \left( t,R\right) \in
E_t^n\mid R\in rot\right\} =\TeXButton{R}{\mathbb{R}}^n\odot rot\quad , 
\end{equation}
\begin{equation}
\label{pp3fo16}N\left( \Gamma \right) =\left\{ \left( t,R\right) \in
E_t^n\mid R\in rot^{\times }\right\} =\TeXButton{R}{\mathbb{R}}^n\odot
rot^{\times }\quad . 
\end{equation}

From now on we focus on the Lorentzian metric, $t=1$, and work in $M=%
\TeXButton{R}{\mathbb{R}}_1^n$. In the following we examine in detail the
extended normalizer of a group $\Gamma =T_{lat}$, whose associated lattice $%
lat$ is given as follows: The basis vectors $\underline{u}=\left(
u_{+},u_1\ldots ,u_{m-1}\right) $ of the lattice contain {\bf one} lightlike
vector, namely $u_{+}$, and $\left( m-1\right) $ spacelike vectors $%
u_1,\ldots ,u_{m-1}$. It is assumed that $u_{+}\perp \left[ u_1,\ldots
,u_{m-1}\right] _{\TeXButton{R}{\mathbb{R}}}$, which is necessary to
guarantee that the $\TeXButton{R}{\mathbb{R}}$-linear span of all basis
vectors, $W\equiv \left[ \underline{u}\right] _{\TeXButton{R}{\mathbb{R}}}$,
is indeed a lightlike vector subspace of $\TeXButton{R}{\mathbb{R}}_1^n$. As
is well known, this means that the restriction $\left. \eta \right| W$ of
the metric to $W$ is degenerate; and furthermore, that $W$ contains a $1$%
-dimensional lightlike vector subspace, in this case given by $\left[
u_{+}\right] _{\TeXButton{R}{\mathbb{R}}}$, but no lightlike vector
otherwise. Applying a suitable Lorentz transformation it can be assumed
without loss of generality that $\left( 1\right) $ the vector subspace $%
U\equiv \left[ u_1,\ldots ,u_{m-1}\right] _{\TeXButton{R}{\mathbb{R}}}$
coincides with the span of the last $\left( m-1\right) $ canonical basis
vectors of $\TeXButton{R}{\mathbb{R}}_1^n$, i.e. $U=\left[ e_{n-m+1},\ldots
,e_{n-1}\right] _{\TeXButton{R}{\mathbb{R}}}$, where $\TeXButton{R}
{\mathbb{R}}_1^n=\left[ e_0,\ldots ,e_{n-1}\right] _{\TeXButton{R}
{\mathbb{R}}}$; and $\left( 2\right) $, that $u_{+}=\frac{e_0+e_1}{\sqrt{2}}$%
. We want to compute $N\left( \Gamma \right) $ and $eN\left( \Gamma \right) $
for this lattice, where $\Gamma =T_{lat}$. According to (\ref{pp3fo15}, \ref
{pp3fo16}) our first task is to identify the sets $rot$ and $rot^{\times }$.
We do this in several steps. Firstly, we identify the subset of $E_1^n$ that
preserves the $1$-dimensional lightlike subspace $\left[ u_{+}\right] _{%
\TeXButton{R}{\mathbb{R}}}$.

\subsection{Preservation of a lightlike $1$-dimensional subspace}

Let $\underline{e}\equiv \left( e_0,e_1,\ldots ,e_{n-1}\right) $ denote the
canonical basis of $\TeXButton{R}{\mathbb{R}}_1^n$. Construct two lightlike
vectors $u_{+,-}\equiv \frac 1{\sqrt{2}}\left( e_0\pm e_1\right) $, and
consider the new basis $\underline{b}\equiv \left( u_{+},u_{-},e_2,\ldots
,e_{n-1}\right) $. The transformation between the two bases is accomplished
by $T=diag\left( \frac 1{\sqrt{2}}\left( 
\begin{array}{cc}
1 & 1 \\ 
1 & -1 
\end{array}
\right) ,{\bf 1}\right) $, with $T^2=1$, so that $\underline{b}=\underline{e}%
T$ and $\underline{e}=\underline{b}T$. In the $\underline{b}$-basis, the
matrix $\eta _{\underline{b}}$ of $\eta $ takes the form $\eta _{\underline{b%
}}=diag\left( \left( 
\begin{array}{cc}
0 & -1 \\ 
-1 & 0 
\end{array}
\right) ,{\bf 1}\right) $.

We want to find the set $rot\left( \left[ u_{+}\right] _{\TeXButton{R}
{\mathbb{R}}};O_{1,n-1}\right) $ of elements $R\in O_{1,n-1}$ that preserve
this subspace, i.e. $\Phi \left( R\right) u_{+}=\lambda \cdot u_{+}$ with $%
\TeXButton{R}{\mathbb{R}}\ni \lambda \neq 0$. Here $\Phi \left( R\right) $
denotes the linear operator associated with $R$, acting according to $\Phi
\left( R\right) \underline{b}=\underline{b}R$. Note the notation
conventions: We write $R$ for the abstract group elemenent as well as for
the matrix representing $\Phi \left( R\right) $ in a particular basis.

From $\Phi \left( R\right) \underline{b}=\underline{b}R$ we see that the
matrix representing $\Phi \left( R\right) $ must take the general form
$
R=\left( 
\begin{array}{cc}
\begin{array}{c}
{\lambda } \\ {\bf 0} 
\end{array}
& * 
\end{array}
\right) $
, where $"*"$ denotes ''something''. We can make an Ansatz for $R$
according to this form, and then impose the condition $R^T\eta _{\underline{b%
}}R=\eta _{\underline{b}}$ that expresses that $R$ is a Lorentz
transformation. This yields a set of matrices $R=\left( V,a,C\right) $
parametrized by $V\in \TeXButton{R}{\mathbb{R}}^{n-2}\subset \TeXButton{R}
{\mathbb{R}}_1^n$ canonically embedded in $\TeXButton{R}{\mathbb{R}}_1^n$
according to $V\leftrightarrow \left( 
\begin{array}{c}
0 \\ 
0 \\ 
V 
\end{array}
\right) $; $a\in \TeXButton{R}{\mathbb{R}}^{\times }$, where $\TeXButton{R}
{\mathbb{R}}^{\times }=\TeXButton{R}{\mathbb{R}}-\left\{ 0\right\} $ denotes
the multiplicative group of units of $\TeXButton{R}{\mathbb{R}}$; and $C\in
O_{n-1}\subset O_{1,n-1}$, canonically embedded according to $%
C\leftrightarrow \left( 
\begin{array}{ccc}
1 & 0 & 0 \\ 
0 & 1 & 0 \\ 
0 & 0 & C 
\end{array}
\right) \in O_{1,n-1}$. The matrices $\left( V,a,C\right) $ take the form 
\begin{equation}
\label{pp3fo17}\left( V,a,C\right) =\left( 
\begin{array}{ccc}
a & \frac{V^2}a & - 
\sqrt{2}V^TC \\ 0 & \frac 1a & 0 \\ 
0 & -\frac{\sqrt{2}V}a & C 
\end{array}
\right) \quad , 
\end{equation}
where $V^2=\sum_{i=2}^{n-1}V_i^2$ denotes the Euclidean quadratic form on $%
\TeXButton{R}{\mathbb{R}}^{n-2}$. These matrices satisfy the group law 
\begin{equation}
\label{pp3fo18}\left( V,a,C\right) \left( V^{\prime },a^{\prime },C^{\prime
}\right) =\left( aCV^{\prime }+V,aa^{\prime },CC^{\prime }\right) \quad , 
\end{equation}
with unit $\left( 0,1,1\right) $, and inverses 
\begin{equation}
\label{pp3fo19}\left( V,a,C\right) ^{-1}=\left( -\frac 1aC^{-1}V,\frac
1a,C^{-1}\right) \quad . 
\end{equation}
Using the group law (\ref{pp3fo18}) we find the standard decomposition of
elements 
\begin{equation}
\label{pp3fo20}\left( V,a,C\right) =\left( V,1,1\right) \left( 0,a,1\right)
\left( 0,1,C\right) \quad , 
\end{equation}
this decomposition being chosen so that factors which form normal subgroups
stand to the left, as it will be shown now. We firstly identify three
subgroups of $G=rot\left( \left[ u_{+}\right] _{\TeXButton{R}{\mathbb{R}}%
};O_{1,n-1}\right) $: The set of all $\left( V,1,1\right) $ forms an Abelian
subgroup of $G$, which is isomorphic to $\TeXButton{R}{\mathbb{R}}^{n-2}$,
as can be seen from the group law 
\begin{equation}
\label{pp3fo21}\left( V,1,1,\right) \left( V^{\prime },1,1\right) =\left(
V+V^{\prime },1,1\right) \quad . 
\end{equation}
The set of all $\left( 0,a,1\right) $ is a subgroup of $G$ with group law $%
\left( 0,a,1\right) \left( 0,a^{\prime },1\right) =\left( 0,aa^{\prime
},1\right) $, which will continue to be denoted by $\TeXButton{R}{\mathbb{R}}%
^{\times }$, and the set of all $\left( 0,1,C\right) $ clearly is a subgroup
isomorphic to $O_{n-2}$. Since $\left( 0,a,1\right) \left( 0,1,C\right)
=\left( 0,1,C\right) \left( 0,a,1\right) $, the last two subgroups form a
direct product subgroup $\TeXButton{R}{\mathbb{R}}^{\times }\otimes O_{n-2}$
of $G$. Furthermore, using the group law (\ref{pp3fo18}) again, we see that
conjugation $I\left( U,a,C\right) $ [where $I\left( g\right) h=ghg^{-1}$] of
an element $\left( V,1,1\right) $ of $\TeXButton{R}{\mathbb{R}}^{n-2}$
yields again a translation, 
\begin{equation}
\label{pp3fo22}I\left( U,a,C\right) \left( V,1,1\right) =\left(
aCV,1,1\right) \quad , 
\end{equation}
from which it follows that $\TeXButton{R}{\mathbb{R}}^{n-2}$ is a normal
subgroup of $G$. This implies that $G$ has the structure of a semidirect
product 
\begin{equation}
\label{pp3fo23}G=rot\left( \left[ u_{+}\right] _{\TeXButton{R}{\mathbb{R}}%
};O_{1,n-1}\right) \;\simeq \;\TeXButton{R}{\mathbb{R}}^{n-2}\odot \left[ 
\TeXButton{R}{\mathbb{R}}^{\times }\otimes O_{n-2}\right] \quad , 
\end{equation}
where $"\odot "$ denotes a semidirect product, and the normal factor $%
\TeXButton{R}{\mathbb{R}}^{n-2}$ stands to the left.

We see that this group has four connected components: They are obtained by
pairing the two connected components $\left( \TeXButton{R}{\mathbb{R}}_{+},%
\TeXButton{R}{\mathbb{R}}_{-}\right) $ of $\TeXButton{R}{\mathbb{R}}^{\times
}$ with the two connected components $\left( SO_{n-2},O_{n-2}^{-}\right) $
of $O_{n-2}$. $\TeXButton{R}{\mathbb{R}}_{-}$ reverses the time direction,
whereas $O_{n-2}^{-}$ reverses spatial orientation. The identity component $%
G_0$ of $G$ is obviously 
\begin{equation}
\label{pp3fo24}G_0=\TeXButton{R}{\mathbb{R}}^{n-2}\odot \left[ \TeXButton{R}
{\mathbb{R}}_{+}\otimes SO_{n-2}\right] \quad . 
\end{equation}
In what follows we shall restrict attention to $G_0$. If we had started this
section with $SO_{1,n-1}^{+}$, then our analysis would naturally render the
identity component $G_0$ for $rot\left( \left[ u_{+}\right] _{\TeXButton{R}
{\mathbb{R}}};O_{1,n-1}\right) $. We adapt our notation to this fact by
denoting as $rot\left( \left[ u_{+}\right] _{\TeXButton{R}{\mathbb{R}}%
};H\right) $ the set of all elements in the group $H\subseteq O_{1,n-1}$
that preserve $\left[ u_{+}\right] $. Then we can conclude this section with
the results 
\begin{equation}
\label{pp3fo25}rot\left( \left[ u_{+}\right] _{\TeXButton{R}{\mathbb{R}}%
};O_{1,n-1}\right) \;\simeq \;\TeXButton{R}{\mathbb{R}}^{n-2}\odot \left[ 
\TeXButton{R}{\mathbb{R}}^{\times }\otimes O_{n-2}\right] \quad , 
\end{equation}
\begin{equation}
\label{pp3fo26}rot\left( \left[ u_{+}\right] _{\TeXButton{R}{\mathbb{R}}%
};SO_{1,n-1}^{+}\right) \;\simeq \;\TeXButton{R}{\mathbb{R}}^{n-2}\odot
\left[ \TeXButton{R}{\mathbb{R}}_{+}\otimes SO_{n-2}\right] \quad . 
\end{equation}

We now give the explicit form of the matrices $\left( V,a,C\right) $ etc. in
the basis $\underline{e}$. Performing a similarity transformation with $T$
then yields, using (\ref{pp3fo17}), 
\begin{equation}
\label{pp3fo27}\left( V,a,C\right) =\left( 
\begin{array}{ccc}
\frac{a+\frac 1a}2+\frac{V^2}{2a} & \frac{a-\frac 1a}2-\frac{V^2}{2a} & 
-V^TC \\ 
\frac{a-\frac 1a}2+\frac{V^2}{2a} & \frac{a+\frac 1a}2-\frac{V^2}{2a} & 
-V^TC \\ 
-\frac Va & \frac Va & {\bf 1}_{n-2} 
\end{array}
\right) \quad . 
\end{equation}
This gives, in particular, 
\begin{equation}
\label{pp3fo28}\left( V,1,1\right) =\left( 
\begin{array}{ccc}
1+\frac{V^2}{2a} & -\frac{V^2}{2a} & -V^T \\ 
\frac{V^2}{2a} & 1-\frac{V^2}{2a} & -V^T \\ 
-V & V & {\bf 1}_{n-2} 
\end{array}
\right) \quad , 
\end{equation}
\begin{equation}
\label{pp3fo29}\left( 0,a,1\right) =\left( 
\begin{array}{cc}
{\rm sgn}\left( a\right) \cdot \left( 
\begin{array}{cc}
\cosh \phi & \sinh \phi \\ 
\sinh \phi & \cosh \phi 
\end{array}
\right) & 0 \\ 
0 & {\bf 1}_{n-2} 
\end{array}
\right) \quad , 
\end{equation}
where $\cosh \phi =\left| \frac{a+\frac 1a}2\right| $ and 
\begin{equation}
\label{pp3fo30}\left( 0,1,C\right) =\left( 
\begin{array}{cc}
{\bf 1}_2 & 0 \\ 
0 & C 
\end{array}
\right) \quad . 
\end{equation}

\subsection{Conformal algebra $cf_{n-2}$}

Before we investigate the Lie algebra of the Lie group $rot\left( \left[
u_{+}\right] _{\TeXButton{R}{\mathbb{R}}};SO_{1,n-1}^{+}\right) $, we
briefly explain the relation between the conformal algebra $cf_{n-2}$ in $%
\left( n-2\right) $ Euclidean dimensions and the Lorentz algebra $so_{1,n-1}$%
. $cf_{n-2}$ is spanned by generators $\left[ \left( L_{ij}\right) _{2\le
i<j\le n-1};\left( K_i,S_j\right) _{i,j=2,\ldots ,n-1};\Delta \right] $,
where $L_{ij}$ and $K_i$ span the Euclidean algebra $so_{n-2}$, $s_j$ are
the generators of special conformal transformations, and $\Delta $ generates
dilations. This basis obeys the relations 
\begin{equation}
\label{pp3fo31}
\begin{array}{c}
\left[ L_{ij},L_{km}\right] =\delta _{ik}\cdot L_{jm}+\delta _{jm}\cdot
L_{ik}-\delta _{im}\cdot L_{jk}-\delta _{jk}\cdot L_{im}\quad , \\ 
\left[ L_{ij},K_k\right] =\delta _{ik}\cdot K_j-\delta _{jk}\cdot K_i\quad ,
\\ 
\left[ L_{ij},S_k\right] =\delta _{ik}\cdot S_j-\delta _{jk}\cdot S_i\quad ,
\\ 
\left[ S_i,K_j\right] =2\left( L_{ij}-\delta _{ij}\cdot \Delta \right) \quad
, \\ 
\left[ K_i,K_j\right] =\left[ S_i,S_j\right] =0\quad , \\ 
\left[ L_{ij},\Delta \right] =0\quad , \\ 
\left[ \Delta ,K_i\right] =K_i\quad , \\ 
\left[ \Delta ,S_j\right] =-S_j\quad . 
\end{array}
\end{equation}
The first two lines contain the $so_{n-2}$ subalgebra.

The generators of $so_{1,n-1}$, on the other hand, are real $\left(
n,n\right) $ matrices $L_{\mu \nu }$ defined by 
\begin{equation}
\label{pp3fo32}\left( L_{\mu \nu }\right) _{\;b}^a=-\delta _\mu ^a\cdot \eta
_{\nu b}+\delta _\nu ^a\cdot \eta _{\mu a}\quad , 
\end{equation}
satisfying 
\begin{equation}
\label{pp3fo33}\left[ L_{\mu \nu },L_{\rho \sigma }\right] =\eta _{\mu \rho
}\cdot L_{\nu \sigma }+\eta _{\nu \sigma }\cdot L_{\mu \rho }-\eta _{\mu
\sigma }\cdot L_{\nu \rho }-\eta _{\nu \rho }\cdot L_{\mu \sigma }\quad . 
\end{equation}
Now we transform the basis $\left( L_{\mu \nu }\right) $ to a new basis%
$$
\left( -L_{01}\,;\,L_{0k}+L_{1k}\,;\,L_{0k}-L_{1i}\,;\,L_{ij}\right) \;= 
$$
\begin{equation}
\label{pp3fo34}=\;\left( \Delta \,;\,K_k\,;\,S_k\,;\,L_{ij}\right) \quad
,\quad k=2,\ldots ,n-1\quad ;\quad 2\le i<j\le n-1\quad . 
\end{equation}
Using (\ref{pp3fo17}) it is easy to verify that this new basis satisfies the
algebra (\ref{pp3fo31}), so that we have the well-known isomorphism of Lie
algebras 
\begin{equation}
\label{pp3fo35}cf_{n-2}\simeq so_{1,n-1}\quad . 
\end{equation}

We now can turn to evaluate the Lie algebra $\hat g_0$ of $rot\left( \left[
u_{+}\right] _{\TeXButton{R}{\mathbb{R}}};SO_{1,n-1}^{+}\right) $. The
generators of the subgroup $\TeXButton{R}{\mathbb{R}}^{n-2}$ with elements $%
\left( V,1,1\right) $ are obtained from (\ref{pp3fo28}) by 
\begin{equation}
\label{pp3fo36}\frac d{dt}\left( t\cdot V,1,1\right)
_{t=0}=\sum_{i=2}^{n-1}V^i\cdot \left( L_{0i}+L_{1i}\right)
=\sum_{i=2}^{n-1}V^i\cdot K_i\quad , 
\end{equation}
with $L_{\mu \nu }$ from (\ref{pp3fo32}), and using the new basis (\ref
{pp3fo34}). Similarly, 
\begin{equation}
\label{pp3fo37}\frac d{dt}\left( 0,t\cdot a,1\right) _{t=0}=-L_{01}=\Delta
\quad . 
\end{equation}
The generators of the $SO_{n-2}$-factor clearly are the elements $\left(
L_{ij}\right) _{2\le i<j\le n-1}$. Thus we see that the Lie algebra $\hat
g_0 $ is a Lie subalgebra of the conformal algebra $cf_{n-2}$ in $\left(
n-2\right) $ Euclidean dimensions; $\hat g_0$ is spanned precisely by those
generators of $cf_{n-2}$, that either annihilate the lightlike vector $u_{+}$%
, or leave it invariant, i.e. 
\begin{equation}
\label{pp3fo38}L_{ij}u_{+}=K_iu_{+}=0\quad ,\quad \Delta u_{+}=u_{+}\quad . 
\end{equation}

\subsection{Preservation of $W=\left[ u_{+}\right] _{\TeXButton{R}
{\mathbb{R}}}\oplus U$}

Consider the subvector spaces $U=\left[ e_{n-m+1},\ldots ,e_{n-1}\right] _{%
\TeXButton{R}{\mathbb{R}}}$ and $W=\left[ u_{+}\right] _{\TeXButton{R}
{\mathbb{R}}}\oplus U$ introduced at the beginning of section \ref{Sc.5}.
Our next question is: Which elements $R$ of $SO_{1,n-1}^{+}$ preserve $W$ in
the sense that $\Phi \left( R\right) W\subset W$? For the sake of
simplicity, we restrict attention to the identity component $SO_{1,n-1}^{+}$
here. This set of elements is again a subgroup and will be denoted by $%
rot\left( W;SO_{1,n-1}^{+}\right) \ $or $rot\left( W\right) $, if no
confusion is likely. Clearly, $rot\left( W\right) \stackunder{\neq }{\subset 
}rot\left( \left[ u_{+}\right] _{\TeXButton{R}{\mathbb{R}}%
};SO_{1,n-1}^{+}\right) $, hence we only need to examine which of the
generators $L_{ij},K_k,\Delta $ discussed in the previous section map $%
e_{n-m+1},\ldots ,e_{n-1}$ into $W$. An easy computation shows that $\Delta $
annihilates $U$, i.e. 
\begin{equation}
\label{pp3fo38a}\Delta U=0\quad , 
\end{equation}
that 
\begin{equation}
\label{pp3fo38b}
\begin{array}{ccc}
K_iU=0 & \text{for} & 2\le i\le n-m \\ 
K_ie_j=-\sqrt{2}\delta _{ij}\cdot u_{+} & \text{for} & n-m+1\le i,j\le n-1 
\end{array}
\quad , 
\end{equation}
and that all generators $L_{ij}$ with $2\le i\le n-m$ but $n-m+1\le j\le n-1$
are broken, so that we the remaining $L$-generators satisfy 
\begin{equation}
\label{pp3fo38c}
\begin{array}{ccc}
L_{ij}U=0 & \text{for} & 2\le i<j\le n-m \\ 
L_{ij}U\stackrel{\text{orth.}}{\subset }U & \text{for} & n-m+1\le i<j\le n-1 
\end{array}
\quad . 
\end{equation}
For the sake of simplicity we now denote indices ranging in $\left\{
2,\ldots ,n-m\right\} $ as $a,b$, etc.; those ranging in $\left\{
n-m+1,\ldots ,n-1\right\} $ as greek $\mu ,\nu $, etc.; and those ranging in 
$\left\{ 2,\ldots ,n-1\right\} $ as $i,j$, etc.\ Then the remaining
generators that preserve $W$ can be written as $\left( \Delta
;K_i;L_{ab};L_{\mu \nu }\right) $; their algebra is 
\begin{equation}
\label{pp3fo39}
\begin{array}{c}
\left[ \Delta ,K_i\right] =K_i\quad . \\ 
\left[ \Delta ,L_{ab}\right] =\left[ \Delta ,L_{\mu \nu }\right] =0\quad .
\\ 
\left[ K_i,K_j\right] =0\quad . \\ 
\left[ L_{ab},K_c\right] =\delta _{ac}\cdot K_b-\delta _{bc}\cdot K_a\quad .
\\ 
\left[ L_{ab},K_\rho \right] =0\quad . \\ 
\left[ L_{\mu \nu },K_a\right] =0\quad . \\ 
\left[ L_{\mu \nu },K_\rho \right] =\delta _{\mu \rho }\cdot K_\nu -\delta
_{\nu \rho }\cdot K_\mu \quad . \\ 
\left[ L_{ab},L_{cd}\right] =\delta _{ac}\cdot L_{bd}+\delta _{bd}\cdot
L_{ac}-\delta _{ad}\cdot L_{bc}-\delta _{bc}\cdot L_{ad}\quad . \\ 
\left[ L_{\mu \nu },L_{\gamma \delta }\right] =\delta _{\mu \gamma }\cdot
L_{\nu \delta }+\delta _{\nu \delta }\cdot L_{\mu \gamma }-\delta _{\mu
\delta }\cdot L_{\nu \gamma }-\delta _{\nu \gamma }\cdot L_{\mu \delta
}\quad . \\ 
\left[ L_{ab},L_{\mu \nu }\right] =0\quad . 
\end{array}
\end{equation}
This is the Lie algebra $Lie\left( rot\left( W\right) \right) $ of $%
rot\left( W\right) $. We see immediately that we have two subalgebras
isomorphic to the Euclidean algebras $Lie\left( E^{n-m-1}\right) $ and $%
Lie\left( E^{m-1}\right) $, which are spanned by $\left( L_{ab};K_c\right) $
and $\left( L_{\mu \nu };K_\rho \right) $, respectively. These subalgebras
commute. Their direct sum $Lie\left( E^{n-m-1}\right) \oplus Lie\left(
E^{m-1}\right) $ is an ideal in the full algebra, in which the dilation
generator $\Delta $ acts non-trivially only on the generators $K_i$. On the
other hand, we can combine the $K$-generators with $\Delta $ to define a
subalgebra $A=\left[ \Delta ,K_i\right] _{\TeXButton{R}{\mathbb{R}}}$, which
is also an ideal in $Lie\left( rot\left( W\right) \right) $.

On exponentiation of this algebra we obtain a covering group of $rot\left(
W\right) $; hence we must have 
\begin{equation}
\label{pp3fo40}rot\left( W\right) \simeq \left[ E^{n-m-1}\otimes
E^{m-1}\right] \odot \TeXButton{R}{\mathbb{R}}_{+}\quad , 
\end{equation}
where the normal factor $E^{n-m-1}\otimes E^{m-1}$ is written to the left of
the multiplicative subgroup $\TeXButton{R}{\mathbb{R}}_{+}$. The group law
can be derived from (\ref{pp3fo18}), if we make a split 
\begin{equation}
\label{pp3fo41}V=\sum_{a=2}^{n-m}V^ae_a\;+\sum_{\mu =n-m+1}^{n-1}V^\mu e_\mu
\;=V_1+V_2\quad , 
\end{equation}
and 
\begin{equation}
\label{pp3fo42}
\begin{array}{c}
C=C_1C_2=C_2C_1\quad ; \\ 
C_1\in SO_{n-m-1}\subset SO_{1,n-1}\quad ;\quad C_2\in SO_{m-1}\subset
SO_{1,n-1}\quad . 
\end{array}
\end{equation}
We observe that, according to the algebra (\ref{pp3fo39}), $C_1$ acts
trivially on $V_2$ and $C_2$ acts trivially on $V_1$, and that $C_1$
commutes with $C_2$. Thus, elements of $rot\left( W\right) $ will be denoted
by 
\begin{equation}
\label{pp3fo43}\left( V,a,C\right) \equiv \left( V_1,C_1,V_2,C_2,a\right)
\quad , 
\end{equation}
With these remarks, the group law for elements (\ref{pp3fo43}) can be
derived from (\ref{pp3fo18}) to be 
$$
\left( V_1,C_1,V_2,C_2,a\right) \left( V_1^{\prime },C_1^{\prime
},V_2^{\prime },C_2^{\prime },a^{\prime }\right) \;= 
$$
\begin{equation}
\label{pp3fo44}=\;\left( aC_1V_1^{\prime }+V_1,C_1C_1^{\prime
},aC_2V_2^{\prime }+V_2,C_2C_2^{\prime },aa^{\prime }\right) \quad . 
\end{equation}
Thus, elements $\left( V_1,C_1,V_2,C_2,a\right) $ decompose according to 
\begin{equation}
\label{pp3fo45}
\begin{array}{c}
\left( V_1,C_1,V_2,C_2,a\right) =\left( V_1,C_1,0,1,1\right) \left(
0,1,V_2,C_2,1\right) \left( 0,1,0,1,a\right) \quad , \\ 
\left( V_1,C_1,0,1,1\right) =\left( V_1,1,0,1,1\right) \left(
0,C_1,0,1,1\right) \quad , \\ 
\left( 0,1,V_2,C_2,1\right) =\left( 0,1,V_2,1,1\right) \left(
0,1,0,C_2,1\right) \quad . 
\end{array}
\end{equation}

We now describe the relationship between exponentiated elements of the Lie
algebra (\ref{pp3fo39}) and the group elements $\left(
V_1,C_1,V_2,C_2,a\right) $. Using straightforward matrix algebra, the
commutation relations (\ref{pp3fo39}) and the decomposition (\ref{pp3fo45})
we find that 
$$
\exp \left[ \sum_{a=2}^{n-m}V_1^a\cdot K_a+\sum_{\mu =n-m+1}^{n-1}V_2^\mu
\cdot K_\mu \right] =\left( V_1,1,V_2,1,1\right) = 
$$
\begin{equation}
\label{pp3fo46}=\left( V_1^2,\ldots ,V_1^{n-m};1;V_2^{n-m+1},\ldots
,V_2^{n-1};1;1\right) \quad , 
\end{equation}
\begin{equation}
\label{pp3fo47}\exp \left( \phi \cdot \Delta \right) =\left( 0,1,0,1,\exp
\phi \right) \quad , 
\end{equation}
and 
$$
\exp \left( \sum_{2\le a<b\le n-m}\omega _1^{ab}\cdot L_{ab}\right) =\left(
0,C_1\left( \omega _1\right) ,0,1,1\right) \quad ; 
$$
\begin{equation}
\label{pp3fo48}\exp \left( \sum_{n-m+1\le \mu <\nu \le n-1}\omega _2^{\mu
\nu }\cdot L_{\mu \nu }\right) =\left( 0,1,0,C_2\left( \omega _2\right)
,1\right) \quad . 
\end{equation}

We finish this subsection with computing the action of group elements $%
\left( V_1,C_1,V_2,C_2,a\right) $ on the transformed basis $\underline{b}$.
We use the relations (\ref{pp3fo38a}-\ref{pp3fo38c}), which we supplement by
the action of the $\left( \Delta ,K_a,K_\mu ,L_{ab},L_{\mu \nu }\right) $%
-basis on the basis vectors of the $\TeXButton{R}{\mathbb{R}}$-linear span 
\begin{equation}
\label{pp3fo49}M^{\prime }\equiv \left[ u_{-},e_2,\ldots ,e_{n-m}\right]
\quad . 
\end{equation}
From the basis transformation introduced at the beginning of this subsection
we see that we have 
\begin{equation}
\label{pp3fo50}M=\left[ u_{+}\right] \oplus M^{\prime }\oplus U\quad , 
\end{equation}
where $M=\TeXButton{R}{\mathbb{R}}_1^n$. The action of $\left( \Delta
,K_a,K_\mu ,L_{ab},L_{\mu \nu }\right) $ on the basis $\underline{b}$ of $M$
is now given by 
\begin{equation}
\label{pp3fo51}
\begin{array}{rclcrclc}
\Delta u_{+} & = & u_{+} & . & \Delta u_{-} & = & -u_{-} & . \\ 
K_au_{+} & = & 0 & . & K_au_{-} & = & -\sqrt{2}e_a & . \\ 
K_\mu u_{+} & = & 0 & . & K_\mu u_{-} & = & -\sqrt{2}e_\mu & . \\ 
L_{ab}u_{+} & = & 0 & . & L_{ab}u_{-} & = & 0 & . \\ 
L_{\mu \nu }u_{+} & = & 0 & . & L_{\mu \nu }u_{-} & = & 0 & . 
\end{array}
\end{equation}
\begin{equation}
\label{pp3fo52}
\begin{array}{rclcrclc}
\Delta e_a & = & 0 & . & \Delta e_\mu & = & 0 & . \\ 
K_ae_b & = & -\sqrt{2}\delta _{ab}\cdot u_{+} & . & K_ae_\mu & = & 0 & . \\ 
K_\mu e_a & = & 0 & . & K_\mu e_\nu & = & -\sqrt{2}\delta _{\mu \nu }\cdot
u_{+} & . \\ 
L_{ab}e_c & = & \delta _{ac}\cdot e_b-\delta _{bc}\cdot e_a & . & 
L_{ab}e_\mu & = & 0 & . \\ 
L_{\mu \nu }e_a & = & 0 & . & L_{\mu \nu }e_\rho & = & \delta _{\mu \rho
}\cdot e_\nu -\delta _{\nu \rho }\cdot e_\mu & . 
\end{array}
\end{equation}

With the help of (\ref{pp3fo46}-\ref{pp3fo48}), the commutation relations (%
\ref{pp3fo39}) and formulas (\ref{pp3fo51}-\ref{pp3fo52}) we can derive the
action of elements $\left( V_1,C_1,V_2,C_2,a\right) $ on $\left[
u_{+}\right] \oplus M^{\prime }\oplus U$. A calculation gives 
\begin{equation}
\label{pp3fo53}
\begin{array}{rclc}
\left( 0,1,0,1,e^\phi \right) u_{+} & = & e^\phi \cdot u_{+} & . \\ 
\left( V_1,1,0,1,1\right) u_{+} & = & u_{+} & . \\ 
\left( 0,1,V_2,1,1\right) u_{+} & = & u_{+} & . \\ 
\left( 0,C_1,0,1,1\right) u_{+} & = & u_{+} & . \\ 
\left( 0,1,0,C_2,1\right) u_{+} & = & u_{+} & . 
\end{array}
\end{equation}
\begin{equation}
\label{pp3fo54}
\begin{array}{rclc}
\left( 0,1,0,1,e^\phi \right) u_{-} & = & e^{-\phi }\cdot u_{-} & . \\ 
\left( V_1,1,0,1,1\right) u_{-} & = & \left( V_1\right) ^2\cdot u_{+}+u_{-}-%
\sqrt{2}V_1 & . \\ 
\left( 0,1,V_2,1,1\right) u_{-} & = & \left( V_2\right) ^2\cdot u_{+}+u_{-}-%
\sqrt{2}V_2 & . \\ 
\left( 0,C_1,0,1,1\right) u_{-} & = & u_{-} & . \\ 
\left( 0,1,0,C_2,1\right) u_{-} & = & u_{-} & . 
\end{array}
\end{equation}
Here $\left( V_1\right) ^2=\sum_{a=1}^{n-m}\left( V_1^a\right) ^2$, $\left(
V_2\right) ^2=\sum_{\mu =n-m+1}^{n-1}\left( V_2^\mu \right) ^2$.
Furthermore, 
\begin{equation}
\label{pp3fo55}
\begin{array}{rclc}
\left( 0,1,0,1,e^\phi \right) e_a & = & e_a & . \\ 
\left( V_1,1,0,1,1\right) e_a & = & -\sqrt{2}V_1^a\cdot u_{+}+e_a & . \\ 
\left( 0,1,V_2,1,1\right) e_a & = & e_a & . \\ 
\left( 0,C_1,0,1,1\right) e_a & = & \sum_{b=2}^{n-m}\left( C_1\right)
_{ab}\cdot e_b & . \\ 
\left( 0,1,0,C_2,1\right) e_a & = & e_a & . 
\end{array}
\end{equation}
\begin{equation}
\label{pp3fo56}
\begin{array}{rclc}
\left( 0,1,0,1,e^\phi \right) e_\mu & = & e_\mu & . \\ 
\left( V_1,1,0,1,1\right) e_\mu & = & e_\mu & . \\ 
\left( 0,1,V_2,1,1\right) e_\mu & = & -\sqrt{2}V_2^\mu \cdot u_{+}+e_\mu & .
\\ 
\left( 0,C_1,0,1,1\right) e_\mu & = & e_\mu & . \\ 
\left( 0,1,0,C_2,1\right) e_\mu & = & \sum_{b=n-m+1}^{n-1}\left( C_2\right)
_{\mu \nu }\cdot e_\nu & . 
\end{array}
\end{equation}

We can decompose $\TeXButton{R}{\mathbb{R}}_1^n$ in the $\underline{b}$%
-basis as 
\begin{equation}
\label{pp3fo57}\TeXButton{R}{\mathbb{R}}_1^n=\left[ u_{+}\right] \oplus
\left[ u_{-}\right] \oplus \,\left[ e_2,\ldots ,e_{n-m}\right] \,\,\oplus
U\quad . 
\end{equation}
Accordingly, we write a general element of $\TeXButton{R}{\mathbb{R}}_1^n$
as $X_{+}+X_{-}+X+Y$, where 
\begin{equation}
\label{pp3fo58}X_{+}=x^{+}\cdot u_{+}\quad ;\quad X_{-}=x^{-}\cdot
u_{-}\quad ;\quad X=\sum_{a=2}^{n-m}x^a\cdot e_a\quad ;\quad Y=\sum_{\mu
=n-m+1}^{n-1}y^\mu \cdot e_\mu \quad . 
\end{equation}

\subsection{Preservation of $lat=\left[ u_{+}\right] _{\TeXButton{Z}
{\mathbb{Z}}}\oplus \left[ u_1,\ldots ,u_{m-1}\right] _{\TeXButton{Z}
{\mathbb{Z}}}$ \label{prlat}}

Having identified the group $rot\left( W;SO_{1,n-1}^{+}\right) $ that
preserves the $\TeXButton{R}{\mathbb{R}}$-linear enveloppe 
\begin{equation}
\label{pp3fo64}\left[ lat\right] =\left[ u_{+}\right] _{\TeXButton{R}
{\mathbb{R}}}\oplus \left[ u_1,\ldots ,u_{m-1}\right] _{\TeXButton{R}
{\mathbb{R}}} 
\end{equation}
of $lat$, we eventually can turn to reduce this group down to the set 
\begin{equation}
\label{pp3fo65}rot\left( lat\right) \cap SO_{1,n-1}^{+}\equiv rot_0\quad , 
\end{equation}
where $rot_0$ is that part of the set $rot$ that lies in the identity
component $SO_{1,n-1}^{+}$ of $O_{1,n-1}$. To this end we must restrict the
enveloppe (\ref{pp3fo64}) to the original lattice points 
\begin{equation}
\label{pp3fo66}lat=\left[ u_{+}\right] _{\TeXButton{Z}{\mathbb{Z}}}\oplus
\left[ u_1,\ldots ,u_{m-1}\right] _{\TeXButton{Z}{\mathbb{Z}}}=\left[ 
\underline{u}\right] _{\TeXButton{Z}{\mathbb{Z}}}\quad . 
\end{equation}
Now we ask, which of the elements (\ref{pp3fo43}) in $rot\left( W\right) $
preserve this set; the answer is found in formulas (\ref{pp3fo53}-\ref
{pp3fo56}) :

Elements $\left( V_1,1,0,1,1\right) $ act as identity on $W$ and hence
preserve $lat$ without further restriction. The same is true for elements $%
\left( 0,C_1,0,1,1\right) $. The set of products $\left(
V_1,C_1,0,1,1\right) $ of these forms a semidirect product subgroup of $%
rot\left( W\right) $ isomorphic to the Euclidean group $E_0^{n-m-1}$ in $%
\left( n-m-1\right) $ dimensions.

Elements $\left( 0,1,V_2,1,1\right) $ map $Y\in U$ into $-\sqrt{2}\left(
Y\bullet V_2\right) \cdot u_{+}+Y$. For $Y\in lat$, this is a lattice vector
if and only if $\sqrt{2}\left( Y\bullet V_2\right) $ is an integer. Since $Y$
now has integer components, we find that the components of $V_2$ must be $%
V_2^\mu =\frac{z^\mu }{\sqrt{2}}$, $z^\mu \in \TeXButton{Z}{\mathbb{Z}}$.

Elements $\left( 0,1,0,C_2,1\right) $ act as identity on $u_{+}$; they must
be further restricted to map the sublattice $lat^{\prime }\equiv \left[
u_1,\ldots ,u_{m-1}\right] _{\TeXButton{Z}{\mathbb{Z}}}$ into itself. Since
the basis lattice vectors are $\TeXButton{R}{\mathbb{R}}$-linearly
independent this will be satisfied only for a finite (hence discrete) subset 
$D\subset SO_{m-1}$. Since the sublattice $lat^{\prime }$ is now {\bf %
spacelike}, theorem \ref{Bedingung} implies that $D$ must be a group [indeed 
$D$ now coincides what in crystallography is called the {\it maximal point
group} of the sublattice $lat^{\prime }$].

The set of products $\left( 0,1,V_2,C_2,1\right) $ forms a discrete subgroup 
$G_{discr}$ of the subgroup $E_0^{m-1}\subset rot\left( W\right) $, where $%
E_0^{m-1}$ is isomorphic to the Euclidean group in $\left( m-1\right) $
dimensions.

The main point comes now: Elements $\left( 0,1,0,1,e^\phi \right) $ must be
restricted to $\left( 0,1,0,1,k\right) $, $k\in \TeXButton{N}{\mathbb{N}}$,
in order to satisfy $\left( 0,k,1\right) u_{+}=k\cdot u_{+}\in lat$.
Although the original set of $\left( 0,1,0,1,e^\phi \right) $ with $e^\phi
\in \TeXButton{R}{\mathbb{R}}_{+}$ was a group, the set of all $\left(
0,1,0,1,k\right) $ is a group no longer, but a semigroup isomorphic to the
semigroup $\left( \TeXButton{N}{\mathbb{N}},\cdot \right) $ of all natural
numbers with multiplicative composition $\left( k,k^{\prime }\right) \mapsto
k\cdot k^{\prime }$, and $1$ as unit. Clearly, $\left( 0,1,0,1,k\right) $ is
still an invertible element of $rot_0\subset SO_{1,n-1}^{+}$; however, as
mentioned above, it has no inverse \underline{in $rot_0$}, since $\left(
0,1,0,1,k\right) ^{-1}=\left( 0,1,0,1,\frac 1k\right) $ is {\bf not}
lattice-preserving, as it maps $u_{+}\mapsto \frac 1k\cdot u_{+}\not \in lat$%
, for $k>1$.

The multiplicative structure of $rot_0$ clearly is the same as that of $%
rot\left( W\right) $, and is given by the group law (\ref{pp3fo44}). Hence
we see that $E^{n-m-1}\otimes G_{discr}$ forms a proper subgroup of $rot_0$.

\subsection{The structure and the Lie algebra of $\TeXButton{R}{\mathbb{R}}%
^n\odot rot\left( W\right) $}

According to (\ref{pp3fo15}, \ref{pp3fo16}), the normalizer and extended
normalizer of $\Gamma $ are given as semidirect products of the
translational group $\TeXButton{R}{\mathbb{R}}^n$ and $rot^{\times }$, $rot$%
, respectively. In order to understand better the Lie algebra of these
normalizing sets, we first examine the Lie algebra of the semidirect product
group $\TeXButton{R}{\mathbb{R}}^n\odot rot\left( W\right) $, in which $%
eN\left( \Gamma \right) =\TeXButton{R}{\mathbb{R}}^n\odot rot$ is embedded.
To this end we again restrict attention to elements lying in $SO_{1,n-1}^{+}$%
, which means that we define 
\begin{equation}
\label{pp3fo67}rot\left( W\right) _0\equiv rot\left( W\right) \cap
SO_{1,n-1}^{+}\quad , 
\end{equation}
and now study the group and Lie algebra 
\begin{equation}
\label{pp3fo68}\TeXButton{R}{\mathbb{R}}^n\odot rot\left( W\right) _0\quad
,\quad Lie\left[ \TeXButton{R}{\mathbb{R}}^n\odot rot\left( W\right)
_0\right] \quad . 
\end{equation}
The elements of the full group $\TeXButton{R}{\mathbb{R}}^n\odot rot\left(
W\right) _0$ now must take the form $\left[ T\mid \Lambda \right] $, where $%
T\in \TeXButton{R}{\mathbb{R}}^n$ and $\Lambda \in rot\left( W\right) _0$.
The group law is the same as that in $E_1^n$, $\left[ T\mid \Lambda \right]
\left[ T^{\prime }\mid \Lambda ^{\prime }\right] =\left[ \Lambda T^{\prime
}+T\mid \Lambda \Lambda ^{\prime }\right] $. In order to determine in detail
how the elements of $rot\left( W\right) _0$ act on the translational factor $%
\TeXButton{R}{\mathbb{R}}^n$, i.e. on elements $\left[ T\mid 1\right] $, we
transform the orthogonal basis $\left( P_0,P_1,\ldots ,P_{n-1}\right) $ of
the Lie algebra $\TeXButton{R}{\mathbb{R}}^n$ of the translational group $%
\TeXButton{R}{\mathbb{R}}^n$ into the new basis 
\begin{equation}
\label{pp3fo69}\left( P_{+},P_{-},P_2,\ldots ,P_{n-m},P_{n-m+1},\ldots
,P_{n-1}\right) \quad ;\quad P_{\pm }=\frac{P_0\pm P_1}{\sqrt{2}}\quad , 
\end{equation}
which is defined in analogy with the $\underline{b}$-basis given above, so
that 
\begin{equation}
\label{pp3fo70}\TeXButton{R}{\mathbb{R}}^n=\left[ P_{+}\right] \oplus \left[
P_{-}\right] \oplus \left[ P_2,\ldots ,P_{n-m}\right] \oplus \left[
P_{n-m+1},\ldots ,P_{n-1}\right] \quad . 
\end{equation}
There is a natural (vector space) isomorphism between $M=\TeXButton{R}
{\mathbb{R}}_1^n$ and the translation algebra $\TeXButton{R}{\mathbb{R}}^n$
defined by 
\begin{equation}
\label{pp3fo71}u_{+}\simeq P_{+}\quad ;\quad u_{-}\simeq P_{-}\quad ;\quad
e_a\simeq P_a\quad ;\quad e_\mu \simeq P_\mu \quad . 
\end{equation}
Accordingly, we write a general element $\left[ Tr\mid 1\right] $ of $%
\TeXButton{R}{\mathbb{R}}^n$ as $Tr=T_{+}+T_{-}+T+T_2\equiv \left[
T_{+},T_{-},T,T_2\mid 1\right] $, where 
\begin{equation}
\label{pp3fo72}T_{+}=t^{+}\cdot P_{+}\quad ;\quad T_{-}=t^{-}\cdot
P_{-}\quad ;\quad T=\sum_{a=2}^{n-m}t^a\cdot P_a\quad ;\quad T_2=\sum_{\mu
=n-m+1}^{n-1}t_2^\mu \cdot P_\mu \quad . 
\end{equation}

In the following, conjugation of a group element $g$ by another group
element $h$ will be denoted by $\left( h,g\right) \mapsto cj\left( h\right)
\cdot g\equiv hgh^{-1}$.

We now present the action of elements $\left[ 0\mid \Lambda \right] $ on $%
\left[ Tr\mid 0\right] $ by conjugation, according to 
\begin{equation}
\label{pp3fo73}\left( \left[ 0\mid \Lambda \right] ,\left[ Tr\mid 0\right]
\right) \mapsto cj\left( \left[ 0\mid \Lambda \right] \right) \cdot \left[
Tr\mid 0\right] \equiv \left[ 0\mid \Lambda \right] \left[ Tr\mid 0\right]
\left[ 0\mid \Lambda \right] ^{-1}=\left[ \Lambda Tr\mid 0\right] \quad , 
\end{equation}
which expresses how $rot\left( W\right) _0$ acts on the (normal)
translational factor $\TeXButton{R}{\mathbb{R}}^n$. These relations can be
directly derived from formulas (\ref{pp3fo53}-\ref{pp3fo56}):

The action of $\left[ 0\mid \Lambda \right] $ on translations $\left[
T_{+},0,0,0\mid 1\right] $ along the lightlike direction $u_{+}$ is given by 
\begin{equation}
\label{pp3fo74}
\begin{array}{rclc}
cj\left( \left[ 0\mid V_1,1,0,1,1\right] \right) \,\cdot \left[
T_{+},0,0,0\mid 1\right] & = & \left[ T_{+},0,0,0\mid 1\right] & \quad . \\ 
cj\left( \left[ 0\mid 0,C_1,0,1,1\right] \right) \cdot \left[
T_{+},0,0,0\mid 1\right] & = & \left[ T_{+},0,0,0\mid 1\right] & \quad . \\ 
cj\left( \left[ 0\mid 0,1,V_2,1,1\right] \right) \cdot \left[
T_{+},0,0,0\mid 1\right] & = & \left[ T_{+},0,0,0\mid 1\right] & \quad . \\ 
cj\left( \left[ 0\mid 0,1,0,C_2,1\right] \right) \cdot \left[
T_{+},0,0,0\mid 1\right] & = & \left[ T_{+},0,0,0\mid 1\right] & \quad . \\ 
cj\left( \left[ 0\mid 0,1,0,1,e^\phi \right] \right) \cdot \left[
T_{+},0,0,0\mid 1\right] & = & \left[ e^\phi \cdot T_{+},0,0,0\mid 1\right]
& \quad . 
\end{array}
\end{equation}
The action of $\left[ 0\mid \Lambda \right] $ on time translations $\left[
0,T_{-},0,0\mid 1\right] $ is given by 
\begin{equation}
\label{pp3fo75}
\begin{array}{rclc}
cj\left( \left[ 0\mid V_1,1,0,1,1\right] \right) \cdot \left[
0,T_{-},0,0\mid 1\right] & = & \left[ t^{-}\left( V_1\right) ^2,T_{-},-t^{-}%
\sqrt{2}V_1,0\mid 1\right] & \quad . \\ 
cj\left( \left[ 0\mid 0,C_1,0,1,1\right] \right) \cdot \left[
0,T_{-},0,0\mid 1\right] & = & \left[ 0,T_{-},0,0\mid 1\right] & \quad . \\ 
cj\left( \left[ 0\mid 0,1,V_2,1,1\right] \right) \cdot \left[
0,T_{-},0,0\mid 1\right] & = & \left[ t^{-}\left( V_2\right)
^2,T_{-},0,-t^{-}\sqrt{2}V_2\mid 1\right] & \quad . \\ 
cj\left( \left[ 0\mid 0,1,0,C_2,1\right] \right) \cdot \left[
0,T_{-},0,0\mid 1\right] & = & \left[ 0,T_{-},0,0\mid 1\right] & \quad . \\ 
cj\left( \left[ 0\mid 0,1,0,1,e^\phi \right] \right) \cdot \left[
0,T_{-},0,0\mid 1\right] & = & \left[ 0,e^{-\phi }\cdot T_{-},0,0\mid
1\right] & \quad . 
\end{array}
\end{equation}
The action of $\left[ 0\mid \Lambda \right] $ on spacelike translations $%
\left[ 0,0,T,0\mid 1\right] $ on the space part of the subspace $M^{\prime }$
is given by 
\begin{equation}
\label{pp3fo76}
\begin{array}{rclc}
cj\left( \left[ 0\mid V_1,1,0,1,1\right] \right) \cdot \left[ 0,0,T,0\mid
1\right] & = & \left[ -\sqrt{2}\left( T\bullet V_1\right) ,0,T,0\mid
1\right] & \quad . \\ 
cj\left( \left[ 0\mid 0,C_1,0,1,1\right] \right) \cdot \left[ 0,0,T,0\mid
1\right] & = & \left[ 0,0,C_1T,0\mid 1\right] & \quad . \\ 
cj\left( \left[ 0\mid 0,1,V_2,1,1\right] \right) \cdot \left[ 0,0,T,0\mid
1\right] & = & \left[ 0,0,T,0\mid 1\right] & \quad . \\ 
cj\left( \left[ 0\mid 0,1,0,C_2,1\right] \right) \cdot \left[ 0,0,T,0\mid
1\right] & = & \left[ 0,0,T,0\mid 1\right] & \quad . \\ 
cj\left( \left[ 0\mid 0,1,0,1,e^\phi \right] \right) \cdot \left[
0,0,T,0\mid 1\right] & = & \left[ 0,0,T,0\mid 1\right] & \quad . 
\end{array}
\end{equation}
The action of $\left[ 0\mid \Lambda \right] $ on translations $\left[
0,0,0,T_2\mid 1\right] $ on the enveloppe $U$ of the sublattice $lat^{\prime
}$ is given by 
\begin{equation}
\label{pp3fo77}
\begin{array}{rclc}
cj\left( \left[ 0\mid V_1,1,0,1,1\right] \right) \cdot \left[ 0,0,0,T_2\mid
1\right] & = & \left[ 0,0,0,T_2\mid 1\right] & \quad . \\ 
cj\left( \left[ 0\mid 0,C_1,0,1,1\right] \right) \cdot \left[ 0,0,0,T_2\mid
1\right] & = & \left[ 0,0,0,T_2\mid 1\right] & \quad . \\ 
cj\left( \left[ 0\mid 0,1,V_2,1,1\right] \right) \cdot \left[ 0,0,0,T_2\mid
1\right] & = & \left[ -\sqrt{2}\left( T_2\bullet V_2\right) ,0,0,T_2\mid
1\right] & \quad . \\ 
cj\left( \left[ 0\mid 0,1,0,C_2,1\right] \right) \cdot \left[ 0,0,0,T_2\mid
1\right] & = & \left[ 0,0,0,C_2T_2\mid 1\right] & \quad . \\ 
cj\left( \left[ 0\mid 0,1,0,1,e^\phi \right] \right) \cdot \left[
0,0,0,T_2\mid 1\right] & = & \left[ 0,0,0,T_2\mid 1\right] & \quad . 
\end{array}
\end{equation}

From formulas (\ref{pp3fo74}-\ref{pp3fo77}) we now can derive the Lie
algebra $Lie\left[ \TeXButton{R}{\mathbb{R}}^n\odot rot\left( W\right)
_0\right] $, using the fact that the conjugating elements $\left[ 0\mid
\Lambda \right] $ occuring on the left hand side of equations (\ref{pp3fo74}-%
\ref{pp3fo77}) are exponentials, as follows from (\ref{pp3fo46}-\ref{pp3fo48}%
). We employ standard Lie algebra machinery, 
\begin{equation}
\label{pp3fo78}\frac d{ds}\left. cj\left( g\right) \cdot \exp sX\right|
_{s=0}=cj\left( g\right) _{*}X=Ad\left( g\right) X\quad , 
\end{equation}
and 
\begin{equation}
\label{pp3fo79}\frac d{ds}\left. Ad\left( \exp sY\right) X\right|
_{s=0}=ad\left( Y\right) X=\left[ Y,X\right] \quad , 
\end{equation}
for elements $X,Y$ in the Lie algebra $Lie\left[ \TeXButton{R}{\mathbb{R}}%
^n\odot rot\left( W\right) _0\right] $. The subalgebra of the generators of $%
rot\left( W\right) _0$ has been derived in (\ref{pp3fo39}) already; hence we
consider commutators $\left[ Y,X\right] $, where $Y\in Lie\left[ rot\left(
W\right) _0\right] =Lie\left[ rot\left( W\right) \right] $, and $X\in 
\TeXButton{R}{\mathbb{R}}^n=Lie\left( \TeXButton{R}{\mathbb{R}}^n\right) $,
this space being decomposed according to (\ref{pp3fo70}): 
\begin{equation}
\label{pp3fo80}
\begin{array}{rclcrclc}
\left[ K_a,P_{+}\right] & = & 0 & \quad .\quad & \left[ K_a,P_{-}\right] & =
& -\sqrt{2}P_a & \quad . \\ 
\left[ L_{ab},P_{+}\right] & = & 0 & \quad .\quad & \left[
L_{ab},P_{-}\right] & = & 0 & \quad . \\ 
\left[ K_\mu ,P_{+}\right] & = & 0 & \quad .\quad & \left[ K_\mu
,P_{-}\right] & = & -\sqrt{2}P_\mu & \quad . \\ 
\left[ L_{\mu \nu },P_{+}\right] & = & 0 & \quad .\quad & \left[ L_{\mu \nu
},P_{-}\right] & = & 0 & \quad . \\ 
\left[ \Delta ,P_{+}\right] & = & P_{+} & \quad .\quad & \left[ \Delta
,P_{-}\right] & = & -P_{-} & \quad . 
\end{array}
\end{equation}
\begin{equation}
\label{pp3fo81}
\begin{array}{rclcrclc}
\left[ K_a,P_b\right] & = & -\sqrt{2}\delta _{ab}\cdot P_{+} & \quad .\quad
& \left[ K_a,P_\mu \right] & = & 0 & \quad . \\ 
\left[ L_{ab},P_c\right] & = & \delta _{ac}\cdot P_b-\delta _{bc}\cdot P_a & 
\quad .\quad & \left[ L_{ab},P_\mu \right] & = & 0 & \quad . \\ 
\left[ K_\mu ,P_a\right] & = & 0 & \quad .\quad & \left[ K_\mu ,P_\nu
\right] & = & -\sqrt{2}\delta _{\mu \nu }\cdot P_{+} & \quad . \\ 
\left[ L_{\mu \nu },P_a\right] & = & 0 & \quad .\quad & \left[ L_{\mu \nu
},P_\rho \right] & = & \delta _{\mu \rho }\cdot P_\nu -\delta _{\nu \rho
}\cdot P_\mu & \quad . \\ 
\left[ \Delta ,P_a\right] & = & 0 & \quad .\quad & \left[ \Delta ,P_\mu
\right] & = & 0 & \quad . 
\end{array}
\end{equation}
For better comparison, we present the algebra of the $rot\left( W\right) $%
-factor again: 
\begin{equation}
\label{pp3fo82}
\begin{array}{c}
\left[ \Delta ,K_i\right] =K_i\quad . \\ 
\left[ \Delta ,L_{ab}\right] =\left[ \Delta ,L_{\mu \nu }\right] =0\quad .
\\ 
\left[ K_i,K_j\right] =0\quad . \\ 
\left[ L_{ab},K_c\right] =\delta _{ac}\cdot K_b-\delta _{bc}\cdot K_a\quad .
\\ 
\left[ L_{ab},K_\rho \right] =0\quad . \\ 
\left[ L_{\mu \nu },K_a\right] =0\quad . \\ 
\left[ L_{\mu \nu },K_\rho \right] =\delta _{\mu \rho }\cdot K_\nu -\delta
_{\nu \rho }\cdot K_\mu \quad . \\ 
\left[ L_{ab},L_{cd}\right] =\delta _{ac}\cdot L_{bd}+\delta _{bd}\cdot
L_{ac}-\delta _{ad}\cdot L_{bc}-\delta _{bc}\cdot L_{ad}\quad . \\ 
\left[ L_{\mu \nu },L_{\gamma \delta }\right] =\delta _{\mu \gamma }\cdot
L_{\nu \delta }+\delta _{\nu \delta }\cdot L_{\mu \gamma }-\delta _{\mu
\delta }\cdot L_{\nu \gamma }-\delta _{\nu \gamma }\cdot L_{\mu \delta
}\quad . \\ 
\left[ L_{ab},L_{\mu \nu }\right] =0\quad . 
\end{array}
\end{equation}

\subsection{The structure and the Lie algebra of $rot_0$}

From (\ref{pp3fo40}) we can now read off the form of $rot_0$, 
\begin{equation}
\label{pp3fo83}rot_0\simeq \left[ E^{n-m-1}\otimes G_{discr}\right] \odot
\left( \TeXButton{N}{\mathbb{N}},\cdot \right) \quad , 
\end{equation}
\begin{equation}
\label{pp3fo84}E^{n-m-1}=\TeXButton{R}{\mathbb{R}}^{n-m-1}\odot
O_{n-m-1}\quad . 
\end{equation}
In analogy with (\ref{pp3fo65}) we furthermore introduce the subgroup $%
rot_0^{\times }$ of the $lat$-invertible transformations that lie in $%
SO_{1,n-1}^{+}$ as 
\begin{equation}
\label{pp3fo85}rot_0^{\times }\equiv rot^{\times }\cap SO_{1,n-1}^{+}\quad . 
\end{equation}
This set contains all Lorentz transformations $R$ belonging to the identity
component of $SO_{1,n-1}$ that preserve the lattice $lat$, such that the
same is true for $R^{-1}$. From the analysis above it is now clear that $%
rot_0^{\times }$ is isomorphic to 
\begin{equation}
\label{pp3fo86}rot_0^{\times }\simeq E^{n-m-1}\otimes G_{discr}\quad . 
\end{equation}
Hence it is the dilations in $\left( \TeXButton{N}{\mathbb{N}},\cdot \right) 
$ that constitute the extension from $rot_0^{\times }$ to $rot_0$, and we
have 
\begin{equation}
\label{pp3fo87}rot_0=rot_0^{\times }\odot \left( \TeXButton{N}{\mathbb{N}}%
,\cdot \right) 
\end{equation}
in this case; i.e., a semidirect product of a group and a semigroup.

The connected component $rot_{00}$ of $rot_0$ can be read off from (\ref
{pp3fo83}); it coincides with the connected component $\left( rot_0^{\times
}\right) _0$ of $rot_0^{\times }$, and is given by 
\begin{equation}
\label{pp3fo88}rot_{00}=\left( rot_0^{\times }\right) _0\simeq
E_0^{n-m-1}\quad . 
\end{equation}
Hence, we have the Lie algebras%
$$
Lie\left( rot\right) =Lie\left( rot_0\right) =Lie\left( rot_{00}\right)
=Lie\left( rot^{\times }\right) =Lie\left( rot_0^{\times }\right) =Lie\left(
rot_0^{\times }\right) _0\simeq 
$$
\begin{equation}
\label{pp3fo89}\simeq euc_0^{n-m-1}\quad . 
\end{equation}

We now can turn eventually to the extended normalizer $eN\left( \Gamma
\right) =\TeXButton{R}{\mathbb{R}}^n\odot rot$, and $N\left( \Gamma \right) =%
\TeXButton{R}{\mathbb{R}}^n\odot rot^{\times }$, as given in (\ref{pp3fo15},%
\ref{pp3fo16}). However, as in the previous subsections, we want to focus on
elements that are in the connected component $SO_{1,n-1}^{+}$. Following (%
\ref{pp3fo65},\ref{pp3fo85}), we accordingly define 
\begin{equation}
\label{pp3fo90}eN\left( \Gamma \right) _0\equiv eN\left( \Gamma \right) \cap
SO_{1,n-1}^{+}=\TeXButton{R}{\mathbb{R}}^n\odot rot_0\quad , 
\end{equation}
\begin{equation}
\label{pp3fo91}N\left( \Gamma \right) _0\equiv N\left( \Gamma \right) \cap
SO_{1,n-1}^{+}=\TeXButton{R}{\mathbb{R}}^n\odot rot_0^{\times }\quad , 
\end{equation}
with identity components 
\begin{equation}
\label{pp3fo92}\left[ eN\left( \Gamma \right) _0\right] _0=\left[ N\left(
\Gamma \right) _0\right] _0=\TeXButton{R}{\mathbb{R}}^n\odot rot_{00}\simeq 
\TeXButton{R}{\mathbb{R}}^n\odot E_0^{n-m-1}\quad . 
\end{equation}
The quotient $eN\left( \Gamma \right) _0/\Gamma $ then takes the form 
\begin{equation}
\label{pp3fo94}eN\left( \Gamma \right) _0/\Gamma =\TeXButton{R}{\mathbb{R}}%
^n/\Gamma \odot rot_0=\left[ \TeXButton{R}{\mathbb{R}}^{n-m}\otimes
T^m\left( R_{+},R_{n-m+1},\ldots ,R_{n-1}\right) \right] \odot rot_0\quad , 
\end{equation}
where $\TeXButton{R}{\mathbb{R}}^{n-m}=\left[ P_{-},P_2,\ldots
,P_{n-m}\right] $, and%
$$
T^m\left( R_{+},R_\mu \right) =T^m\left( R_{+},R_{n-m+1},\ldots
,R_{n-1}\right) \equiv 
$$
\begin{equation}
\label{pp3fo96}=\left[ u_{+}\right] _{\TeXButton{R}{\mathbb{R}}}/\left[
u_{+}\right] _{\TeXButton{Z}{\mathbb{Z}}}\otimes \left[ u_{n-m+1}\right] _{%
\TeXButton{R}{\mathbb{R}}}/\left[ u_{n-m+1}\right] _{\TeXButton{Z}
{\mathbb{Z}}}\cdots \otimes \left[ u_{n-1}\right] _{\TeXButton{R}{\mathbb{R}}%
}/\left[ u_{n-1}\right] _{\TeXButton{Z}{\mathbb{Z}}}\quad , 
\end{equation}
where we have written 
\begin{equation}
\label{pp3fo97}R_A\equiv \frac{\left\| u_A\right\| _E}{2\pi }\quad ,\quad
\left\| u_A\right\| _E\equiv \sqrt{\sum_{i=0}^{n-1}\left| u_A^i\right| }%
\quad ,\quad A\in \left\{ 0,n-m+1,\ldots ,n-1\right\} 
\end{equation}
i.e. $\left\| u_A\right\| _E$ here is the {\bf Euclidean} norm of $u_A$, and 
$R_A$ are the radii of the associated circles $\left[ u_A\right] _{%
\TeXButton{R}{\mathbb{R}}}/\left[ u_A\right] _{\TeXButton{Z}{\mathbb{Z}}}$. $%
T^m$ is the translation group of a torus $\TeXButton{R}{\mathbb{R}}^m/\Gamma 
$ which we will denote by the same symbol $T^m\left( R_{+},R_\mu \right) $;
coordinates $\left( x^{+},x^{-},x^a,y^\mu \right) $ on this latter torus can
be obtained from (\ref{pp3fo58}). The quotient $\TeXButton{R}{\mathbb{R}}%
^m\rightarrow \TeXButton{R}{\mathbb{R}}^m/\Gamma $ clearly preserves Lie
algebras, so that $Lie\left( T^m\right) =\TeXButton{R}{\mathbb{R}}^m$. $T^m$
acts on $T^m$ as%
$$
\exp \left( t^{+}\cdot P_{+}+t^{-}\cdot P_{-}+\sum_{a=2}^{n-m}t^a\cdot
P_a+\sum_{\mu =n-m+1}^{n-1}t_2^\mu \cdot P_\mu \right) \left(
x^{+},x^{-},x^a,y^\mu \right) = 
$$
\begin{equation}
\label{pp3fo98}=\left( t^{+}+x^{+},t^{-}+x^{-},t^a+x^a,y^\mu +t_2^\mu
\right) \;{\rm mod}\;lat\quad . 
\end{equation}
The full quotient $M/\Gamma $ is $M/\Gamma =\TeXButton{R}{\mathbb{R}}%
^{n-m}\times T^m$. As was explained in section \ref{Sc.5}, the metric on $%
T^{m-1}\left( R^\mu \right) $ is positive definite, whereas it is
identically zero on $T\left( R_{+}\right) =\left[ u_{+}\right] _{%
\TeXButton{R}{\mathbb{R}}}/\left[ u_{+}\right] _{\TeXButton{Z}{\mathbb{Z}}}$.

We now turn to the Lie algebra of $\left[ eN\left( \Gamma \right) _0\right]
_0/\Gamma $. Since $\Gamma $ is a discrete normal subgroup of the connected
component $\left[ eN\left( \Gamma \right) _0\right] _0$, the groups $\left[
eN\left( \Gamma \right) _0\right] _0$ and $\left[ eN\left( \Gamma \right)
_0\right] _0/\Gamma $ are locally isomorphic, hence \cite{SagleWald} they
possess isomorphic Lie algebras. This implies%
$$
Lie\left\{ \left[ eN\left( \Gamma \right) _0\right] _0/\Gamma \right\}
\simeq Lie\left\{ eN\left( \Gamma \right) _0/\Gamma \right\} \simeq
Lie\left\{ N\left( \Gamma \right) _0/\Gamma \right\} \simeq Lie\left\{
N\left( \Gamma \right) /\Gamma \right\} \simeq 
$$
\begin{equation}
\label{pp3fo99}\simeq Lie\left\{ \left[ eN\left( \Gamma \right) _0\right]
_0\right\} \simeq Lie\left\{ N\left( \Gamma \right) \right\} \simeq
Lie\left\{ I\left( M/\Gamma \right) \right\} \simeq Lie\left\{ \TeXButton{R}
{\mathbb{R}}^n\odot E_0^{n-m-1}\right\} \quad . 
\end{equation}
This algebra is spanned by 
\begin{equation}
\label{pp3fo100}\left( P_{+},P_{-},P_a,P_\mu \mid K^a,L_{ab}\right) 
\end{equation}
subject to the relations (\ref{pp3fo80}-\ref{pp3fo82}). We rewrite these
relations in a slightly different form; to this end we redefine 
\begin{equation}
\label{pp3fo101}K_a\mapsto \sqrt{2}K_a\quad ;\quad P_{-}\mapsto -P_{-}\quad
. 
\end{equation}
Then 
\begin{equation}
\label{pp3fo102}\left[ K_a,P_\mu \right] =\left[ L_{ab},P_\mu \right]
=0\quad , 
\end{equation}
and 
\begin{equation}
\label{pp3fo103}
\begin{array}{rclcrclc}
\left[ K_a,P_{+}\right] & = & 0 & \quad .\quad & \left[ K_a,P_{-}\right] & =
& P_a & \quad . \\ 
\left[ L_{ab},P_{+}\right] & = & 0 & \quad .\quad & \left[
L_{ab},P_{-}\right] & = & 0 & \quad . \\ 
\left[ K_a,P_b\right] & = & -\delta _{ab}\cdot P_{+} & \quad .\quad & \left[
L_{ab},P_c\right] & = & \delta _{ac}\cdot P_b-\delta _{bc}\cdot P_a & \quad
. 
\end{array}
\end{equation}
\begin{equation}
\label{pp3fo104}
\begin{array}{c}
\left[ K_i,K_j\right] =0\quad . \\ 
\left[ L_{ab},K_c\right] =\delta _{ac}\cdot K_b-\delta _{bc}\cdot K_a\quad .
\\ 
\left[ L_{ab},L_{cd}\right] =\delta _{ac}\cdot L_{bd}+\delta _{bd}\cdot
L_{ac}-\delta _{ad}\cdot L_{bc}-\delta _{bc}\cdot L_{ad}\quad . 
\end{array}
\end{equation}
This defines a direct sum of Lie algebras,
\begin{equation}
\label{pp3fo104a}A\oplus gal_{ce}^{n-m-1}\quad .
\end{equation}
Here $A$ is an Abelian Lie algebra isomorphic to $\TeXButton{R}{\mathbb{R}}%
^{m-1}$ with generators $\left( P_\mu \right) $. This algebra will play no
further role in what we discuss in the remainder of the paper. The second
algebra has generators $\left( P_{+},P_{-},P_a\mid K^a,L_{ab}\right) $ and
is isomorphic to the centrally extended Galilean algebra $gal_{ce}^{n-m-1}$
in $\left( n-m-1\right) $ dimensions, where $P_{-}$
generates ''time'' translations, $P_a$ generate ''space'' translations, $K_a$
generate Galilei boosts, $L_{ab}$ generate rotations, and $P_{+}$ is the
''mass generator'' spanning a $1$-dimensional central extension, in the
following denoted by $gal_{ce}^{n-m-1}$, of the unextended Galilei algebra $%
gal^{n-m-1}$. . If the generators for $\tau $-, $T$-, $V$-, $R$%
-transformations in the Galilei algebra are denoted by $P_{-}$, $P_a$, $K_a$%
, $L_{ab}$, respectively, we find that these generators satisfy (\ref
{pp3fo102}-\ref{pp3fo104}). Furthermore, if $P_{+}$ denotes the generator of
the central extension $gal_{ce}$ of the Noether charge algebra carried by a
point particle of mass $m$, then the relations of $P_{+}$ with the remaining
generators are given by the left block of formulas (\ref{pp3fo103}), where
the last equation is to be replaced by 
\begin{equation}
\label{pp3fo105}\left[ K_a,P_b\right] =-m\delta _{ab}\cdot P_{+}\quad , 
\end{equation}
which is why $P_{+}$ is called a mass generator (see, e.g. \cite{Azcarraga}).

\section{The effect of semigroup transformations \label{Sc.6}}

Finally, we briefly discuss how the semigroup elements $\Phi \equiv \left(
0,1,0,1,k\right) $ act on the compactified spacetime $\TeXButton{R}
{\mathbb{R}}^{n-m}\times T^m$, and on quantum fields defined on such a
spacetime. From subsection \ref{prlat} we see that the map $\Phi $ acts
trivially on $x^a$-coordinates; on the other hand, the pair $\left(
x^{+},x^{-}\right) $ of coordinates in the $u_{\pm }$-direction is mapped
into $\left( kx^{+},\frac 1kx^{-}\right) $; it corresponds to contraction in
the ''time'' coordinate $x^{-}$, and, more important, to a dilation $%
x^{+}\mapsto kx^{+}$. As $x^{+}$ takes values in $\left[ 0,2\pi R_{+}\right] 
$, its image under $\Phi $ therefore winds $k$ times around the lightlike $%
S_1$-factor. The $2$-cylinder $\left[ u_{-}\right] _{\TeXButton{R}
{\mathbb{R}}}\times \left[ u_{+}\right] _{\TeXButton{R}{\mathbb{R}}}/\left[
u_{+}\right] _{\TeXButton{Z}{\mathbb{Z}}}$ therefore gets contracted and $k$
times wound around itself; this means that the mass generator $P_{+}$ should
correspond to a topological mass term in this context.

If $\phi :\left[ 0,2\pi R_{+}\right] \times \TeXButton{R}{\mathbb{R}}%
^{n-m-1}\times T^{m-1}\rightarrow tgt$ is a field on the compactified
spacetime taking values in some target space $tgt$, and ${\cal L}={\cal L}%
\left[ \phi \right] $ is a Lagrangian governing its dynamics, then 
\begin{equation}
\label{pp3fo106}\int\limits_{\left[ 0,2\pi R_{+}\right] }dx^{+}\int
dx^{-}dx^i\cdot {\cal L}\left[ \phi \right] \;=\;\int\limits_{\left[ 0,\frac{%
2\pi R_{+}}k\right] }dx^{+}\int dx^{-}dx^i\cdot \Phi ^{*}{\cal L}\left[ \phi
\right] \quad , 
\end{equation}
where $\Phi ^{*}{\cal L}\left[ \phi \right] $ is the pullback of ${\cal L}$
to the space $\left[ 0,\frac{2\pi R_{+}}k\right] \times \TeXButton{R}
{\mathbb{R}}^{n-m-1}\times T^{m-1}$. (\ref{pp3fo106}) therefore shows the
important result that, although $\Phi $ is originally a map that preserves
the lattice $lat$, and hence the spacetime $\left[ 0,2\pi R_{+}\right]
\times \TeXButton{R}{\mathbb{R}}^{n-m-1}\times T^{m-1}$, $\Phi $
nevertheless induces a map on actions so as to map a theory on a spacetime
with lightlike compactification radius $R_{+}$ to a theory with the smaller
radius $\frac{R_{+}}k$. This operation corresponds to finite discrete
transformations associated with the ''mass'' generator $P_{+}$, which
commutes with all observables in the Galilei algebra and hence is a
superselection operator for the (spacetime degrees of freedom of the)
theory. This means that it labels different, non-coherent, subspaces of
physical states in the overall Hilbert space of the system; amongst these
different superselection sectors, the superposition principle is no longer
valid. It therefore would seem that the $\Phi $-map, when applied to
actions, relates different superselection sectors of the theory. From the
non-invertibility of $\Phi $ on the lattice we deduce that this is a one-way
operation.

\chapter{Covering Spaces and Moment Maps}
\label{MomentMaps}

\section{Introduction}

For a physical system whose Lagrangian is invariant under the action of the
isometry group $G$ of a background spacetime, the algebra of conserved
charges coincides with the Noether charge algebra of the system associated
with the isometry group. In recent years, it was observed by many authors
working in the field of high energy physics, that when classical or quantum
fields propagate in background spacetimes which are topologically
non-trivial then the associated algebra of conserved charges may reflect
this non-triviality by exhibiting extensions of the Noether charge algebra
which measure the topology of the background, provided the system is only
semi-invariant under the isometry group of the spacetime (see, e.g., \cite
{Azca}). On the other hand, the algebra of conserved charges defines a
partition of the underlying phase space by distinguishing those subsets of
the phase space on which the values of all conserved charges involved are
constant (this is just the first step to a Marsden-Weinstein reduction).
These subsets then could be called ''elements of classical $G$-multiplets'',
a multiplet being defined as a $G$-orbit of such subsets. On every such
multiplet, the symmetry group $G$ acts transitively, thus exhibiting a
property which is the classical analogue of an irreducible representation in
the quantum theory \cite{Woodhouse}. Now, if dynamical systems are described
in the framework of symplectic formalism, the quantity determining this
phase space partition is what we call a global moment map on the phase
space. However, this global moment map need not exist in the case that the
phase space is not simply connected. This observation was the starting point
for the investigations in this work. For systems with finite degrees of
freedom, we generalize the concept of a global moment map to multiply
connected phase spaces and general symplectic manifolds. We show that the
appropriate generalization involves a locally defined multi-valued moment
map, whose different branches are labelled by the fundamental group of the
phase space. Furthermore, it is shown how the different local branches can
be smoothly glued together by a glueing condition, which is expressed by
certain \u Cech cocycles on the underlying symplectic manifold. These
constructions are intimately related with the existence of a universal
symplectic covering manifold of the original phase space. On the covering,
global moment maps defining $G$-multiplets always exist. At the end of this
work we show how these multiplets on the covering space can be related to $G$%
-multiplets on the original symplectic manifold by an identification map
which derives from the covering projection.

In order to formulate these ideas rigourosly we first had to examine the
question of liftability of symplectic group actions on a symplectic manifold
to a covering space. In the course of this, we proved a series of theorems
investigating the existence and uniqueness of such lifts, and to which
extent the lifted action preserves the group structure of the original
symmetry group $G$.

The plan of this chapter is as follows: In section \ref{BasicsAboutCoverings}
we collect basic statements about covering spaces on which the rest of this
work relies. In section \ref{CoveringsAndDiffForms} and \ref{MultiValFunc}
we examine multi-valued potential functions for closed but not exact
differential forms on a multiply connected manifold. These considerations
will be needed later on when examining local moment maps. In section \ref
{NotationSymplecticManifolds}--\ref{CotangentBundles} we collect notation
conventions and basic facts about symplectic manifolds, Hamiltonian vector
fields, and cotangent bundles. In section \ref{CoveringCotBundle} we study
how covering projections can be extended to local symplectomorphisms of
coverings of cotangent bundles. Sections \ref{LiftOfGroupActions} and \ref
{PreservationOfTheGroupLaw} examine the conditions under which an action of
a Lie group $G$ on a manifold can be lifted to an action on a covering
manifold, and when such a lift preserves the group law of $G$. Section \ref
{GSpacesCoveringMaps} examines the relation between group actions on
covering spaces and equivariance of the covering map. In \ref
{SymplecticGActionsMoMaps} we introduce our notation for symplectic $G$%
-actions on a symplectic manifold. Section \ref{GlobalMomentMaps}
recapitulates the notion of global moment maps as they are usually defined,
while this concept is generalized to local moment maps in the subsequent
section \ref{LocalMomentMaps}. Equivariance of global and local moment maps
is discussed in section \ref{EquivarianceOfMomentMaps}. Section \ref
{NonSimplyConnected} discusses the relation of local moment maps to covering
spaces which are themselves multiply connected. In \ref{GStateSpaces} we
introduce the concept of $G$-states, while the subsequent section \ref
{SplittingOfMultiplets} shows how a splitting of $G$-states on multiply
connected symplectic manifold arises.

\section{Basic facts about coverings \label{BasicsAboutCoverings}}

In this section we quote the main results on coverings and lifting theorems
on which the rest of this work is built; it is mainly based on
\cite{Wolf,Fulton,Jaehnich,ONeill}. In this work we are interested in
covering spaces that are manifolds. Consequently, some of the definitions
and quotations of theorems to follow are not presented in their full
generality, as appropriate for general topological spaces, but rather we
give working definitions and formulations pertaining to manifolds and the
fact that these are special topological spaces (locally homeomorphic to $%
\TeXButton{R}{\mathbb{R}}^n$).

--- A covering of manifolds is a triple $\left( p,X,Y\right) $, where $%
p:X\rightarrow Y$ is a smooth surjective map of smooth manifolds $X,Y$,
where $Y$ is connected, such that for every $y\in Y$ there exists an open
neighbourhood $V\subset Y$ of $y$ for which $p^{-1}\left( V\right) $ is a
disjoint union of open sets $U$ in $X$, on each of which the restriction $%
\left. p\right| U:U\rightarrow Y$ is a diffeomorphism. Every open $V\subset
Y $ for which this is true is called admissible (with respect to $p$). This
means that, for each $y\in Y$, the inverse image $p^{-1}\left( y\right)
\subset X$, called the fibre over{\it \ }$y$, is discrete. Since $Y$ is
connected, all fibres have the same cardinality. $p$ is called projection or
covering map. A diffeomorphism $\phi :X\rightarrow X$ such that $p\circ \phi
=p$ is called a deck{\it \ }transformation of the covering. The set ${\cal D}
$ of all deck transformations of the covering is a group under composition
of maps. Since every $\phi \in {\cal D}$ permutes the elements in the fibres 
$p^{-1}\left( y\right) $, the group ${\cal D}$ of all $\phi $ is discrete.
If $X$ is connected, deck transformations are uniquely determined by their
value at a given point $x\in X$.

--- Let $x\in p^{-1}\left( y\right) $, and let $\pi _1\left( X,x\right) $, $%
\pi _1\left( Y,y\right) $ denote the fundamental groups of $X$ and $Y$ based
at $x$ and $y$, respectively. The projection $p$ induces a homomorphism $%
p_{\#}:\pi _1\left( X,x\right) \rightarrow \pi _1\left( Y,y\right) $ of
fundamental groups, such that $p_{\#}\pi _1\left( X,x^{\prime }\right) $
ranges through the set of all conjugates of $p_{\#}\pi _1\left( X,x\right) $
in $\pi _1\left( Y,y\right) $, as $x^{\prime }$ ranges through the elements
in the fibre $p^{-1}\left( y\right) $. A covering $p:X\rightarrow Y$ is
called normal if $p_{\#}\pi _1\left( X,x^{\prime }\right) $ is normal in $%
\pi _1\left( Y,y\right) $ for some (hence any) $x^{\prime }\in p^{-1}\left(
y\right) $. One can show that a covering is normal if and only if the deck
transformation group ${\cal D}$ acts transitively on the fibres, i.e. for
all $x,x^{\prime }\in X$ with $p\left( x\right) =p\left( x^{\prime }\right) $
there exists a unique deck transformation $\phi \in {\cal D}$ with $%
x^{\prime }=\phi \left( x\right) $. This is certainly true when $X$ is
simply connected; in this case, the deck transformation group ${\cal D}$ is
isomorphic to the fundamental group $\pi _1\left( Y,y\right) $.

--- Let $\Gamma $ be a group of diffeomorphisms acting on the manifold $X$,
and let $\Gamma x$ denote the orbit of $x$ under $\Gamma $. The set of all
orbits is denoted by $X/\Gamma $, and is called an orbit space. The natural
projection $pr:X\rightarrow X/\Gamma $ sends each $x\in X$ to its orbit, $%
pr\left( x\right) =\Gamma x$. The topology on $X/\Gamma $ is the quotient
topology, for which $p$ is continuous, and an open map. $\Gamma $ acts
properly discontinuously on $X$ if every $x\in X$ has a neighbourhood $U$
such that the set $\left\{ \gamma \in \Gamma \mid \gamma U\cap U\neq
\emptyset \right\} $ is finite. $\Gamma $ acts freely if no $\gamma \neq e$
has a fixed point in $x$. $\Gamma $ acts properly discontinuously and freely
if each $x\in X$ has a neighbourhood $U$ such that $e\neq \gamma \in \Gamma $
implies $\gamma U\cap U=\emptyset $.

--- Given a covering $p:X\rightarrow Y$ and a map $f:V\rightarrow Y$ defined
on some manifold $V$, a map $\tilde f:V\rightarrow X$ is called a lift of $f$%
\ through $p$ if $p\circ \tilde f=f$. Two maps $f,g:V\rightarrow Y$ are
called homotopic if there exists a continuous map $G:I\times V\rightarrow Y$%
, $\left( t,v\right) \mapsto G_t\left( v\right) $, such that $G_0=f$ and $%
G_1=g$. $G$ is called a homotopy of $f$\ and $g$.

\TeXButton{Abst}{\vspace{0.8ex}}--- We now quote without proof a couple of
theorems from \cite{Wolf} which we will make us of frequently. These
theorems are actually proven in every textbook on Algebraic Topology.

{\bf Theorem:} \label{TheoWolf1}\quad If $\Gamma $ acts properly
discontinuously and freely on the connected manifold $X$, the natural
projection $pr:X\rightarrow X/\Gamma $ onto the orbit space is a covering
map, and the covering is normal. Furthermore, the deck transformation group $%
{\cal D}$ of this covering is $\Gamma $.

\TeXButton{Abst}{\vspace{0.8ex}}As a converse, we have

{\bf Theorem:} \label{TheoWolf2}\quad If $p:X\rightarrow Y$ is a covering
and ${\cal D}$ is the group of deck transformations, then ${\cal D}$ acts
properly discontinuously and freely on $X$. --- If $X$ is a simply connected
manifold, every covering $p:X\rightarrow Y$ is a natural projection $%
pr:X\rightarrow X/\Gamma $ for some discrete group $\Gamma $ of
diffeomorphisms acting properly discontinuously and freely on $X$.

\TeXButton{Abst}{\vspace{0.8ex}}As a consequence, we have

{\bf Corollary:} \label{CoroWolf3}\quad Let $Y$ be a connected manifold.
Then $Y$ is diffeomorphic to an orbit space $X/\Gamma $, where $X$ is a
simply connected covering manifold $X$, and $\Gamma $ is a group of
diffeomorphisms acting properly discontinuously and freely on $X$. In this
case, $\Gamma ={\cal D}$ coincides with the deck transformation group of the
covering.

\TeXButton{Abst}{\vspace{0.8ex}}--- Existence and uniqueness of lifts are
determined by the following lifting theorems:

{\bf Theorem (''Unique Lifting Theorem''):} \label{UniqueLiftingTheorem}%
\quad Let $p:X\rightarrow Y$ be a covering of manifolds. If $V$ is a
connected manifold, $f:V\rightarrow Y$ is continuous, and $g_1$ and $g_2$
are lifts of $f$ through $p$ that coincide in one point, $g_1\left( v\right)
=g_2\left( v\right) $, then $g_1=g_2$.

\TeXButton{Abst}{\vspace{0.8ex}}{\bf Theorem (''Covering Homotopy Theorem''):%
} \label{CoveringHomotopyTheorem}\quad Let $p:X\rightarrow Y$ be a covering.
Let $f,g:V\rightarrow Y$ be continuous, let $\tilde f:V\rightarrow X$ be a
lift of $f$, and let $G$ be a homotopy of $f$ and $g$. Then there is a
unique pair $\left( \tilde G,\tilde g\right) $, where $\tilde g$ is a lift
of $g$, $\tilde G$ is a lift of $\tilde G$, and $\tilde G$ is a homotopy of $%
\tilde f$ and $\tilde g$.

\TeXButton{Abst}{\vspace{0.8ex}}{\bf Theorem (''Lifting Map Theorem''):} 
\label{LiftingMapTheorem}\quad Let $p:X\rightarrow Y$ be a covering, let $V$
be a connected manifold, let $f:V\rightarrow Y$ be continuous. Let $x\in X$, 
$y\in Y$, $v\in V$ such that $p\left( x\right) =y=f\left( v\right) $. Then $%
f $ has a lift $\tilde f:V\rightarrow X$ with $\tilde f\left( v\right) =x$
if and only if $f_{\#}\pi _1\left( V,v\right) \subset p_{\#}\pi _1\left(
X,x\right) $, where $f_{\#}$ denotes the homomorphism of fundamental groups
induced by $f$.

\section{Covering spaces and differential forms \label{CoveringsAndDiffForms}
}

Let $p:X\rightarrow Y$ be a covering of smooth manifolds. In this case the
covering projection $p$ is a local diffeomorphism. In this section we
examine the relation between differential forms on $X$ and $Y$ which are
related by pull-back through the covering map $p$, and where $X$ is simply
connected. The results will be needed to formulate the concept of local
moment maps on non-simply connected symplectic manifolds in section \ref
{LocalMomentMaps}.

\subsection{Differential forms on $X$ and $Y$}

First consider a smooth $q$-form $\omega $ on $Y$. The pull-back $\Omega
\equiv p^{*}\omega $ is a $q$-form on $X$, and $\Omega $ is invariant under
the deck transformation group of the covering: For, let $\gamma \in {\cal D}$%
, then $\gamma ^{*}\Omega =\left( p\circ \gamma \right) ^{*}\omega =\Omega $%
. Conversely, let $\Omega $ be a $q$-form on $X$. Then, locally, $\Omega $
can be pulled back to $Y$ to give a multi-valued $q$-form on $Y$, since the
covering map $p$ is a local diffeomorphism: To see this, let $V$ be an
admissible neighbourhood of a point $y\in Y$, such that the inverse image $%
p^{-1}\left( V\right) $ is a disjoint union of neighbourhoods $U_i\,$in $X$.
On each $U_i$, the restriction $\left. p\right| U_i:U_i\rightarrow Y$ is a
diffeomorphism, so that we can pull back $\left. \Omega \right| U_i\mapsto
\left( \left. p\right| U_i\right) ^{-1*}\Omega $ on $V$. This gives a
multi-valued $q$-form on $V$, each branch being labelled by some $i$. We ask
under which condition all branches coincide. If this happens to be, we have $%
\left( \left. p\right| U_i\right) ^{-1*}\Omega =\left( \left. p\right|
U_j\right) ^{-1*}\Omega $ for all $i,j$ labelling different neighbourhoods $%
U_i$, $U_j$; but this means that 
\begin{equation}
\label{pp6t1form1}\left[ \left( \left. p\right| U_i\right) ^{-1}\circ \left(
\left. p\right| U_j\right) \right] ^{*}\left( \left. \Omega \right|
U_i\right) =\left( \left. \Omega \right| U_j\right) \quad . 
\end{equation}
For a general covering we can proceed no further, since there need not exist
a deck transformation mapping the neighbourhoods $U_i$ and $U_j$ into each
other. Such a deck transformation exists, however, if $X$ is simply
connected. If $x_i\in U_i$ and $x_j\in U_j$ such that $p\left( x_j\right)
=p\left( x_i\right) =y$, then there is a unique deck transformation $\gamma $
with $x_j=\gamma \left( x_i\right) $. On the other hand, the map on the LHS
of (\ref{pp6t1form1}) satisfies 
\begin{equation}
\label{pp6t1form2}\left( \left. p\right| U_i\right) \circ \left[ \left(
\left. p\right| U_i\right) ^{-1}\circ \left( \left. p\right| U_j\right)
\right] =\left( \left. p\right| U_j\right) 
\end{equation}
and maps $x_j$ to $x_i$; hence it coincides with the restriction $\left(
\gamma \mid U_j,U_i\right) $. This argument can be performed for any
neighbourhood $U$ of some point $x\in X$; in turn, this implies that $\gamma
^{*}\Omega =\Omega $. The fact that ${\cal D}$ now acts transitively on the
fibres of the covering means that this relation must hold for all deck
transformations $\gamma \in {\cal D}$.

As a consequence, if a $q$-form $\Omega $ on $X$ is invariant under ${\cal D}
$, then there is a uniquely determined $q$-form $\omega $ on $Y$ such that $%
\Omega =p^{*}\omega $; for, $\omega $ is defined by mapping any vector $W\in
T_yY$ to $\left( p_{loc}\right) _{*}W$, where $p_{loc}$ denotes the
restriction of $p$ to one of the connected components of the inverse image
of an admissible neighbourhood of $y$ in $Y$, and subsequently performing
the pairing $\left\langle \Omega ,\left( p_{loc}\right) _{*}W\right\rangle $%
. Because of the ${\cal D}$-invariance of $\Omega $, it does not matter
which connected component we choose, and hence this construction is
well-defined.

Altogether, we have shown

\subsection{Proposition \label{DeckInvariant}}

Let $p:X\rightarrow Y$ be a covering of connected manifolds, where $X$ is
simply connected. Then a smooth $q$-form $\Omega $ on $X$ is the pull-back
of a $q$-form $\omega $ on $Y$, $\Omega =p^{*}\omega $, if and only if $%
\gamma ^{*}\Omega =\Omega $ for all $\gamma \in {\cal D}$.

\TeXButton{Abst}{\vspace{0.8ex}}We now discuss the closure and exactness of
forms on $X$ and $Y$ that are related by the covering map according to $%
\Omega =p^{*}\omega $. Since closure of a differential form is a local
property, $\Omega $ is closed if and only if $\omega $ is closed. For if $%
d\omega =0$, then $dp^{*}\omega =p^{*}d\omega =0$; and conversely, if $V$ is
an admissible neighbourhood in $Y$, $d\left( \left. \omega \right| V\right)
=d\left[ \left( \left. p\right| U_i\right) ^{-1*}\Omega \right] =0$, if $%
d\Omega =0$; here $U_i$ is any neighbourhood in $X$ that projects down to $V$%
. This result is actually independent of whether $X$ is simply connected or
not, and makes use only of the existence of a form $\Omega $ such that $%
\Omega =p^{*}\omega $.

Now we examine exactness. Trivially, if the form $\omega $ on $Y$ is exact,
then $\Omega $ is exact, since then $\Omega =p^{*}d\alpha =d\left(
p^{*}\alpha \right) $. The converse is not true in general, however. For
assume that $\Omega =d\eta $ for a $\left( q-1\right) $-form $\eta $ on $X$.
Assuming that $\Omega =p^{*}\omega $, proposition \ref{DeckInvariant} says
that $d\left( \gamma ^{*}\eta -\eta \right) =0$. Thus $\gamma ^{*}\eta -\eta 
$ is closed, and since $X$ is simply connected, it is also exact. This means
that there exists a $\left( q-2\right) $-form $\chi \left( \gamma \right) $
on $X$ such that 
\begin{equation}
\label{pp6t1form3}\gamma ^{*}\eta =\eta +d\chi \left( \gamma \right) \quad . 
\end{equation}
Now unless $d\chi =0$, we see from proposition \ref{DeckInvariant} that $%
\eta $ can{\bf not} be the pull-back of a $\left( q-1\right) $-form on $Y$
under $p$, as this requires $\eta $ to be ${\cal D}$-invariant. Thus,
although $\Omega $ is exact, $\omega $ need not be exact; it is closed,
however, and hence defines an element $\left[ \omega \right] \in
H_{deRham}^q\left( Y\right) $.

The content of the last two paragraphs can be cast into a convenient form by
introducing ${\cal D}${\it -invariant cohomology classes }of forms on $X$:

\subsection{Definition}

Let $\Lambda _{{\cal D}I}^q\left( X\right) $ denote the subspace of all $%
{\cal D}$-invariant $q$-forms on $X$, i.e. $\gamma ^{*}\Omega =\Omega $ for
all $\gamma \in {\cal D}$. Let $Z_{{\cal D}I}^q\left( X\right) $ denote the
class of all elements $\Omega $ in $\Lambda _{{\cal D}I}^q\left( X\right) $
which are closed under $d$, $d\Omega =0$. Let $B_{{\cal D}I}^q\left(
X\right) $ denote the class of forms $\Omega $ in $\Lambda _{{\cal D}%
I}^q\left( X\right) $ which are exact under $d$, i.e. there exists a $\eta
\in \Lambda _{{\cal DI}}^{q-1}\left( X\right) $ such that $\Omega =d\eta $.
Now define the $q$-th ${\cal D}${\it -invariant cohomology group} $H_{{\cal D%
}I}^q\left( X\right) $ on $X$ as the quotient 
\begin{equation}
\label{pp6t1form4}H_{{\cal D}I}^q\left( X\right) \equiv Z_{{\cal D}%
I}^q\left( X\right) /B_{{\cal D}I}^q\left( X\right) \quad . 
\end{equation}
Formula (\ref{pp6t1form3}) shows that $B_{{\cal D}I}^q\left( X\right) 
\stackunder{\neq }{\subset }Z_{{\cal D}I}^q\left( X\right) $ in general, and
so $\dim H_{{\cal D}I}^q\left( X\right) $ can be non-vanishing although $X$
has trivial de Rham cohomology groups. This is expressed in the next
proposition, which is a consequence of proposition \ref{DeckInvariant} and
the discussion in the last paragraph:

\subsection{Proposition \label{DInvariantPullBack}}

Assume that $X$ is simply connected. Then the pull-back $p^{*}\Lambda
^q\left( Y\right) $ of the space of $q$-forms $\Lambda ^q\left( Y\right) $
on $Y$ by the covering map $p$ coincides with the subspace of all ${\cal D}$%
-invariant $q$-forms $\Lambda _{{\cal D}I}^q\left( X\right) $ on $X$, and $%
p^{*}$ is a group isomorphism onto $\Lambda _{{\cal D}I}^q\left( X\right) $.
Furthermore, $Z_{{\cal D}I}^q\left( X\right) =p^{*}Z^q\left( Y\right) $ and $%
B_{{\cal D}I}^q\left( X\right) =p^{*}B^q\left( Y\right) $, and therefore 
\begin{equation}
\label{pp6t1form5}\dim H_{{\cal D}I}^q\left( X\right) =\dim \frac{%
p^{*}Z^q\left( Y\right) }{p^{*}B^q\left( Y\right) }=\dim \frac{Z^q\left(
Y\right) }{B^q\left( Y\right) }=\dim H^q\left( Y\right) \;>\;0 
\end{equation}
in general.

\TeXButton{Abst}{\vspace{0.8ex}}There is a cohomological description of
formula (\ref{pp6t1form3}) in terms of special $\Lambda ^{*}\left( X\right) $%
-valued cochains, where $\Lambda ^{*}\left( X\right) $ denotes the ring of
differential forms on $X$; the associated cohomology is defined and
described in the appendix, chapter \ref{FormValuedCohomologyOnD}. For the
work pursued here the general case has no immediate application, but the
case when $X$ is simply connected and the forms involved are $1$-forms is
important. To start, let $\alpha $ be a closed $1$-form on $Y$; then $%
p^{*}\alpha $ is closed on $X$, hence exact, since $X$ is simply connected.
Thus there exists a smooth function $F:X\rightarrow \TeXButton{R}{\mathbb{R}}
$ with $dF=p^{*}\alpha $. The discussion in proposition \ref{DeckInvariant}
has shown that $dF$ is ${\cal D}$-invariant; therefore, $d\left( \gamma
^{*}F-F\right) =0$, or 
\begin{equation}
\label{pp6t1form6}F\circ \gamma -F\equiv c\left( \gamma \right) \in 
\TeXButton{R}{\mathbb{R}} 
\end{equation}
is a real constant on $X$, depending only on $\gamma $. In particular, $%
c\left( \gamma \right) \circ \gamma ^{\prime }=c\left( \gamma \right) $, and
it follows that 
\begin{equation}
\label{pp6t1form7}c\left( \gamma \gamma ^{\prime }\right) =F\circ \left(
\gamma \gamma ^{\prime }\right) -F=\left[ F+c\left( \gamma \right) \right]
\circ \gamma ^{\prime }-F=c\left( \gamma \right) +c\left( \gamma ^{\prime
}\right) \quad ; 
\end{equation}
hence $c:{\cal D}\rightarrow \TeXButton{R}{\mathbb{R}}$, $\gamma \mapsto
c\left( \gamma \right) $ is a real $1$-dimensional representation of ${\cal D%
}$. We have proven:

\subsection{Proposition \label{Sheets}}

Let $p:X\rightarrow Y$ be a covering of smooth manifolds, where $X$ is
simply connected. Let $\alpha $ be a closed $1$-form on $Y$. Then the
pull-back $p^{*}\alpha $ is an exact $\gamma $-invariant $1$-form on $X$,
with $p^{*}\alpha =dF$, where $F\in {\cal F}\left( X\right) $ is a smooth
function on $X$. Under ${\cal D}$-transformations, $F$ is invariant up to a
real $1$-dimensional ${\cal D}$-representation $c:{\cal D}\rightarrow 
\TeXButton{R}{\mathbb{R}}$, i.e. 
\begin{equation}
\label{pp6t1form8}F\circ \gamma =F+c\left( \gamma \right) \quad . 
\end{equation}

\TeXButton{Abst}{\vspace{0.8ex}}Now we see how the function $F$ gives rise
to multi-valued locally defined functions $f_\gamma $ on $Y$, which
represent local potentials, i.e. $0$-forms, for the closed $1$-form $\alpha $%
: Given an admissible open neighbourhood $V\subset Y$, choose a connected
component $U\subset X$ of $p^{-1}\left( V\right) $; then $\left. p\right| U$
is a diffeomorphism onto $V$, and every other connected component in $%
p^{-1}\left( V\right) $ is obtained as the image of $U$ under a deck
transformation $\gamma $. Since the sets $\gamma U$, $\gamma \in {\cal D}$,
are disjoint, this determines a collection $\left( f_\gamma \right) $ of
local potentials for $\alpha $ on $V$, each $f_\gamma $ being defined as 
\begin{equation}
\label{pp6t1pot1}f_\gamma =F\circ \left( \left. p\right| \gamma U\right)
^{-1}\quad . 
\end{equation}
By construction, we have $df_\gamma =\alpha $ for every $\gamma \in {\cal D}$%
. Furthermore, since $\left( \left. p\right| \gamma U\right) \circ \gamma
=\left. p\right| U$, it follows that 
\begin{equation}
\label{pp6t1pot2}f_\gamma =F\circ \left[ \gamma \circ \left( \left. p\right|
U\right) ^{-1}\right] =\left[ F\circ \gamma \right] \circ \left( \left.
p\right| U\right) ^{-1}=\left[ F+c\left( \gamma \right) \right] \circ \left(
\left. p\right| U\right) ^{-1}=f_e+c\left( \gamma \right) \quad , 
\end{equation}
where we have used (\ref{pp6t1form8}).

\TeXButton{Abst}{\vspace{0.8ex}}We now prove the important result, that $%
c\left( \gamma \right) $ can be expressed as an integral of $\alpha $ over
certain $1$-cycles or loops in $Y$: To this end we recall that the deck
transformation group ${\cal D}$ acts properly discontinuously and freely on
the simply connected covering manifold $X$, and that $Y$ is the orbit space $%
X/{\cal D}$. Furthermore, if $x,y$ are base points of $X,Y$, with $p\left(
x\right) =y$, then the fundamental group $\pi _1\left( Y,y\right) $ of $Y$
at $y$ is isomorphic to ${\cal D}$. This isomorphism is defined as follows:
If $\gamma \in {\cal D}$, let $\lambda $ be an arbitrary path in $X$
connecting the base point $x$ with its image $\gamma \left( x\right) $; then 
$\lambda $ projects into a loop $p\circ \lambda $ at $y$, whose associated
homotopy class $\left[ p\circ \lambda \right] $ represents $\gamma \in \pi
_1\left( Y,y\right) $. Any other choice $\lambda ^{\prime }$ of path is
homotopic to $\lambda $ due to $X$ being simply connected, hence the loops $%
\left[ p\circ \lambda \right] $ and $\left[ p\circ \lambda ^{\prime }\right] 
$ both represent the same homotopy class. Conversely, given a loop $l$ at $y$
representing $\gamma $, there exists a unique lift $\tilde l$ of $l$ through 
$p$ with initial point $x\in p^{-1}\left( y\right) $ (as follows from
theorem \ref{UniqueLiftingTheorem}); this means that $p\circ \tilde l=l$,
and $\tilde l\left( 1\right) =\gamma \left( x\right) $. Now we can prove

\subsection{Theorem \label{Cozykel}}

Let $y\in Y$ be the base point of $Y$, let $l$ be a loop at $y$ with $\left[
l\right] =\gamma \in \pi _1\left( Y,y\right) \simeq {\cal D}$. Then 
\begin{equation}
\label{pp6t1fo9}c\left( \gamma \right) =\int\limits_l\alpha \quad . 
\end{equation}
The integral depends only on the homotopy class $\left[ l\right] $ of $l$.

\TeXButton{Beweis}{\raisebox{-1ex}{\it Proof :}
\vspace{1ex}}

The lift $\tilde l$ of $l$ to the base point $x$ of $X$ satisfies $\tilde
l\left( 1\right) =\gamma \left( x\right) $. Since $p\circ \tilde l=l$, we
have%
$$
\int\limits_l\alpha =\int\limits_{p\circ \tilde l}\alpha
=\int\limits_{\tilde l}p^{*}\alpha \quad , 
$$
but $p^{*}\alpha =dF$, and since $\tilde l$ connects $x$ and $\gamma \left(
x\right) $, the last integral in the above equation is%
$$
\int\limits_{\tilde l}p^{*}\alpha =\int\limits_x^{\gamma \left( x\right)
}dF=F\circ \gamma \left( x\right) -F\left( x\right) =\left[ F\circ \gamma
-F\right] \left( x\right) =c\left( \gamma \right) \quad , 
$$
according to the definition (\ref{pp6t1form8}) of $c\left( \gamma \right) $.
This proves (\ref{pp6t1fo9}). Any other loop $l^{\prime }$ homotopic to $l$
lifts to a path $\tilde l^{\prime }$ homotopic to $\tilde l$; hence the
difference between the associated integrals is an integral of $dF$ over a
loop, which must vanish, as $dF$ is exact. \TeXButton{BWE}
{\hfill
\vspace{2ex}
$\blacksquare$}

\subsection{Corollary}

Any element $\gamma $ of ${\cal D}$ that has torsion lies in the kernel of $%
c $; in other words, if there is a $k\in \TeXButton{N}{\mathbb{N}}$ with $%
\gamma ^k=e$ then $c\left( \gamma \right) =0$.

\TeXButton{Beweis}{\raisebox{-1ex}{\it Proof :}
\vspace{1ex}}

Insert $e=\gamma ^k$ into $c\left( e\right) =0$, which gives $0=c\left(
\gamma ^k\right) =k\cdot c\left( \gamma \right) $ due to (\ref{pp6t1form7}).
Since $k>0$, $c\left( \gamma \right) =0$. \TeXButton{BWE}
{\hfill
\vspace{2ex}
$\blacksquare$}

\section{\u Cech cohomology and multi-valued functions \label{MultiValFunc}}

We now want to make more precise the notion of multi-valued functions that
serve as local potentials for closed $1$-forms on the non-simply connected
manifold $Y$. We want to express the local potentials, as they were defined
in formula (\ref{pp6t1pot1}), and their mutual relations, in terms of
locally defined quantities and glueing conditions without explicit reference
to a specific covering manifold. It is clear that we must make up for the
information that is lost by discarding the covering space from consideration
by some additional structure on the manifold $Y$. It turns out that the
necessary ingredients are

\begin{description}
\item[1.)]  \quad a countable simply connected path-connected open cover $%
{\cal V=}\left\{ V_a\subset Y\mid a\in A\right\} $ of $Y$, i.e. a collection
of countably many open sets $V_a\subset Y$ whose union gives $Y$, and such
that every loop in $V_a$ is homotopic in $Y$ to a constant loop, and all $%
V_a $ are path-connected.
\end{description}

Since $Y$ is a manifold, a cover of the type just described always exists.
We note that, as every element $V_a\in {\cal V}$ is simply connected (in $Y$%
), it is automatically admissible with respect to the covering map $%
p:X\rightarrow Y$; for, assume it were not admissible; then the inverse
image $p^{-1}\left( V_a\right) $ contained a connected, hence
path-connected, component $U$ on which the restriction of $p$ is not
injective. In particular, there are points $x,x^{\prime }\in U$, $x\neq
x^{\prime }$, but $p\left( x\right) =p\left( x^{\prime }\right) $. Choose a
path $\lambda $ connecting $x$ and $x^{\prime }$ in $U$; then this projects
into a loop in $V_a$ at $p\left( x\right) $, which is non-contractible in $Y$%
. This contradicts the assumption of $V_a$ being simply connected. --- The
second ingredient is

\begin{description}
\item[2.)]  \quad a ${\cal D}$-valued $1$-\u Cech-cocycle $\left( g_{ab}\in 
{\cal D}\right) $, $a,b\in A$, on ${\cal V}$, satisfying a certain condition
which expresses that the class of ${\cal D}$-isomorphic covering spaces to
which it refers is simply connected (elements of \u Cech cohomology and its
relation to covering spaces are explained in the appendix, chapters \ref
{CechCohomology}, \ref{DCoverings}, \ref{CechAndGlueing}).
\end{description}

We now explain what this condition means. Let $p:X\rightarrow Y$ be a
universal covering manifold of $Y$ with base points $x\in X$ and $y\in Y$
such that $y=p\left( x\right) $. The deck transformation group ${\cal D}$ of
such a covering is isomorphic to the fundamental group $\pi _1\left(
Y,y\right) $, the isomorphism being defined as in the discussion preceding
theorem \ref{Cozykel}: If $\left[ \lambda \right] $ is any loop class in $%
\pi _1\left( Y,y\right) $, let $\tilde \lambda $ denote the unique lift of $%
\lambda $ to $x$; then there exists a unique deck transformation $\gamma $
such that $\gamma x=\tilde \lambda \left( 1\right) $. This deck
transformation is the image of $\left[ \lambda \right] $ under the
above-mentioned isomorphism. The manifold $Y$ can be considered as the orbit
space $X/{\cal D}$ with the quotient topology. We now identify ${\cal D}$
with the fundamental group $\pi _1\left( Y,y\right) $, so that $Y$ can be
regarded as the orbit space $X/\pi _1\left( Y,y\right) $, $\pi _1$ acting
via deck transformations $\gamma $ on $X$, and $p:X\rightarrow X/{\cal D}=Y$
is a ${\cal D}$-covering.

Next, we note that, although a simply connected covering manifold $X$ of $Y$
is uniquely defined up to ${\cal D}$-isomorphisms, in general there are also 
${\cal D}$-coverings $q:Z\rightarrow Y=Z/{\cal D}$ of $Y$ which are {\bf not}
connected. This means that $Y$ can be expressed as an orbit space of
different, not necessarily connected, manifolds $Z$ (with base point $z$),
under an action of ${\cal D}$. These manifolds can be assembled into
equivalence classes, equivalence being expressed by ${\cal D}$-isomorphism
(see appendix, chapter \ref{DCoverings}), and a universal covering manifold $%
X$ determines just one class in this collection. As explained in the
appendix, chapter \ref{CechAndGlueing}, there is a bijection between these
equivalence classes and the classes of cohomologous $1$-\u Cech-cocycles on $%
{\cal V}$, in other words, the elements of $H^1\left( {\cal V;D}\right) $;
hence the bijection 
\begin{equation}
\label{pp6t1multValFunc10}\left\{ {\cal D}\text{-coverings with base point}%
\right\} /\text{isomorphism\ }\leftrightarrow \;H^1\left( {\cal V;D}\right)
\quad . 
\end{equation}
Furthermore, a result in the theory of covering spaces (see, e.g., \cite
{Fulton}) states that there is a bijection between $\left\{ {\cal D}\text{%
-coverings with base point}\right\} /$isomorphism and the set $Hom\left( 
{\cal D,D}\right) $ of homomorphisms from ${\cal D}=\pi _1\left( Y,y\right)
\rightarrow {\cal D}$. Hence we also have a bijection 
\begin{equation}
\label{pp6t1multValFunc11}H^1\left( {\cal V;D}\right) \;\leftrightarrow
\;Hom\left( {\cal D,D}\right) \quad . 
\end{equation}
We explain (LHS$\rightarrow $RHS) of this bijection. Let the \u Cech cocycle 
$\left( g_{ab}\right) $ be given. We first show that a ${\cal D}$-covering $%
Z $ of $Y$ exists such that the \u Cech cocycle determined by a collection $%
\left( i_a\right) $ of trivializations is the given one. To this end,
consider the topological sum of all ${\cal D}\times V_a$ (i.e. the
underlying set is a disjoint union) and define the relation $\left(
d,y\right) \sim \left( d^{\prime },y^{\prime }\right) $ for elements $\left(
d,y\right) \in {\cal D}\times V_a$, $\left( d^{\prime },y^{\prime }\right)
\in {\cal D}\times V_b$ to be true if and only if $\left( d^{\prime
},y^{\prime }\right) =\left( d\cdot g_{ab},y\right) $; then properties
(Trans1-Trans3) in the appendix, chapter \ref{CechAndGlueing}, guarantee
that $"\sim "$ is an equivalence relation. Now define $Z$ to be the quotient 
$Z\equiv \sqcup _{a\in A}{\cal D}\times V_a/\sim $, endowed with the final
topology (quotient topology). It is easy to see that $Z$ is a ${\cal D}$%
-covering of $Y$, i.e. $Y=Z/{\cal D}$. The set of maps $\left( i_a\right) $
that send elements $\left( d,y\right) \in {\cal D}\times V_a$ to the
equivalence classes $i_a\left( d,y\right) $ to which they belong provides
the natural collection of trivializations for this ${\cal D}$-space; it
follows that $i_b^{-1}\circ i_a\left( d,y\right) =\left( d\cdot
g_{ab},y\right) $, so that the associated \u Cech cocycle is the one we have
started with. Choose base points $z\in Z$, $y=q\left( z\right) $. Now
observe that the group ${\cal D}\simeq \pi _1\left( Y,y\right) $ enters this
construction in two different ways: Firstly, the elements of ${\cal D}$
locally label the different sheets of the covering in a trivialization.
Secondly, ${\cal D}$ is the set of homotopy classes of loops at $y$ on the
base manifold $Y$. We now construct a homomorphism $\rho $ from ${\cal D}$
as the set of homotopic loops to ${\cal D}$ as the labelling space for the
sheets of the covering: Choose a homotopy class $\left[ \gamma \right] $,
where $\gamma $ is a loop at the base point $y\in Y$. The unit interval $%
\left[ 0,1\right] $ can be divided \cite{Fulton} into $0=t_0<t_1<\cdots
<t_n=1$ such that the image of each interval $\left[ t_{i-1},t_i\right] $
lies in the open set $V_{a\left( i\right) }$. Then every point $\gamma
\left( t_i\right) $ lies in $V_{a\left( i\right) }\cap V_{a\left( i+1\right)
}$; on this domain, the cocycle $g_{a\left( i\right) a\left( i+1\right) }$
is constant. We lift the loop $\gamma $ to a curve $\tilde \gamma $ starting
at the base point $z$ in $Z$. If $i_{a\left( 0\right) }\left( d,y\right)
=z=\tilde \gamma \left( 0\right) $, then we find that 
\begin{equation}
\label{pp6t1muVaFu12}i_{a\left( 0\right) }^{-1}\tilde \gamma \left( 1\right)
=\left( d\cdot g_{a\left( 0\right) a\left( 1\right) }\cdot \cdots \cdot
g_{a\left( n\right) a\left( 0\right) },y\right) \equiv \left( d\cdot \rho
\left[ \gamma \right] ,y\right) \quad , 
\end{equation}
which defines an element $\rho \left[ \gamma \right] \equiv g_{a\left(
0\right) a\left( 1\right) }\cdot \cdots \cdot g_{a\left( n\right) a\left(
0\right) }\in {\cal D}$. It can be shown that this is independent of the
representative $\gamma $ of the homotopy class $\left[ \gamma \right] $, and
furthermore, that the assignment $\left[ \gamma \right] \mapsto \rho \left[
\gamma \right] $ is a homomorphism.

Thus, ${\cal D}$-coverings of $Y$, or equivalently, \u Cech cocycles on $%
{\cal V}$, are characterized by, and in turn characterize, homomorphisms $%
{\cal D}\rightarrow {\cal D}$. Cohomologous cocycles $g_{ab}^{\prime
}=h_a^{-1}g_{ab}h_b$ give rise to homomorphisms $\rho $, $\rho ^{\prime }$
that differ by conjugation, i.e. an inner automorphism of ${\cal D}$, $\rho
^{\prime }\left[ \gamma \right] =h_{a\left( 0\right) }^{-1}\cdot \rho \left[
\gamma \right] \cdot h_{a\left( 0\right) }$. For example, the homomorphism $%
{\cal D}\rightarrow {\cal D}$ is the trivial one, i.e. $\left[ \gamma
\right] \mapsto e\in {\cal D}$ for all elements $\left[ \gamma \right] $ in $%
{\cal D}$, if and only if the associated ${\cal D}$-covering of $Y$ is
(isomorphic to) the trivial $\#{\cal D}$-sheeted covering ${\cal D}\times
Y\rightarrow Y$ of $Y$, ${\cal D}$ acting on ${\cal D}\times Y$ by left
multiplication on the first factor. On the other hand, we now show that the
class of ${\cal D}$-isomorphic simply connected ${\cal D}$-coverings $%
p:X\rightarrow Y=X/{\cal D}$ is characterized by homomorphisms $\rho $ which
are inner automorphisms of ${\cal D}$: To see this, we first examine the
simply connected covering space $X$ consisting of all homotopy classes $%
\left[ \gamma \right] $ of curves $\gamma $ in $Y$ with initial point $%
\gamma \left( 0\right) =y$, where $y$ is the base point of $Y$. Choose
trivializations $i_a$ on ${\cal D}\times V_a$ so that the image $%
i_a^{-1}\left[ c\right] $ of the constant loop $c$ at $y$ is represented by $%
\left( \left[ c\right] ,y\right) $ for all $a$ for which $y\in V_a$. Then an
arbitrary loop class $\left[ \gamma \right] \in \pi _1\left( Y,y\right) $ is
represented by $\left[ \gamma \right] \equiv i_a\left( \left[ \gamma \right]
,y\right) \in X$; but this element is just the endpoint $\tilde \gamma
\left( 1\right) $ of the lift $\tilde \gamma $ of $\gamma $ to $\left[
c\right] \in p^{-1}\left( y\right) $, which implies by formula (\ref
{pp6t1muVaFu12}) that $\rho \left[ \gamma \right] =\left[ \gamma \right] $.
Since this holds for all $\left[ \gamma \right] $, we have $\rho =\left.
id\right| {\cal D}$ in this case. Now, if $p^{\prime }:X^{\prime
}\rightarrow Y$ is another simply connected covering, we have seen above
that the associated cocycles are cohomologous, hence the associated ${\cal D}
$-homomorphisms differ by an inner automorphism; but since $\rho $ is the
identity, this means that every homomorphism $\rho ^{\prime }:{\cal D}%
\rightarrow {\cal D}$ must be an inner automorphism of ${\cal D}$.

The developments of the last paragraph together with the content of theorems 
\ref{Sheets} and \ref{Cozykel} are summarized in

\subsection{Theorem: Multi-valued potentials \label{MultiValuedPotential}}

Let $\alpha $ be a closed $1$-form on the smooth manifold $Y$ with base
point $y$. Let ${\cal V}=\left\{ V_a\mid a\in A\right\} $ be a simply
connected path-connected open cover of $Y$. Let ${\cal D}\equiv \pi _1\left(
Y,y\right) $. Then

\begin{description}
\item[(A)]  \quad for every ${\cal D}$-valued $1$-\u Cech-cocycle $\left(
g_{ab}\right) $, $a,b\in A$, on ${\cal V}$ whose associated homomorphism $%
\rho :{\cal D}\rightarrow {\cal D}$ is an {\it inner} automorphism of ${\cal %
D}$, i.e. $\rho \left( d^{\prime }\right) =d\cdot d^{\prime }\cdot d^{-1}$
for some fixed $d\in {\cal D}$, there exists a collection of functions $%
f_{a,d}:V_a\rightarrow \TeXButton{R}{\mathbb{R}}$ for $a\in A$, $d\in {\cal D%
}$, such that

\begin{enumerate}
\item  \quad $f_{a,d}$ is a local potential for $\alpha $, i.e. $%
df_{a,d}=\alpha $ on $V_a$, for all $a\in A$ and $d\in {\cal D}$;

\item  \quad let $\lambda $ be a loop at $y$ with $\left[ \lambda \right]
=d\in \pi _1\left( Y,y\right) \simeq {\cal D}$. Then 
\begin{equation}
\label{pp6t1SheetsOfPotential}f_{a,d}=f_{a,e}+\int\limits_\lambda \alpha
\quad , 
\end{equation}
where $e$ is the identity in ${\cal D}$.

\item  \quad the $f_{a,d}$ satisfy a {\it glueing condition}, expressed by 
\begin{equation}
\label{pp6t1MultiValFunc1}f_{a,d}=f_{b,d\cdot g_{ab}} 
\end{equation}
on $V_a\cap V_b\neq \emptyset $.
\end{enumerate}

\item[(B)]  \quad Let $\left( g_{ab}^{\prime }\right) $ be a cocycle
cohomologous to $\left( g_{ab}\right) $, and let $\left( f_{a,d}^{\prime
}\right) $ be a collection of functions on ${\cal V}$ satisfying properties
(A1--A3) with respect to $\left( g_{ab}^{\prime }\right) $. Then there
exists a real constant $c$ and a ${\cal D}$-valued $0$-\u Cech cochain $%
\left( k_a:V_a\rightarrow {\cal D}\right) $ on ${\cal V}$ such that 
\begin{equation}
\label{pp6t1MultiValFunc2}f_{a,d}^{\prime }=f_{a,d\cdot k_a}+c 
\end{equation}
for all $a\in A$, $d\in {\cal D}$. The $0$-cochain $\left( k_a\right) $ is
determined by the cocycles $\left( g_{ab}\right) $ and $\left(
g_{ab}^{\prime }\right) $ up to its value $k_{a_0}$ on the open set $%
V_{a_0}\in {\cal V}$ which contains the base point $y$; on $V_{a_0}$, $%
k_{a_0}$ can range arbitrarily in the coset $h_{a_0}^{-1}\cdot {\cal D}%
_{center}$, where $h_{a_0}$ is the value of the \u Cech cochain which
relates the cocycles $\left( g_{ab}\right) $ and $\left( g_{ab}^{\prime
}\right) $ on $V_{a_0}$, and ${\cal D}_{center}$ is the center of ${\cal D}$.

\item[(C)]  \quad {\bf Definition:\quad }A collection $\left(
g_{ab};f_{a,d}\right) $ satisfying properties (A1--A3) will be called a {\it %
multi-valued potential function} for the closed $1$-form $\alpha $ on $Y$.
\end{description}

\TeXButton{Beweis}{\raisebox{-1ex}{\it Proof :}
\vspace{1ex}}

\underline{Ad (A) :}\quad Let $X$ be the identification space $i_a:{\cal D}%
\times V_a\rightarrow X\equiv \bigsqcup\limits_{a\in A}{\cal D}\times
V_a/\sim $, where the relation $\left( d,y\right) \sim \left( d^{\prime
},y^{\prime }\right) $ for $\left( d,y\right) \in {\cal D}\times V_a$, $%
\left( d^{\prime },y^{\prime }\right) \in {\cal D}\times V_b$ is defined to
be true if and only if $\left( d^{\prime },y^{\prime }\right) =\left( d\cdot
g_{ab},y\right) $, in which case these elements are identified according to $%
i_a\left( d,y\right) =i_b\left( d^{\prime },y^{\prime }\right) $. Then $X$
is a covering space of $Y$, and $Y$ is the space of orbits on $X$ under the
action of ${\cal D}$ on $X$ according to $\left( d^{\prime },\left(
d,y\right) \right) \mapsto \left( d^{\prime }\cdot d,y\right) $. Since the
homomorphism $\rho :{\cal D\rightarrow D}$ associated with $\left(
g_{ab}\right) $ is an inner automorphims by assumption, it follows from the
discussion at the beginning of this section that $X$ is simply connected.
Since $Y$ is a smooth manifold, $X$ is a smooth manifold. Let $%
p:X\rightarrow Y$ be the projection, which is a local diffeomorphism. Then $%
p^{*}\alpha $ is a closed, hence exact, $1$-form on $X$, and has a potential 
$F$ with $dF=p^{*}\alpha $. The identification maps $i_a:{\cal D}\times
V_a\rightarrow p^{-1}\left( V_a\right) $ are the natural trivializations for
this covering. If we write $i_a\left( d,y\right) \equiv i_{a,d}\left(
y\right) $, then $i_{a,d}$ is the inverse of the restriction $\left.
p\right| i_a\left( \left\{ d\right\} \times V_a\right) $. Now define $%
f_{a,d}\left( y\right) \equiv F\circ i_a\left( d,y\right) $ for $y\in V_a$.
By construction, $df_{a,d}=i_{a,d}^{*}dF$, and since $dF=\left[ \left.
p\right| i_a\left( \left\{ d\right\} \times V_a\right) \right] ^{*}\alpha $,
it follows that $df_{a,d}=\alpha $ on $V_a$. Furthermore, if also $y\in V_b$%
, then $F\circ i_a\left( d,y\right) =\left( F\circ i_b\right) \circ \left(
i_b^{-1}\circ i_a\right) \left( d,y\right) =\left( F\circ i_b\right) \left(
d\cdot g_{ab},y\right) =f_{b,d\cdot g_{ab}}\left( y\right) $. Formula (\ref
{pp6t1SheetsOfPotential}) is a consequence of theorems \ref{Sheets} and \ref
{Cozykel}. This proves (A).

\underline{Ad (B) :}\quad From the cocycles $\left( g_{ab}\right) $, $\left(
g_{ab}^{\prime }\right) $, construct coverings $p:X\rightarrow Y$, $%
q:Z\rightarrow Y$ as in the proof of (A), with trivializations $i_a:{\cal D}%
\times V_a\rightarrow X$, $j_a:{\cal D}\times V_a\rightarrow Z$. Then both $%
X $ and $Z$ are smooth, simply connected manifolds. The glueing condition $%
f_{a,d}=f_{b,d\cdot g_{ab}}$ for $\left( f_{a,d}\right) $ implies that there
exists a smooth function $F$ on $X$ such that $F\circ i_{a,d}=f_{a,d}$: For,
we have $f_{a,d}\left( y\right) =f_{b,d^{\prime }}\left( y^{\prime }\right) $
whenever $i_a\left( d,y\right) =i_b\left( d^{\prime },y^{\prime }\right) $;
the universal property of the identification space \cite{Brown} $i_a:{\cal D}%
\times V_a\rightarrow X$ guarantees the existence of a smooth $F$ with the
desired property. A similar function $F^{\prime }$ with $f^{\prime }\circ
j_{a,d}=f_{a,d}^{\prime }$ exists on $Z$. Since $X$ and $Z$ are ${\cal D}$%
-isomorphic, there exists a diffeomorphism $\phi :Z\rightarrow X$ preserving
fibres, i.e. $p\circ \phi =q$, and being ${\cal D}$-equivariant, i.e. $\phi
\left( d\cdot z\right) =d\cdot \phi \left( z\right) $. In the
trivializations employed above we have 
\begin{equation}
\label{pp6t1zwi20}i_a^{-1}\circ \phi \circ j_a\left( d,y\right) =\left(
d\cdot k_a,y\right) \quad , 
\end{equation}
where the collection $\left( k_a:V_a\rightarrow {\cal D}\right) $ defines a $%
0$-\u Cech cochain on ${\cal V}$.

Since $d\left( F\circ \phi \right) =\phi ^{*}dF=\phi ^{*}p^{*}\alpha
=q^{*}\alpha $, we see that both $F\circ \phi $ and $F^{\prime }$ are
potentials for $q^{*}\alpha $ on $Z$; since $Z$ is simply connected, it
follows that $F^{\prime }=F\circ \phi +c$ with $c\in \TeXButton{R}
{\mathbb{R}}$. Then%
$$
f_{a,d}^{\prime }\left( y\right) =F^{\prime }\circ j_a\left( d,y\right)
=F\circ \left( \phi \circ j_a\right) \left( d,y\right) +c=\left( F\circ
i_a\right) \left( d\cdot k_a,y\right) +c=f_{a,d\cdot k_a}\left( y\right)
+c\quad , 
$$
where we have used (\ref{pp6t1zwi20}); thus (\ref{pp6t1MultiValFunc2})
follows.

Furthermore, assume that $V_a\cap V_b\neq \emptyset $, then the analogue of (%
\ref{pp6t1zwi20}) on $V_b$ reads $i_b^{-1}\circ \phi \circ j_b\left( d\cdot
g_{ab}^{\prime },y\right) =\left( d\cdot g_{ab}^{\prime }\cdot k_b,y\right) $%
, which implies%
$$
\left( i_b^{-1}\circ i_a\right) \circ \left( i_a^{-1}\circ \phi \circ
j_a\right) \circ \left( j_a^{-1}\circ j_b\right) \left( d\cdot
g_{ab}^{\prime },y\right) =\left( d\cdot g_{ab}^{\prime }\cdot k_b,y\right)
\quad , 
$$
from which it follows that 
\begin{equation}
\label{pp6t1zwi40}g_{ab}^{\prime }\cdot k_b=k_a\cdot g_{ab}\quad . 
\end{equation}
This formula says that the cochain $\left( k_a\right) $ is not arbitrary,
but is completely determined by the cocycles $\left( g_{ab}\right) $ and $%
\left( g_{ab}^{\prime }\right) $, once a choice has been made for the value
of $k_a$ on one selected $V_a$ (e.g. the $V_{a_0}$ which contains the base
point $y$ of $Y$). This follows from path-connectedness of $Y$; for, if $%
V_{a_n}$ is any open set in ${\cal V}$, there exist finite sequences $%
g_{a_0a_1}^{\prime }$, $\ldots $, $g_{a_{n-1}a_n}^{\prime }$, and $%
g_{a_0a_1} $, $\ldots $, $g_{a_{n-1}a_n}$ so that%
$$
k_{a_n}=g_{a_{n-1}a_n}^{\prime -1}\cdots g_{a_0a_1}^{\prime -1}\cdot
k_{a_0}\cdot g_{a_0a_1}\cdots g_{a_{n-1}a_n}\quad . 
$$
Since the selected $k_{a_0}$ is constant on $V_{a_0}$, a choice of $k_{a_0}$
is just a choice of an element of ${\cal D}$. Furthermore, if $l$ is any
loop in $Y$ at the base point $y$ representing the element $\delta \in {\cal %
D}$, then from (\ref{pp6t1muVaFu12}) we see that the associated series $%
\left( g_{a_ia_{i+1}}\right) $ and $\left( g_{a_ia_{i+1}}^{\prime }\right) $
represent $\rho \left[ l\right] \equiv g_{a\left( 0\right) a\left( 1\right)
}\cdots g_{a\left( n\right) a\left( 0\right) }$, $\rho ^{\prime }\left[
l\right] \equiv g_{a\left( 0\right) a\left( 1\right) }^{\prime }\cdots
g_{a\left( n\right) a\left( 0\right) }^{\prime }$, respectively. Thus, we
must have%
$$
k_{a_0}=\rho ^{\prime }\left[ l\right] ^{-1}\cdot k_{a_0}\cdot \rho \left[
l\right] 
$$
for all homotopy classes $\left[ l\right] \in \pi _1\left( Y,y\right) \equiv 
{\cal D}$. However, as ${\cal D}$ acts transitively on each fibre of the
covering $p:X\rightarrow Y$, it follows that the elements $\rho ^{\prime
}\left[ l\right] $, $\rho \left[ l\right] $ take any value in ${\cal D}$, as 
$\left[ l\right] $ ranges in ${\cal D}$. Using $\rho ^{\prime }\left[
l\right] =h_{a_0}^{-1}\cdot \rho \left[ l\right] \cdot h_{a_0}$ it follows
that $k_{a_0}=h_{a_0}^{-1}\cdot \rho \left[ l\right] ^{-1}\cdot h_{a_0}\cdot
k_{a_0}\cdot \rho \left[ l\right] $, or%
$$
\delta \cdot \left( h_{a_0}k_{a_0}\right) =\left( h_{a_0}k_{a_0}\right)
\cdot \delta 
$$
for all $\delta \in {\cal D}$. But this implies that $h_{a_0}k_{a_0}$ must
lie in the center of ${\cal D}$, which is a normal subgroup of ${\cal D}$.
Therefore $k_{a_0}$ can range in the coset $h_{a_0}^{-1}\cdot {\cal D}%
_{center}$. Hence the collection $\left( f_{a,d}\right) $ is determined up
to a real constant and an arbitrary element in $h_{a_0}^{-1}\cdot {\cal D}%
_{center}$. \TeXButton{BWE}{\hfill
\vspace{2ex}
$\blacksquare$}

\TeXButton{Abst}{\vspace{0.8ex}}In the sequel we apply the covering
techniques discussed so far to coverings of symplectic manifolds. We first
present our notational conventions:

\section{Notation and conventions for symplectic manifolds \label
{NotationSymplecticManifolds}}

Sections \ref{NotationSymplecticManifolds}--\ref{HamiltonianVectorFields}
are based on \cite{Woodhouse,cram/pir,GuillStern}.

--- We recall that a symplectic form $\omega $ on a manifold $M$ is a
closed, nondegenerate $2$-form on $M$. In this case, the pair $\left(
M,\omega \right) $ is called a symplectic manifold.

--- By ${\cal F}\left( M\right) $ we denote the set of all smooth functions $%
f:M\rightarrow \TeXButton{R}{\mathbb{R}}$. On a symplectic manifold we can
make ${\cal F}\left( M\right) $ into a real Lie algebra using Poisson
brackets.

--- By $\chi \left( M\right) $ we denote the set of all smooth vector fields
on $M$.

--- If $G$ is a Lie group, we will frequently denote its Lie algebra by $%
\hat g$, and the coalgebra, i.e. the space dual to $\hat g$, by $g^{*}$.

--- Given an action $\phi :G\times M\rightarrow M$ of a Lie group $G$ on a
manifold $M$, we will frequently denote the components of its tangent map $%
\phi _{*}$ by $\phi _{*}=\left( \frac{\partial \phi }{\partial G},\frac{%
\partial \phi }{\partial M}\right) $. If $A\in \hat g$, the induced vector
field on $M$ will be denoted by $\frac{\partial \phi }{\partial G}A$ or $%
\tilde A$.

--- Interior multiplication of a vector $V$ with a $q$-form $\omega $ will
be denoted by $V\TeXButton{i}{\intmul}\omega $.

--- A diffeomorphism $f:M\rightarrow M$ on a symplectic manifold $\left(
M,\omega \right) $ is called canonical transformation, if $f^{*}\omega
=\omega $. An action $\phi :G\times M\rightarrow M$ of a Lie group $G$ on $M$
is called symplectic if every $\phi _g$ is a canonical transformation.

--- $I$ generally denotes the closed interval $I=\left[ 0,1\right] \subset 
\TeXButton{R}{\mathbb{R}}$.

\section{Hamiltonian and locally Hamiltonian vector fields \label
{HamiltonianVectorFields}}

On a symplectic manifold, the symplectic form $\omega $ provides a
non-natural isomorphism between tangent spaces $T_xM$ and cotangent spaces $%
T_x^{*}M$ at every point $x\in M$, since $\omega $ is non-degenerate. In
particular, for every $f\in {\cal F}\left( M\right) $ there exists a unique
vector field $\rho f\in \chi \left( M\right) $ such that 
\begin{equation}
\label{pp6t1fo1}\rho f\TeXButton{i}{\intmul}\omega +df=0\quad . 
\end{equation}
This gives us a well-defined map $\rho :{\cal F}\left( M\right) \rightarrow
\chi \left( M\right) $. A vector field $V\in \chi \left( M\right) $ which is
the image of a function $f\in {\cal F}\left( M\right) $ under $\rho $, $%
V=\rho f$, is called Hamiltonian. The set of all (smooth) Hamiltonian vector
fields on $M$ is denoted by $\chi _H\left( M\right) $, and is a real vector
space.

On the other hand, the set $\chi _{LH}\left( M\right) $ of vector fields $V$
on $M$ which satisfy 
\begin{equation}
\label{pp6t1fo2}{\cal L}_V\omega =0\quad , 
\end{equation}
where ${\cal L}_V$ denotes a Lie derivative, is called the set of locally
Hamiltonian vector fields. This means that on every simply connected open
neighbourhood $U\subset M$, the $1$-form $V\TeXButton{i}{\intmul}\omega $ is
exact, hence there exists a smooth function $f\in {\cal F}\left( U\right) $
such that 
\begin{equation}
\label{pp6t1fo4}V\TeXButton{i}{\intmul}\omega +df=0\quad \text{on }U\quad . 
\end{equation}
As (\ref{pp6t1fo4}) holds in a neighbourhood of every point, we refer to $V$
as a locally Hamiltonian vector field. The functions $f\in {\cal F}\left(
U\right) $ need not be globally defined. If $M$ is simply connected, then
every locally Hamiltonian vector field is Hamiltonian, and $\chi _{LH}\left(
M\right) =\chi _H\left( M\right) $. This will not be true for the manifolds
we are interested in in this work.

\section{Cotangent bundles \label{CotangentBundles}}

In this section we compile some standard facts about cotangent bundles we
shall use throughout this paper. This material is discussed in standard
textbooks on Symplectic Geometry (e.g. \cite{GuillStern}), Mechanics (e.g. 
\cite{AbrahamMarsden}), Differential Geometry (e.g. \cite{cram/pir}), and
Algebraic Topology (e.g. \cite{DodsonParker}).

--- On the cotangent bundle $T^{*}M$ of a manifold $M$ we have a projection $%
\tau :T^{*}M\rightarrow M$, and a natural symplectic $2$-form being given as
the differential $\omega =d\theta $ of the canonical $1$-form \cite
{GuillStern,cram/pir} $\theta $ on $T^{*}M$, which is defined as
follows: For $V\in T_{\left( m,p\right) }T^{*}M$, the action of $\theta $ on 
$V$ is defined by $\left\langle \theta ,V\right\rangle \left( m,p\right)
\equiv \left\langle p,\tau _{*}V\right\rangle $.

--- The homotopy groups of $T^{*}M$ are determined by those of $M$; in fact
we have 
\begin{equation}
\label{pp6t1homotopyGroups}\pi _n\left( T^{*}M\right) \simeq \pi _n\left(
M\right) 
\end{equation}
for all $n\ge 0$. This follows from the exact homotopy sequence for
fibrations (see any textbook on Algebraic Topology, e.g. \cite{DodsonParker}%
), 
\begin{equation}
\label{pp6t1homSeq}\cdots \rightarrow \pi _{n+1}\left( B\right) \stackrel{%
\partial }{\longrightarrow }\pi _n\left( F\right) \stackrel{i_{\#}}{%
\longrightarrow }\pi _n\left( E\right) \stackrel{p_{\#}}{\longrightarrow }%
\pi _n\left( B\right) \rightarrow \cdots \quad , 
\end{equation}
where $E\stackrel{p}{\longrightarrow }B$ is a fibration with standard fibre $%
F$. For a vector bundle with $F\simeq \TeXButton{R}{\mathbb{R}}^k$, the
homotopy groups $\pi _n\left( \TeXButton{R}{\mathbb{R}}^k\right) $ are
trivial, hence%
$$
0\rightarrow \pi _n\left( E\right) \stackrel{p_{\#}}{\longrightarrow }\pi
_n\left( B\right) \rightarrow 0 
$$
is an exact sequence, which says that $p_{\#}$ is an isomorphism in this
case. As a consequence, (\ref{pp6t1homotopyGroups}) follows.

A diffeomorphism $f:M\rightarrow M$ can be extended to a diffeomorphism $%
^{*}f:T^{*}M\rightarrow T^{*}M$ as follows \cite{GuillStern}: For $\left(
m,p\right) \in T^{*}M$, let 
\begin{equation}
\label{pp6t1cot1}\left( ^{*}f\right) \left( m,p\right) \equiv \left( f\left(
m\right) ,\left( f^{-1}\right) ^{*}p\right) \quad . 
\end{equation}
$^{*}f$ is fibre-preserving and hence a bundle map. Given two
diffeomorphisms $f,g$ we find 
\begin{equation}
\label{pp6t1cot1a}^{*}\left( fg\right) =\left( ^{*}f\right) \left(
^{*}g\right) \quad . 
\end{equation}
From definition (\ref{pp6t1cot1}) it follows immediately that every $^{*}f$
preserves the canonical $1$-form $\theta $, 
\begin{equation}
\label{pp6t1cot3}\left( ^{*}f\right) ^{*}\theta =\theta \quad . 
\end{equation}

\section{Coverings of cotangent bundles \label{CoveringCotBundle}}

In this section we start with symplectic manifolds $\left( T^{*}Y,d\theta
\right) $ which are the cotangent bundles $T^{*}Y$ of non-simply connected
configuration spaces $Y$, and are endowed with the natural symplectic $2$%
-form $d\theta $ which is associated with the canonical $1$-form $\theta $
on the cotangent bundle. Then we extend a simply connected covering manifold 
$X$ of $Y$ to a covering manifold $T^{*}X$ of $T^{*}Y$ and show how the
projection map $p:X\rightarrow Y$ can be extended to give a local
symplectomorphism between $T^{*}X$ and $T^{*}Y$ which is also a covering map.

\subsection{Covering spaces and their cotangent bundles}

Let $p:X\rightarrow Y$ be a covering of manifolds. Let $\Theta ,\theta $
denote the canonical $1$-forms on the cotangent bundles $T^{*}X$, $T^{*}Y$,
respectively (see section \ref{CotangentBundles}). The bundle projections
are written as $\sigma :T^{*}X\rightarrow X$ and $\tau :T^{*}Y\rightarrow Y$%
; the canonical symplectic $2$-forms on $T^{*}X$, $T^{*}Y$ are $\Omega
=d\Theta $, $\omega =d\theta $, respectively.

Central to our developments is the observation that we can extend the
projection $p$ to a covering map of cotangent bundles; for all coverings it
is understood that they are smooth:

\subsection{Theorem \label{PStarIsACovering}}

Let $p:X\rightarrow Y$ be a covering. Then $p$ can be extended to a bundle
map $^{*}p:T^{*}X\rightarrow T^{*}Y$ such that $^{*}p:T^{*}X\rightarrow
T^{*}Y$ is a covering.

\TeXButton{Beweis}{\raisebox{-1ex}{\it Proof :}
\vspace{1ex}}

We first have to show that a well-defined extension $^{*}p$ exists. To this
end, let $V\subset Y$ be an admissible open neighbourhood in $Y$, and let $%
U\subset X$ denote a connected component of $p^{-1}\left( V\right) $. On
every $U$, the restriction $\left. p\right| U:U\rightarrow Y$ is a
diffeomorphism, hence its extension 
\begin{equation}
\label{pp6t1pstarisacovering}^{*}\left( \left. p\right| U\right) \equiv
\left( \left. p\right| U,\left[ \left. p\right| U\right] ^{-1*}\right) 
\end{equation}
is well-defined. The collection of all $U$, being the collection of all
inverse images of admissible open neighbourhoods $V$ in $Y$ forms an open
cover of $X$. On the intersection of any two of the $U,U^{\prime }$, the
locally defined maps (\ref{pp6t1pstarisacovering}) coincide. Now define $%
^{*}p$ to be the uniquely determined function on $T^{*}X$ whose restriction
to any of the $U$ coincides with $^{*}\left( \left. p\right| U\right) $.

From this it follows that $^{*}p$ is well-defined, and is smooth, provided $%
p $ is smooth. Its local form (\ref{pp6t1pstarisacovering}) shows that $%
^{*}p $ preserves fibres and hence is a bundle map. If $V$ is any admissible
neighbourhood in $Y$, then $\bigcup\limits_{y\in V}T_y^{*}Y$ is a
neighbourhood in $T^{*}Y$ whose inverse image under $^{*}p$ is a disjoint
union of neighbourhoods in $T^{*}X$. This says that $^{*}p$ is a covering
map. \TeXButton{BWE}{\hfill
\vspace{2ex}
$\blacksquare$}

\TeXButton{Abst}{\vspace{0.8ex}}As a consequence of the last theorem, we can
pull back $q$-forms $\omega $ on $T^{*}Y$ to $q$-forms $\left( ^{*}p\right)
^{*}\omega $ on $T^{*}X$ via $^{*}p$. Hence the pull-back $\left(
^{*}p\right) ^{*}\theta $ is well-defined. The next theorem explains the
relation between this pull-back and the canonical symplectic potential $%
\Theta $ on $T^{*}X$. To this end we note that

\subsection{Lemma}

\begin{equation}
\label{pp6t1cover2}\tau \circ \left( ^{*}p\right) =p\circ \sigma \quad . 
\end{equation}
\TeXButton{Beweis}{\raisebox{-1ex}{\it Proof :}
\vspace{1ex}}

This follows from the definition $\left( ^{*}p\right) \left( x,\alpha
\right) =\left( p\left( x\right) ,\left( \left. p\right| U\right)
^{-1*}\alpha \right) $ for $\left( x,\alpha \right) \in T^{*}X$. 
\TeXButton{BWE}{\hfill
\vspace{2ex}
$\blacksquare$}

\subsection{Theorem}

The pull-back of the canonical $1$-form $\theta $ on $Y$ under $^{*}p$
coincides with the canonical $1$-form $\Theta $ on $X$,

\begin{equation}
\label{pp6t1cover3}\left( ^{*}p\right) ^{*}\theta =\Theta \quad . 
\end{equation}
\TeXButton{Beweis}{\raisebox{-1ex}{\it Proof :}
\vspace{1ex}}

Let $\left( x,\alpha \right) \in T^{*}X$, let $V\in T_{\left( x,\alpha
\right) }T^{*}X$. Then $\left( ^{*}p\right) \left( x,\alpha \right) =\left(
p\left( x\right) ,\left( \left. p\right| U\right) ^{-1*}\alpha \right) $,
where $U$ is a neighbourhood of $x$ in $X$ on which the restriction of $p$
is injective. Now if $\left( ^{*}p\right) _{*}V$ is paired with $\theta $ at
the point $\left( ^{*}p\right) \left( x,\alpha \right) \in T^{*}Y$, we
obtain $\left\langle \theta ,\left( ^{*}p\right) _{*}V\right\rangle
=\left\langle \left( \left. p\right| U\right) ^{-1*}\alpha ,\tau _{*}\left(
^{*}p\right) _{*}V\right\rangle $; however, from (\ref{pp6t1cover2}) we have
that $\tau _{*}\left( ^{*}p\right) _{*}=p_{*}\sigma _{*}$, and hence%
$$
\left\langle \left( \left. p\right| U\right) ^{-1*}\alpha ,\tau _{*}\left(
^{*}p\right) _{*}V\right\rangle =\left\langle \alpha ,\sigma
_{*}V\right\rangle =\left\langle \Theta ,V\right\rangle \quad , 
$$
by the definition of $\Theta $; this proves the theorem. \TeXButton{BWE}
{\hfill
\vspace{2ex}
$\blacksquare$}

\TeXButton{Abst}{\vspace{0.8ex}}There is an immediate important consequence:

\subsection{Corollary \label{LocalSymplectomorphism}}

The canonical symplectic $2$-form $d\Theta $ on $T^{*}X$ is the pull-back of
the canonical symplectic $2$-form $d\theta $ on $T^{*}Y$ under $^{*}p$, and
hence the map $^{*}p:T^{*}X\rightarrow T^{*}Y$ is a {\bf local
symplectomorphism}.

\TeXButton{Beweis}{\raisebox{-1ex}{\it Proof :}
\vspace{1ex}}

This follows from $d\Theta =d\left( ^{*}p\right) \theta =\left( ^{*}p\right)
d\theta $. \TeXButton{BWE}{\hfill
\vspace{2ex}
$\blacksquare$}

\TeXButton{Abst}{\vspace{0.8ex}}Hence we can relate, and locally identify,
the dynamics taking place on $T^{*}Y$ to an associated dynamical system on
the symplectic covering manifold $T^{*}X$; this is one of the major
statements of this work. As was shown in (\ref{pp6t1homotopyGroups}), the
homotopy groups of a cotangent bundle are isomorphic to those of its base
space. In particular, if $X$ is simply connected, the fundamental groups
obey the relations 
\begin{equation}
\label{pp6t1fndGrp}\pi _1\left( T^{*}X\right) =\pi _1\left( X\right) =0\quad
,\quad \pi _1\left( T^{*}Y\right) =\pi _1\left( Y\right) ={\cal D}\quad , 
\end{equation}
and it follows that the deck transformation group ${\cal D}\left(
T^{*}X\right) $ of the covering $^{*}p:T^{*}X\rightarrow T^{*}Y$ is just $%
\pi _1\left( T^{*}Y\right) ={\cal D}$. This will enable us to remove the
multi-valuedness of a local moment map given on a cotangent space $T^{*}Y$
by constructing the local symplectomorphism $^{*}p$, and then studying the
associated dynamics on the simply connected symplectic covering space $%
T^{*}X $, on which every locally Hamiltonian vector field has a globally
defined charge, and hence every symplectic group action has a global moment
map; see section \ref{LocalMomentMaps} for the details. Our next task
therefore is to study how (Lie) group actions defined on $Y$ (and, in turn, $%
T^{*}Y$) can be lifted to a covering space space $X$ (and $T^{*}X$), in
particular, when $X$ is simply connected; this is done in sections \ref
{LiftOfGroupActions} and \ref{PreservationOfTheGroupLaw} for general
symplectic manifolds which are not necessarily cotangent bundles.

\section{Lift of group actions under covering maps \label{LiftOfGroupActions}
}

In this section we prove a couple of theorems about the lifting of a left
action $\phi $ of a Lie group $G$ on a manifold $Y$ to a covering manifold $%
X $. Here we examine existence and uniqueness of lifts; in the next section
we examine under which conditions the group law of $G$ is preserved under
the lift.

Let $p:X\rightarrow Y$ be a covering, where $X$, $Y$ are connected, let $V$
be a manifold, let $f:V\rightarrow Y$. As explained in section \ref
{BasicsAboutCoverings}, one calls a map $\tilde f:V\rightarrow X$ a {\it %
lift of }$f${\it \ through }$p$, if $p\circ \tilde f=f$. For the sake of
convenience, we establish a similar phrase for a related construction which
will frequently appear in the following: If $f:Y\rightarrow Y$, we call $%
\hat f:X\rightarrow X$ a {\it lift of }$f${\it \ to }$X$ if 
\begin{equation}
\label{pp6t1lift01}p\circ \hat f=f\circ p\quad . 
\end{equation}
We enhance this condition for the case that $G$ is a group and $\phi
:G\times Y\rightarrow Y$ is a left action of $G$ on $Y$. In this case, we
call a smooth map $\hat \phi :G\times X\rightarrow X$ a {\it lift of }$\phi $%
{\it \ to }$X$ if

\begin{description}
\item[(L1)]  \quad $p\circ \hat \phi =\phi \circ \left( id_G\times p\right) $%
, and

\item[(L2)]  $\quad \hat \phi _e=id_X$,
\end{description}

where $\hat \phi _e$ denotes the map $x\mapsto \hat \phi _e\left( x\right)
\equiv \hat \phi \left( e,x\right) $.

We remark that (L1) does {\bf not} imply that $p$ is $G$-equivariant (or a $%
G $-morphism). This is because $G$-equivariance requires $X$ to be a $G$%
-space, i.e. a manifold with a smooth left {\bf action of }$G$ on it. This
in turn means that the lift $\hat \phi $ must preserve the group law of $G$,
i.e. $\widehat{\phi _{gh}}=\hat \phi _g\hat \phi _h$. We will see shortly
that this is guaranteed only if $G$ is connected. If $G$ has several
connected components, the lift can give rise to an extension $\tilde G$ of
the original group by the deck transformation group ${\cal D}$; this is
described in theorem \ref{ExtensionOfG}. The matter of equivariance of the
covering map $p$ is taken up in section \ref{GSpacesCoveringMaps}.

We note that if a map $\hat \phi $ satisfying (L1) exists, it is determined
only up to a deck transformation $\gamma $;\ for if we define $\hat \phi
^{\prime }\equiv \hat \phi \left( id_G\times \gamma \right) $, $\hat \phi
^{\prime }$ also satisfies (L1); this is why we have to impose (L2)
additionally. However, we show that if $\hat \phi $ exists, then it can
always be assumed that it satisfies (L2):

\subsection{Proposition \label{Smoothness}}

\begin{enumerate}
\item  \quad Let $\phi :G\times Y\rightarrow Y$ be a smooth left action of $%
G $ on $Y$. If a smooth map $\hat \phi :G\times X\rightarrow X$ satisfying
(L1) exists, it can always be redefined so that 
\begin{equation}
\label{pp6t1lift2}\hat \phi _e=id_X\quad . 
\end{equation}

\item  \quad Every $\hat \phi _g$ is a diffeomorphism, and the assignment $%
\left( g,x\right) \mapsto \hat \phi _g^{-1}\left( x\right) $ is smooth.
\end{enumerate}

\TeXButton{Beweis}{\raisebox{-1ex}{\it Proof :}
\vspace{1ex}}

\underline{Ad (1) :}\quad By assumption, $p\hat \phi _e=\phi _ep=p$, hence $%
\hat \phi _e$ is a deck transformation $\gamma $ of the covering. Now
redefine $\hat \phi \mapsto \hat \phi ^{\prime }=\hat \phi \left( id_G\times
\gamma ^{-1}\right) $, then $\hat \phi ^{\prime }$ satisfies $p\hat \phi
^{\prime }=\phi \left( id_G\times p\right) $ and $\hat \phi _e^{\prime
}=id_X $.

\underline{Ad (2) :}\quad The map $\hat \phi _g\hat \phi _{g-1}:X\rightarrow
X$ projects into $p$, hence is a deck transformation $\gamma $. Thus $\hat
\phi _g^{-1}=\widehat{\phi _{g^{-1}}}\circ \gamma ^{-1}$ is smooth, since $%
\gamma ^{-1}$ is smooth, hence $\hat \phi _g$ is a diffeomorphism. The
assignment $\left( g,x\right) \mapsto \hat \phi _g^{-1}\left( x\right) =%
\widehat{\phi _{g^{-1}}}\circ \gamma ^{-1}\left( x\right) $ is smooth, since
the map $G\rightarrow G$, $g\mapsto g^{-1}$ is smooth. \TeXButton{BWE}
{\hfill
\vspace{2ex}
$\blacksquare$}

\TeXButton{Abst}{\vspace{0.8ex}}In the following we assume that $X$ and $Y$
are connected manifolds, and $Y$ is not simply connected. We note that

\subsection{Remark}

The fundamental group of the Lie group $G$ coincides with the fundamental
group of its identity component, 
\begin{equation}
\label{pp6t1remark1}\pi _n\left( G\right) =\pi _n\left( G_0\right) \text{%
\quad }. 
\end{equation}
\TeXButton{Beweis}{\raisebox{-1ex}{\it Proof :}
\vspace{1ex}}

This follows from the exact homotopy sequence (\ref{pp6t1homSeq})%
$$
\cdots \rightarrow \pi _{n+1}\left( B\right) \stackrel{\partial }{%
\longrightarrow }\pi _n\left( F\right) \stackrel{i_{\#}}{\longrightarrow }%
\pi _n\left( E\right) \stackrel{p_{\#}}{\longrightarrow }\pi _n\left(
B\right) \rightarrow \cdots \quad , 
$$
for the fibration $pr:G\rightarrow G/G_0=Ds$. Put $B=Ds$, $F=G_0$, $E=G$, $%
\pi _n\left( Ds\right) =0$ to obtain (\ref{pp6t1remark1}). \TeXButton{BWE}
{\hfill
\vspace{2ex}
$\blacksquare$}

\TeXButton{Abst}{\vspace{0.8ex}}Now we derive a necessary and sufficient
condition for the existence of a lift satisfying (L1, L2) in terms of the
fundamental group $\pi _1\left( G\right) $ of $G$. To prove this, we first
need a

\subsection{Lemma \label{HomotopyLemma}}

Let $X,Y$ be topological spaces, let $\lambda \times \mu $ be a loop in $%
X\times Y$ at $\left( x,y\right) $, where $\lambda $ is a loop in $X$ at $x$%
, and $\mu $ is a loop in $Y$ at $y$. Then $\lambda \times \mu $ is
homotopic to a product of loops 
\begin{equation}
\label{pp6t1homolemma1}\lambda \times \mu \sim \left( \left\{ x\right\}
\times \mu \right) *\left( \lambda \times \left\{ y\right\} \right) \quad , 
\end{equation}
where $"\sim "$ means ''homotopic'', and $"*"$ denotes a product of paths.

\TeXButton{Beweis}{\raisebox{-1ex}{\it Proof :}
\vspace{1ex}}

We explicitly give a homotopy effecting (\ref{pp6t1homolemma1}). Define $%
h:I\times I\rightarrow X\times Y$, $\left( s,t\right) \mapsto h\left(
s,t\right) =h_s\left( t\right) $. Here $s$ labels the loops $t\mapsto
h_s\left( t\right) $, and $t$ is the loop parameter. Define 
\begin{equation}
\label{pp6t1homolemma2}h\left( s,t\right) \equiv \left\{ 
\begin{array}{ccl}
\left( x,\mu \left( \frac t{1-\frac s2}\right) \right) & \text{for} & t\in
\left[ 0,\frac s2\right) \quad . \\ 
\left( \lambda \left( \frac{t-\frac s2}{1-\frac s2}\right) ,\mu \left( \frac
t{1-\frac s2}\right) \right) & \text{for} & t\in \left[ \frac s2,1-\frac
s2\right) \quad . \\ 
\left( \lambda \left( \frac{t-\frac s2}{1-\frac s2}\right) ,y\right) & \text{%
for} & t\in \left[ 1-\frac s2,1\right] \quad . 
\end{array}
\right. 
\end{equation}
This is the required homotopy. \TeXButton{BWE}
{\hfill
\vspace{2ex}
$\blacksquare$}

\TeXButton{Abst}{\vspace{0.8ex}}Now we can turn our theorem. To this end,
consider a left action $\phi :G\times Y\rightarrow Y$, and choose $y\in Y$
fixed. By $\phi _y$ we denote the map $\phi _y:G\rightarrow Y$, $g\mapsto
\phi _y\left( g\right) \equiv \phi \left( g,y\right) $. $\phi _y$ induces
the homomorphism $\phi _{y\#}:\pi _1\left( G,e\right) \rightarrow \pi
_1\left( Y,y\right) $. Furthermore, if $x\in p^{-1}\left( y\right) $ is any
point in the fibre over $y$, then $p$ induces a map $p_{\#}:\pi _1\left(
X,x\right) \rightarrow \pi _1\left( Y,y\right) $. We now can state:

\subsection{Theorem \label{ConditionForGroupLift}}

Let $x\in X$ arbitrary, let $y=p\left( x\right) $. Let $G$ be connected.
Then the action $\phi :G\times Y\rightarrow Y$ possesses a unique lift $\hat
\phi :G\times X\rightarrow X$ satisfying (L1, L2) if and only if 
\begin{equation}
\label{pp6t1lift3}\phi _{y\#}\,\pi _1\left( G,e\right) \subset p_{\#}\,\pi
_1\left( X,x\right) \quad . 
\end{equation}
\TeXButton{Beweis}{\raisebox{-1ex}{\it Proof :}
\vspace{1ex}}

Assume a lift $\hat \phi $ exists. Then $\hat \phi $ is the unique lift of
the map $G\times X\rightarrow Y$, $\left( g,x\right) \mapsto \phi \left(
g,p\left( x\right) \right) $ through $p$ with the property $\hat \phi \left(
e,x\right) =x$. According to the ''Lifting Map Theorem'' \ref
{LiftingMapTheorem} it follows that 
\begin{equation}
\label{pp6t1lift4}\left[ \phi \left( id_G\times p\right) \right] _{\#}\,\pi
_1\left( G\times X,\left( e,x\right) \right) \subset p_{\#}\pi _1\left(
X,x\right) \quad . 
\end{equation}
Conversely, if (\ref{pp6t1lift4}) holds, then $\phi \left( id_G\times
p\right) $ lifts uniquely to $\hat \phi $ satisfying (L1, L2). Hence we need
only show the equivalence (\ref{pp6t1lift3}) $\Leftrightarrow $ (\ref
{pp6t1lift4}).

\underline{(\ref{pp6t1lift3})\ $\Leftarrow $\ (\ref{pp6t1lift4}) :}$\quad $%
Let $\lambda $ be a loop in $G$ at $e$. Then $\lambda \times \left\{
x\right\} $ is a loop in $G\times X$ at $\left( e,x\right) $, hence $%
t\mapsto \phi \left( \lambda \left( t\right) ,p\left( x\right) \right) =\phi
_y\circ \lambda \left( t\right) $ is a loop in $Y$ at $y$ giving rise to the
class%
$$
\phi _{y\#}\left[ \lambda \right] =\left[ \phi \left( id_G\times p\right)
\right] _{\#}\left[ \lambda \times \left\{ x\right\} \right] \in 
$$
$$
\in \left[ \phi \left( id_G\times p\right) \right] _{\#}\pi _1\left( G\times
X,\left( e,x\right) \right) \subset p_{\#}\pi _1\left( X,x\right) \quad , 
$$
where the last inclusion follows from assumption, and hence (\ref{pp6t1lift3}%
) follows.

\underline{(\ref{pp6t1lift3})\ $\Rightarrow $\ (\ref{pp6t1lift4}) :}$\quad $%
Let $\lambda \times \mu $ be a loop in $G\times X$ at $\left( e,x\right) $.
Then lemma \ref{HomotopyLemma} says that%
$$
\lambda \times \mu \sim \left( \left\{ e\right\} \times \mu \right) *\left(
\lambda \times \left\{ x\right\} \right) \quad , 
$$
hence%
$$
\left[ \phi \left( id_G\times p\right) \right] \lambda \times \mu \sim
\left[ \phi \left( id_G\times p\right) \right] \left( \left\{ e\right\}
\times \mu \right) *\left[ \phi \left( id_G\times p\right) \right] \left(
\lambda \times \left\{ x\right\} \right) = 
$$
$$
=\phi \left( e,p\mu \right) *\phi \left( \lambda ,p\left( x\right) \right)
=\left( p\mu \right) *\left( \phi _y\lambda \right) \quad . 
$$
But by assumption (\ref{pp6t1lift3}), there exists a loop $\rho $ in $X$ at $%
x$ such that $\phi _y\lambda \sim p\rho $. Then the last expression in the
last line above becomes%
$$
\left( p\mu \right) *\left( \phi _y\lambda \right) \sim p\left( \mu *\rho
\right) \quad , 
$$
where $\mu *\rho $ is a loop in $X$ at $x$. But this proves%
$$
\left[ \phi \left( id_G\times p\right) \right] \left[ \lambda \times \mu
\right] \in p_{\#}\,\pi _1\left( X,x\right) \quad , 
$$
where $\left[ \lambda \times \mu \right] $ is the homotopy class of $\lambda
\times \mu $, and hence (\ref{pp6t1lift4}) follows. \TeXButton{BWE}
{\hfill
\vspace{2ex}
$\blacksquare$}

\section{Preservation of the group law \label{PreservationOfTheGroupLaw}}

In this section we show that a lift $\hat \phi $ to a simply connected
covering space is not unique, if $G$ is not connected. Every lift gives rise
to an extension of the original group $G$ by the group ${\cal D}$ of deck
transformations of the covering. If ${\cal D}$ is Abelian, all these
extensions are equivalent.

\TeXButton{Abst}{\vspace{0.8ex}}We start with theorem \ref
{ConditionForGroupLift}: This gives a condition for the existence of a
smooth map $\hat \phi $ satisfying (L1,L2) under the assumption that $G$ is
connected. We now temporarily relax the last condition and allow $G$ to be a
Lie group with several connected components. In this case we simply assume
that a lift $\hat \phi $ exists. As mentioned above, the lift $\hat \phi $
need not preserve the group law on $G$, in which case it is not an action of 
$G$ on $X$. In particular, this means, that the set 
\begin{equation}
\label{pp6t1liftset}\left\{ \hat \phi _g\mid g\in G\right\} 
\end{equation}
of diffeomorphisms $\hat \phi _g:X\rightarrow X$ is no longer a group.
Clearly, we want to know under which circumstances the lift {\bf is} an
action of some group $\tilde G$. We examine this under the assumption that $%
G $ is a semidirect product $G=G_0\odot Ds$ of its identity component and a
discrete factor $Ds$, where $G/G_0\equiv Ds$, as $G_0$ is normal in $G$. In
this case, every element of $G$ has a unique expression $\left( g,\kappa
\right) $, where $g\in G_0$, $\kappa \in Ds$. The group law is 
\begin{equation}
\label{pp6t1lift5}\left( g,\kappa \right) \left( g^{\prime },\kappa ^{\prime
}\right) =\left( g\cdot a\left( \kappa \right) g^{\prime },\kappa \kappa
^{\prime }\right) \quad , 
\end{equation}
where $a\left( \kappa \right) :G_0\rightarrow G_0$ is an outer automorphism
of $G_0$, and $a:Ds\rightarrow Aut\left( G_0\right) $ is a representation of 
$Ds$ in the automorphism group of $G_0$, as $a\left( \kappa \kappa ^{\prime
}\right) =a\left( \kappa \right) \circ a\left( \kappa ^{\prime }\right) $,
and $a\left( e\right) =id$. We first need a

\subsection{Lemma \label{DeckObjects}}

Let $\gamma \in {\cal D}$, $g,g^{\prime }\in G$. Then the maps 
\begin{equation}
\label{pp6t1lift6}\hat \phi _g\circ \gamma \circ \hat \phi _g^{-1}\quad
,\quad \hat \phi _g\circ \hat \phi _{g^{\prime }}\circ \widehat{\phi
_{gg^{\prime }}}^{-1} 
\end{equation}
are deck transformations.

\TeXButton{Beweis}{\raisebox{-1ex}{\it Proof :}
\vspace{1ex}}

This follows immediately by applying $p$ and using (L1,L2). \TeXButton{BWE}
{\hfill
\vspace{2ex}
$\blacksquare$}

\TeXButton{Abst}{\vspace{0.8ex}}Thus, for a given $g\in G$, (\ref{pp6t1lift6}%
) defines a map 
\begin{equation}
\label{pp6t1lift7}b\left( g\right) :{\cal D}\rightarrow {\cal D}\quad ,\quad
\gamma \mapsto b\left( g\right) \gamma \equiv \hat \phi _g\circ \gamma \circ
\hat \phi _g^{-1}\quad , 
\end{equation}
which implies that $b\left( g\right) \in Aut\left( {\cal D}\right) $ is a $%
{\cal D}$-automorphism. $b:G\rightarrow Aut\left( {\cal D}\right) $ need not
be a representation, however! The second expression in (\ref{pp6t1lift6})
defines a map 
\begin{equation}
\label{pp6t1lift8}\Gamma :G\times G\rightarrow {\cal D}\quad ,\quad \left(
g,g^{\prime }\right) \mapsto \Gamma \left( g,g^{\prime }\right) \equiv \hat
\phi _g\circ \hat \phi _{g^{\prime }}\circ \widehat{\phi _{gg^{\prime }}}%
^{-1}\quad . 
\end{equation}
These maps determine how the group law of $G$ is changed under the lift $%
\hat \phi $; in particular (\ref{pp6t1lift8}) shows that $\Gamma $ expresses
the deviation of the lifted diffeomorphisms $\hat \phi _g$ from forming a
group isomorphic to $G$. This is the content of the next

\subsection{Theorem \label{ExtensionOfG}}

Let $\phi :G\times Y\rightarrow Y$ be a smooth action of $G$ on $Y$. Assume
that a smooth lift $\hat \phi $ satisfying (L1,L2) exists. Then

\begin{enumerate}
\item  \quad the set $\left\{ \hat \phi _g\mid g\in G\right\} $ is no longer
a group in general; instead, the lift $\hat \phi $ provides an {\it %
extension }$\tilde G${\it \ of }$G${\it \ by the deck transformation group }$%
{\cal D}$, i.e. ${\cal D}$ is normal in $\tilde G$, and $\tilde G/{\cal D}=G$%
. The extended group is given by the set 
\begin{equation}
\label{pp6t1lift9}\tilde G=\left\{ \gamma \circ \hat \phi _g\mid \gamma \in 
{\cal D\;};\;g\in G\right\} \quad . 
\end{equation}
If the elements of this set are denoted as pairs, $\gamma \circ \hat \phi
_g\leftrightarrow \left( \gamma ,g\right) $, then the group law of $\tilde G$
is given by 
\begin{equation}
\label{pp6t1lift10}\left( \gamma ,g\right) \cdot \left( \gamma ,g^{\prime
}\right) =\left( \gamma \cdot b\left( g\right) \gamma ^{\prime }\cdot \Gamma
\left( g,g^{\prime }\right) ,gg^{\prime }\right) \quad , 
\end{equation}
and inverses are 
\begin{equation}
\label{pp6t1lift10a}\left( \gamma ,g\right) ^{-1}=\left( b\left( g\right)
^{-1}\left[ \gamma ^{-1}\cdot \Gamma ^{-1}\left( g,g^{-1}\right) \right]
,g^{-1}\right) = 
\end{equation}
\begin{equation}
\label{pp6t1lift10b}=\left( \left[ b\left( g^{-1}\right) \gamma \cdot \Gamma
\left( g^{-1},g\right) \right] ^{-1},g^{-1}\right) \quad . 
\end{equation}
The group law (\ref{pp6t1lift10}) expresses the non-closure of the set $%
\left\{ \hat \phi _g\mid g\in G\right\} $ as discussed in (\ref{pp6t1liftset}%
), since now 
\begin{equation}
\label{pp6t1lift10c}\left( e,g\right) \cdot \left( e,g^{\prime }\right)
=\left( \Gamma \left( g,g^{\prime }\right) ,gg^{\prime }\right) \quad . 
\end{equation}

\item  \quad If $H,K\subset G$ are any connected subsets of $G$, then the
restrictions $\left. b\right| H$ and $\left. \Gamma \right| H\times K$ are
constant. Hence both $b$ and $\Gamma $ descend to the quotient $G/G_0=Ds$, 
\begin{equation}
\label{pp6t1lift11}b:Ds\rightarrow {\cal D}\quad ,\quad \Gamma :Ds\times
Ds\rightarrow {\cal D}\quad . 
\end{equation}

\item  \quad In particular, on the identity component $G_0$ we have $\left.
b\right| G_0=id$ and $\left. \Gamma \right| G_0\times G_0=e$. Hence, for
elements $\left( \gamma ,g\right) \in {\cal D}\times G_0$ we have the group
law 
\begin{equation}
\label{pp6t1lift12}\left( \gamma ,g\right) \cdot \left( \gamma ^{\prime
},g^{\prime }\right) =\left( \gamma \gamma ^{\prime },gg^{\prime }\right)
\quad . 
\end{equation}
As a consequence, the identity component $G_0$ can be regarded as a subgroup
of the extension $\tilde G$, and can be identified with the set of all
elements of the form $\left( e,g\right) $, $g\in G_0$, so that (cf. \ref
{pp6t1lift10c}) 
\begin{equation}
\label{pp6t1lift12a}\left( e,g\right) \cdot \left( e,g^{\prime }\right)
=\left( e,gg^{\prime }\right) \quad . 
\end{equation}
Also, (\ref{pp6t1lift12}) says that ${\cal D}$ and $G_0$ commute. As a
consequence of all that, the identity component $\tilde G_0$ of $\tilde G$
coincides (up to isomorphism) with the identity component of $G$, $\tilde
G_0=G_0$, and hence we have an isomorphism of Lie algebras 
\begin{equation}
\label{pp6t1lift13}Lie\left( \tilde G\right) =Lie\left( G\right) =\hat
g\quad . 
\end{equation}
\end{enumerate}

\TeXButton{Beweis}{\raisebox{-1ex}{\it Proof :}
\vspace{1ex}}

\underline{Ad (1) :}\quad We first must verify that the set $\tilde G$ as
defined in (\ref{pp6t1lift9}) is indeed a group of diffeomorphisms on $X$.
Since $\hat \phi _e=id_X$ by (L2), we have that $\tilde G$ contains $e\hat
\phi _e=id_X$. Next, if $\gamma ,\gamma ^{\prime }\in {\cal D}$ and $%
g,g^{\prime }\in G$, then $\gamma \hat \phi _g\gamma ^{\prime }\hat \phi
_{g^{\prime }}=\gamma \circ b\left( g\right) \gamma ^{\prime }\circ \hat
\phi _g\circ \hat \phi _{g^{\prime }}$ according to (\ref{pp6t1lift7}); but $%
\hat \phi _g\circ \hat \phi _{g^{\prime }}=\Gamma \left( g,g^{\prime
}\right) \circ \widehat{\phi _{gg^{\prime }}}$ by (\ref{pp6t1lift8}), which
gives 
\begin{equation}
\label{pp6t1lift14}\gamma \circ \hat \phi _g\circ \gamma ^{\prime }\circ
\hat \phi _{g^{\prime }}=\left[ \gamma \circ b\left( g\right) \gamma
^{\prime }\circ \Gamma \left( g,g^{\prime }\right) \right] \circ \left[ 
\widehat{\phi _{gg^{\prime }}}\right] \quad , 
\end{equation}
where the first factor in square brackets on the RHS is an element of ${\cal %
D}$, and the second factor is a lifted diffeomorphism, and therefore the LHS
is an element of the set $\tilde G$. Furthermore, when using the pair
notation $\left( \gamma ,g\right) $ for elements of $\tilde G$, formula (\ref
{pp6t1lift14}) yields the group law (\ref{pp6t1lift10}). Finally, we must
show that inverses exist in $\tilde G$; it is easy to use (\ref{pp6t1lift10}%
) to arrive at (\ref{pp6t1lift10a}) for inverses, which means that inverses
have the form $\gamma \circ \hat \phi _g$ as required. Furthermore, from (%
\ref{pp6t1lift10}) it follows that ${\cal D}$ is normal in $\tilde G$, and
that the cosets ${\cal D}\cdot \left( \gamma ,g\right) $ obey the group law
of $G$, since%
$$
\left[ {\cal D}\cdot \left( e,g\right) \right] \left[ {\cal D}\cdot \left(
e,g^{\prime }\right) \right] ={\cal D}\cdot \left[ \left( e,g\right) \left(
e,g^{\prime }\right) \right] ={\cal D}\cdot \left( \Gamma \left( g,g^{\prime
}\right) ,gg^{\prime }\right) ={\cal D}\cdot \left( e,gg^{\prime }\right)
\quad , 
$$
thus $\tilde G/{\cal D}=G$.

\underline{Ad (2+3) :}\quad Let $\gamma \in {\cal D}$ be arbitrary, and
consider the map $G\rightarrow p^{-1}\left( p\left( x\right) \right) $, $%
g\mapsto \left[ b\left( g\right) \gamma \right] x$, where the fibre $%
p^{-1}\left( p\left( x\right) \right) $ is a discrete space for arbitrary
fixed $x\in X$. From the definition (\ref{pp6t1lift7}) and proposition \ref
{Smoothness} we see that this map is smooth; since the target space is
discrete, it must therefore be constant on every connected subset of the
domain. Since $X$ is connected, every deck transformation of the covering $%
p:X\rightarrow Y$ is uniquely determined by its value at a single point $%
x\in X$. Hence $b\left( g\right) \gamma =const.$ for all $g$ within a
connected component of $G$. On the identity component $G_0$, $b\left(
g\right) =b\left( e\right) =id_{{\cal D}}$. -- A similar argument applies to 
$\Gamma $, since (\ref{pp6t1lift8}) shows that all maps involved in the
definition of $\Gamma $ are smooth in all arguments. In particular, $\Gamma
\left( e,g\right) =\Gamma \left( g,e\right) =e_{{\cal D}}$. (\ref
{pp6t1lift12}) is a consequence of (\ref{pp6t1lift10}). \TeXButton{BWE}
{\hfill
\vspace{2ex}
$\blacksquare$}

\TeXButton{Abst}{\vspace{0.8ex}}If ${\cal D}$ is Abelian, the map $\Gamma $
defined in (\ref{pp6t1lift8}) has a special significance:

\subsection{Theorem}

If ${\cal D}$ is Abelian, then $\Gamma :G\times G\rightarrow {\cal D}$
defines a $2$-cocycle in the ${\cal D}$-valued cohomology on $G$ as defined
in appendix, chapter \ref{DeckTransformationValuedCohomOnG}.

\TeXButton{Beweis}{\raisebox{-1ex}{\it Proof :}
\vspace{1ex}}

The proof is a standard argument: Compute $\left( \hat \phi _g\hat \phi
_h\right) \hat \phi _k=\hat \phi _g\left( \hat \phi _h\hat \phi _k\right) $
using (\ref{pp6t1lift8}); this leads to%
$$
b\left( g\right) \Gamma \left( h,k\right) +\Gamma \left( g,hk\right) -\Gamma
\left( gh,k\right) -\Gamma \left( g,h\right) =\left( \delta \Gamma \right)
\left( g,h,k\right) =0\quad , 
$$
where we have used (\ref{pp6t1co14}). \TeXButton{BWE}
{\hfill
\vspace{2ex}
$\blacksquare$}

\TeXButton{Abst}{\vspace{0.8ex}}--- Any two lifts $\hat \phi $ and $\hat
\phi ^{\prime }$ must coincide on the identity component $G_0$, by
uniqueness. However, they may differ on the components $G_0\cdot \kappa $, $%
\kappa \in Ds$, by deck transformations. This gives rise to different
cocycles (in case that ${\cal D}$ is Abelian) $\Gamma $ and $\Gamma ^{\prime
}$. We now show that $\Gamma $ and $\Gamma ^{\prime }$ must differ by a
coboundary:

\subsection{Theorem}

Let the deck transformation group ${\cal D}$ of the covering $p:X\rightarrow
Y$ be Abelian. Given two lifts $\hat \phi $ and $\hat \phi ^{\prime }$ of
the action $\phi :G\times Y\rightarrow Y$ to $X$, with associated cocycles $%
\Gamma $ and $\Gamma ^{\prime }$, there exists a $1$-cochain $\eta \in
C^1\left( G,{\cal D}\right) $ in the ${\cal D}$-valued cohomology on $G$ as
defined in Appendix, section \ref{DeckTransformationValuedCohomOnG}, such
that 
\begin{equation}
\label{pp6t1lift15}\Gamma ^{\prime }=\Gamma +\delta \eta \quad . 
\end{equation}
On the connected component $G_0$, $\eta =e$.

\TeXButton{Beweis}{\raisebox{-1ex}{\it Proof :}
\vspace{1ex}}

Both $\hat \phi $ and $\hat \phi ^{\prime }$ project down to $\phi $ by
(L1), which implies that $\hat \phi _g^{\prime }\hat \phi _g^{-1}$ is a deck
transformation $\eta \left( g\right) $, hence 
\begin{equation}
\label{pp6t1lift151}\hat \phi _g^{\prime }=\eta \left( g\right) \hat \phi
_g\quad . 
\end{equation}
$\eta $ is constant on connected components, hence is trivial on $G_0$. We
have $\hat \phi _g^{\prime }\hat \phi _h^{\prime }=\Gamma ^{\prime }\left(
g,h\right) \widehat{\phi _{gh}^{\prime }}$, and inserting (\ref{pp6t1lift151}%
) gives%
$$
\eta \left( g\right) \cdot \left[ b\left( g\right) \eta \left( h\right)
\right] \cdot \Gamma \left( g,h\right) \cdot \widetilde{\phi _{gh}}=\Gamma
^{\prime }\left( g,h\right) \cdot \eta \left( gh\right) \cdot \widetilde{%
\phi _{gh}}\quad ; 
$$
hence, on using additive notation for the group composition in ${\cal D}$, 
\begin{equation}
\label{pp6t1lift16}\Gamma ^{\prime }\left( g,h\right) =\Gamma \left(
g,h\right) +\,\stackunder{=\left( \delta \eta \right) \left( g,h\right) }{%
\underbrace{b\left( g\right) \eta \left( h\right) -\eta \left( gh\right)
+\eta \left( g\right) }}\quad , 
\end{equation}
using (\ref{pp6t1co14}). \TeXButton{BWE}
{\hfill
\vspace{2ex}
$\blacksquare$}

\TeXButton{Abst}{\vspace{0.8ex}}Thus, all the different lifts $\hat \phi $
of the group action $\phi $ to $X$ give rise to the same cohomology class $%
\left[ \Gamma \right] \in H^2\left( G,{\cal D}\right) $ in the ${\cal D}$%
-valued cohomology on $G$, which also implies, that the possible group
extensions $\tilde G$ of $G$ associated with these lifts are all equivalent 
\cite{Azcarraga}. Thus, up to equivalence, there is only one group extension 
$\tilde G$ of $G$ by ${\cal D}$, and this is basically determined by the
geometry of the covering $p:X\rightarrow Y$.

\section{$\tilde G$-spaces and equivariant covering maps \label
{GSpacesCoveringMaps}}

Although property (L1), $p\hat \phi _g=\phi _gp$, seems to suggest that
every covering map $p$ is a $G$-morphism, this is not true in general, if $G$
is not connected; for in this case, $X$ is not a $G$-space, but only a $%
\tilde G$-space, as explained above. However, we can make $p$ equivariant
with respect to the larger group $\tilde G$: To this end, we first note that
this requires $X$ and $Y$ to be $\tilde G$-spaces. This can be accomplished
by introducing the projection $pr:\tilde G\rightarrow \tilde G/{\cal D}=G$
and observing that $Y$ is trivially a $\tilde G$-space by defining the
action of $\tilde G$ on $Y$ as 
\begin{equation}
\label{pp6t1lift17a}\Phi :\tilde G\times Y\rightarrow Y\quad ,\quad \Phi
\equiv \phi \left( pr\times id_Y\right) \quad . 
\end{equation}
Now we must define a suitable action $\hat \Phi $ of $\tilde G$ on $X$; this
is accomplished by the assignment 
\begin{equation}
\label{pp6t1lift17}\hat \Phi :\tilde G\times X\rightarrow X\quad ,\quad
\left( \left( \gamma ,g\right) ,x\right) \mapsto \hat \Phi \left( \left(
\gamma ,g\right) ,x\right) \equiv \gamma \circ \hat \phi _g\left( x\right)
\quad . 
\end{equation}
Using the group law (\ref{pp6t1lift10}) for the case that ${\cal D}$ is
Abelian we see that 
\begin{equation}
\label{pp6t1lift18}\hat \Phi _{\left( \gamma ,g\right) }\hat \Phi _{\left(
\gamma ^{\prime },g^{\prime }\right) }=\hat \Phi _{\left( \gamma +b\left(
g\right) \gamma ^{\prime }+\Gamma \left( g,g^{\prime }\right) ,gg^{\prime
}\right) }=\hat \Phi _{\left( \gamma ,g\right) \left( \gamma ^{\prime
},g^{\prime }\right) }\quad , 
\end{equation}
i.e. $\hat \Phi $ is indeed a left action. Furthermore, under the projection 
$p$ we have%
$$
p\hat \Phi \left( \left( \gamma ,g\right) ,x\right) =p\hat \phi _g\left(
x\right) =\phi \left( g,p\left( x\right) \right) =\phi \left( pr\times
p\right) \left( \left( \gamma ,g\right) ,x\right) \quad , 
$$
or 
\begin{equation}
\label{pp6t1lift19}p\circ \hat \Phi =\phi \circ \left( pr\times p\right)
=\Phi \circ \left( id_{\tilde G}\times p\right) \quad . 
\end{equation}
But this equation now says that the covering projection $p$ is a $\tilde G$%
-morphism with respect to the group $\tilde G$, where $X$ and $Y$ are now
regarded as $\tilde G$-spaces. Let us summarize:

\subsection{Theorem: Equivariance of the covering map}

Define actions $\Phi ,\hat \Phi $ of the extended group $\tilde G$ on $Y,X$
according to formulas (\ref{pp6t1lift17a}), (\ref{pp6t1lift17}). With this
definition, $X$ and $Y$ become $\tilde G$-spaces, and the covering map $%
p:X\rightarrow Y$ is $\tilde G$-equivariant (a $\tilde G$-morphism).

\TeXButton{Abst}{\vspace{0.8ex}}--- We now discuss a situation in which a
lift $\hat \phi $ as introduced in section \ref{LiftOfGroupActions} always
exists. A glance at formula (\ref{pp6t1lift3}) in section \ref
{ConditionForGroupLift} shows that the lift exists if the fundamental groups
of both $G$ and $X$ are trivial. Now we perform the following construction:
We assume that $G$ is connected and $X$ is simply connected. If $G$ acts on $%
Y$ via $\phi $, then so does the universal simply connected covering group $%
CG$ of $G$; just let $pro:CG\rightarrow G$ be the natural projection (which
is a homomorphism), then $c\phi :CG\times Y\rightarrow Y$, $c\phi \equiv
\phi \left( pro\times id_Y\right) $ defines a smooth left action of $CG$ on $%
Y$. Therefore, in this case there always exists a lift $\widehat{c\phi }%
:CG\times X\rightarrow X$ of $c\phi $ to $X$ satisfying (L1,L2). Since $G$
is connected, formula (\ref{pp6t1lift12a}) in theorem \ref{ExtensionOfG}
tells us that the lift preserves the group law of $CG$, and hence $\widehat{%
c\phi }$ is an action of $CG$ on $X$. Thus,

\subsection{Theorem}

Let $p:X\rightarrow Y$ be a covering, where $X$ is simply connected. Let the
connected Lie group $G$ act on $Y$ via $\phi $, and let $CG$ denote the
universal covering group of $G$ with projection homomorphism $%
pro:CG\rightarrow G$. Then $c\phi \equiv \phi \left( pro\times id_Y\right) $
defines an action of $CG$ on $Y$, and there exists a lift $\widehat{c\phi }%
:CG\times X\rightarrow X$ such that $\widehat{c\phi }$ preserves the group
law on $CG$, 
\begin{equation}
\label{pp6t1lift20}\widehat{c\phi _{gh}}=\widehat{c\phi _g}\circ \widehat{%
c\phi _h}\quad . 
\end{equation}
Thus, the projection map $p$ is a $CG$-morphism of $CG$-spaces, 
\begin{equation}
\label{pp6t1lift21}p\circ \widehat{c\phi _g}=c\phi _g\circ p\quad , 
\end{equation}
for $g\in CG$.

\TeXButton{Abst}{\vspace{0.8ex}}--- A consequence of the developments in
this section is this: Assume that $p:X\rightarrow Y$ is a smooth covering of
manifolds, where $Y$ is a symplectic manifold with symplectic $2$-form $%
\omega $. Since $p$ is a local diffeomorphism, the $2$-form $\Omega \equiv
p^{*}\omega $ is closed and non-degenerate, and hence is a valid symplectic $%
2$-form on $X$. Now assume that a connected Lie group $G$ acts on $Y$ via $%
\Phi :G\times Y\rightarrow Y$ such that all diffeomorphisms $\Phi _g$ are
canonical transformations, $\Phi ^{*}\omega =\omega $. Assume that a lift $%
\tilde \Phi $ to $X$ exists; as $G$ is connected, the group law is then
preserved. Since the lift obeys $p\circ \tilde \Phi _g=\Phi _g\circ p$, it
follows that $\tilde \Phi _g^{*}p^{*}\omega =p^{*}\Phi _g^{*}\omega $, or $%
\tilde \Phi _g^{*}\Omega =\Omega $; hence all diffeomorphisms $\tilde \Phi
_g $ are canonical transformations on the symplectic covering manifold $X$.
In summary,

\subsection{Theorem \label{CanonicalTransOnCovering}}

Let $p:X\rightarrow Y$ be a covering of smooth manifolds, where $Y$ is a
symplectic manifold with symplectic $2$-form $\omega $. Assume the connected
Lie group $G$ acts via $\Phi $ on $Y$ from the left, such that all $\Phi _g$
are canonical transformations with respect to $\omega $. Assume a lift $%
\tilde \Phi $ to $X$ exists. Then all diffeomorphisms $\tilde \Phi _g$ are
canonical transformations with respect to the symplectic $2$-form $\Omega
\equiv p^{*}\omega $ on $X$.

\TeXButton{Abst}{\vspace{0.8ex}}--- Finally, we wish to study the following
problem: Given a non-simply connected manifold $Y$, thought of as the
''configuration space'' of a dynamical system (or rather, a class of
dynamical systems), and the left action $\phi :G\times Y\rightarrow Y$ of a
(Lie) group $G$ on $Y$; if a lift $\hat \phi $ exists, we can extend it to a
map $^{*}\hat \phi :G\times T^{*}X\rightarrow T^{*}X$ on the cotangent
bundle of $X$ using formula (\ref{pp6t1cot1}). On the other hand, from the
discussion in section \ref{PStarIsACovering} we know that $T^{*}X$ is itself
a covering space of $T^{*}Y$, and $^{*}p$ as defined in equation (\ref
{pp6t1pstarisacovering}) is the covering map. Hence we can extend the action 
$\phi $ first to an action $^{*}\phi $ on the cotangent bundle $T^{*}Y$, and
then ask whether a lift $\widehat{^{*}\phi }$ of $^{*}\phi $ to $T^{*}X$
exists, and if yes, whether it coincides with the extension $^{*}\hat \phi $
of the lift $\hat \phi $ of $\phi $. The next theorem gives the answer for
connected $G$:

\subsection{Theorem \label{LiftsAndExtensions}}

Let $G$ be a connected Lie group; let $p:X\rightarrow Y$ be a covering,
where $X$ is connected. Let $\phi :G\times Y\rightarrow Y$ be a smooth
action of $G$ on $Y$, which possesses a lift $\hat \phi :G\times
X\rightarrow X$ satisfying (L1, L2). Then $\left( 1\right) $ a lift $%
\widehat{^{*}\phi }:G\times T^{*}X\rightarrow T^{*}X$ of the extension $%
^{*}\phi :G\times T^{*}Y\rightarrow T^{*}Y$ of the action $\phi $ to the
cotangent bundle $T^{*}X$ of $X$ exists; and $\left( 2\right) $ this lift
coincides with the uniquely determined extension $^{*}\widehat{\phi }%
:G\times T^{*}X\rightarrow T^{*}X$ of the lift $\hat \phi :G\times
X\rightarrow X$ of $\phi $ to the cotangent bundle $T^{*}X$ of $X$; i.e. 
\begin{equation}
\label{pp6t1lift22}\widehat{^{*}\phi }=\,^{*}\widehat{\phi }\quad , 
\end{equation}
yielding a commutative diagram.

\TeXButton{Beweis}{\raisebox{-1ex}{\it Proof :}
\vspace{1ex}}

Let $\hat \phi $ be the lift of $\phi $ satisfying (L1,L2). Its extension to
the cotangent bundle $T^{*}X$ takes the form $^{*}\widehat{\phi }=\left(
\hat \phi ,\hat \phi ^{-1*}\right) $. This map satisfies 
\begin{equation}
\label{pp6t1lift23a}^{*}\widehat{\phi _e}=id_{T^{*}X}\quad , 
\end{equation}
and under the extended projection $^{*}p$ it behaves as 
\begin{equation}
\label{pp6t1lift23}^{*}p\circ \,^{*}\hat \phi =\,^{*}\phi \circ \,\left(
id_G\times \,^{*}p\right) \quad . 
\end{equation}
But formulas (\ref{pp6t1lift23a},\ref{pp6t1lift23}) are precisely the
conditions (L1, L2) for a lift $\widehat{^{*}\phi }$ of the extended action $%
^{*}\phi $ under the covering $^{*}p:T^{*}X\rightarrow T^{*}Y$. This means
that $\left( 1\right) $ $^{*}\hat \phi $ is a lift of $^{*}\phi $; and,
since $G$ and $X$, and hence $G\times T^{*}X$, are connected, this lift is
unique as a consequence of the ''Unique Lifting Theorem'' \ref
{UniqueLiftingTheorem}, and hence coincides with $\widehat{^{*}\phi }$. 
\TeXButton{BWE}{\hfill
\vspace{2ex}
$\blacksquare$}

\section{Symplectic $G$-actions and moment maps \label
{SymplecticGActionsMoMaps}}

In the following sections we generalize the usual definition of a moment map
to a construction we call local moment map. The standard definitions found
in the literature generally involve that a moment map is a {\bf globally}
defined function from a symplectic manifold $M$ into the coalgebra of the
Lie algebra of a Lie group $G$ which acts on $M$ symplectically (e.g. \cite
{AbrahamMarsden}). Other authors (e.g. \cite{Woodhouse,GuillStern})
give even more restrictions by introducing moment maps only together with
the condition that the first and second Chevalley-Eilenberg cohomology
groups of the Lie group $G$ are trivial, or equivalently, that the
associated Lie algebra cohomoloy groups $H_0^1\left( \hat g,\TeXButton{R}
{\mathbb{R}}\right) $ and $H_0^2\left( \hat g,\TeXButton{R}{\mathbb{R}}%
\right) $ are trivial. In this work we make no assumptions about cohomologies on the group $G$,
nor do we assume that moment maps exist globally; on the contrary, it is the
purpose of this work to generalize moment maps to situations where the
underlying symplectic manifold is non-simply connected, and hence in general
does not admit a global moment map.

Let $M$ be a symplectic manifold with symplectic form $\omega $. Let $G$ be
a Lie group, let $\Phi :G\times M\rightarrow M$ be a smooth symplectic left
action of $G$ on $M$, i.e. $\Phi \left( g,x\right) =\Phi _g\left( x\right) $%
, with $\Phi _{gg^{\prime }}=\Phi _g\Phi _{g^{\prime }}$, $\Phi _e=id_M$,
and $\Phi _g^{*}\omega =\omega $ for all $g$. Let $\hat g$ be the Lie
algebra of $G$, and $g^{*}$ denote the coalgebra. If $A\in \hat g$, then the
vector field induced by $A$ on $M$ is denoted $\frac{\partial \Phi }{%
\partial G}A\equiv \tilde A$. Since $\Phi $ preserves the symplectic form,
the Lie derivative of $\omega $ with respect to $\tilde A$ vanishes, hence $%
\tilde A$ is a locally Hamiltonian vector field according to (\ref{pp6t1fo2}%
). From this it follows that the $1$-form $\frac{\partial \Phi }{\partial G}A%
\TeXButton{i}{\intmul}\omega $ is closed.

\section{Global moment maps \label{GlobalMomentMaps}}

First we assume that $M$ is simply connected. By a standard argument, an $%
\TeXButton{R}{\mathbb{R}}$-linear map $f$ from a vector space $\hat g$ to
the space of smooth closed $1$-forms $Z_{deRham}^1\left( M\right) $ on $M$
can always be lifted to an $\TeXButton{R}{\mathbb{R}}$-linear map $h:\hat
g\rightarrow {\cal F}\left( M\right) $, $A\mapsto h_A$ such that $%
dh_A=f\left( A\right) $: For, since $M$ is simply connected, every closed $1$%
-form $f\left( A\right) $ has a potential $h_A$ with $dh_A=f\left( A\right) $%
; the assignment $\left( A,x\right) \mapsto h_A\left( x\right) $ can be
assumed to be smooth in $x$, but need not be smooth in $A$. Now choose an
arbitrary fixed point $x_0\in M$, and replace $h_A$ by $h_A-h_A\left(
x_0\right) $; then $A\mapsto h_A\left( x\right) -h_A\left( x_0\right) $ is
linear in $A$.

In particular, the map $\hat g\ni A\mapsto \frac{\partial \Phi }{\partial G}A%
\TeXButton{i}{\intmul}\omega \in Z^1\left( M\right) $ can be lifted to an $%
\TeXButton{R}{\mathbb{R}}$-linear map $h:\hat g\rightarrow {\cal F}\left(
M\right) $, $A\mapsto -h_A$, with 
\begin{equation}
\label{pp6t1sym1}\frac{\partial \Phi }{\partial G}A\TeXButton{i}{\intmul}%
\omega +dh_A=0\quad . 
\end{equation}
Since $A\mapsto h_A\left( x\right) $ is linear for every fixed $x\in M$, $h$
defines a map $J:M\rightarrow g^{*}$, $\left\langle J\left( x\right)
,A\right\rangle \equiv h_A\left( x\right) $. $J$ is called a {\it moment map}
associated with the action $\Phi $. From its definition via $h$ we see that $%
J$ is determined up to addition $J\mapsto J+L$ of an $M$-constant, $%
\TeXButton{R}{\mathbb{R}}$-linear map $L:\hat g\rightarrow \TeXButton{R}
{\mathbb{R}}$, with $dL=0$.

\TeXButton{Abst}{\vspace{0.8ex}}$h$ as defined above is a homomorphisms of
vector spaces by linearity; in general it is not a homomorphism of Lie
algebras, however. Rather, the algebra of Poisson brackets can provide a
central extension of the Lie algebra of the Hamiltonian vector fields \cite
{GuillStern}.

\section{Local moment maps \label{LocalMomentMaps}}

Now we assume that $Y$ is connected, but not simply connected. If $y$ is a
base point of $Y$, set ${\cal D}\equiv \pi _1\left( Y,y\right) $. Let ${\cal %
V}=\left\{ V_a\mid a\in A\right\} $ be a countable simply connected open
cover of $Y$ as introduced in section \ref{MultiValFunc}. For every $A\in
\hat g$, $\frac{\partial \Phi }{\partial G}A\TeXButton{i}{\intmul}\omega $
is a closed $1$-form on $Y$; hence a multi-valued potential function $\left(
h_{A,a,d}\right) $, $d\in {\cal D}$, exists for every \u Cech cocycle $%
\left( g_{ab}\right) $ associated with a simply connected cover of $Y$,
according to theorem \ref{MultiValuedPotential}. However, here we can no
longer be sure whether all $h_{A,a,d}$ can be made linear in $A$
simultaneously, without spoiling the glueing conditions $h_{A,a,d}=h_{A,b,d%
\cdot g_{ab}}$. We therefore have to formulate the problem in terms of an
appropriate covering space $X$, which we take, as in the proof of theorem 
\ref{MultiValuedPotential}, to be the identification space $i_a:{\cal D}%
\times V_a\rightarrow X\equiv \bigsqcup\limits_{a\in A}{\cal D}\times
V_a/\sim $, where $\sim $ relates elements $\left( d,y\right) $ and $\left(
d^{\prime },y^{\prime }\right) =\left( d\cdot g_{ab},y\right) $ on ${\cal D}%
\times V_a$ and ${\cal D}\times V_b$ that are identified as $i_a\left(
d,y\right) =i_b\left( d^{\prime },y^{\prime }\right) $. Then $X$ is a
universal covering space of $Y$, such that the projection $p:X\rightarrow Y$
is a local diffeomorphism. $Y$ is the space of orbits under the action of $%
{\cal D}$ on $X$. If $\omega $ denotes the symplectic $2$-form on $Y$, $%
\Omega \equiv p^{*}\omega $ is a symplectic $2$-form on $X$.

We must ask whether $\Phi $ can be lifted to $X$, i.e. whether there exists
a map $\hat \Phi :G\times X\rightarrow X$ satisfying $p\circ \hat \Phi =\hat
\Phi \circ \left( id_G\times p\right) $, and $\hat \Phi _e=id_X$.
Furthermore, one has to examine whether the group law is preserved by the
lift, i.e. whether $\widehat{\Phi _{gh}}=\hat \Phi _g\hat \Phi _h$. The
question of existence of a lift is examined in theorem \ref
{ConditionForGroupLift}; in theorem \ref{ExtensionOfG} it is proven that,
for connected $G$, the lift preserves the group law of $G$. In theorem \ref
{CanonicalTransOnCovering} it is proven that, for connected $G$, the
diffeomorphisms $\hat \Phi _g$ are canonical transformations with respect to 
$\Omega $, hence $\hat \Phi $ is a symplectic action. In this case we have a
left action $\hat \Phi $ of $G$ on the simply connected manifold $X$, so
that the results from section \ref{GlobalMomentMaps} can be applied: There
exists a lift of the $\TeXButton{R}{\mathbb{R}}$-linear map $\hat g\ni
A\mapsto \frac{\partial \hat \Phi }{\partial G}A\TeXButton{i}{\intmul}\Omega
\in Z_{deRham}^1\left( X\right) $ to an $\TeXButton{R}{\mathbb{R}}$-linear
map $\hat h:\hat g\rightarrow {\cal F}\left( X\right) $, $A\mapsto -\hat h_A$%
, with 
\begin{equation}
\label{pp6t1LocMom1}\frac{\partial \hat \Phi }{\partial G}A\TeXButton{i}
{\intmul}\Omega +d\hat h_A=0\quad , 
\end{equation}
and there exists a global moment map $\hat J:X\rightarrow g^{*}$, $%
\left\langle \hat J\left( x\right) ,A\right\rangle \equiv \hat h_A\left(
x\right) $. $\hat J$ is determined up to addition $\hat J\mapsto \hat J+L$
of an $X$-constant, $\TeXButton{R}{\mathbb{R}}$-linear map $L:\hat
g\rightarrow \TeXButton{R}{\mathbb{R}}$, with $dL=0$. Now we can set $%
h_{a,d}\left( y\right) \equiv \hat h\circ i_a\left( d,y\right) $, and $%
J_{a,d}\left( y\right) \equiv \hat J\circ i_a\left( d,y\right) $ for $y\in
V_a$, $d\in {\cal D}$. Then 
\begin{equation}
\label{pp6t1LocMom2}\frac{\partial \Phi }{\partial G}A\TeXButton{i}{\intmul}%
\omega +d\left\langle J_{a,d},A\right\rangle =0 
\end{equation}
on $V_a$, and for all $d\in {\cal D}$. The glueing condition is easily
derived as%
$$
J_{a,d}\left( y\right) =\hat J\circ i_a\left( d,y\right) =\left( \hat J\circ
i_b\right) \circ \left( i_b^{-1}\circ i_a\right) \left( d,y\right) = 
$$
\begin{equation}
\label{pp6t1LocMom3}=\left( \hat J\circ i_b\right) \left( d\cdot
g_{ab},y\right) =J_{b,d\cdot g_{ab}}\left( y\right) \quad . 
\end{equation}

--- Now we examine to which extent a collection $\left( J_{a,d}\right) $ is
determined by the action $\Phi $ and a cocycle: Let $\left( g_{ab}\right) $, 
$\left( g_{ab}^{\prime }\right) $ be cocycles such that the associated
homomorphisms $\rho $, $\rho ^{\prime }$ are inner automorphisms of ${\cal D}
$, hence give rise to simply connected coverings; and let $\left(
J_{a,d}\right) $, $\left( J_{a,d}^{\prime }\right) $ be collections
satisfying relations (\ref{pp6t1LocMom2},\ref{pp6t1LocMom3}), respectively.
Then $\left( g_{ab}\right) $ and $\left( g_{ab}^{\prime }\right) $ are
cohomologous. As above, construct smooth simply connected covering manifolds 
$p:X\rightarrow Y$, $q:Z\rightarrow Y$ as identification spaces; the
trivializations $\left( i_a\right) $ with respect to $X$ identify $i_a\left(
d,y\right) =i_b\left( d\cdot g_{ab},y\right) $, and the trivializations $%
\left( j_a\right) $ with respect to $Z$ identify $j_a\left( d,y\right)
=j_b\left( d\cdot g_{ab}^{\prime },y\right) $. A lift $\tilde \Phi $ of $%
\Phi $ to $Z$ exists precisely when a lift $\hat \Phi $ of $\Phi $ to $X$
exists. The glueing conditions (\ref{pp6t1LocMom3}) for $\left(
J_{a,d}\right) $, $\left( J_{a,d}^{\prime }\right) $ guarantee that there
exist smooth functions $\hat h_A$, $\tilde h_A$ on $X$, $Z$, obeying 
\begin{equation}
\label{pp6t1LocMom70}\hat h_A\circ i_{a,d}=\left\langle
J_{a,d},A\right\rangle \text{\quad and\quad }\tilde h_A\circ
j_{a,d}=\left\langle J_{a,d}^{\prime },A\right\rangle \quad . 
\end{equation}
These functions define global moment maps $\left\langle \tilde
J,A\right\rangle =\tilde h_A$ on $Z$ and $\left\langle \hat J,A\right\rangle
=\hat h_A$ on $X$. $\tilde J$ satisfies the analogue of (\ref{pp6t1LocMom1}%
), 
\begin{equation}
\label{pp6t1LocMom7}\frac{\partial \tilde \Phi }{\partial G}A\TeXButton{i}
{\intmul}\Omega ^{\prime }+d\tilde h_A=0 
\end{equation}
on $Z$, where $\Omega ^{\prime }\equiv q^{*}\omega $. Furthermore, $Z$ and $%
X $ are ${\cal D}$-isomorphic, the isomorphism being effected by a
diffeomorphism $\phi :Z\rightarrow X$, where $\phi $ preserves fibres, $%
p\circ \phi =q$, and is ${\cal D}$-equivariant. We examine the relation
between the lifts $\hat \Phi $ and $\tilde \Phi $: Define a map $\psi
:G\times Z\rightarrow Z$, 
\begin{equation}
\label{pp6t1LocMom4}\left( g,z\right) \mapsto \psi \left( g,z\right) \equiv
\phi ^{-1}\circ \hat \Phi \left( g,\phi \left( z\right) \right) \quad . 
\end{equation}
A calculation shows that $\psi $ satisfies $q\circ \psi =\Phi \circ \left(
id_G\times q\right) $, and $\psi _e=id_Z$. But, as discussed in section \ref
{PreservationOfTheGroupLaw}, there exists precisely one function $G\times
Z\rightarrow Z$ with these two properties, and this is just the lift $\tilde
\Phi $! Hence we deduce that $\psi =\tilde \Phi $, or 
\begin{equation}
\label{pp6t1LocMom5}\phi \circ \tilde \Phi =\hat \Phi \circ \left(
id_G\times \phi \right) \quad , 
\end{equation}
which gives a commutative diagram. It follows that 
\begin{equation}
\label{pp6t1LocMom6}\frac{\partial \hat \Phi }{\partial G}=\phi _{*}\frac{%
\partial \tilde \Phi }{\partial G}\quad . 
\end{equation}

The symplectic $2$-forms on $X$ and $Z$ are $\Omega \equiv p^{*}\omega $ and 
$\Omega ^{\prime }\equiv q^{*}\omega $, where $\omega $ is the symplectic $2$%
-form on $Y$. From $p\phi =q$ we infer that $\Omega ^{\prime }=\phi
^{*}\Omega $, hence $\phi $ is a symplectomorphism. From theorem \ref
{CanonicalTransOnCovering} we know that $\hat \Phi $, $\tilde \Phi $ are
symplectic actions with respect to $\Omega $, $\Omega ^{\prime }$.

If we insert (\ref{pp6t1LocMom6}) into (\ref{pp6t1LocMom1}) we obtain 
\begin{equation}
\label{pp6t1LocMom8}\left[ \phi _{*}\frac{\partial \tilde \Phi }{\partial G}%
A\right] \TeXButton{i}{\intmul}\Omega +d\hat h_A=0\quad ; 
\end{equation}
using the relation $\Omega ^{\prime }=\phi ^{*}\Omega $ and (\ref
{pp6t1LocMom7}) we deduce $\left[ \phi _{*}\frac{\partial \tilde \Phi }{%
\partial G}A\right] \TeXButton{i}{\intmul}\Omega =\phi ^{-1*}\left[ \frac{%
\partial \tilde \Phi }{\partial G}A\TeXButton{i}{\intmul}\Omega ^{\prime
}\right] $, which, together with (\ref{pp6t1LocMom8}), yields $d\left\langle
\tilde J,A\right\rangle =\phi ^{*}d\left\langle \hat J,A\right\rangle $, or 
\begin{equation}
\label{pp6t1LocMom9}\tilde J=\hat J\circ \phi +L\quad , 
\end{equation}
where $L$ is a $Z$-constant linear map $\hat g\rightarrow \TeXButton{R}
{\mathbb{R}}$. Using the trivializations $\left( i_a\right) $, $\left(
j_a\right) $, (\ref{pp6t1LocMom70}) and (\ref{pp6t1LocMom9}) we have%
$$
J_{a,d}^{\prime }\left( y\right) =\tilde J\circ j_{a,d}=\left[ \hat J\circ
\phi +L\right] \circ j_{a,d}=\left( \hat J\circ i_a\right) \circ \left(
i_a^{-1}\circ \phi \circ j_a\right) \left( d,y\right) +L\quad . 
$$
As in the proof of theorem \ref{MultiValuedPotential}, $\left( i_a^{-1}\circ
\phi \circ j_a\right) \left( d,y\right) =\left( d\cdot k_a,y\right) $ for a $%
0$-\u Cech cochain $\left( k_a\right) $ which is determined by the cocycles $%
\left( g_{ab}\right) $, $\left( g_{ab}^{\prime }\right) $ up to its value in
a coset of the center of ${\cal D}$ in ${\cal D}$. Then the last equation
gives%
$$
J_{a,d}^{\prime }=J_{a,d\cdot k_a}+L\quad . 
$$
In summary, we have proven:

\subsection{Theorem and Definition: Local moment map \label
{TheoremLocalMomentMap}}

Let $Y$ be a connected, but not simply connected symplectic manifold with
base point $y$, with ${\cal D}\equiv \pi _1\left( Y,y\right) $, and $\omega $
is the symplectic $2$-form on $Y$. Let $\Phi :G\times Y\rightarrow Y$ be a
symplectic left action of a connected Lie group $G$ on $Y$ with respect to $%
\omega $. Let ${\cal V}=\left\{ V_a\mid a\in A\right\} $ be a simply
connected path-connected open cover of $Y$. Then

\begin{description}
\item[(A)]  \quad for every ${\cal D}$-valued $1$-\u Cech-cocycle $\left(
g_{ab}\right) $, $a,b\in A$, on ${\cal V}$, whose associated homomorphism $%
\rho :{\cal D}\rightarrow {\cal D}$ is an {\it inner} automorphism of ${\cal %
D}$, there exists a collection $\left( J_{a,d}\right) $ of coalgebra-valued
functions $J_{a,d}:V_a\rightarrow g^{*}$ for $a\in A$, $d\in {\cal D}$, such
that

\begin{enumerate}
\item  
\begin{equation}
\label{pp6t1LocMom10}\frac{\partial \Phi }{\partial G}A\TeXButton{i}{\intmul}%
\omega +d\left\langle J_{a,d},A\right\rangle =0 
\end{equation}
on $V_a$, and for all $d\in {\cal D}$;

\item  \quad let $\lambda $ be a loop at $y$ with $\left[ \lambda \right]
=d\in \pi _1\left( Y,y\right) \simeq {\cal D}$. Then 
\begin{equation}
\label{pp6t1SheetsOfMoment}\left\langle J_{a,d},A\right\rangle =\left\langle
J_{a,e},A\right\rangle -\int\limits_\lambda \frac{\partial \Phi }{\partial G}%
A\TeXButton{i}{\intmul}\omega 
\end{equation}
for all $A\in \hat g$, where $e$ is the identity in ${\cal D}$.

\item  \quad the $J_{a,d}$ satisfy a {\it glueing condition}, expressed by 
\begin{equation}
\label{pp6t1LocMom11}J_{a,d}=J_{b,d\cdot g_{ab}} 
\end{equation}
on $V_a\cap V_b\neq \emptyset $.
\end{enumerate}

\item[(B)]  \quad Let $\left( g_{ab}^{\prime }\right) $ be another cocycle
giving rise to a simply connected cover of $Y$, and let $\left(
J_{a,d}^{\prime }\right) $ be another collection of functions on ${\cal V}$
satisfying properties (A1--A3) with respect to $\left( g_{ab}^{\prime
}\right) $ and the action $\Phi $. Then there exists a $Y$-constant linear
map $L:\hat g\rightarrow \TeXButton{R}{\mathbb{R}}$ (i.e. $dL=0$) and a $%
{\cal D}$-valued $0$-\u Cech cochain $\left( k_a:V_a\rightarrow {\cal D}%
\right) $ on ${\cal V}$ such that 
\begin{equation}
\label{pp6t1LocMom12}J_{a,d}^{\prime }=J_{a,d\cdot k_a}+L 
\end{equation}
for all $a\in A$, $d\in {\cal D}$. The $0$-cochain $\left( k_a\right) $ is
determined by the cocycles $\left( g_{ab}\right) $ and $\left(
g_{ab}^{\prime }\right) $ as expressed in theorem \ref{MultiValuedPotential}.

\item[(C)]  \quad {\bf Definition:\quad }A collection $\left(
g_{ab};J_{a,d}\right) $ satisfying properties (A1--A3) will be called a {\it %
local moment map} for the action $\Phi $ on the symplectic manifold $\left(
Y,\omega \right) $.
\end{description}

\section{Equivariance of moment maps \label{EquivarianceOfMomentMaps}}

Usually, moment maps are introduced in a more restricted context. For
example, conditions are imposed from the start so as to guarantee the
existence of a uniquely determined single-valued globally defined moment
map. Furthermore, it is often assumed that the first and second
Chevalley-Eilenberg cohomology groups of the group $G$ vanish, which then
provides a sufficient condition for the moment map to transform as a $G$%
-morphism \cite{Woodhouse,GuillStern}. Our approach will be slightly
more general. The first generalization has been made above, allowing for
moment maps to be only locally defined. The second
one is, that we do not want to enforce the moment maps to behave as strict $%
G $-morphisms; rather, the deviation from transforming equivariantly is
determined by a cocycle in a certain cohomology on $\hat g$, which in turn
gives rise to a central extension of the original Lie algebra $\hat g$,
which is interesting in its own right and also physically relevant.

Let the conditions of theorem \ref{TheoremLocalMomentMap} be given.
Reconstruct a simply connected cover $p:X\rightarrow Y$ from ${\cal V}$ and $%
\left( g_{ab}\right) $ as in section \ref{LocalMomentMaps}, together with
trivializations $\left( i_a\right) $ such that $i_b^{-1}\circ i_a\left(
d,y\right) =\left( d\cdot g_{ab},y\right) $. Assume a lift $\hat \Phi _g$
exists. Assume that $p\circ \hat \Phi _g\left( x\right) \in V_b$, where $%
x=i_a\left( d,y\right) $; then there exists a unique $d^{\prime }\in {\cal D}
$ with $i_b^{-1}\circ \hat \Phi _g\left( x\right) =\left( d^{\prime },\Phi
_g\left( y\right) \right) $, with $\Phi _g\left( y\right) \in V_b$. Here $%
d^{\prime }$ is a function of $d$, $g$, and $y$; its structure can be
understood from the following consideration: Let $\lambda $ be a path in $G$
connecting $e$ with $g$. Then $t\mapsto \Phi \left( \lambda \left( t\right)
,y\right) $ is a path in $Y$ connecting $y$ with $\Phi _g\left( y\right) $,
which has a unique lift to $i_a\left( d,y\right) $, whose endpoint is just $%
\hat \Phi _g\circ i_a\left( d,y\right) $, by the definition of the lift $%
\hat \Phi $. The condition $\Phi _{y\#}\pi _1\left( G,e\right) \subset \pi
_1\left( Y,y\right) $ guarantees that this construction is independent of
the path $\lambda $. The interval $\left[ 0,1\right] $ can be partitioned
into subintervals $\left[ a_{i-1},a_i\right] $ so that $\Phi \left( \lambda
\left( t\right) ,y\right) \in V_{a_i}$ for $t\in \left[ a_{i-1},a_i\right] $%
, and $\lambda \left( t_n\right) =g$. It follows that 
\begin{equation}
\label{pp6t1equi2}i_{a_n}^{-1}\circ \hat \Phi _g\left( x\right) =\left(
d\cdot g_{a_0a_1}\cdots g_{a_{n-1}a_n},\Phi _g\left( y\right) \right) \quad
, 
\end{equation}
and hence $d^{\prime }=d\cdot g_{a_0a_1}\cdots g_{a_{n-1}a_n}\equiv d\cdot
\psi _{a_n}\left( g,y\right) $. Since ${\cal D}$ is discrete, $\psi _{a_n}$
is locally constant. Altogether we have shown that if $p\circ \hat \Phi
_g\left( x\right) \in V_b$, then 
\begin{equation}
\label{pp6t1equi3}i_b^{-1}\circ \hat \Phi _g\left( x\right) =\left( d\cdot
\psi _b,\Phi _g\left( y\right) \right) \quad . 
\end{equation}

\TeXButton{Abst}{\vspace{0.8ex}}--- Next we recall without proof the $G$%
-transformation behaviour for global moment maps (see \cite{Azcarraga}):

\subsection{Theorem: $G$-transformation of global moment maps \label
{GTransGlobalMomentMap}}

Let $\left( M,\omega \right) $ be a simply connected symplectic manifold,
let $\Phi :G\times M\rightarrow M$ be a symplectic left action of the Lie
group $G$ on $M$. Let $J$ be a global moment map for the action $\Phi $.
Then 
\begin{equation}
\label{pp6t1equi4}J\circ \Phi _g=Ad^{*}\left( g\right) \cdot J+\alpha \left(
g\right) \quad , 
\end{equation}
where $\alpha :G\rightarrow g^{*}$ is a $1$-cocycle in the $g^{*}$-valued
cohomology on $G$ as defined in section \ref{gStarOnGroup}, i.e. $\alpha \in
Z^1\left( G,g^{*}\right) $. This means that 
\begin{equation}
\label{pp6t1equi5}\left( \delta \alpha \right) \left( g,h\right)
=Ad^{*}\left( g\right) \cdot \alpha \left( h\right) -\alpha \left( gh\right)
+\alpha \left( g\right) =0 
\end{equation}
for all $g,h\in G$. Thus, (\ref{pp6t1equi4}) says that $J$ transforms
equivariantly under $G$ up to a cocycle in the $g^{*}$-valued cohomology.

\TeXButton{Abst}{\vspace{0.8ex}}--- Now we generalize this result to local
moment maps. We prove:

\subsection{Theorem: $G$-transformation behaviour of local moment maps \label
{GTransLocalMomentMap}}

Let $Y$ be a connected symplectic manifold, with ${\cal D}\equiv \pi
_1\left( Y,y\right) $. Let $\Phi :G\times Y\rightarrow Y$ be a symplectic
left action of a connected Lie group $G$ on $Y$. Let ${\cal V}=\left\{
V_a\mid a\in A\right\} $ be a simply connected open cover of $Y$. Let $%
\left( g_{ab}\right) $, $a,b\in A$, be a ${\cal D}$-valued $1$-\u
Cech-cocycle on ${\cal V}$ describing a simply connected ${\cal D}$-covering
space of $Y$, and let the collection $\left( J_{a,d};g_{ab}\right) $ be the
associated local moment map with respect to the action $\Phi $.

Let $y\in Y$, and assume that $p\circ \Phi _g\left( y\right) \in V_b$. Then 
\begin{equation}
\label{pp6t1GTrans1}J_{b,d\cdot \psi _b}\circ \Phi _g=Ad^{*}\left( g\right)
\cdot J_{a,d}\;+\;\alpha \left( g\right) \quad , 
\end{equation}
where $\alpha :G\rightarrow g^{*}$ is a $1$-cocycle in the $g^{*}$-valued
cohomology on $G$ as defined in appendix, chapter \ref{gStarOnGroup}, and $%
\psi _g$ is defined in formula (\ref{pp6t1equi3}).

\TeXButton{Beweis}{\raisebox{-1ex}{\it Proof :}
\vspace{1ex}}

Let the conditions of theorem \ref{TheoremLocalMomentMap} be given, and
reconstruct a simply connected cover $p:X\rightarrow Y$ from ${\cal V}$ and $%
\left( g_{ab}\right) $ as in section \ref{LocalMomentMaps}, together with
trivializations $\left( i_a\right) $ such that $i_b^{-1}\circ i_a\left(
d,y\right) =\left( d\cdot g_{ab},y\right) $. Assume a lift $\hat \Phi _g$
exists. By $\hat J$ we denote the global moment map on $X$ with respect to $%
\hat \Phi $. This satisfies 
\begin{equation}
\label{pp6t1GTrans2}\hat J\circ \hat \Phi _g=Ad^{*}\left( g\right) \cdot
\hat J+\alpha \left( g\right) \quad , 
\end{equation}
according to (\ref{pp6t1equi4}). From (\ref{pp6t1equi3}) it follows that 
\begin{equation}
\label{pp6t1GTrans3}i_b^{-1}\circ \hat \Phi _g\circ i_a\left( d,y\right)
=\left( d\cdot \psi _b,\Phi _g\left( y\right) \right) \quad . 
\end{equation}
Using $J_{a,d}=\hat J\circ i_{a,d}$ in (\ref{pp6t1GTrans2}), (\ref
{pp6t1GTrans3}) implies%
$$
\left[ \hat J\circ i_b\right] \circ \left[ i_b^{-1}\circ \hat \Phi _g\circ
i_a\right] \left( d,y\right) =J_{b,d\cdot \psi _b}\left( y\right) 
$$
for the LHS, and hence the result (\ref{pp6t1GTrans1}). \TeXButton{BWE}
{\hfill
\vspace{2ex}
$\blacksquare$}

\section{Non-simply connected coverings \label{NonSimplyConnected}}

In this section we study the relation between local moment maps on
symplectic manifolds $Z$, $Y$ where $q:Z\rightarrow Y$ is a covering of
manifolds, but $Z$ is not necessarily simply connected:

Let $\zeta \in Z$, $\eta \in Y$ be base points with $q\left( \zeta \right)
=\eta $; let ${\cal D}\equiv \pi _1\left( Y,\eta \right) $, and ${\cal H}%
\equiv \pi _1\left( Z,\zeta \right) $. Let $X$ be a universal covering
manifold $p:X\rightarrow Y$ of $Y$ such that $Y$ is the orbit space $Y=X/%
{\cal D}$. Then $X$ is also a universal cover of $Z$, and $Z$ is isomorphic
to the orbit space $X/{\cal H}$; for the sake of simplicity we ignore this
isomorphism and identify $Z=X/{\cal H}$. There is a covering projection $%
r:X\rightarrow Z$, taking the base point $\xi \in X$ to $r\left( \xi \right)
=\zeta $, and $p=q\circ r$. Since $q_{\#}\pi _1\left( Z,\zeta \right) $ is
an injective image of ${\cal H}$ in ${\cal D}$, we identify ${\cal H}$ with
its image under $q_{\#}$, and thus can regard ${\cal H}$ as a subgroup of $%
{\cal D}$.

If $\omega $ is a symplectic form on $Y$, then $\Omega \equiv q^{*}\omega $
and $\hat \Omega \equiv p^{*}\omega $ are the natural symplectic forms on $Z$
and $X$, respectively, and $\hat \Omega =r^{*}\Omega $.

Let the Lie group $G$ act on $Y$ from the left via $\Phi :G\times
Y\rightarrow Y$. We assume that the lift $\hat \Phi $ of $\Phi $ to $X$
exists, which is true if and only if $\Phi _{\eta \#}\pi _1\left( G,e\right)
=\left\{ e\right\} $. It is easy to show that in this case the lift $\tilde
\Phi $ of $\Phi $ to $Z$ exists; and furthermore, that the lift of $\tilde
\Phi $ to $X$ coincides with $\hat \Phi $.

We now introduce trivializations for $p:X\rightarrow Y$, and subsequently,
construct preferred trivializations of the coverings $r:X\rightarrow Z$ and $%
q:Z\rightarrow Y$ based on this. Firstly, let ${\cal V}$ be a simply
connected open cover of $Y$ as above, and let $\left( i_a:{\cal D}\times
V_a\rightarrow X\right) $ be a trivialization of the covering $%
p:X\rightarrow Y$. Given an element $d\in {\cal D}$, let $\left[ d\right] $
denote the coset $\left[ d\right] \equiv {\cal H}\cdot d$, where ${\cal H}%
\subset {\cal D}$ is regarded as a subgroup of ${\cal D}$, as above. For $%
V_a\in {\cal V}$ and $d\in {\cal D}$, define sets $U_{a,\left[ d\right]
}\equiv r\circ i_a\left( \left\{ d\right\} \times V_a\right) \subset Z$. The
sets $U_{a,\left[ d\right] }$ are open and simply connected by construction,
and their totality 
\begin{equation}
\label{pp6t1nonsim1}{\cal U}\equiv \left\{ U_{a,\left[ d\right] }\mid a\in
A\,;\;\left[ d\right] \in {\cal D/H}\right\} 
\end{equation}
covers $Z$, since the sets $i_a\left( d,V_a\right) $ cover $X$. The $%
U_{a,\left[ d\right] }$ are just the connected components of the inverse
image $q^{-1}\left( V_a\right) \subset Z$, hence we have $q\left(
U_{a,\left[ d\right] }\right) =V_a$ for all $\left[ d\right] \in {\cal D}/%
{\cal H}$, and ${\cal U}$ is a simply connected open (countable) cover of $Z$%
. We define a trivialization $\left( k_a:{\cal D}/{\cal H}\times
V_a\rightarrow Z\right) $ of the covering $q:Z\rightarrow Y$ by $k_a\left(
\left[ d\right] ,y\right) \equiv r\circ i_a\left( d,y\right) $. Since $r$
maps points $x\in X$ into orbits $r\left( x\right) =\hat \Phi \left( {\cal H}%
,x\right) $, this definition is independent of the representative $d$ of $%
\left[ d\right] $. Now we can construct trivializations of $r:X\rightarrow Z$
based on the cover ${\cal U}$ of $Z$; in particular, by specifying
representatives $d_0$ of the various cosets $\left[ d_0\right] $, we see
that there exists a trivialization $\left( j_{a,\left[ d_0\right] }:{\cal H}%
\times U_{a,\left[ d_0\right] }\rightarrow X\right) $ such that 
\begin{equation}
\label{pp6t1nonsim2}j_{a,\left[ d_0\right] }\left( h,k_a\left( \left[
d_0\right] ,y\right) \right) =i_a\left( h\cdot d_0,y\right) 
\end{equation}
for all arguments.

As $X$ is simply connected, the action $\hat \Phi $ has a global moment map $%
\hat J$ satisfying 
\begin{equation}
\label{pp6t1nonsim3}\frac{\partial \hat \Phi }{\partial G}A\TeXButton{i}
{\intmul}p^{*}\omega +d\left\langle \hat J,A\right\rangle =0\quad . 
\end{equation}
Using the arguments in section \ref{LocalMomentMaps} we find that 
\begin{equation}
\label{pp6t1nonsim4}\frac{\partial \tilde \Phi }{\partial G}A\TeXButton{i}
{\intmul}q^{*}\omega +d\left\langle \hat J\circ j_{a,\left[ d_0\right]
,h},A\right\rangle =0\quad . 
\end{equation}
Hence, introducing the quantities $\tilde J_{a,\left[ d_0\right] ,h}\equiv
\hat J\circ j_{a,\left[ d_0\right] ,h}$, and taking into account that the
trivializations $\left( j_{a,\left[ d_0\right] }\right) $ define an ${\cal H}
$-valued $1$-\u Cech cocycle $\left( \hat g_{a,\left[ d_0\right]
\,;\,b,\left[ d_0^{\prime }\right] }\right) $ by 
\begin{equation}
\label{pp6t1nonsim5}j_{b,\left[ d_0^{\prime }\right] }^{-1}\circ j_{a,\left[
d_0\right] }\left( h,z\right) =\left( h\cdot \hat g_{a,\left[ d_0\right]
\,;\,b,\left[ d_0^{\prime }\right] },z\right) \quad , 
\end{equation}
we see that the collection $\left( \tilde J_{a,\left[ d_0\right]
,h}\,;\,\hat g_{a,\left[ d_0\right] \,;\,b,\left[ d_0^{\prime }\right]
}\right) $ defines a local moment map for the action $\tilde \Phi $ with
respect to $q^{*}\omega $. Similarly, the equation 
\begin{equation}
\label{pp6t1nonsim6}\frac{\partial \Phi }{\partial G}A\TeXButton{i}{\intmul}%
\omega +d\left\langle \hat J\circ i_{a,d},A\right\rangle =0 
\end{equation}
shows that the collection $\left( J_{a,d}\,;\,g_{ab}\right) $, where $%
J_{a,d}\equiv \hat J\circ i_{a,d}$, and $\left( g_{ab}\right) $ is the $%
{\cal D}$-valued $1$-\u Cech cocycle with respect to $\left( i_a\right) $,
is a local moment map for the action $\Phi $ with respect to $\omega $. The
relation between these two local moment maps is easily found using (\ref
{pp6t1nonsim2}) to be 
\begin{equation}
\label{pp6t1nonsim7}\tilde J_{a,\left[ d_0\right] ,h}\circ k_{a,\left[
d_0\right] }=J_{a,h\cdot d_0}\quad . 
\end{equation}

\section{$G$-state spaces and moment maps \label{GStateSpaces}}

Finally, in this section we discuss our concept of a $G$-state space. This
is an identification space based on a partitioning of a symplectic manifold
into connected subsets, on each of which a given global moment map is
constant. These connected subsets are then invariant under the Hamiltonian
flow associated with every Hamiltonian $h$ that commutes with the \thinspace 
$G$-action $\hat \Phi $. This construction coincides with the first step in
a Marsden-Weinstein reduction of the symplectic manifold. We first consider
the case where the symplectic manifold is simply connected:

Let $X$ be a simply connected symplectic manifold with symplectic $2$-form $%
\Omega $. Let $\hat \Phi $ be a symplectic action of a Lie group $G$ on $X$.
There exists a global moment map $J$ associated with $\hat \Phi $. To every $%
x\in X$ we now assign the connected component $s\left( x\right) $ of $%
J^{-1}\left( J\left( x\right) \right) $ that contains $x$; i.e., $s\left(
x\right) \subset J^{-1}\left( J\left( x\right) \right) $ is connected (in
the induced topology), and $x\in s\left( x\right) $. The collection of all $%
s\left( x\right) $, as $x$ ranges through $X$, is denoted as $\Sigma _X$.
Then $\Sigma _X$ is an identification space, where $s:X\rightarrow \Sigma _X$
is the identification map. We endow $\Sigma _X$ with the quotient topology
inherited from $X$. One can assume further technical conditions in order to
guarantee that the sets $s\left( x\right) $ are presymplectic manifolds
which give rise to reduced phase spaces; such a reduction is called
Marsden-Weinstein reduction \cite{Woodhouse}. We do not make these
assumptions here, since they are not necessary for our purposes.

By construction, the moment map $J$ is constant on every connected component
of $J^{-1}\left( J\left( x\right) \right) $, and hence descends to the space 
$\Sigma _X$; i.e., there exists a unique map $\iota :\Sigma _X\rightarrow
g^{*}$ satisfying $\iota \circ s=J$. Also, every diffeomorphism $\hat \Phi
_g $ maps connected components of $J^{-1}\left( J\left( x\right) \right) $
into connected components; this follows from formula (\ref{pp6t1equi4}).
Hence, $\hat \Phi $ descends to an action $\hat \phi :G\times \Sigma
_X\rightarrow \Sigma _X$, satisfying 
\begin{equation}
\label{pp6t1GState0}s\circ \hat \Phi _g=\hat \phi _g\circ s\quad , 
\end{equation}
which gives rise to an analogue of formula (\ref{pp6t1equi4}) on $\Sigma _X$%
, 
\begin{equation}
\label{pp6t1GState1}\iota \circ \hat \phi _g=Ad^{*}\left( g\right) \cdot
\iota +\alpha \left( g\right) \quad . 
\end{equation}

In this construction, we have identified all states $x\in X$ which are
mapped into the same value under the moment map, and which can be connected
by a path on which the moment map is constant. $\Sigma _X$ is a $G$-space
with action $\hat \phi $. There is a semi-equivariant map $\iota $ from $%
\Sigma _X$ to the $G$-space $g^{*}$. Furthermore, the connected components $%
s\left( x\right) $ are preserved by any Hamiltonian that commutes with $G$.
This discussion can be summarized in the

\subsection{Theorem and definition}

Let $X$ be a simply connected symplectic manifold, let $\hat \Phi $ be a
symplectic action of a connected Lie group $G$ with Lie algebra $\hat g$ on $%
X$, let $J$ be a global moment map associated with $\hat \Phi $. Then

\begin{description}
\item[(A)]  \quad there exists a space $\Sigma _X$ with a $G$-action $\hat
\phi $, a projection $s:X\rightarrow \Sigma _X$, and a semi-equivariant map $%
\iota :\Sigma _X\rightarrow g^{*}$ satisfying $\iota \circ s=J$ such that (%
\ref{pp6t1GState1}) holds.

\item[(B)]  \quad If $h$ is any Hamiltonian on $X$ satisfying the
Poisson-bracket relations $\left\{ h,\left\langle J,A\right\rangle \right\}
=0$ for all $A\in \hat g$, then the associated Hamiltonian flow $f_t\left(
x\right) $ preserves the sets $s^{-1}\left( \sigma \right) $, $\sigma \in
\Sigma $. In other words, 
\begin{equation}
\label{pp6t1GState2}J\circ f_t\left( x\right) =J\left( x\right) 
\end{equation}
for all $x\in s^{-1}\left( \sigma \right) $ and $t\in \TeXButton{R}
{\mathbb{R}}$.

\item[(C)]  \quad $\Sigma _X$ will be called a $G${\it -state space} for the
pair $\left( X,\hat \Phi \right) $.
\end{description}

\section{The splitting of multiplets \label{SplittingOfMultiplets}}

Now we turn to investigate the relation of the objects defined above to a
similar construction on a non-simply connected manifold $Y$, where $%
p:X\rightarrow Y$ is a universal covering of $Y$, and $p$ is a local
symplectomorphism of symplectic forms $\omega $ on $Y$ and $\Omega \equiv
p^{*}\omega $ on $X$. The first thing to observe is that the diffeomorphisms 
$\gamma $ of the deck transformation group ${\cal D}$ of the covering
descend to $\Sigma _X$: To see this, let $\lambda $ be a path lying entirely
in one of the connected components $s^{-1}\left( \sigma \right) \subset
J^{-1}\left( J\left( x\right) \right) $ [Here we assume that connectedness
implies path-connectedness]. If $\dot \lambda $ denotes its tangent, we have%
$$
\frac d{dt}\left\langle J\circ \gamma \circ \lambda ,A\right\rangle =\dot
\lambda \TeXButton{i}{\intmul}\gamma ^{*}d\left\langle J,A\right\rangle
=-\Omega \left( \frac{\partial \hat \Phi }{\partial G}A,\gamma _{*}\dot
\lambda \right) = 
$$
$$
=-\left( \gamma ^{*}\Omega \right) \left( \gamma _{*}^{-1}\frac{\partial
\hat \Phi }{\partial G}A,\dot \lambda \right) =-\Omega \left( \frac{\partial
\hat \Phi }{\partial G}A,\dot \lambda \right) =\frac d{dt}\left\langle
J\circ \lambda ,A\right\rangle =0\quad . 
$$
Here, the last equation follows from the fact that $\lambda $ lies in a
connected component of $J^{-1}\left( J\left( x\right) \right) $;
furthermore, we have used that $\Omega $ is ${\cal D}$-invariant, and the
action $\hat \Phi $ commutes with ${\cal D}$, since $G$ is connected (this
follows from theorem \ref{ExtensionOfG}). But this result says that $J$ is
constant on the $\gamma $-image of every connected component $J^{-1}\left(
J\left( x\right) \right) $; since this image is connected itself, it must
lie in one of the connected components of $J^{-1}\left( J\circ \gamma \left(
x\right) \right) $. As $\gamma $ is invertible, it follows that $\gamma $
maps connected components onto connected components, and hence descends to a
map $\bar \gamma :\Sigma _X\rightarrow \Sigma _X$ such that 
\begin{equation}
\label{pp6t1GState3}\bar \gamma \circ s=s\circ \gamma \quad . 
\end{equation}
Thus, we have a well-defined action of ${\cal D}$ on $\Sigma _X$. We now
construct a space $\Sigma _Y$ analogous to $\Sigma _X$: Define $\Sigma _Y$
as the quotient $\Sigma _X/{\cal D}$, with projection $q:\Sigma
_X\rightarrow \Sigma _Y$. We note that this is not a covering space in
general, since the action of ${\cal D}$ on $\Sigma _X$ need not necessarily
be free; for example, if $\gamma $ maps one of the connected components $%
s^{-1}\left( \sigma \right) $ onto itself, then $\bar \gamma $ has a
fixpoint on $\Sigma _X$. However, formula (\ref{pp6t1GState3}) implies that
the map $s$ descends to the quotient $\Sigma _Y=\Sigma _X/{\cal D}$, which
means that there exists a unique map $\bar s:Y\rightarrow \Sigma _Y$ such
that 
\begin{equation}
\label{pp6t1GState4}\bar s\circ p=q\circ s\quad . 
\end{equation}
Using (\ref{pp6t1GState3}) and (\ref{pp6t1GState4}) it is easy to see that
the action $\Phi $ of $G$ on $Y$ preserves the $G$-states on $Y$, i.e. the
images $p\circ s^{-1}\left( \sigma \right) =\bar s^{-1}\circ q\left( \sigma
\right) $ of the connected components of $J^{-1}\left( J\left( x\right)
\right) $, where $x\in s^{-1}\left( \sigma \right) $, under $p$: For, let $%
y\in p\circ s^{-1}\left( \sigma \right) $, then there exists an $x\in
s^{-1}\left( \sigma \right) $ with $y=p\left( x\right) $. Then%
$$
\Phi _g\left( y\right) =\Phi _g\circ p\left( x\right) =p\circ \hat \Phi
_g\left( x\right) \in p\circ \hat \Phi _g\circ s^{-1}\left( \sigma \right)
=p\circ s^{-1}\circ \hat \phi _g\left( \sigma \right) =\bar s^{-1}\circ
q\circ \hat \phi _g\left( \sigma \right) \quad , 
$$
which implies that $\bar s\circ \Phi _g\left( y\right) \in q\circ \hat \phi
_g\left( \sigma \right) $ for all $y$ in $\bar s^{-1}\left( q\left( \sigma
\right) \right) $. Hence, $\Phi $ descends to an action $\phi :G\times
\Sigma _Y\rightarrow \Sigma _Y$ with 
\begin{equation}
\label{pp6t1GState5}\phi _g\circ \bar s=\bar s\circ \Phi _g\quad , 
\end{equation}
which is the analogue of (\ref{pp6t1GState0}).

The orbits $\hat \phi _G\left( \sigma \right) $, $\sigma \in \Sigma _X$, and 
$\phi _G\left( \tau \right) $, $\tau \in \Sigma _Y$, are the classical
analogue of carrier spaces of irreducible $G$-representations in the quantum
context \cite{Woodhouse}. However, for every $G$-state $\tau \in \Sigma _Y$
there exists a collection $q^{-1}\left( \tau \right) $ of $G$-states in $%
\Sigma _X$ which are identified under $q$. The elements in the collection $%
q^{-1}\left( \tau \right) $ are labelled by the elements $d$ of the
fundamental group ${\cal D}\simeq \pi _1\left( Y,y\right) $. We call this
phenomenon the ''splitting of (classical) multiplets'' on account of the
multiple-connectedness of the background $Y$. It is basically a consequence
of the fact that the group $G$, when lifted to the covering space $X$, is
extended to a group $\tilde G$ by the deck transformation group ${\cal D}$,
whose group law in the case under consideration is determined by formula (%
\ref{pp6t1lift12}) in theorem \ref{ExtensionOfG}.

\begin{appendix}

\chapter{Form-valued cohomology on the deck transformation group \label
{FormValuedCohomologyOnD}}

Here we compile the cohomologies that are used in this work. References are 
\cite{Fulton,Azcarraga,DodsonParker}.

\vspace{0.8ex}

--- Let $p:X\rightarrow Y$ be a covering of manifolds, where $X$ is simply
connected. The deck transformation group of the covering is ${\cal D}$. Our
analysis is based on formula (\ref{pp6t1form3}), 
\begin{equation}
\label{pp6t1cohomFormVal1}\gamma ^{*}\eta =\eta +d\chi \left( \gamma \right)
\quad , 
\end{equation}
where $\eta $ is the potential for a $q$-form $d\eta $ which is the
pull-back of a closed $q$-form $\omega $ on $Y$, i.e. $d\eta =p^{*}\omega $.
Then $d\eta $ is ${\cal D}$-invariant, as follows from proposition \ref
{DeckInvariant}, but $\eta $ is not, as follows from the last equation. In
particular this means that for $\chi \neq 0$, $\eta $ is not the pull-back
under $p^{*}$ of a form on $Y$. Now (\ref{pp6t1cohomFormVal1}) defines a
cochain in a cohomology on ${\cal D}$ defined as follows (our notation
conventions are those of \cite{Azcarraga}): An $n$-cochain $\alpha _n$ is a
map $\alpha _n:{\cal D}^n\rightarrow \Lambda ^{*}\left( X\right) $, where $%
\Lambda ^{*}\left( X\right) =\bigoplus\limits_{q\ge 0}\Lambda ^q\left(
X\right) $ denotes the ring of differential forms on $X$. The deck
transformation group ${\cal D}$ acts via pull-back of elements $\gamma $ on
forms: ${\cal D}\times \Lambda ^{*}\left( X\right) \ni \left( \gamma ,\alpha
\right) \mapsto \gamma ^{*}\alpha $. This is a {\bf right} action, in
contrast to the the cohomologies to be discussed below. A zero-cochain $%
\alpha _0$ is an element of $\Lambda ^{*}\left( X\right) $. The coboundary
operator $\delta $ in this cohomology is defined to act on $0$-, $1$-, $2$%
-cochains according to 
\begin{equation}
\label{pp6t1co6}
\begin{array}{rcl}
\left( \delta \alpha _0\right) \left( \gamma \right) & = & \gamma ^{*}\alpha
_0-\alpha _0\quad , \\ 
\left( \delta \alpha _1\right) \left( \gamma _1,\gamma _2\right) & = & 
\gamma _2^{*}\alpha _1\left( \gamma _1\right) -\alpha _1\left( \gamma
_1\gamma _2\right) +\alpha _1\left( \gamma _2\right) \quad , \\ 
\left( \delta \alpha _2\right) \left( \gamma _1,\gamma _2,\gamma _3\right) & 
= & \gamma _3^{*}\alpha _2\left( \gamma _1,\gamma _2\right) +\alpha _2\left(
\gamma _1\gamma _2,\gamma _3\right) - \\  
&  & -\alpha _2\left( \gamma _1,\gamma _2\gamma _3\right) -\alpha _2\left(
\gamma _2,\gamma _3\right) \quad , 
\end{array}
\end{equation}
and $\delta $ is nilpotent, $\delta \circ \delta =0$, as usual. Now consider
the $0$-cochain $\eta $ in equation (\ref{pp6t1cohomFormVal1}). With the
help of (\ref{pp6t1co6}), equation (\ref{pp6t1cohomFormVal1}) can be
expressed as 
\begin{equation}
\label{pp6t1co7}\delta \eta =d\chi \quad , 
\end{equation}
where it is understood that $\chi $ is a function of arguments in the set $%
{\cal D}\times X$. The commutativity $\left[ \delta ,d\right] =0$ and the
nilpotency $\delta ^2=0$ and $d^2=0$ of the coboundary operators give rise
to a chain of equations similar to (\ref{pp6t1co7}): Applying $\delta $ to (%
\ref{pp6t1co7}) gives 
\begin{equation}
\label{pp6t1co8}d\left( \delta \chi \right) =0\quad . 
\end{equation}
Since $X$ is simply connected 
\begin{equation}
\label{pp6t1co9}\delta \chi =d\chi ^{\prime } 
\end{equation}
for some $\chi ^{\prime }$. Now the process can be repeated with the last
equation, etc.

\chapter{\u Cech cohomology on an open cover \label{CechCohomology}}

The definitions in this chapter are based on \cite{Fulton,DodsonParker}.

Let $Y$ be a topological space and ${\cal D}$ be a discrete group. Let $%
{\cal V}\equiv \left\{ V_a\mid a\in A\right\} $ be an open cover of $Y$ such
that every $V_a$ is admissible. [If $Y$ is a manifold, we can assume that $A$
is countable, and every $V_a$ is simply connected.] A function $%
g:Y\rightarrow {\cal D}$ on the topological space $Y$ is called {\it locally
constant} if every point $y\in Y$ possesses a neighbourhood $V$ on which the
restriction of $g$ is constant.

Let ${\cal S}^0$ denote the sum ${\cal S}^0\equiv \bigsqcup\limits_{a\in
A}V_a$; let ${\cal S}^1$ denote the sum ${\cal S}^1\equiv
\bigsqcup\limits_{a,b}V_a\cap V_b$, for all $a,b$ for which $V_a\cap V_b\neq
\emptyset $, allowing for $a=b$. We denote the images of $V_a$, $V_a\cap V_b$%
, ..., under the associated injections simply by $\left( a\right) $, $\left(
a,b\right) $, etc. [We recall that the set underlying a sum $\Sigma =B\sqcup
C$ is the disjoint union of $B$ and $C$. Furthermore, if $i:B\rightarrow
\Sigma $, $j:C\rightarrow \Sigma $ are the injections, then $i\left(
B\right) $ and $j\left( C\right) $ are both open and closed in $\Sigma $,
which means that $\Sigma $ is disconnected. Hence if each $V_a$ is connected
in $Y$, then a locally constant function on ${\cal S}^0$ is constant on the
images of all $V_a$ under the appropriate injection; similarly, a locally
constant function on ${\cal S}^1$ is constant on the images of $V_a\cap V_b$
under injection. Therefore in this case, a locally constant function on $%
{\cal S}^0,{\cal S}^1$ is constant on all $\left( a\right) $, and $\left(
a,b\right) $, respectively]. A locally constant function $f_0:{\cal S}%
^0\rightarrow {\cal D}$ is called a $0$-\u Cech-cochain (with respect to $%
{\cal V}$). A locally constant function $f_1:{\cal S}^1\rightarrow {\cal D}$
is called a $1$-\u Cech-cochain (with respect to ${\cal V}$). A $1$-\u
Cech-cochain $f_1$ is called a $1${\it -\u Cech-cocycle} if

\begin{description}
\item[(Coc1)]  \ $\left. f_1\right| \left( a,a\right) =e$, where $e$ is the
identity in ${\cal D}$,

\item[(Coc2)]  \ $\left. f_1\right| \left( b,a\right) =\left.
f_1^{-1}\right| \left( a,b\right) $, $f_1^{-1}$ denoting the inverse of $f_1$
in ${\cal D}$;
\end{description}

and for all $V_a,V_b,V_c$ for which $V_a\cap V_b\cap V_c\neq \emptyset $ it
is true that

\begin{description}
\item[(Coc3)]  \ $\left. f_1\right| \left( a,c\right) =\left[ \left.
f_1\right| \left( a,b\right) \right] \cdot \left[ \left. f_1\right| \left(
b,c\right) \right] $,
\end{description}

where $\left( a,b\right) $, $\left( b,a\right) $, $\left( a,c\right) $,
etc., are to be regarded as disjoint subsets of ${\cal S}^1$.

Two $1$-\u Cech-cocycles $f,f^{\prime }$ are said to be {\it cohomologous}
if there exists a $0$-\u Cech-cochain $h$ such that

\begin{description}
\item[(Coh)]  \ $\left. f^{\prime }\right| \left( a,b\right) =\left[ \left.
h^{-1}\right| \left( a\right) \right] \cdot \left[ \left. f\right| \left(
a,b\right) \right] \cdot \left[ \left. h\right| \left( b\right) \right] $.
\end{description}

The property of being cohomologous defines an equivalence relation amongst
all $1$-\u Cech-cocycles with respect to ${\cal V}$; the equivalence classes
are called {\it first \u Cech cohomology classes on }${\cal V}$ with
coefficients in ${\cal D}$. The set of these classes is denoted as $%
H^1\left( {\cal V};{\cal D}\right) $.

For $n>1$, the $n$-th cohomology class is described more readily when ${\cal %
D}$ is Abelian. Assuming this, an $n$-{\it \u Cech-cochain} $f_n$ with
coefficients in ${\cal D}$ is a locally constant map $f_n:{\cal S}%
^n\rightarrow {\cal D}$, where ${\cal S}^n$ is the topological sum ${\cal S}%
^n\equiv \bigsqcup\limits_{a_0,\ldots ,a_n}V_{a_0}\cap V_{a_1}\cdots \cap
V_{a_n}$, for all $a_0,\ldots ,a_n$ for which $V_{a_0}\cap V_{a_1}\cdots
\cap V_{a_n}\neq \emptyset $. The coboundary operator $\delta $ sends $n$%
-cochains $f_n$ to $\left( n+1\right) $-cochains $\delta f_n$ defined by 
\cite{DodsonParker} 
\begin{equation}
\label{pp6t1cechcohom1}\left. \delta f_n\right| \left( a_0,\ldots
,a_{n+1}\right) =\sum_{k=0}^{n+1}\left( -1\right) ^k\cdot \left. f_n\right|
\left( a_0,\ldots ,\widehat{a_k},\ldots ,a_{n+1}\right) \quad , 
\end{equation}
where $\widehat{a_k}$ means that this argument has to be omitted, and $-f_n$
denotes the inverse of $f_n$ in the Abelian group ${\cal D}$. $\delta $ is
nilpotent, $\delta _{n+1}\circ \delta _n=0$. As usual, $n$-cocycles are
elements in $\ker \delta _n$, $n$-coboundaries are elements in $\left. {\rm %
im\,}\delta _{n-1}\right. $, and the $n$-th \u Cech cohomology group on $%
{\cal V}$ is the quotient $H^n\left( {\cal V};{\cal D}\right) =\ker \delta
_n/\left. {\rm im\,}\delta _{n-1}\right. $. Obviously, statements
(Coc1-Coc3) and (Coh) above generalize this pattern to non-Abelian groups $%
{\cal D}$.

\chapter{${\cal D}$-coverings \label{DCoverings}}

This and the next chapter are mainly based on \cite{Fulton}.

Let ${\cal D}$ denote a discrete group. Let $X$ be a topological space on
which ${\cal D}$ acts properly discontinuously and freely. Let $%
p:X\rightarrow X/{\cal D}$ denote the projection onto the space of orbits,
endowed with the quotient (final) topology. Then $p$ is a covering map.

Generally, if a covering $p:X\rightarrow Y$ arises in this way from a
properly discontinuous and free action ${\cal D}$ on $X$, we call the
covering a ${\cal D}${\it -covering}. Given two coverings $p:X\rightarrow Y$
and $p^{\prime }:X^{\prime }\rightarrow Y^{\prime }$ of topological spaces,
a homeomorphism $\phi :X\rightarrow X^{\prime }$ is called {\it isomorphism
of coverings} if $\phi $ is fibre-preserving, $p^{\prime }\circ \phi =p$. An 
{\it isomorphism of ${\cal D}$-coverings} is an isomorphism of coverings
that commutes with the actions of ${\cal D}$, i.e. $\phi \left( d\cdot
x\right) =d\cdot \phi \left( x\right) $; in this case we also say that $\phi 
$ is ${\cal D}${\it -equivariant}, and we say that the spaces involved are $%
{\cal D}${\it -isomorphic}. The {\it trivial} ${\cal D}${\it -covering} of $%
Y $ is the Cartesian product ${\cal D}\times Y$ together with projection
onto the second factor as covering map, and ${\cal D}$ acts on $\left(
d,y\right) \in {\cal D}\times Y$ by left multiplication on the first factor, 
$\left( d^{\prime },\left( d,y\right) \right) \mapsto \left( d^{\prime
}d,y\right) $. An isomorphism $i:{\cal D}\times Y\rightarrow X$ of the
trivial ${\cal D}$-covering onto a ${\cal D}$-covering $X$ is called a {\it %
trivialization of }$X$.

Let $p:X\rightarrow X/{\cal D}=Y$ be a ${\cal D}$-covering. Let $V\subset Y$
be an admissible connected open set in $Y$. A choice of a connected
component $U\subset p^{-1}\left( V\right) $ defines a {\it trivialization} $%
i:{\cal D}\times V\rightarrow p^{-1}\left( V\right) $ of the ${\cal D}$%
-covering $p:p^{-1}\left( V\right) \rightarrow V$ as follows: For $\left(
d,y\right) \in {\cal D}\times V$, let $i\left( d,y\right) \equiv d\cdot
\left( \left. p\right| U\right) ^{-1}\left( y\right) $. Then $i$ is
evidently a bijection and hence a homeomorphism; furthermore it is
fibre-preserving, since projection onto the second factor of $\left(
d,y\right) $ yields the same as $p\circ i$ applied to $\left( d,y\right) $;
and it is ${\cal D}$-equivariant by definition. This says that a ${\cal D}$%
-covering is trivial over each admissible neighbourhood $V\subset Y$. A
different choice $U^{\prime }$ of connected components in $p^{-1}\left(
V\right) $ defines a trivialization $i^{\prime }:{\cal D}\times V\rightarrow
p^{-1}\left( V\right) $ with $i^{\prime -1}\circ i\left( d,y\right) =\left(
d\cdot g\left( y\right) ,y\right) $, where $g:V\rightarrow {\cal D}$ is
continuous, and hence constant on every connected subset of $V$, since $%
{\cal D}$ is discrete.

\chapter{\u Cech cohomology and the glueing of ${\cal D}$-coverings \label
{CechAndGlueing}}

Let $p:X\rightarrow X/{\cal D}=Y$ be a ${\cal D}$-covering; let ${\cal V}$
be an open cover of $Y$ by admissible subsets $V\subset Y$. For sufficiently
simple spaces such as manifolds it can be assumed that every $V\in {\cal V}$
is simply connected in $Y$ and path-connected. As explained in the last
chapter, a choice of connected component $U_a\subset p^{-1}\left( V_a\right) 
$ in the inverse image of every $V_a$, $a\in A$, gives rise to a set of
local trivializations $i_a:{\cal D}\times V_a\rightarrow p^{-1}\left(
V_a\right) $, which, in turn, define a collection $\left( g_{ab}\right) $ of
transition functions $g_{ab}:V_a\cap V_b\rightarrow {\cal D}$, $%
i_b^{-1}\circ i_a\left( d,y\right) =\left( d\cdot g_{ab}\left( y\right)
,y\right) $. If all $V_a$ are connected, the transition functions $g_{ab}$
are constant due to continuity. It is easily seen that the $\left(
g_{ab}\right) $ satisfy

\begin{description}
\item[(Trans1)]  \ $g_{aa}=e$,

\item[(Trans2)]  \ $g_{ba}=g_{ab}^{-1}$,

\item[(Trans3)]  \ $g_{ac}=g_{ab}\cdot g_{bc}$,
\end{description}

the last equation following from $i_a^{-1}i_ci_c^{-1}i_bi_b^{-1}i_a=id$,
whenever $V_a\cap V_b\cap V_c\neq \emptyset $. Comparison with (Coc1-Coc3)
in chapter \ref{CechCohomology} shows that the collection $\left(
g_{ab}\right) $ defines a $1$-\u Cech-cocycle on ${\cal V}$. Now suppose we
choose different trivializations $i_a^{\prime }$. Then the trivializations
are related by $i_a^{\prime -1}\circ i_a\left( d,y\right) =\left( d\cdot
h_a\left( y\right) ,y\right) $, with a collection $\left( h_a\right) $ of
locally constant functions $h_a:V_a\rightarrow {\cal D}$, which defines a $0$%
-\u Cech-cochain on ${\cal V}$, as explained in chapter \ref{CechCohomology}%
. The transition functions $\left( g_{ab}^{\prime }\right) $ associated with 
$\left( i_a^{\prime }\right) $ are defined by $i_b^{\prime -1}\circ
i_a^{\prime }\left( d,y\right) =\left( d\cdot g_{ab}^{\prime },y\right) $;
on the other hand, from the definition of $\left( h_a\right) $, we find that 
$i_a^{\prime }\left( d,y\right) =i_a\left( d\cdot h_a^{-1},y\right) $, which
implies that $i_b^{\prime -1}\circ i_a^{\prime }\left( d,y\right) =\left(
d\cdot h_a^{-1}g_{ab}h_b,y\right) $. Thus, 
\begin{equation}
\label{pp6t1appCechCoh}g_{ab}^{\prime }=h_a^{-1}\cdot g_{ab}\cdot h_b\quad . 
\end{equation}
Statement (Coh) in chapter \ref{CechCohomology} now shows that the cocycles $%
\left( g_{ab}\right) $ and $\left( g_{ab}^{\prime }\right) $ are
cohomologous. This means that a ${\cal D}$-covering $p:X\rightarrow X/{\cal D%
}=Y$ determines a unique \u Cech cohomology class in $H^1\left( {\cal V;D}%
\right) $. -- Furthermore, let $q:Z\rightarrow Z/{\cal D}=Y$ be another $%
{\cal D}$-covering of $Y$ such that there exists an isomorphism $\phi
:X\rightarrow Z$ of ${\cal D}$-coverings. Assume that $i_a:{\cal D}\times
V_a\rightarrow p^{-1}\left( V_a\right) \subset X$ is the trivialization over 
$V_a$ in the ${\cal D}$-covering $p:X\rightarrow Y$. Then $\left( d,y\right)
\mapsto \phi \circ i_a\left( d,y\right) $ is a trivialization over $V_a$ in
the covering $q:Z\rightarrow Y$ with transition functions $g_{ab}^{\prime
}=g_{ab}$. Any other trivialization of $q:Z\rightarrow Y$ gives a
cohomologous cocycle. Thus, we have found that ${\cal D}$-coverings which
are ${\cal D}$-isomorphic define a unique \u Cech cohomology class in $%
H^1\left( {\cal V;D}\right) $.

-- Conversely, we want to show that a cohomology class represented by $%
\left( g_{ab}\right) $ defines a ${\cal D}$-covering of $Y$ up to ${\cal D}$%
-isomorphisms. Let $p:X\rightarrow Y=X/{\cal D}$, $p^{\prime }:X^{\prime
}\rightarrow Y=X^{\prime }/{\cal D}$ be two ${\cal D}$-coverings with
trivializations $i_a:{\cal D}\times V_a\rightarrow p^{-1}\left( V_a\right) $%
, $i_a^{\prime }:{\cal D}\times V_a\rightarrow p^{\prime -1}\left(
V_a\right) $ and associated cocycles $\left( g_{ab}\right) $, $\left(
g_{ab}^{\prime }\right) $ defined by $i_b^{-1}\circ i_a\left( d,y\right)
=\left( d\cdot g_{ab},y\right) $, $i_b^{\prime -1}\circ i_a^{\prime }\left(
d,y\right) =\left( d\cdot g_{ab}^{\prime },y\right) $, such that $%
g_{ab}^{\prime }=h_a^{-1}\cdot g_{ab}\cdot h_b$, where $\left(
h_a:V_a\rightarrow {\cal D}\right) $ is a $0$-\u Cech-cochain on ${\cal V}$.
It follows that $i_a^{\prime }\left( d\cdot h_a,y\right) =i_b^{\prime
}\left( d\cdot g_{ab}\cdot h_b,y\right) $. On the sets ${\cal D}\times V_a$
we now define a collection of functions $k_a:{\cal D}\times V_a\rightarrow
X^{\prime }$ determined by $k_a\left( d,y\right) =i_a^{\prime }\left( d\cdot
h_a,y\right) $. By construction, all $k_a$ are continuous. Furthermore, if $%
\left( d,y\right) \in {\cal D}\times V_a$ and $\left( d^{\prime },y^{\prime
}\right) \in {\cal D}\times V_b$ are identified under the trivializations $%
i_a$, $i_b$, so that $i_a\left( d,y\right) =i_b\left( d^{\prime },y^{\prime
}\right) $, then $k_a\left( d,y\right) $ and $k_b\left( d^{\prime
},y^{\prime }\right) $ coincide; for, in this case, we must have $\left(
d^{\prime },y^{\prime }\right) =\left( d\cdot g_{ab},y\right) $, and hence%
$$
k_b\left( d^{\prime },y^{\prime }\right) =i_b^{\prime }\left( d^{\prime
}\cdot h_b,y^{\prime }\right) =i_b^{\prime }\left( d\cdot g_{ab}\cdot
h_b,y\right) =i_b^{\prime }\left( dh_a\cdot g_{ab}^{\prime },y\right) = 
$$
$$
=i_a^{\prime }\left( d\cdot h_a,y\right) =k_a\left( d,y\right) \quad . 
$$
Now the set of trivializations $\left( i_a:{\cal D}\times V_a\rightarrow
X\right) $ as defined here can be regarded as a collection of identification
maps $\left( i_a\right) $ on the sets ${\cal D}\times V_a$, where ${\cal D}$
has the discrete topology, and the $V_a$ have the topology induced from $Y$.
The identification space $X$ has the $\phi $-universal property \cite{Brown}
that for every topological space $X^{\prime }$ and any collection of
continuous functions $\left( k_a:{\cal D}\times V_a\rightarrow X^{\prime
}\right) $ which coincide on elements $\left( d,y\right) $, $\left(
d^{\prime },y^{\prime }\right) $ which are identified in the identification
space $\left( i_a:{\cal D}\times V_a\rightarrow X\right) $, there exists a
unique continuous map $\psi :X\rightarrow X^{\prime }$ such that $\psi \circ
i_a=k_a$. Since the arguments leading to this result can be reversed, it
follows that $\psi $ has a continuous inverse, and hence is a homeomorphism.
Locally, we have $i_a^{\prime -1}\circ \psi \circ i_a\left( d,y\right)
=\left( d\cdot h_a,y\right) $, which says that $\psi $ is fibre-preserving,
and hence is a covering isomorphism. The same formula shows that $\psi $ is $%
{\cal D}$-equivariant. Altogether, therefore, $\psi $ is an isomorphism of $%
{\cal D}$-coverings.

The last two paragraphs therefore prove that there is a 1--1 correspondence
between ${\cal D}$-coverings of $Y$ which are ${\cal D}$-isomorphic, and
cohomology classes in the first \u Cech cohomology group $H^1\left( {\cal V;D%
}\right) $ on ${\cal V}$ with values in ${\cal D}$.

\chapter{Deck-transformation-valued cohomology on $G$ \label
{DeckTransformationValuedCohomOnG}}

Consider the scenario of theorem \ref{ExtensionOfG}; there it was shown that
when the action $\phi :G\times Y\rightarrow Y$ of the Lie group $G$ is
lifted to a smooth map $\hat \phi :G\times X\rightarrow X$ satisfying
(L1,L2), then the set $\left\{ \hat \phi _g\mid g\in G\right\} $ usually no
longer closes into a group, but is extended to a larger group $\tilde G$
which contains ${\cal D}$ as a normal subgroup such that $\tilde G/{\cal D}%
=G $. The deviation from closure was measured by the map 
\begin{equation}
\label{pp6t1co10}\Gamma :G\times G\rightarrow {\cal D}\quad ,\quad \left(
g,h\right) \mapsto \Gamma \left( g,h\right) \equiv \hat \phi _g\hat \phi _h%
\widehat{\phi _{gh}}^{-1}\quad , 
\end{equation}
see (\ref{pp6t1lift8}). Furthermore, we have a map 
\begin{equation}
\label{pp6t1co11}b\left( g\right) :{\cal D}\rightarrow {\cal D}\quad ,\quad
\gamma \mapsto b\left( g\right) \gamma \equiv \hat \phi _g\circ \gamma \circ
\hat \phi _g^{-1}\quad . 
\end{equation}
In the discussion following formula (\ref{pp6t1lift7}) it was pointed out
that $b:G\rightarrow Aut\left( {\cal D}\right) $ usually is not a
representation; here we show that if ${\cal D}$ is Abelian, then $b$ {\bf is}
a representation, and hence defines a left action 
\begin{equation}
\label{pp6t1co12}G\times {\cal D}\rightarrow {\cal D\quad },\quad \left(
g,\gamma \right) \mapsto b\left( g\right) \gamma \quad , 
\end{equation}
of $G$ on ${\cal D}$. To see this, consider the expression%
$$
b\left( gg^{\prime }\right) \gamma =\hat \phi _{gg^{\prime }}\gamma \hat
\phi _{gg^{\prime }}^{-1}\quad ; 
$$
using (\ref{pp6t1co10},\ref{pp6t1co11}) this becomes%
$$
\hat \phi _{gg^{\prime }}\gamma \hat \phi _{gg^{\prime }}^{-1}=\Gamma
^{-1}\left( g,g^{\prime }\right) \left[ b\left( g\right) \circ b\left(
g^{\prime }\right) \gamma \right] \Gamma \left( g,g^{\prime }\right) \quad ; 
$$
but the expression $b\left( g\right) \circ b\left( g^{\prime }\right) \gamma 
$ in square brackets is an element of ${\cal D}$, as are the $\Gamma $'s.
Hence, since ${\cal D}$ is Abelian, the last expression is%
$$
\Gamma ^{-1}\left( g,g^{\prime }\right) \left[ b\left( g\right) \circ
b\left( g^{\prime }\right) \gamma \right] \Gamma \left( g,g^{\prime }\right)
=b\left( g\right) \circ b\left( g^{\prime }\right) \gamma \quad , 
$$
which proves 
\begin{equation}
\label{pp6t1co13}b\left( gg^{\prime }\right) =b\left( g\right) \circ b\left(
g^{\prime }\right) \quad . 
\end{equation}

In the sequel we use an additive notation for the group law in ${\cal D}$;
i.e. $\left( \gamma ,\gamma ^{\prime }\right) \mapsto \gamma +\gamma
^{\prime }\in {\cal D}$. We now introduce a ${\cal D}$-valued cohomology on $%
G$ as follows: $n$-cochains $\alpha _n$ are maps $G^n\rightarrow {\cal D}$; $%
0$-cochains are elements of ${\cal D}$. The coboundary operator $\delta $ is
defined to act on $0$-, $1$-, $2$-cochains according to 
\begin{equation}
\label{pp6t1co14}
\begin{array}{rcl}
\left( \delta \alpha _0\right) \left( g\right) & = & b\left( g\right) \alpha
_0-\alpha _0\quad , \\ 
\left( \delta \alpha _1\right) \left( g,h\right) & = & b\left( g\right)
\alpha _1\left( h\right) -\alpha _1\left( gh\right) +\alpha _1\left(
g\right) \quad , \\ 
\left( \delta \alpha _2\right) \left( g,h,k\right) & = & b\left( g\right)
\alpha _2\left( h,k\right) +\alpha _2\left( g,hk\right) - \\  
&  & -\alpha _2\left( gh,k\right) -\alpha _2\left( g,h\right) \quad . 
\end{array}
\end{equation}
This is well-defined, since (\ref{pp6t1co13}) says that $b$ is now an
action. We denote the sets of $n$-cochains, -cocycles, -coboundaries, and $n$%
-cohomology groups by $C^n\left( G,{\cal D}\right) $, $Z^n\left( G,{\cal D}%
\right) $, $B^n\left( G,{\cal D}\right) $, and $H^n\left( G,{\cal D}\right)
=Z^n\left( G,{\cal D}\right) /B^n\left( G,{\cal D}\right) $.

\TeXButton{Abst}{\vspace{0.8ex}}--- Another cohomology on $G$ and $\hat g$
that occurs in studying moment maps is the

\chapter{$g^{*}$-valued cohomology on $G$ \label{gStarOnGroup}}

$n${\it -}cochains{\it \ }are smooth maps $\alpha _n:G^n\rightarrow g^{*}$. $%
G$ acts on $g^{*}$ via the coadjoint representation $Ad^{*}$ of $G$ on $%
g^{*} $; this is a left action, see the beginning of the appendix. The set
of all $g^{*}$-valued $n$-cochains is denoted by $C^n\left( G,g^{*}\right) $%
. The coboundary operator $\delta :C^n\rightarrow C^{n+1}$ acts on $0$-, $1$%
-, $2$-cochains $\alpha _0$, $\alpha _1$, $\alpha _2$ according to 
\begin{equation}
\label{pp6t1co4}
\begin{array}{rcl}
\left( \delta \alpha _0\right) \left( g\right) & = & Ad^{*}\left( g\right)
\alpha _0-\alpha _0\quad , \\ 
\left( \delta \alpha _1\right) \left( g,h\right) & = & Ad^{*}\left( g\right)
\alpha _1\left( h\right) -\alpha _1\left( gh\right) +\alpha _1\left(
g\right) \quad , \\ 
\left( \delta \alpha _2\right) \left( g,h,k\right) & = & Ad^{*}\left(
g\right) \alpha _2\left( h,k\right) +\alpha _2\left( g,hk\right) - \\  
&  & -\alpha _2\left( gh,k\right) -\alpha _2\left( g,h\right) \quad , 
\end{array}
\end{equation}
etc.. The set of all $n$-cocycles is denoted by $Z^n\left( G,g^{*}\right) $,
the set of all $n$-coboundaries is denoted as $B^n\left( G,g^{*}\right) $.
The $n$-th cohomology group of $G$ with values in $g^{*}$ is the quotient $%
H^n\left( G,g^{*}\right) =Z^n\left( G,g^{*}\right) /B^n\left( G,g^{*}\right) 
$.

\end{appendix}


\begin{thebibliography}{99}


\bibitem{Wolf}  J.A. Wolf, ''Spaces of Constant Curvature''. McGraw Hill,
1967.

\bibitem{Hull}   C.M. Hull and B. Julia. Nucl.Phys. B534 (1998) 250-260.

\bibitem{Woodhouse}  N.M.J. Woodhouse, ''Geometric Quantization''. Oxford
Science Publications, 1991.

\bibitem{Azca}  J.A. de Azcarraga, J.P. Gauntlett, J.M. Izquierdo, and P.K.
Townsend, Phys. Rev. Lett. 63, no. 22 (1989) 2443.

\bibitem{Soro}  Dmitri Sorokin and Paul K. Townsend, Phys.Lett. B412 (1997) 265-273.

\bibitem{Achu}  A. Achucarro, J.M. Evans, P.K. Townsend and D.L. Wiltshire,
Phys. Lett. B 198 (1987) 441.

\bibitem{PKT1}   P.K. Townsend, University of Cambridge preprint R-88 11.

\bibitem{cram/pir}  M. Crampin and F.A.E. Pirani, Applicable Differential
Geometry, 1986.

\bibitem{Govaerts}  J. Govaerts, Hamiltonian Dynamics and Constrained
Systems, Leuven University Press, 1991.

\bibitem{BergTown1}  E. Bergshoeff and P.K. Townsend, Nucl.Phys. B490 (1997) 145-162.

\bibitem{Thor}  L. Thorlacius, Nucl.Phys.Proc.Suppl. 61A (1998) 86-98.

\bibitem{Berg2}  E. Bergshoeff and M. de Roo and M.B. Green, G. Papadopoulos
and P.K. Townsend, Nucl.Phys. B470 (1996) 113-135.

\bibitem{Berg3}  E. Bergshoeff and P.M. Cowdall and P.K. Townsend, Phys.Lett. B410 (1997) 13-21.

\bibitem{Ceder1}  Martin Cederwall, Alexander von Gussich, Bengt E.W.
Nilsson, Per Sundell and Anders Westerberg, Nucl.Phys. B490 (1997) 179-201.

\bibitem{Nakahara}  M. Nakahara, Geometry, Topology and Physics, 1990.

\bibitem{Corn1}  J.F. Cornwell, ''Group Theory in Physics'', vol.1. 
Academic Press, 1984.

\bibitem{Fulton}  William Fulton, ''Algebraic Topology''. Springer Verlag,
1995.

\bibitem{Jaehnich}  Klaus J\"ahnich, ''Topology''. Springer-Verlag, 1980.

\bibitem{Massey}  William S. Massey, ''A Basic Course in Algebraic
Topology''. Springer-Verlag, 1991.

\bibitem{ONeill}  Barrett O'Neill, ''Semi-Riemannian Geometry with
Applications to Relativity''. Academic Press, 1983.

\bibitem{SagleWald}  Arthur A. Sagle, Ralph E. Walde, ''Introduction to Lie
Groups and Lie Algebras''. Academic Press, 1973.

\bibitem{Warner}  Frank W. Warner, ''Foundations of Differentiable Manifolds
and Lie Groups''. Scott, Foresman and Company, 1971.

\bibitem{Azcarraga}  Jose de Azcarraga and Jose M. Izquierdo, ''Lie groups,
Lie algebras, cohomology and some applications in physics''. Cambridge
University Press, 1995.

\bibitem{Brown}  Ronald Brown, ''Topology''. Ellis Harwood Limited, 1988.

\bibitem{GuillStern}  Victor Guillemin, Shlomo Sternberg, ''Symplectic
techniques in physics''. Cambridge University Press, 1984.

\bibitem{AbrahamMarsden}  Ralph Abraham, Jerrold E. Marsden. ''Foundations
of Mechanics''. The Benjamin/Cummings Publishing Company, Inc., 1978.

\bibitem{DodsonParker}  C.T.J. Dodson and Phillip E. Parker, ''A User's
Guide to Algebraic Topology''. Kluwer Academic Publishers.

\end{thebibliography}
\end{document}